\documentclass[reprint,aps,prb,twocolumn,amsmath,amssymb,mathrsfs,showpacs,superscriptaddress,notitlepage,longbibliography]{revtex4-2}
\usepackage{graphicx}% Include figure files
\usepackage{dcolumn}% Align table columns on decimal point
\usepackage{bm}% bold math

\usepackage{framed}
\usepackage{tabularx}
\usepackage{booktabs}
\usepackage{bm}
\usepackage{graphicx}
\usepackage{color}
\usepackage{multirow}
\usepackage{amsmath}
\usepackage{physics}
\usepackage{float}
\usepackage{hyperref}
\hypersetup{colorlinks=true,linkcolor=red,anchorcolor=red,citecolor=blue,urlcolor=blue}
\usepackage{ulem}
\usepackage{subcaption}
\usepackage[justification=raggedright,singlelinecheck=false]{caption}

\newcommand{\gscaep}[0]{Graduate School of China Academy of Engineering Physics, Beijing 100193, China}

\begin{document}

% \preprint{APS/123-QED}

\title{Real-space construction and classification for time-reversal symmetric crystalline superconductors in 2D interacting fermionic systems   }
% Force line breaks with \\

\author{Yi-Ming Liu }
\affiliation{ Department of Physics, Southern University of Science and Technology, Shenzhen 518055, China
}
\affiliation{\gscaep}
\author{Wei-Qiang Chen}
\email{chenwq@sustech.edu.cn}
\affiliation{ Department of Physics, Southern University of Science and Technology, Shenzhen 518055, China
}

\author{Zheng-Cheng Gu}
\email{zcgu@phy.cuhk.edu.hk}
\affiliation{ Department of Physics, The Chinese University of Hong Kong, Shatin, New Territories, Hong Kong, China
}

\date{\today}

\begin{abstract}
Crystalline symmetry and time-reversal symmetry are commonly present in real superconducting materials. However, the topological classification of systems respecting these symmetries—particularly in the context of interacting fermions—remains incomplete. In this work, we systematically classify time-reversal symmetry-protected crystalline topological superconductors in two-dimensional interacting fermionic systems, using an explicit real-space construction. Among the resulting phases, we identify \textit{intrinsically interacting fermionic topological superconductors}, i.e. phases that cannot be realized in either free-fermion systems or interacting bosonic systems. In particular, for spinless fermions with protecting symmetry group $C_4 \times \mathbb{Z}_2^T \times \mathbb{Z}_2^f$ or $D_4 \times \mathbb{Z}_2^T \times \mathbb{Z}_2^f$ (here $C_4$ and $D_4$ denote the rotation and dihedral groups, respectively; $\mathbb{Z}_2^T$ and $\mathbb{Z}_2^f$ denote time-reversal and fermion parity symmetry groups), the intrinsic sector has a $\mathbb{Z}_4$ classification. The corresponding root phase generating this $\mathbb{Z}_4$ classification admits a transparent real-space construction in terms of decorated $1$D blocks. For $C_4$, the root phase is a one-dimensional fermionic symmetry-protected topological (FSPT) phase protected by $\mathbb{Z}_2^T  \times \mathbb{Z}_2^f$, realizable as double Majorana chains. For $D_4$, the root phase is a one-dimensional FSPT phase protected by $\mathbb{Z}_2^T \times \mathbb{Z}_2^M \times \mathbb{Z}_2^f$, where $\mathbb{Z}_2^M$ denotes an effective on-site symmetry associated with reflection, and also can be realized by double Majorana chains.  We further find the corresponding $\mathbb{Z}_4$ spinless intrinsic phases in examples with wallpaper groups $p4$, $p4m$, and $p4g$. We also find an additional $\mathbb{Z}_2$ intrinsically interacting phase for spinless fermions with wallpaper group $pm$, which is absent with the corresponding point-group symmetry only.  Moreover, these intrinsic phases naturally give rise to higher-order FSPT phases that support corner zero modes in the presence of boundaries. Finally, we verify the widely-conjectured \textit{crystalline equivalence principle} for generic 2D interacting FSPT systems in the presence of both crystalline and internal symmetries.
\end{abstract}

\maketitle

\tableofcontents

\section{ INTRODUCTION}

\subsection{Overview}
According to the Landau--Ginzburg paradigm~\cite{Ginzburg:1950sr}, phases of matter and phase transitions are traditionally understood through the mechanism of spontaneous symmetry breaking. However, for gapped quantum systems at zero temperature, quantum entanglement introduces an additional and essential organizing principle to classify quantum phases~\cite{LU}. Within this framework, gapped quantum phases can be classified based on their entanglement structure under finite-depth local unitary (LU) transformations. Specifically, a quantum state is said to possess short-range entanglement if and only if it can be transformed into a trivial product state via such LU evolution. In contrast, states that cannot be disentangled in this way exhibit long-range entanglement, a hallmark of topological order~\cite{TO}. Surprisingly, if systems with global symmetry are considered, even the so-called ``short-range entangled" phases can also have many different classes. Among them, one class is the conventional symmetry breaking phases described by the Landau paradigm. However, it turns out that there exists a new class of topological phases -- the so called symmetry-protected topological (SPT) phases~\cite{Gu_2009,Pollmann}.

An SPT state features a gapped, unique ground state in the bulk (on a closed manifold), yet they universally host protected gapless states at their boundaries (on an open manifold). Two SPT states are considered to belong to the same phase if they can be connected by a symmetric local unitary (SLU) circuit—i.e., a finite-depth quantum circuit that preserves both the symmetry and the bulk energy gap~\cite{zeng2018quantuminformationmeetsquantum}. For systems with on-site symmetries, the classification of SPT phases is well-developed. On one hand, group cohomology and group supercohomology provide explicit classification data, as well as discretized non-linear sigma model topological terms~\cite{chenSymmetryProtectedTopological2013,Chen_2011, sigmaModel}. On the other hand, cobordism theory and related invariants from invertible quantum field theory offer a framework to classify low-energy effective response actions~\cite{kapustin2014symmetry,Kapustin_2015,kapustin2015equivariant,Freed_2021}. In particular, decorated domain wall construction, together with the Atiyah–Hirzebruch spectral sequence method, provides a clear and physically intuitive approach to understanding both the construction and classification of interacting fermionic SPT phases~\cite{Chen_2014,Wang2018, Wang2020,ren2023stacking}. Experimentally, well-known realizations of SPT phases include topological insulators and topological superconductors~\cite{TIreview,TSC,Sato_2017,Sasaki_2012,FU_TSC_2010,FU_proximity_2008}, which have been extensively studied and observed in condensed matter systems.

Crystalline symmetries can also protect nontrivial short-range entangled phases, giving rise to crystalline SPT phases~\cite{CryTI,CryTI2,Tanaka_2012,Dziawa_2012,Okada_2013,ma2017experimental,hsieh2012topological,Isobe_2015,Zou_2018,Po_2017,Song_2020_beyond,Jiang_2017,Kruthoff_2017,wire_2022,Shiozaki_2022,Rasmussen_2020,rasmussen2018intrinsicallyinteractingtopologicalcrystalline,Cheng_Lieb_2019,Huang_Surface_2018,Ono_2021,lee2024crystallineequivalenttopologicalphasesmanybody,lee2024connectionfreefermioninteractingcrystalline}. For bosonic systems, two main approaches enable their classification: topological crystal constructions~\cite{SongHao,pointgroup}, and topological field theory where spatial symmetries are coupled to background gauge fields~\cite{Thorngren}. The topological crystal construction approach admits a unified formulation via spectral sequences~\cite{Song_2020}. Extending these methods to fermionic systems is conceptually and computationally more involved, especially in the presence of time-reversal symmetry. Moreover, for bosonic systems, the crystalline equivalence principle has been established~\cite{Thorngren,defect_networks}: SPT phases protected by a space group $G$ correspond to SPT phases protected by an internal symmetry $G$, with orientation-reversing elements interpreted as time-reversal (antiunitary) operations. For fermions, the so-called fermionic crystalline equivalence principle has also been proposed~\cite{debray2021invertible,Nonperturbativeconstraints}, but a fully rigorous proof is still missing.

Given the prevalence of time reversal symmetry in most superconducting materials, we delve into a systematic exploration of 2D fermionic SPT phases protected by both crystalline symmetries (point groups and space groups) and time reversal symmetry. Using recently introduced fermionic topological crystal constructions~\cite{Song_2020,RotationSPT,pointgroup,2Dcrystalline,zhang2022construction}, we employ a real-space approach to construct and classify these phases. Our study extends to comparing the results with the classification of 2D FSPTs safeguarded by corresponding internal symmetries, thus validating the fermionic crystalline equivalence principle.

\subsection{General strategy}
\label{sec:classification strategy}

\begin{figure}[tb]
    \centering
    \includegraphics[width=0.46\textwidth]{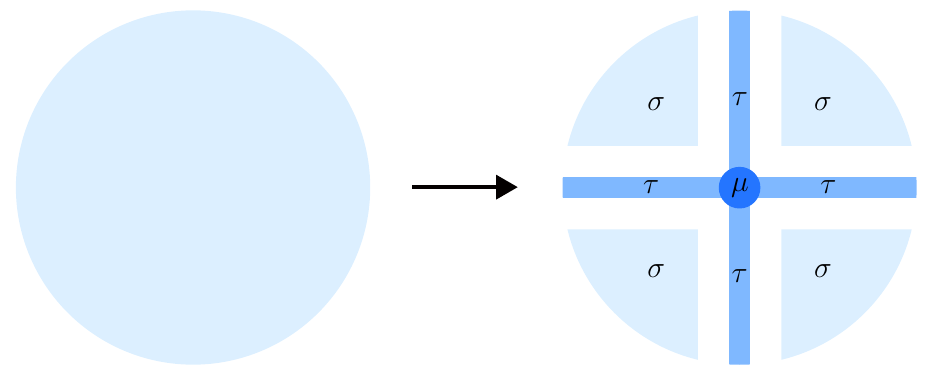}
    \caption{Cell decomposition for the point group $SG=C_4$. The 2D system (left) is decomposed into symmetry-related blocks (right). Any two-dimensional block $\sigma$ is mapped to any other $2$D block through a finite number of $\pi/2$ rotations. The $1$D blocks $\tau$ are also connected by $\pi/2$ rotations.}
    \label{C4 cell}
\end{figure}

In this work, we adopt the real-space topological crystal construction approach introduced in Ref. \cite{pointgroup, 2Dcrystalline}. To make this paper self-contained, we briefly review their framework in this section, which provides a general construction and classification for FSPT states with crystalline symmetry. It contains the following four major steps:
\begin{description}
    \item[Cell decomposition]
    For a lattice system with symmetry group $G_f$, we decompose space into symmetry-related blocks of different dimensions, arranged to respect the spatial subgroup $SG\subset G_f$. Figure~\ref{C4 cell} shows an example for $SG=C_4$.
    \item[Block-state decoration]  
    We assign to each $n$D-block a SPT state (block-state) protected by the subgroup of symmetries that fix the block pointwise.  Symmetry-related blocks must carry the same block-state. When applicable, we also allow invertible topological order states (e.g., a Majorana chain).
     A decoration is obstruction-free if one can add symmetry-preserving interactions on each $(n-1)$-dimensional boundary ($n=1,2$) that fully gap out the boundary modes from adjacent $n$D-blocks and leave a unique ground state. 
    \item[Bubble equivalence] 
   
    Generally, if the decorated block-states contain topologically non-trivial states, the resulting system constitutes a non-trivial crystalline SPT phase. However, in certain cases, these non-trivial block-states may still be trivialized. Later, we will introduce a special class of trivial block-states called bubbles. Two crystalline SPT phases differing only by such bubbles are considered topologically equivalent, and this is called the ``bubble equivalence”. 
    \item[Extension problem]
     After imposing obstruction-free condition and bubble equivalence, one obtains candidate classification groups in each dimension. The full SPT classification is obtained by stacking decorations of different dimensions, which may realize non-trivial (central) group extensions rather than a direct product.

\end{description}

We elaborate on the general real-space construction of fermionic SPT phases protected by a symmetry group $G_f$ (abbreviated as $G_f$-FSPTs), focusing on the case where the spatial subgroup $SG\subset G_f$ is a two-dimensional point group. The generalization to space groups is straightforward.

We partition the system into open, mutually disjoint blocks that are related by point-group symmetries. In two dimensions this consists of one $0$D block $M^{0\mathrm{D}}$, several $1$D blocks $M^{1\mathrm{D}}_i$, and several $2$D blocks $M^{2\mathrm{D}}_j$. We denote the resulting division by
\[
Y=\{M^{0\mathrm{D}},\, M^{1\mathrm{D}}_i,\, M^{2\mathrm{D}}_j\}.
\]
Among the many possible divisions, we restrict to a specific class~\cite{Jiang2021}: for any block $\Delta\in Y$ with site-symmetry group $SG_\Delta\subset SG$ (the subgroup that maps $\Delta$ to itself), every $g\in SG_\Delta$ acts pointwise on $\Delta$. Equivalently, $SG_\Delta$ acts as an on-site internal symmetry when restricted to $\Delta$.

Then, consider a block $M_j$. To make the system topological, we decorate $M_j$ with a FSPT block-state compatible with the symmetry acting on $M_j$. 
Because our division $Y$ is chosen such that the site symmetry group $SG_{M_j}\subset SG$ acts pointwise on $M_j$, the spatial symmetries in $SG_{M_j}$ can be regarded as on-site symmetries when restricted to this block. 
Together, $SG_{M_j}$ (acting on-site on $M_j$) and the internal symmetry in $G_f$ impose the symmetry constraint on the block-state on $M_j$.

In addition, we allow $M_j$ to host an invertible topological order phase that does not require any symmetry for protection, e.g., the 1D Majorana chain. Due to point group symmetry, rotation-/reflection-related blocks must be decorated by the same phase. In this way, by appropriately decorating the blocks, we assign a block-state to every block in $Y$. However, since the blocks are disconnected, gapless modes generally appear on the boundaries between adjacent blocks.

To obtain a fully gapped state, there must exist symmetry-preserving interaction terms defined on the $(n-1)$-dimensional interfaces ($n=1,2$) such that:
(1) they preserve the internal symmetry in $G_f$, and are also invariant under the site-symmetry subgroup of the interface block, and
(2) they can fully gap out the modes originating from adjacent $n$-dimensional blocks, leaving a unique ground state.
Physically, this indicates that anomalies contributed by all adjacent block-states cancel each other~\cite{Song_2020}. A gapped assembly of block-states is called an obstruction-free decoration. Only obstruction-free decorations are allowed, which is referred to as the no-open-edge condition.

To obtain a non-trivial  obstruction-free block-state decoration, we need to additionally consider potential trivializations. 
For $n$-dimensional blocks ($n\geq 1$), we can further decorate certain $(n-1)$-dimensional degrees of freedom, which may be trivialized if they adiabatically shrink to a point. This construction is
called bubble equivalence. 

In particular, for 2D space groups, a $2$D block has no nontrivial site-symmetry group, bubbles come entirely from the internal symmetry. For $2$D blocks with only fermion parity $\mathbb{Z}_2^f$ symmetry, a $2$D bubble is a closed Majorana chain (or an integer multiple thereof). By definition of bubble equivalence, we only keep Majorana chains with anti-periodic boundary conditions (anti-PBC), since such a loop can be trivialized upon shrinking to a point without breaking any symmetry~\cite{2Dcrystalline}. 
In the examples considered in this paper, the internal symmetry further includes time-reversal symmetry $\mathbb{Z}_2^T=\{I,T\}$, leading to two distinct cases. (i) If $T^2=1$, the $2$D bubble is a single Majorana chain with anti-PBC (or an integer multiple thereof). (ii) If $T^2=P_f$, the $2$D bubble consists of two Majorana chains with anti-PBC (or an integer multiple thereof).

For $1$D blocks, we consider two types of $1$D bubble states, both of which must be trivialized when shrunk to a point. The first type is the fermionic $1$D bubble: we decorate each edge of a $1$D block with a complex fermion, and upon fusion, they form an atomic insulating state with even fermion parity. These fermions transform under a linear representation of the group \( G_f \). The second type is the bosonic $1$D bubble: we decorate each edge with a boson, which linearly transforms under the physical symmetry group \( G_b = G_f / \mathbb{Z}_2^f \).  
Furthermore, we need not consider 3D bubbles in our 2D systems.

In particular, we note that the block-states are constructed layer-by-layer. We denote obstruction-free states and trivialization-free states in $d$-dimension as $\{\mathrm{OFBS}\}^{dD}$ and $\{\mathrm{TBS}\}^{dD}$, respectively. Then, we  define the $d$-dimensional obstruction-free and trivialization-free block-states as $E^{dD}$:
\begin{align}
    E^{d\mathrm{D}}=\{\mathrm{OFBS}\}^{d\mathrm{D}}/\{\mathrm{TBS}\}^{d\mathrm{D}}
\end{align}

$E^{d\mathrm{D}}$ can only be treated as a group within the context of $d$-dimensional block-states.
In order to obtain the ultimate classification of SPT phases, we should further consider the possible stacking of block-states with different dimensions. 

Considering all obstruction-free and trivialization-free block-states across all dimensions, the ultimate classification with exact group structure $\mathcal{G}$ of 2D crystalline FSPT phases is a group extension among $E^{0\mathrm{D}}$, $E^{1\mathrm{D}}$, and $E^{2\mathrm{D}}$:
\begin{equation}
    \begin{gathered}
        1\rightarrow E^{0\mathrm{D}} \rightarrow E^{\leq 1\mathrm{D}}\rightarrow E^{1\mathrm{D}}\rightarrow 1\\
        1\rightarrow E^{{\leq 1}\mathrm{D}}\rightarrow \mathcal{G}\rightarrow E^{2\mathrm{D}}\rightarrow 1,
    \end{gathered}
    \label{eq:extension}
\end{equation}
where $E^{\leq 1 \mathrm{D}}$ denotes the set of all obstruction-free and non-trivial states consisting of 0D and $1$D block-states. Both group extensions are central \cite{RotationSPT}. See Appendix~\ref{sec:group extension} for a brief review of the group extension.

\subsection{Symmetry properties} \label{sec:sym and cell}

In this section, we provide a detailed definition of the symmetry group $G_f$
for the  systems considered in this study.
First, we will outline the general structure of symmetries in fermionic systems.

In a fermionic system with a total symmetry group $G_f$, there always exists a subgroup $\mathbb{Z}_{2}^{f}=\{1,P_{f}=(-1)^{F}\}$, where $F$ denotes the fermion number operator counting the number of fermions in a state. The subgroup $\mathbb{Z}_{2}^{f}$ resides in the center of $G_f$, as all physical symmetry transformations  preserve the fermion parity and thus commute with $P_f$. 
Specifically, in systems lacking charge conservation symmetry $U^f(1)$, we define the physical (bosonic) symmetry group  as the quotient group $G_{b}=G_{f}/\mathbb{Z}_{2}^{f}$. Conversely, for a given physical symmetry group $G_b$, there exist multiple total symmetry groups $G_{f}$ that serve as central extensions of $G_b$ by $\mathbb{Z}_{2}^{f}$. 
This relationship can be expressed through the following short exact sequence:\begin{equation}
    1 \rightarrow\mathbb{Z}_{2}^{f}\rightarrow G_{f}\rightarrow G_{b}\rightarrow 1,
\end{equation}  where different extensions of $G_f$ are distinguished by factor systems represented as 2-cocycles $\omega_{2}\in   \mathcal{H}^{2}(G_{b},\mathbb{Z}_{2})$. Accordingly, $G_{f}$ is denoted as $\mathbb{Z}_{2}^{f}\times_{\omega_{2}}G_{b}$. The spin of fermions (spinless or spin-1/2) is characterized by different choices of 2-cocycles $\omega_2$, i.e., the spinless corresponds to a trivial $\omega_2$  while spin-1/2 (spinful) fermion corresponds to specific choice of nontrivial $\omega_2$.

In this work, we study 2D time-reversal symmetric crystalline superconductors in interacting fermionic systems. Since these systems lack charge conservation symmetry $U^f(1)$, as discussed earlier, this imposes a specific structure on the total symmetry group, which can be written as $G_f = \mathbb{Z}_2^f \times_{\omega_2} G_b$. Furthermore, for the systems we consider, the physical symmetry group $G_b$ is given by $G_b = SG \times \mathbb{Z}_2^T$, where $SG$ denotes the 2D point group or wallpaper group, and $\mathbb{Z}_2^T$ corresponds to the time-reversal symmetry group. Therefore, to fully specify $G_f$, it suffices to provide the 2-cocycles $\omega_2$.

In 2D lattice systems, there are ten point groups, classified as rotation (cyclic) groups $C_n$ and dihedral groups $D_n$ ($n =1,2,3,4,6$). Mathematically, each dihedral group is a semidirect product of a rotation and a reflection symmetry group, expressed as $D_n = C_n \rtimes \mathbb{Z}^M_2$.

For the even-fold symmetry group $G_b = C_{2n} \times \mathbb{Z}_2^T$, the distinct equivalence classes of 2-cocycles $\omega_2$ are characterized by the group cohomology groups:
\begin{equation}
    \mathcal{H}^2(C_{2n} \times \mathbb{Z}_2^T, \mathbb{Z}_2) = \mathbb{Z}_2^3,
\end{equation}

Similarly, for the even-fold symmetry group $G_b = D_{2n} \times \mathbb{Z}_2^T$, the equivalence classes of 2-cocycles $\omega_2$ are characterized by group cohomology groups:
\begin{equation}
  \mathcal{H}^2(D_{2n} \times \mathbb{Z}_2^T, \mathbb{Z}_2) = \mathbb{Z}_2^6,  
\end{equation}

For odd-fold symmetry groups $G_b=C_{2n-1} \times \mathbb{Z}_2^T$ and $G_b=D_{2n-1} \times \mathbb{Z}_2^T$, different equivalent classes of 2-cocycles $\omega_2$ are characterized by group cohomology groups:
\begin{align}
    &\mathcal{H}^2(C_{2n-1} \times \mathbb{Z}_2^T,\mathbb{Z}_2)=\mathbb{Z}_2, \\
   & \mathcal{H}^2(D_{2n-1} \times \mathbb{Z}_2^T,\mathbb{Z}_2)=\mathbb{Z}_2^3.
\end{align}

In this paper, we focus on the most physically relevant scenarios involving spinless or spin-$1/2$ (spinful) fermionic systems. Accordingly, our attention is restricted to specific 2-cocycles that govern the topological spin rotation properties around different axes~\cite{pointgroup}. For convenience, we denote ${R}$, ${M}$, and ${T}$ as the rotation, reflection, and time-reversal generators in the respective groups.

For spinless fermions, all 2-cocycles are trivial, with $G_f$ taking the form: $G_f=C_n \times \mathbb{Z}_2^T\times \mathbb{Z}_2^f$ or $G_f=D_n \times \mathbb{Z}_2^T\times \mathbb{Z}_2^f$. 

For spinful fermions,  generators of  even-fold cyclic group $C_{2n}$ should satisfy \begin{equation}T^2=P_f,R^{2n}=P_f,TR=RT.\end{equation} The corresponding 2-cocycles take the form:
\begin{equation} \omega_2(g_1h_2,g_2h_2):=\delta_2(g_1,g_2)\epsilon_2(h_1,h_2) \label{eq:extension cocycles of cyclic groups}
\end{equation} where $g_j \in \mathbb{Z}_2^T \subset G_b, h_j \in C_{2n} \subset G_b$. Here the $\epsilon_2$ is given by:
\begin{equation}
    \epsilon_2(R^a,R^b)=e^{i\pi \lfloor \frac{[a]_{2n}+[b]_{2n}}{2n}\rfloor} 
\end{equation} where we define $[x]_n \equiv x\pmod{n}$, $\lfloor x \rfloor$ as the greatest integer less than or equal to $x$. The factor $\delta_2$ represents the nontrivial 2-cocycle in $ \mathcal{H}^2(\mathbb{Z}_2^T,\mathbb{Z}_2)$.
For even-fold dihedral group $D_{2n}$, the generators must satisfy:
\begin{equation}T^2=P_f,M^2=P_f,R^{2n}=P_f,MRM^{-1}=R^{-1}\end{equation} with $T$ commuting with both $M$ and $R$. The corresponding 2-cocycles are: \begin{equation}
\omega_2(g_1h_2,g_2h_2):=\delta_2(g_1,g_2)\varepsilon_2(h_1,h_2)  \label{2-cocycle}
\end{equation} where $g_j \in \mathbb{Z}_2^T \subset G_b, h_j \in D_{2n} \subset G_b$. The factor $\varepsilon_2$ was defined in Eq.~(8) of Ref.~\cite{2Dcrystalline}, while $\delta_2$ is the nontrivial 2-cocycle in $ \mathcal{H}^2(\mathbb{Z}_2^T,\mathbb{Z}_2)$.

\begin{table}[tb]
    \centering
    \renewcommand\arraystretch{1.2}

    \begin{tabular}{ccccc}
        \toprule
        $SG$  & $E_0^{2D}$                      & $E_0^{1D}$                                                            & $E_0^{0D}$         & $\mathcal{G}_0$                                                                                       \\
        \midrule
        $C_2$ & ${\mathbb{Z}_1}$              & ${\color{blue}\mathbb{Z}_{2}}$                                      & ${\mathbb{Z}_{1}}$ & ${\color{blue}\mathbb{Z}_{2}}$                                                                        \\
        % \hline
        $C_3$ & ${\mathbb{Z}_1}$              & $\mathbb{Z}_{1}$                                                    & ${\mathbb{Z}_{1}}$ & ${\mathbb{Z}_{1}}$                                                                                      \\
        % \hline
        $C_4$ & ${\mathbb{Z}_1}$              & ${\color{red}\mathbb{Z}_{4}}$                                       & ${\mathbb{Z}_{1}}$ & ${\color{red}\mathbb{Z}_4}$                                                                           \\
        % \hline
        $C_6$ & ${\mathbb{Z}_1}$              & ${\color{blue}\mathbb{Z}_{2}}$                                      & ${\mathbb{Z}_{1}}$ & ${\color{blue}\mathbb{Z}_{2}}$                                                                        \\
        % \hline
        $D_2$ & ${\mathbb{Z}_1}$              & ${\color{blue}\mathbb{Z}^3_{2}}$                                    & ${\color{red}\mathbb{Z}_{2}}$ & ${\color{blue}\mathbb{Z}^3_2}\times{\color{red}\mathbb{Z}_{2}}  $                                     \\
        % \hline
        $D_3$ & ${\mathbb{Z}_1}$              & ${\color{red}\mathbb{Z}_8}$                                         & ${ \mathbb{Z}_{1}}$ & ${\color{red}\mathbb{Z}_8}$                                                                           \\
        % \hline
        $D_4$ & ${\mathbb{Z}_1}$              & ${\color{blue} \mathbb{Z}_{2}^2 }\times{\color{red}\mathbb{Z}_{4}}$ & ${\color{red}\mathbb{Z}_{2}}$ & ${\color{blue} \mathbb{Z}_{2}^2 }\times{\color{red}\mathbb{Z}_{4}} \times{\color{red}\mathbb{Z}_{2}}$ \\
        % \hline
        $D_6$ & ${\mathbb{Z}_1}$              & ${\color{blue}\mathbb{Z}^3_{2}}$                                    & ${\color{red}\mathbb{Z}_{2}}$ & ${\color{blue}\mathbb{Z}_2^3}\times{\color{red}\mathbb{Z}_{2}}$                                       \\
        \bottomrule
    \end{tabular}
    \caption{Interacting classification of 2D topological superconductors in spinless fermionic systems protected by point-group and time-reversal symmetries. The results are organized in a layer-by-layer fashion: \(E^{nD}_0\) denotes the group of obstruction-free, nontrivial \(n\)D block-states, and \(\mathcal{G}_0\) denotes the final classification group. Classification indices associated with fermionic and bosonic root phases are highlighted in red and blue, respectively.}
    \label{table II}
\end{table}

\begin{table}[tb]
    \centering
    \renewcommand\arraystretch{1.2}
    \begin{tabular}{ccccc}
        \toprule
        $SG$  & $E_{1/2}^{2D}$                      & $E_{1/2}^{1D}$                      & $E^{0D}_{1/2}$                       & $\mathcal{G}_{1/2}$                                                                     \\ 
        \midrule
        $C_2$ & ${\color{red}\mathbb{Z}_{2}}$     & ${\color{red}\mathbb{Z}_{2}}$ & ${\color{blue}\mathbb{Z}_{2}}$ & ${\color{blue}\mathbb{Z}_2} \times {\color{red}\mathbb{Z}_2^2}$                       \\ 
        % \hline
        $C_3$ & ${\color{red}\mathbb{Z}_2}$   & $\mathbb{Z}_{1}$              & $\mathbb{Z}_{1}$               & ${\color{red}\mathbb{Z}_2}$                                                         \\ 
        % \hline
        $C_4$ & ${\color{red}\mathbb{Z}_{2}}$  & ${\color{red}\mathbb{Z}_{2}}$ & ${\color{blue}\mathbb{Z}_{2}}$ & ${\color{blue}\mathbb{Z}_2} \times {\color{red}\mathbb{Z}_2^2}$                       \\ 
        % \hline
        $C_6$ & ${\color{red}\mathbb{Z}_{2}}$  & ${\color{red}\mathbb{Z}_{2}}$ & ${\color{blue}\mathbb{Z}_{2}}$ & ${\color{blue}\mathbb{Z}_2} \times {\color{red}\mathbb{Z}_2^2}$                       \\ 
        % \hline
        $D_2$ & ${\color{red}\mathbb{Z}_2}$    & ${\color{red}\mathbb{Z}_4^2}$ & ${\color{blue}\mathbb{Z}_2^2}$  & ${\color{blue}\mathbb{Z}_2^2} \times {\color{red}\mathbb{Z}_4 \times \mathbb{Z}_8}$ \\ 
        % \hline
        $D_3$ & ${\color{red}\mathbb{Z}_{2}}$   & ${\color{red}\mathbb{Z}_{4}}$ & ${\mathbb{Z}_1}$               & ${\color{red}\mathbb{Z}_8}$                                                         \\ 
        % \hline
        $D_4$ & ${\color{red}\mathbb{Z}_{2}}$   & ${\color{red}\mathbb{Z}_4^2}$   & ${\color{blue}\mathbb{Z}_2^2}$ & ${\color{blue}\mathbb{Z}_2^2} \times {\color{red}\mathbb{Z}_4 \times \mathbb{Z}_8}$ \\ 
        % \hline
        $D_6$ & ${\color{red}\mathbb{Z}_{2}}$   & ${\color{red}\mathbb{Z}_4^2}$   & ${\color{blue}\mathbb{Z}_2^2}$ & ${\color{blue}\mathbb{Z}_2^2 }\times {\color{red}\mathbb{Z}_4 \times \mathbb{Z}_8}$ \\ 
        \bottomrule
    \end{tabular}
    \caption{ Interacting classification of 2D topological superconductors in spinful fermionic systems protected by point-group and time-reversal symmetries. Notations and color conventions follow Table~\ref{table II}.}
    \label{table I}
\end{table}

For odd-fold groups, i.e. groups $C_n$ and $D_n$ with $n=3$, we directly present the fermionic symmetry groups $G_f$:
\begin{align}
    G_b=C_3:&\quad G_f=\{T,R,P_f \mid \begin{array}{l}
                                R^3=T^2=P_f, \\
                                RT=TR, \\
                                P_f^2=I 
                              \end{array}\} \notag\\
           &\cong\{r=TR^{-1}\mid r^{12}=I \} = \mathbb{Z}^T_{12}, \notag\\                   
    G_b=D_3:& \quad G_f=\{T,R,M,P_f \mid \begin{array}{l}
                                  R^3=M^2=T^2=P_f, \\
                                  MRM^{-1}=R^{-1}, \\
                                  RT=TR, \\
                                  MT=TM, \\
                                  P_f^2=I
                                \end{array}\}\notag \\
        &\cong\{T,r=R^2,m=MT \mid \begin{array}{l}
                                           T^4=m^2=r^3=I, \\ 
                                           mrm^{-1}=r^{-1}
                                         \end{array}\} \\
        & = \mathbb{Z}^T_{4}\times D^T_3 \notag 
\end{align}

After the discussion of point group, we now turn our attention to wallpaper groups. In two-dimensional systems, there are 17 wallpaper groups (e.g., $p4m$, $p2$, $pgg$). We consider systems with symmetry group $G_f = \mathbb{Z}_2^f \times_{\bar{\omega}_2} (SG \times \mathbb{Z}_2^T)$, where $SG$ represents a wallpaper group. We restrict our analysis to phases where translation symmetry is not fractionalized (i.e., no magnetic flux). Mathematically, this implies that the extension is trivial regarding the translation subgroup $\mathcal{T}\subset SG$. For spinless fermions we take the cocycles $\bar{\omega}_2$ to be trivial. But for spinful fermions, the 2-cocycle $\bar{\omega}_2$ for the full group $G_f$ is then defined as the pullback of the cocycle $\omega_2$ introduced above (see Eq.~\eqref{2-cocycle} and Eq.~\eqref{eq:extension cocycles of cyclic groups}):
\begin{equation}
\bar{\omega}_2(g_1, g_2) := \omega_2(\pi(g_1), \pi(g_2)), \quad \forall g_1, g_2 \in G_b.
\end{equation}
Here $G_b=SG \times \mathbb{Z}_2^T$, and $\pi: G_b \to G_b/\mathcal{T} \cong PG \times \mathbb{Z}_2^T$ is the natural projection homomorphism that maps group elements of $SG \times \mathbb{Z}_2^T$ to their point-group and time-reversal components $PG\times \mathbb{Z}_2^T$; $PG$ here specifically denotes the point group.

\begin{table}[htb]
    \centering
    \renewcommand\arraystretch{1.2}
    \begin{tabular}{ccccc}
        \toprule
        SG   & $ E_{0}^{2D}$ & $ E_{0}^{1D}$                                                                                                                                                                                                                            & $ E_{0}^{0D}$                                                                                                                                                     & $ \mathcal{G}_{0}$                                                                                                                                                                                                                                                                                                                  \\
         \midrule
        $p1$   & $ \mathbb{Z}_{1}$      & {\color{red}$ \mathbb{Z}_{8}^{2}$} & $ \mathbb{Z}_{1}$        & {\color{red}$ \mathbb{Z}_{8}^{2}$}                                                                                                                                                                                                                                                                                       \\
        % \hline
        $p2$   & $ \mathbb{Z}_{1}$      & {\color{blue}$ \mathbb{Z}_{2}^{3}$}                                                                                                                                                                                                & $ {\color{blue}\mathbb{Z}}{\color{blue}_{2}} \times {\color{red}\mathbb{Z}}{\color{red}_{2}^{3}}$     & $ {\color{blue}\mathbb{Z}}{\color{blue}_{2}^{4}} \times {\color{red}\mathbb{Z}}{\color{red}_{2}^{3}}$                                                                                                                                                                   \\
        % \hline
        $pm$   & $ \mathbb{Z}_{1}$      & $ {\color{blue}\mathbb{Z}}{\color{blue}_{2}} \times {\color{red}\mathbb{Z}}{\color{red}_{8}^{2}} \times {\color{red}\mathbb{Z}}{\color{red}_{4}}$ & $ {\color{blue}\mathbb{Z}}{\color{blue}_{2}} \times {\color{red}\mathbb{Z}}{\color{red}_{2}^{2}}$     & $ {\color{blue}\mathbb{Z}}{\color{blue}_{2}^{2}} \times {\color{red}\mathbb{Z}}{\color{red}_{2}^{2}} \times {\color{red}\mathbb{Z}}{\color{red}_{8}^{2}} \times {\color{red}\mathbb{Z}}{\color{red}_{4}}$ \\
        % \hline
        $pg$   & $ \mathbb{Z}_{1}$      & $ {\color{red}\mathbb{Z}}{\color{red}_{8}} \times {\color{red}\mathbb{Z}}{\color{red}_{2}}$                                                                                    & {\color{red}$ \mathbb{Z}_{2}$}                                                                                                                         & {\color{red}$\mathbb{Z}_8\times \mathbb{Z}_2^2$}                                                                                                                                               \\
        % \hline
        $cm$   & $ \mathbb{Z}_{1}$      & {\color{red}$ \mathbb{Z}_{8}^{2}$}                                                                                                                                                                                                & $ {\color{blue}\mathbb{Z}}{\color{blue}_{2}} \times {\color{red}\mathbb{Z}}{\color{red}_{2}}$         & {\color{red}$ {\color{blue}\mathbb{Z}}{\color{blue}_{2}} \times {\color{red}\mathbb{Z}}{\color{red}_{2}} \times \mathbb{Z}_{8}^{2}$}                                                                                                                         \\
        % \hline
        $pmm$  & $ \mathbb{Z}_{1}$      & {\color{blue}$ \mathbb{Z}_{2}^{7}$}                                                                                                                                                                                                & $ {\color{blue}\mathbb{Z}}{\color{blue}_{2}^{4}} \times {\color{red}\mathbb{Z}}{\color{red}_{2}^{4}}$ & $ {\color{blue}\mathbb{Z}}{\color{blue}_{2}^{11}} \times {\color{red}\mathbb{Z}}{\color{red}_{2}^{4}}$                                                                                                                                                                  \\
        % \hline
        $pmg$  & $ \mathbb{Z}_{1}$      & $ {\color{blue}\mathbb{Z}}{\color{blue}_{2}^{2}} \times {\color{red}\mathbb{Z}}{\color{red}_{8}}$                                                                                & $ {\color{blue}\mathbb{Z}}{\color{blue}_{2}} \times {\color{red}\mathbb{Z}}{\color{red}_{2}^{3}}$     & $ {\color{blue}\mathbb{Z}}{\color{blue}_{2}^{3}} \times {\color{red}\mathbb{Z}}{\color{red}_{8}} \times {\color{red}\mathbb{Z}}{\color{red}_{2}^{3}}$                                                                                    \\
        % \hline
        $pgg$  & $ \mathbb{Z}_{1}$      & $ {\color{blue}\mathbb{Z}}{\color{blue}_{2}} \times {\color{red}\mathbb{Z}}{\color{red}_{2}}$                                                                                    & $ {\color{blue}\mathbb{Z}}{\color{blue}_{2}} \times {\color{red}\mathbb{Z}}{\color{red}_{2}^2}$       & $  {\color{blue}\mathbb{Z}}{\color{blue}_{2}}\times {\color{red}\mathbb{Z}}{\color{red}_4} \times {\color{red}\mathbb{Z}}{\color{red}_2^2}  $                                                                                                                            \\
        % \hline
        $cmm$  & $ \mathbb{Z}_{1}$      & {\color{blue}$ \mathbb{Z}_{2}^{4}$}                                                                                                                                                                                                & $ {\color{blue}\mathbb{Z}}{\color{blue}_{2}^{2}} \times {\color{red}\mathbb{Z}}{\color{red}_{2}^{3}}$ & $ {\color{blue}\mathbb{Z}}{\color{blue}_{2}^{6}} \times {\color{red}\mathbb{Z}}{\color{red}_{2}^{3}}$                                                                                                                                                                   \\
        % \hline
        $p4$   & $ \mathbb{Z}_{1}$      & {\color{blue}$ \mathbb{Z}_{2} \times {\color{red}\mathbb{Z}}{\color{red}_{4}}$}                                                                                                                     & $ {\color{blue}\mathbb{Z}}{\color{blue}_{2}} \times {\color{red}\mathbb{Z}}{\color{red}_{2}^{2}}$     & $ {\color{blue}\mathbb{Z}}{\color{blue}_{2}^{2}} \times {\color{red}\mathbb{Z}}{\color{red}_{4}}{\color{red}\times \mathbb{Z}}{\color{red}_{2}^{2}}$                                                                                     \\
        % \hline
        $p4m$  & $ \mathbb{Z}_{1}$      & $ {\color{blue}\mathbb{Z}}{\color{blue}_{2}^{4}} \times {\color{red}\mathbb{Z}}{\color{red}_{4}}$                                                                                & $ {\color{blue}\mathbb{Z}}{\color{blue}_{2}^{3}} \times {\color{red}\mathbb{Z}}{\color{red}_{2}^{3}}$ & $ {\color{blue}\mathbb{Z}}{\color{blue}_{2}^{7}} \times {\color{red}\mathbb{Z}}{\color{red}_{2}^{3}}{\color{red}\times \mathbb{Z}}{\color{red}_{4}}$                                                                                     \\
        % \hline
        $p4g$  & $ \mathbb{Z}_{1}$      & $ {\color{blue}\mathbb{Z}}{\color{blue}_{2}} \times {\color{red}\mathbb{Z}}{\color{red}_{4}}$                                                                                    & $ {\color{blue}\mathbb{Z}}{\color{blue}_{2}} \times {\color{red}\mathbb{Z}}{\color{red}_{2}^{2}}$     & $ {\color{blue}\mathbb{Z}}{\color{blue}_{2}^{2}} \times {\color{red}\mathbb{Z}}{\color{red}_{2}^{2}}{\color{red}\times \mathbb{Z}}{\color{red}_{4}}$                                                                                     \\
        % \hline
        $p3$   & $ \mathbb{Z}_{1}$      & $ \mathbb{Z}_{1}$                                                                                                                                                                                                                                     & {\color{red}$ \mathbb{Z}_{2}$}                                                                                                                         & {\color{red}$ \mathbb{Z}_{2}$}                                                                                                                                                                                                                                                                                           \\
        % \hline
        $p3m1$ & $ \mathbb{Z}_{1}$      & {\color{red}$ \mathbb{Z}_{8}$}                                                                                                                                                                                                    & $ {\color{blue}\mathbb{Z}}{\color{blue}_{2}} \times {\color{red}\mathbb{Z}}{\color{red}_{2}}$         & $ {\color{blue}\mathbb{Z}}{\color{blue}_{2}} \times {\color{red}\mathbb{Z}}{\color{red}_{2}}{\color{red}\times \mathbb{Z}}{\color{red}_{8}}$                                                                                             \\
        % \hline
        $p31m$ & $ \mathbb{Z}_{1}$      & {\color{red}$ \mathbb{Z}_{8}$}                                                                                                                                                                                                    & $ {\color{blue}\mathbb{Z}}{\color{blue}_{2}} \times {\color{red}\mathbb{Z}}{\color{red}_{2}}$         & $ {\color{blue}\mathbb{Z}}{\color{blue}_{2}} \times {\color{red}\mathbb{Z}}{\color{red}_{2}}{\color{red}\times \mathbb{Z}}{\color{red}_{8}}$                                                                                             \\
        % \hline
        $p6$   & $ \mathbb{Z}_{1}$      & {\color{blue}$ \mathbb{Z}_{2}$}                                                                                                                                                                                                    & $ {\color{blue}\mathbb{Z}}{\color{blue}_{2}} \times {\color{red}\mathbb{Z}}{\color{red}_{2}}$         & $ {\color{blue}\mathbb{Z}}{\color{blue}_{2}^{2}} \times {\color{red}\mathbb{Z}}{\color{red}_{2}}$                                                                                                                                                                       \\
        % \hline
        $p6m$  & $ \mathbb{Z}_{1}$      & {\color{blue}$ \mathbb{Z}_{2}^{3}$}                                                                                                                                                                                                & $ {\color{blue}\mathbb{Z}}{\color{blue}_{2}^{3}} \times {\color{red}\mathbb{Z}}{\color{red}_{2}}$     & $ {\color{blue}\mathbb{Z}}{\color{blue}_{2}^{6}} \times {\color{red}\mathbb{Z}}{\color{red}_{2}}$                                                                                                                                                                       \\
        \bottomrule
    \end{tabular}
    \caption{Interacting classification of 2D topological superconductors in spinless fermionic systems protected by wallpaper-group and time-reversal symmetries. Here \(E^{nD}_0\) denotes the group of obstruction-free, nontrivial \(n\)D block-states, and the final classification—including possible group-extension structure—is denoted by \(\mathcal{G}_0\). Classification indices corresponding to fermionic and bosonic root phases are highlighted in red and blue, respectively.}
    \label{plane group spinless}
\end{table}
\begin{table}[htb]
    \centering

    \renewcommand\arraystretch{1.2}
    \begin{tabular}{cccccc}
    \toprule
    $~~SG~~$&$~~~E_{1/2}^{\mathrm{2D}}~~~$&$~~~E_{1/2}^{\mathrm{1D}}~~~$&$~~~E_{1/2}^{\mathrm{0D}}~~~$&$~~~\mathcal{G}_{1/2}~~~$\\
    \midrule
    $p1$&${\color{red}\mathbb{Z}_2}$&${\color{red}\mathbb{Z}_2^2}$&${\color{black}\mathbb{Z}_1}$&$
    {\color{red}\mathbb{Z}_2^3}$\\
   %  \hline
    $p2$&${\color{red}\mathbb{Z}_2}$&${\color{red}\mathbb{Z}_2^3}$&${\color{blue}\mathbb{Z}_2^4}$&$
    {\color{red}\mathbb{Z}_2^4}\times
    {\color{blue}\mathbb{Z}_2^4}$\\
   %  \hline
    $pm$&${\color{red}\mathbb{Z}_2}$&${\color{red}\mathbb{Z}_2 \times \mathbb{Z}_4^2}$&${\color{blue}\mathbb{Z}_2^2}$&${\color{red}\mathbb{Z}_2}\times
    {\color{red}\mathbb{Z}_4 }\times{\color{red} \mathbb{Z}_8}\times
    {\color{blue}\mathbb{Z}_2^2}$\\
   %  \hline
    $pg$&${\color{red}\mathbb{Z}_2}$&${\color{red}\mathbb{Z}_2^2}$&$~{\color{black}\mathbb{Z}_1}~$&$
    {\color{red}\mathbb{Z}_4\times\mathbb{Z}_2}$\\
   %  \hline
    $cm$&${\color{red}\mathbb{Z}_2}$&${\color{red}\mathbb{Z}_2 \times \mathbb{Z}_4}$&${\color{blue}\mathbb{Z}_2}$&${\color{red}\mathbb{Z}_2 \times \mathbb{Z}_8}\times
    {\color{blue}\mathbb{Z}_2}$\\
   %  \hline
    $pmm$&${\color{red}\mathbb{Z}_2}$&${~\color{red}\mathbb{Z}_4^4~}$&${\color{blue}\mathbb{Z}_2^8}$&${\color{red}\mathbb{Z}_8}\times{~\color{red}\mathbb{Z}_4^3~}\times{\color{blue}\mathbb{Z}_2^8}$\\
   %  \hline
    $pmg$&${\color{red}\mathbb{Z}_2}$&${\color{red}\mathbb{Z}_2^2 \times \mathbb{Z}_4}$&${\color{blue}\mathbb{Z}_2^3}$&${\color{red}\mathbb{Z}_2^2 \times \mathbb{Z}_8}\times{\color{blue}\mathbb{Z}_2^3}$\\
   %  \hline
    $pgg$&${\color{red}\mathbb{Z}_2}$&${\color{red}\mathbb{Z}_2^2}$&${\color{blue}\mathbb{Z}_2^2}$&${\color{red}\mathbb{Z}_4 \times \mathbb{Z}_2}\times{\color{blue}\mathbb{Z}_2^2}$\\
   %  \hline
    $cmm$&${\color{red}\mathbb{Z}_2}$&${\color{red}\mathbb{Z}_2 \times \mathbb{Z}_4^2}$&${\color{blue}\mathbb{Z}_2^5}$&${\color{red}\mathbb{Z}_2}\times{\color{red}\mathbb{Z}_4 \times \mathbb{Z}_8}\times{\color{blue}\mathbb{Z}_2^5}$\\
   %  \hline
    $p4$&${\color{red}\mathbb{Z}_2}$&${\color{red}\mathbb{Z}_2^2}$&${\color{blue}\mathbb{Z}_2^3}$&${\color{red}\mathbb{Z}_2^3}\times{\color{blue}\mathbb{Z}_2^3}$\\
   %  \hline
    $p4m$&${\color{red}\mathbb{Z}_2}$&${\color{red}\mathbb{Z}_4^3}$&${\color{blue}\mathbb{Z}_2^6}$&${\color{red}\mathbb{Z}_8}\times{\color{red}\mathbb{Z}_4^2}\times{\color{blue}\mathbb{Z}_2^6}$\\
   %  \hline
    $p4g$&${\color{red}\mathbb{Z}_2}$&${\color{red}\mathbb{Z}_2 \times \mathbb{Z}_4}$&${\color{blue}\mathbb{Z}_2^3}$&${\color{red}\mathbb{Z}_2 \times \mathbb{Z}_8}\times{\color{blue}\mathbb{Z}_2^3}$\\
   %  \hline
    $p3$&${\color{red}\mathbb{Z}_2}$&$\mathbb{Z}_1$&${\color{black}\mathbb{Z}_1}$&${\color{red}\mathbb{Z}_2}$\\
   %  \hline
    $~~~p3m1~~~$&${\color{red}\mathbb{Z}_2}$&${\color{red}\mathbb{Z}_4}$&${\color{blue}\mathbb{Z}_2}$&${\color{red}\mathbb{Z}_8}\times{\color{blue}\mathbb{Z}_2}$\\
   %  \hline
    $p31m$&${\color{red}\mathbb{Z}_2}$&${\color{red}\mathbb{Z}_4}$&${\color{blue}\mathbb{Z}_2}$&${\color{red}\mathbb{Z}_8}\times {\color{blue}\mathbb{Z}_2}$\\
   %  \hline
    $p6$&${\color{red}\mathbb{Z}_2}$&${\color{red}\mathbb{Z}_2}$&${\color{blue}\mathbb{Z}_2^2}$&${\color{red}\mathbb{Z}_2^2}\times {\color{blue}\mathbb{Z}_2^2}$\\
   %  \hline
    $p6m$&${\color{red}\mathbb{Z}_2}$&${\color{red}\mathbb{Z}_4^2}$&${\color{blue}\mathbb{Z}_2^4}$&${\color{red}\mathbb{Z}_8}\times{\color{red}\mathbb{Z}_4}\times{\color{blue}\mathbb{Z}_2^4}$\\
    \bottomrule
    \end{tabular} 
     \caption{Interacting classification of 2D topological superconductors in spinful fermionic systems protected by wallpaper-group and time-reversal symmetries. Notations and color conventions follow Table~\ref{plane group spinless}.
    }
    \label{plane group spin-1/2}
    \end{table}

\subsection{Main results} 
The classification results are summarized in the following tables. Table~\ref{table II} provides the classification results for point group and time-reversal symmetry-protected topological superconductors (TSC) in spinless fermionic systems, while Table~\ref{table I} presents the corresponding results for spinful fermionic systems. These TSCs exhibit physical symmetry groups  of the same form as \( G_b = C_n \times \mathbb{Z}_2^T \) or \( G_b = D_n \times \mathbb{Z}_2^T \). In these tables, the left column specifies the point group components $SG$ of $G_b$, corresponding to $C_n$ or $D_n$. The classification group associated with $d$-dimensional obstruction-free and non-trivial block-state decorations is denoted as $E^{d\mathrm{D}}$, with the data organized layer by layer, corresponding to contributions from 0D, 1D, and $2$D blocks, respectively. Finally, the ultimate group structures, represented by $\mathcal{G}$, of the classifications are analyzed by explicitly examining possible non-trivial stacking relations between different dimensional block-states.
In particular, we label the classification indices with the fermionic root phase by red, and the classification indices with the bosonic root phase by blue. 

Table~\ref{plane group spinless} summarizes all the classification results for wallpaper group and time-reversal symmetry-protected topological superconductors (TSC) for spinless fermions, while Table~\ref{plane group spin-1/2} presents the results for spinful fermions. The leftmost column in all the tables indicates the space group component $SG$ of \( G_b \). Again, we label the classification group attributed to the $d$-dimensional obstruction-free and non-trivial block-state decorations by $E^{d\mathrm{D}}$. And $\mathcal{G}$ denotes the ultimate classification.
Compared with the classification of 2D FSPTs protected by the corresponding internal symmetries \cite{ren2023stacking}, the above classification results validate the fermionic crystalline equivalence principle~\cite{debray2021invertible,Nonperturbativeconstraints}.

\section{CONSTRUCTION AND CLASSIFICATION OF POINT GROUP AND TIME-REVERSAL SYMMETRY-PROTECTED TOPOLOGICAL SUPERCONDUCTORS}
\label{sec:point_group}

In this section, we discuss the real-space construction for FSPT phases in 2D interacting fermionic systems protected by both point group symmetry and time-reversal. For 2D FSPT systems protected by point group symmetries only, a complete classification has been provided in previous works~\cite{RotationSPT,pointgroup}, which also employed the real-space construction. In this paper, we extend the discussion to systems with additional time-reversal symmetry.

\begin{figure*}[tb]
    \centering
    \begin{subfigure}{0.24\textwidth}
        \includegraphics[width=\linewidth]{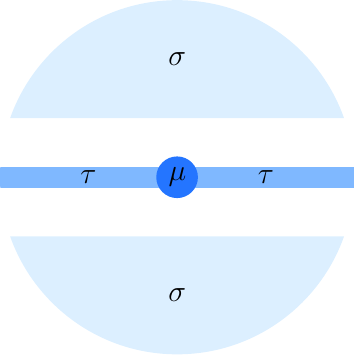}
        \caption{ $C_2$}
    \end{subfigure}\hfill
    \begin{subfigure}{0.24\textwidth}
        \includegraphics[width=\linewidth]{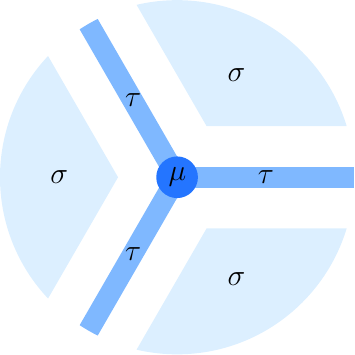}
        \caption{ $C_3$}
    \end{subfigure}\hfill
    \begin{subfigure}{0.24\textwidth}
        \includegraphics[width=\linewidth]{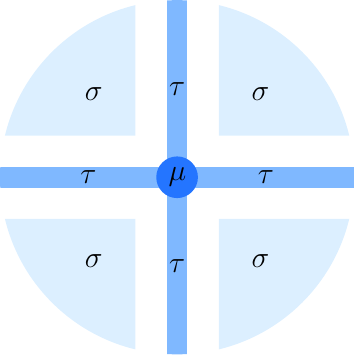}
        \caption{ $C_4$}
    \end{subfigure}\hfill
    \begin{subfigure}{0.24\textwidth}
        \includegraphics[width=\linewidth]{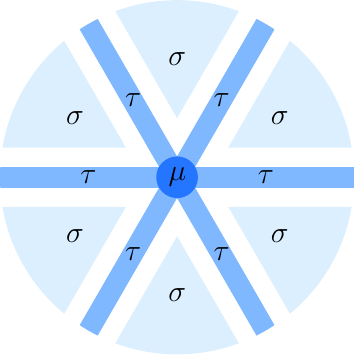}
        \caption{ $C_6$}
    \end{subfigure}
    \caption{Cell decompositions for cyclic groups $C_n$  with $n=2,3,4,6$, arranged from left to right. Each cell decomposition contains a single symmetrically independent block in each dimension ($0$D, $1$D, and $2$D), denoted as $\mu$, $\tau$, and $\sigma$, respectively. }
    \label{Cn cell}
\end{figure*}

\begin{figure*}[tb]
    \centering
    \begin{subfigure}{0.24\textwidth}
        \includegraphics[width=\linewidth]{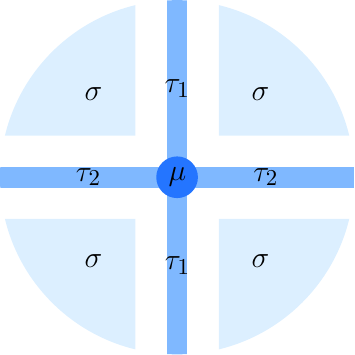}
        \caption{ $D_2$}
    \end{subfigure}\hfill
    \begin{subfigure}{0.24\textwidth}
        \includegraphics[width=\linewidth]{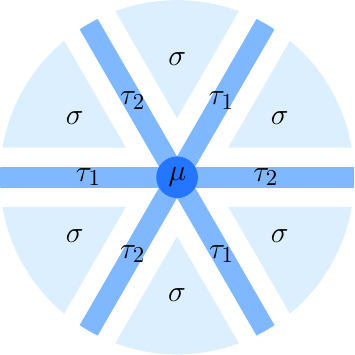}
        \caption{ $D_3$}
    \end{subfigure}\hfill
    \begin{subfigure}{0.24\textwidth}
        \includegraphics[width=\linewidth]{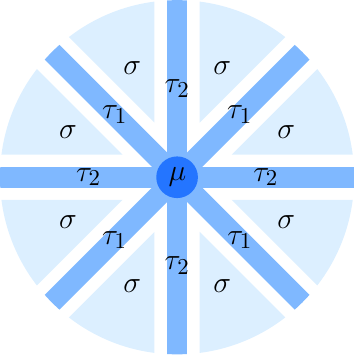}
        \caption{ $D_4$}
    \end{subfigure}\hfill
    \begin{subfigure}{0.24\textwidth}
        \includegraphics[width=\linewidth]{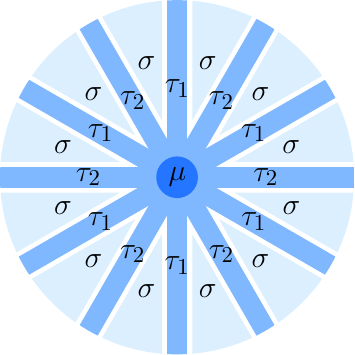}
        \caption{ $D_6$}
    \end{subfigure}
    \caption{Cell decompositions for dihedral groups $D_n$  with $n=2,3,4,6$, arranged from left to right. Each cell decomposition contains one symmetrically independent block for dimensions $0$D and $2$D, denoted as $\mu$ and $\sigma$, respectively. However, there are two classes of independent $1$D blocks, labeled by $\tau_1$ and $\tau_2$.}
    \label{Dn cell}
\end{figure*}

In 2D lattice systems, there are 10 point groups, classified into rotation (cyclic) groups $C_n$ and dihedral groups $D_n$ ($n = 1, 2, 3, 4, 6$).  
The cell decompositions for the groups $C_n$ and $D_n$ are listed in Figures~\ref{Cn cell},\ref{Dn cell}. The block notations $\mu,\tau,\sigma$ used in the following discussion correspond directly to the labels shown in these figures.

We begin with a fermionic system exhibiting $D_4$ point group symmetry—a paradigmatic non-Abelian point group of even order—and time-reversal symmetry. This system hosts intrinsically interacting phases beyond free fermion realization.

\subsection{$D_4$}

The $D_4$-symmetric cell decomposition (Fig.~\ref{Dn cell}(c)) consists of a $0$D block $\mu$ and two types of $1$D blocks, $\tau_1$ and $\tau_2$. We take $C_4$ (the $\pi/2$ rotation) and $M_{\tau_j}$ (reflection about the $\tau_j$-block axis) as the generators of the group $D_4$. And the cell decomposition remains invariant under the action of these generators.

\subsubsection{Spinless fermions}\label{sec:D4 spinless fermion}
 Consider the $0$D block-state decoration first.
The $0$D block $\mu$ has a physical symmetry group $G_b = D_4 \times \mathbb{Z}_2^T$, where $T$ denotes the time-reversal generator.
Following the method in Ref.~\cite{Wang2020}, we characterize the 0D FSPT phases by an ordered pair $(n_0, \nu_1)$ of cohomology data. Specifically, $n_0 \in \mathcal{H}^0(G_b, \mathbb{Z}_2^f)$ represents the complex fermion decoration, while $\nu_1 \in \mathcal{H}^1(G_b, U_T(1))$ corresponds to the bosonic phase. For the present case, the corresponding $0$D block FSPT phases are characterized by:
\begin{equation}
    \begin{aligned}
&n_0 \in \mathcal{H}^0\left(G_b, \mathbb{Z}_2^f\right) = \mathbb{Z}_2,\\
&\nu_1 \in \mathcal{H}^1\left(G_b, U_T(1)\right) = \mathbb{Z}_2 \times \mathbb{Z}_2.
\end{aligned}
\end{equation}
The two non-trivial representative cocycles of $\nu_1$ are characterized by:
\begin{equation}
\nu_1(g) =
\begin{cases}
-1, & \text{if } g = M_{\tau_1} \text{ or } M_{\tau_2}\\
+1, & \text{otherwise}
\end{cases}
\end{equation}
It should be noted that no non-trivial cocycles are contributed by the subgroup $\mathbb{Z}_2^T$.

As a result, the classification group of 0D FSPT phases is given by the direct product $\mathcal{H}^0(G_b, \mathbb{Z}_2^f) \times \mathcal{H}^1(G_b, U_T(1))$. This leads to a $\mathbb{Z}_2^3$ group structure, where each $\mathbb{Z}_2$ factor corresponds to the possible eigenvalues $\pm 1$ of the reflection operators $M_{\tau_1}$, $M_{\tau_2}$, and the fermion parity operator $P_f$.

Next, we consider the $1$D block-state decoration. All $1$D blocks, denoted as $(\tau_1, \tau_2)$, share the same symmetry group :
\begin{align}
    G_b &= \mathbb{Z}_2^M \times \mathbb{Z}_2^T \nonumber\\
    G_f &= \mathbb{Z}_2^M \times \mathbb{Z}_2^T \times \mathbb{Z}_2^f,
\end{align}
The classification data for 1D FSPT phases is a pair $(n_1, \nu_2)$ that satisfies certain symmetry and consistency conditions~\cite{Wang2020}:
\begin{equation}
    \begin{aligned}
    & n_1 \in \mathcal{H}^1\left(G_b, \mathbb{Z}_2^f\right), \\
    & \nu_2 \in C^2\left(G_b, U_T(1)\right) / B^2\left(G_b, U_T(1)\right) / \Gamma^2, \\
    & \mathrm{d} \nu_2 = (-1)^{\omega_2 \cup n_1}, \\
    & \Gamma^2 = \left\{ (-1)^{\omega_2} \in \mathcal{H}^2\left(G_b, U_T(1)\right) \right\},
\end{aligned} \label{eq:equation of 1D FSPT}
\end{equation}
where \( n_1 \) is a 1-cochain representing fermion decorations, and \( \nu_2 \) is a 2-cochain representing bosonic SPT (BSPT) layer candidates. Here, \(C^2\) denotes the 2-cochain group, and \(B^2\) consists of 2-cochains \(\nu_2\) satisfying the condition \(\mathrm{d} \nu_2 = (-1)^{\omega_2 \cup n_1}\). Notably, these conditions have a solution if and only if \((-1)^{\omega_2 \cup n_1}\) is a coboundary, which imposes an additional constraint on \(n_1\). And any solution $\nu_2$ of $\mathrm{d} \nu_2 = (-1)^{\omega_2 \cup n_1}$ must be excluded from consideration if it belongs to the trivialization subgroup $\Gamma^2$~\cite{sigmaModel}.
For this case, the classification data for the corresponding 1D FSPT phase are:
\begin{equation}
\begin{aligned}
    n_1 & \in \mathcal{H}^1(G_b, \mathbb{Z}_2^f) = \mathbb{Z}_2 \times \mathbb{Z}_2, \\
    \nu_2 & \in \mathcal{H}^2(G_b, U_T(1)) = \mathbb{Z}_2 \times \mathbb{Z}_2.
\end{aligned}
\end{equation}
The two representative non-trivial cocycles for \( n_1 \) take the form:
\begin{equation}
    n_1^q(g) \equiv  \text{sum of powers of $q$ in $g$} \pmod{2}
 \label{eq:definition of n1M}
\end{equation}
 where $q$ represents either $M$ or $T$, and any group element $g\in G_b$ is expressed in its canonical form $g=M^jT^k$ with integer powers $j$ and $k$. For instance, $n_1^M(M)=n_1^M(MT)=1$,  while $n_1^M(T)=0$.
 
      For \( \nu_2 \), the non-trivial representatives are:
\begin{equation}
(-1)^{n_1^T \cup n_1^M} \quad \text{and} \quad (-1)^{n_1^T \cup n_1^T},
\end{equation}
or equivalently:
\begin{equation}
(-1)^{n_1^M \cup n_1^M} \quad \text{and} \quad (-1)^{n_1^T \cup n_1^T}.
\end{equation} 

In our convention, \( n_1 \) takes values in the additive group \( \mathbb{Z}_2 = \{0, 1\} \), whereas \( \nu_2 \) takes values in the multiplicative group \( U(1) \).
For any $n_1$, the obstruction function $O[n_1]=(-1)^{\omega_2 \cup n_1}$ and the trivialization subgroup $\Gamma^2=(-1)^{\omega_2}$  are both trivial (identically $1$), since $\omega_2$ is zero.

Consequently, the stacking law \cite{ren2023stacking} takes the form: 
\begin{equation}
\begin{aligned}
   & (n_1^T, 1) \boxplus (n_1^T, 1) = (0, (-1)^{n_1^{T} \cup n_1^{T}}),\\
&(n_1^M, 1) \boxplus (n_1^M, 1) = (0, (-1)^{n_1^M\cup n_1^M}).
\end{aligned} \label{eq:D4 spinless stacking law}
\end{equation}
 This indicates that stacking two \( n_1^T \) (or \( n_1^M \)) FSPT phases produces a single \( (-1)^{n_1^{T} \cup n_1^{T}} \) (or \( (-1)^{n_1^M\cup n_1^M} \)) BSPT phase.  Therefore, the corresponding 1D \( G_f \)-FSPT has an ultimate classification:
\begin{equation}
\mathbb{Z}_4 \times \mathbb{Z}_4,
\end{equation}
where the two order-four group generators correspond to the root phases \( n_1^M \) and \( n_1^T \).
\begin{figure}[tb]
    \centering
    \includegraphics[width=0.48\textwidth]{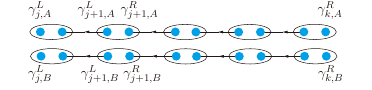}
    \caption{1D $G_f$-FSPT  constructed by two Majorana chains. The arrows indicate pairings between Majorana operators (depicted as blue dots); for example, an arrow pointing from $\gamma^L_{j+1}$ to $\gamma^R_j$ represents a coupling term $i\gamma^R_{j} \gamma^L_{j+1}$. The two chains are labeled by index $A$ and $B$, respectively.  }
    \label{four chain definition}
\end{figure}

Such root phases admit a concrete realization via two Majorana chains. The corresponding ground-state wavefunction, illustrated in Fig.~\ref{four chain definition}, provides a fixed-point representative of these phases. The Hamiltonian is constructed from Majorana operators $\gamma_{j,\sigma}$ and takes the form
\begin{equation}
H = i t \sum_{j, \sigma} \gamma^R_{j, \sigma} \gamma^L_{j+1, \sigma}, \label{eq:D4 spinless definition}
\end{equation}
where integer \( j \) represents the lattice sites, and \( \sigma = A, B \) denotes the chain index.

Distinct $G$-FSPT phases are distinguished by the symmetry transformation properties of their fixed-point ground-state wavefunctions. In the following, we specify the symmetry action for the present model.

For the \( n_1^T \) FSPT phase, the symmetry action is:
\begin{equation}
\begin{aligned}
    T: & \quad
    \left\{
    \begin{aligned}
        & i \mapsto -i, \\
        & \gamma^L_{j, \sigma} \mapsto \gamma^L_{j, \sigma}, \quad \gamma^R_{j, \sigma} \mapsto -\gamma^R_{j, \sigma},
    \end{aligned}
    \right. \\
    M: & \qquad \gamma^L_{j, \sigma} \mapsto \gamma^L_{j, \sigma}, \quad \gamma^R_{j, \sigma} \mapsto \gamma^R_{j, \sigma},
\end{aligned} \label{eq:symm property of n1T}
\end{equation}
% where \( j \) is an integer, and \( \sigma = A, B \). 
These symmetry transformations leave the Hamiltonian invariant and realize the symmetry group $G_f$ of the $1$D blocks.

To determine whether the ground state of the model realizes the $n_1^T$ FSPT phase, a convenient diagnostic is to analyze the phase obtained under stacking. In particular, stacking two identical copies of the model produces a system of four Majorana chains, whose ground state realizes a BSPT phase. This procedure enables an unambiguous identification of the original FSPT phase via the inverse mapping of the stacking law in Eq.~\eqref{eq:D4 spinless stacking law}.
While the symmetry properties in Eq.~\eqref{eq:symm property of n1T} define a specific linear representation of \( G_f \) acting on the Hilbert space of the four-Majorana-chains, further insight can be gained by examining their manifestation at the boundary. Under open boundary conditions, one finds that the induced representations of \( G_f \) in the boundary Hilbert space satisfy \( T^2 = -1 \) and \( M^2 = 1 \), 
with the associated factor system given by \( (-1)^{n_1^{T} \cup n_1^{T}} \). This pattern of symmetry fractionalization signals that the four-chain model realizes the \( (-1)^{n_1^{T} \cup n_1^{T}} \) BSPT phase. Consequently, the original two-chain model is identified as a \( n_1^T \) FSPT phase, in full agreement with the stacking law in Eq.~\eqref{eq:D4 spinless stacking law}. 

In contrast to the $n_1^T$ phase, the \( n_1^M \) phase is characterized by a different set of symmetry transformations:
\begin{equation}
\begin{aligned}
    T: & \quad
    \left\{
    \begin{aligned}
        & i \mapsto -i, \\
        & \gamma^L_{j, A} \mapsto \gamma^L_{j, A},\quad \gamma^L_{j, B} \mapsto -\gamma^L_{j, B}\\
        & \gamma^R_{j, A} \mapsto -\gamma^R_{j, A},\quad \gamma^R_{j, B} \mapsto \gamma^R_{j, B}
    \end{aligned}
    \right. \\
    M: & \quad  \left\{
    \begin{aligned}&   \gamma^L_{j, A} \mapsto \gamma^L_{j, A},\quad\gamma^L_{j, B} \mapsto -\gamma^L_{j, B}\\
        & \gamma^R_{j, A} \mapsto \gamma^R_{j, A},\quad\gamma^R_{j, B} \mapsto -\gamma^R_{j, B}
    \end{aligned} \right.
\end{aligned} \label{eq:symm property of n1M}
\end{equation}
Similarly, one can verify that, for the BSPT four-chain model obtained via stacking, the induced symmetry action on the boundary Hilbert space instead satisfies \(T^{2}=1\) and \(M^{2}=-1\), with the corresponding factor system \( (-1)^{\,n_1^{M}\cup n_1^{M}} \). This confirms that our original model (Eq.~\eqref{eq:D4 spinless definition}), together with the symmetry assignment in Eq.~\eqref{eq:symm property of n1M}, is correctly identified as the \(n_1^{M}\) FSPT phase.

In addition, the $n_1^T+ n_1^M$ phase can be realized by the two Majorana chains in Eq.~\eqref{eq:D4 spinless definition} with the symmetry actions as follows:
\begin{equation}
    \begin{aligned}
    T: & \quad
    \left\{
    \begin{aligned}
        & i \mapsto -i, \\
        & \gamma^L_{j, \sigma} \mapsto \gamma^L_{j, \sigma}, \quad \gamma^R_{j, \sigma} \mapsto -\gamma^R_{j, \sigma}
    \end{aligned}
    \right. \\
    M: & \quad  \left\{
    \begin{aligned}&   \gamma^L_{j, A} \mapsto \gamma^L_{j, A},\quad\gamma^L_{j, B} \mapsto  -\gamma^L_{j, B}\\
        & \gamma^R_{j, A} \mapsto \gamma^R_{j, A}, \quad\gamma^R_{j, B} \mapsto -\gamma^R_{j, B}
    \end{aligned} \right.
\end{aligned} , \label{eq:symm property of n1Mn1T}
\end{equation}
 Similarly, one can verify that, for the corresponding BSPT four-chain model obtained by stacking, the induced representation of \(G_f\) on the boundary Hilbert space satisfies \(T^{2}=-1\) and \(M^{2}=-1\), with the associated factor system
\begin{equation}
    (-1)^{n_1^{T}\cup n_1^{T}+n_1^{M}\cup n_1^{M}}.
    \label{eq:n1mn1t relation}
\end{equation}

Having discussed all possible FSPT phases above, we now turn to the single Majorana chain as a potential candidate for $1$D block-state decoration. Consider the Hamiltonian for a single Majorana chain:
\begin{equation}
H = i t \sum_{j} \gamma^R_{j} \gamma^L_{j+1}, \label{eq:definition of Majorana chain}
\end{equation}
with symmetry actions
\begin{equation}
\begin{aligned}
    T: & \quad 
    \begin{cases}
        i \mapsto -i, \\
        \gamma^L_j \mapsto \gamma^L_j, \quad  \gamma^R_j \mapsto -\gamma^R_j,
    \end{cases} \\
    M: & \qquad \gamma^L_j \mapsto \gamma^L_j, \quad \gamma^R_j \mapsto \gamma^R_j,
\end{aligned} \label{eq:symm majorana chain spinless}
\end{equation}
which leave the Hamiltonian invariant and realize the 1D-block symmetry group \( G_f \). 

The symmetry action here is identical to that of the \(n_1^T\) phase, indicating that the \(n_1^T\) FSPT state can be realized as a stack of two single Majorana chains. As a consequence, the complete set of $1$D block-state candidates forms a group \begin{equation}
    \mathbb{Z}_8 \times \mathbb{Z}_4. \label{eq:Z8}
\end{equation} Here the $\mathbb{Z}_8$ component is generated by a single Majorana chain, while the $\mathbb{Z}_4$ component is generated by the $n_1^M$ root phase. Notably, the $\mathbb{Z}_8$ component aligns with the established classification of one-dimensional topological superconductors~\cite{Kitaev_2001}.

\begin{figure}[b]
    \centering
\includegraphics[width=0.3\textwidth]{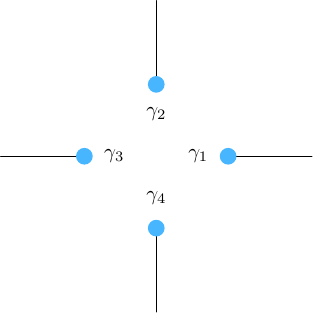}%width for pictures about dangling modes 
    \caption{ Single Majorana chain decoration on $1$D blocks labeled by $\tau_1$ or $\tau_2$, leaving $4$ dangling Majorana modes  (depicted as blue dots) $\gamma_1,\gamma_2,\gamma_3,\gamma_4$ at $0$D block $\mu$.}
    \label{D4 1chain}
\end{figure}

Next, we examine which block-state decorations satisfy the no-open-edge condition. All zero-dimensional decorations automatically meet this requirement. Consequently, the obstruction-free $0$D block-states form the group
\begin{equation}
\{\mathrm{OFBS}\}^{\mathrm{0D}} = \mathbb{Z}_2^3.
\end{equation}

Having classified the obstruction-free $0$D block-states, we now turn to $1$D block decorations and determine when they can satisfy the no-open-edge condition.
Decorating each $1$D block \(\tau_1\) or \(\tau_2\) with a single Majorana chain yields four dangling Majorana modes at \(\mu\) (Fig.~\ref{D4 1chain}). 

Crucially, any $1$D block-state necessarily has two boundaries, with dangling modes of type $\gamma_j^L$ and $\gamma_j^R$ appearing at opposite ends and transforming distinctly under symmetries. Given the freedom in choosing the decoration pattern, specifically the pairing direction of Majorana chains, we can always arrange for one type of boundary modes to appear at $\mu$. Here, we adopt a uniform pairing direction convention, resulting in only $\gamma_j^L$ modes at $\mu$, and henceforth omit the superscript $L$ unless explicitly needed. In later sections, we will also consider cases where different decoration patterns are implemented on the independent blocks $\tau_1$ and $\tau_2$.

In the vicinity of \( \mu \), these Majorana modes transform under the fourfold rotation as (with indices modulo $4$):
\(
C_4: \gamma_j \mapsto \gamma_{j+1}.
\)
The corresponding local fermion-parity operator and its symmetry properties are given by:
\begin{equation}
  P_f = -\gamma_1 \gamma_2 \gamma_3 \gamma_4,~ C_4: P_f \mapsto -P_f.  
\end{equation}
As a result, these four Majorana modes break the fermion parity, implying that the four dangling Majorana modes cannot be symmetrically gapped without breaking fermion parity. Therefore, decorating \( \tau_1 \) or \( \tau_2 \) with a single Majorana chain does not yield a nontrivial crystalline TSC, as it violates the no-open-edge condition.

\begin{figure}[bt]
    \centering
\includegraphics[width=0.3\textwidth]{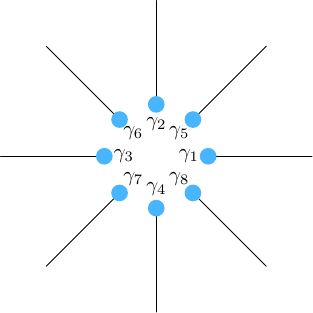}%width for pictures about dangling modes 
    \caption{ Single Majorana chain decoration on $1$D blocks $\tau_1$ and $\tau_2$, which leaves $8$ dangling Majorana modes $\gamma_1,\gamma_2,\gamma_3,\dots,\gamma_8$ at  $0$D block $\mu$.}
    \label{D4 1chain tau1tau2}
\end{figure}

Building on the above result, it is natural to ask whether the obstruction can be removed by decorating both $1$D blocks  $\tau_1$ and $\tau_2$ simultaneously.

Now, consider decorating single Majorana chains on both \( \tau_1 \) and \( \tau_2 \), which leaves eight dangling Majorana modes at the $0$D block \( \mu \) (Fig.~\ref{D4 1chain tau1tau2}). Though the local fermion-parity operator commutes with the rotation operator $C_4$, it anti-commutes with the reflection operator $M$. Specifically, for the fermion-parity operator $ P_f = \prod^8_{j=1}\gamma_j$, it transforms under $M$ as follows: \begin{equation}M: P_f \mapsto \gamma_1 \gamma_4 \gamma_3 \gamma_2 \gamma_8 \gamma_7 \gamma_6 \gamma_5= -P_f.\end{equation} This anti-commutation under $M$ indicates that the dangling Majorana modes cannot be symmetrically gapped while preserving the full $D_4$ symmetry. Therefore, this Majorana chain decoration is still incompatible with the $D_4$ symmetry and remains forbidden by the no-open-edge condition.

\begin{figure}[bt]
    \centering
\includegraphics[width=0.3\textwidth]{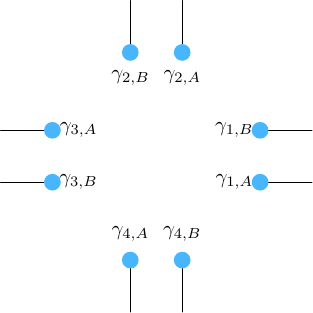}%width for pictures about dangling modes 
    \caption{ 1D FSPT decoration on $1$D blocks $\tau_1$, which leaves $8$ dangling Majorana modes $\gamma_{1,A},\gamma_{1,B},\dots,\gamma_{4,A},\gamma_{4,B}$  at  $0$D block $\mu$.}
    \label{fig:spinless D4 2chain tau1}
\end{figure}

We now analyze the 1D FSPT decorations of $n_1^T$ phase on the $\tau_1$ blocks, implemented via two Majorana chains construction defined in Eq.~\eqref{eq:D4 spinless definition}. The specific configuration of these decorated $\tau_1$ blocks is illustrated in Fig.~\ref{fig:spinless D4 2chain tau1}.

At the $0$D block $\mu$,  the decoration leaves eight dangling zero modes (Fig.~\ref{fig:spinless D4 2chain tau1}). Their transformation properties under the $0$D block symmetry group are:
\begin{align}
    n_1^T:&\left\{\begin{aligned}
        T: &~ i \mapsto -i,\\
            &~\gamma_{j,\sigma} \mapsto \gamma_{j,\sigma}, \\
        M: & ~\gamma_{j,\sigma} \mapsto \gamma_{6-j,\sigma},\\
        C_4:& ~\gamma_{j,\sigma} \mapsto \gamma_{j+1,\sigma},
    \end{aligned}\right.  \label{eq:D4 n1T}
\end{align}
where $j=1,2,3,4$ labels the sites (with periodic  identification $j =j+4$) and $\sigma=A,B$ labels the two chains.

The eight dangling Majorana modes furnish a projective representation of \(G_f\),  as diagnosed by relation \begin{equation}
   (MT)^2=-1.
\end{equation} Here it is important to emphasize that, unlike the symmetry action previously defined at the level of operators, the relation is obtained by evaluating the symmetry action on the state space of the zero modes. Equivalently, although \(M\) and \(T\) act linearly on Majorana operators, their induced action on the many-body states satisfies the group multiplication law only up to a phase, leading to a projective factor.
Since \(\omega_2(MT,MT)\) is a gauge-invariant 2-cocycle, this implies
\begin{equation}
 \omega_2(MT,MT)=-1, 
\end{equation}
confirming a nontrivial projective representation. Such a representation is necessarily multidimensional, so any symmetric local Hamiltonian must retain a ground-state degeneracy at \(\mu\). Therefore, this decoration is obstructed.

\begin{figure}[bt]
    \centering
\includegraphics[width=0.3\textwidth]{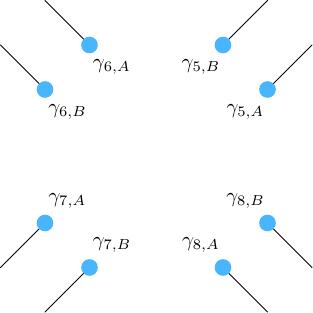}%width for pictures about dangling modes 
    \caption{ 1D FSPT decoration on $1$D blocks labeled by $\tau_2$, which leaves $8$ dangling Majorana modes $\gamma_{5,A},\gamma_{5,B},\dots,\gamma_{8,A},\gamma_{8,B}$  at  $0$D block $\mu$.}
    \label{fig:spinless D4 2chain tau2}
\end{figure}

The above argument is also applicable to the $n_1^T$ FSPT decorations on block $\tau_2$ (see Fig.~\ref{fig:spinless D4 2chain tau2}). There also exist eight dangling Majorana zero modes at the $0$D block $\mu$, forming a new Hilbert space. The symmetry properties are consistent with those presented in Eq.~\eqref{eq:D4 n1T}, with the exception of the reflection operation, which takes the modified form:\begin{equation}M:\gamma_{5,\sigma} \leftrightarrow \gamma_{6,\sigma},~\gamma_{7,\sigma} \leftrightarrow \gamma_{8,\sigma}.\end{equation}
Maintaining the previously established reflection axis and rotation direction, we observe that within this Hilbert space, the symmetry operators generate an operator algebra characterized by the relation $(MC_4T)^2 = -1$. 
Since the corresponding cohomological invariant has a non-trivial value, that is, $\omega_2(MC_4T, MC_4T) = -1$,
the 0D block symmetry group possesses a nontrivial projective representation. Moreover, the symmetry operators also satisfy the relations $(MT)^2 = 1$. 
This indicates that the representation in this case belongs to a different cohomology class from the representation discussed in the last paragraph.
Since the projective representations of group $G_f = D_4 \times \mathbb{Z}_2^T \times \mathbb{Z}_2^f$ are classified by $ \mathcal{H}^2(G_f, U_T(1)) = \mathbb{Z}_2^7$, the direct sum of two non-equivalent non-trivial representations cannot be reduced to a trivial representation (i.e. a linear representation).
Therefore, the simultaneous two-chain decorations of the $n_1^T$ phase on the $1$D blocks $\tau_1$ and $\tau_2$ are also obstructed.

With the case of $n_1^T$ decoration understood, we next consider $n_1^M$ phase decoration on $\tau_1$. The $0$D block symmetry group $G_f$ acts on Majorana operators as follows:
\begin{align}
    n_1^M:&\left\{\begin{aligned}
    T: & ~i \mapsto -i,\\
    &~\gamma_{j,A} \mapsto \gamma_{j,A}, \\
    &~\gamma_{j,B} \mapsto -\gamma_{j,B}, \\
M: & ~\gamma_{j,A} \mapsto \gamma_{6-j,A},\\
& ~\gamma_{j,B} \mapsto -\gamma_{6-j,B},\\
C_4:& ~\gamma_{j,\sigma} \mapsto \gamma_{j+1,\sigma},
\end{aligned}\right.\label{eq:D4 n1M}
\end{align} 
This symmetry definition uses the same reflection axis and rotation direction as discussed in the $n_1^T$ phase cases above. Similarly, in the Hilbert space formed by eight dangling modes on $\mu$, the $G_f$ symmetry operators satisfy $(MT)^2 = -1,(MC_4T)^2 = 1$. 
In addition, the $n_1^M$ decoration on $\tau_2$ gives a representation on $\mu$ with $(MT)^2 = 1,(MC_4T)^2 = -1$.  The symmetry properties are also consistent with those presented in Eq.~\eqref{eq:D4 n1M}, with the exception of the reflection operation, which takes the modified form:
\begin{align}
    M:\left\{\begin{aligned}
        &\gamma_{5,A} \leftrightarrow \gamma_{6,A},~\gamma_{7,A} \leftrightarrow \gamma_{8,A}\\
    &\gamma_{5,B} \leftrightarrow -\gamma_{6,B},~\gamma_{7,B} \leftrightarrow -\gamma_{8,B}
    \end{aligned} \right .
\end{align}
Therefore, by the same reason as for the $n_1^T$ decoration, any two-chain decorations of $n_1^M$ phase— whether on block $\tau_1$ (or $\tau_2$) alone or simultaneously on both $\tau_1$ and $\tau_2$- is obstructed.

As established earlier, $ \mathcal{H}^2(G_f, U_T(1)) = \mathbb{Z}_2^7$. This implies that applying the above 1D decorations twice automatically removes the obstruction and enforces the no-open-edge condition.  More precisely, the decorations on block $\tau_1$ (or $\tau_2$) corresponding to the following BSPT phases are obstruction-free:
\begin{equation}
(-1)^{n_1^T \cup n_1^T} \quad \text{or} \quad (-1)^{n_1^M \cup n_1^M}.
\end{equation}
Following Ref.~\cite{Song_2020}, verification of the no-open-edge condition for such BSPT decorations requires only the computation of anomalies on $0$D blocks through direct summation of block-states on adjacent $1$D blocks. Let $[\alpha],[\beta] \in  \mathcal{H}^2(\mathbb{Z}_2^M\times \mathbb{Z}_2^T \times \mathbb{Z}_2^f,U_T(1))=\mathbb{Z}_2^4$ denote the wavefunctions of these two BSPT phases. And note that rotation (or reflection) operations do not act on the local degrees of freedom of each wavefunction. Following Eq.~(6) in Ref.~\cite{Song_2020}, we have \begin{equation}[\alpha]+C_4[\alpha]+C_4^2[\alpha]+C_4^3[\alpha]=4[\alpha]=0,\end{equation} and an analogous relation holding for $[\beta]$. 
Consequently, decorating $\tau_1$ (or $\tau_2$) with the \( (-1)^{n_1^T \cup n_1^T} \) or \( (-1)^{n_1^M \cup n_1^M} \) BSPT phase satisfies the no-open-edge condition.

Although the $n_1^T$ and $n_1^M$ phases originate from distinct decorations, their corresponding projective representations belong to identical cohomology classes.
Hence, the decorations of phase $n_1^T + n_1^M$  on $\tau_1$ (or $\tau_2$)  are obstruction-free due to the structure of $ \mathcal{H}^2(G_f, U_T(1))$. We will explicitly present the interaction terms that open the corresponding gapless modes in the following.

As introduced earlier, this decoration  can likewise be realized through the two Majorana chains construction (with symmetry properties summarized in Eq.~\eqref{eq:symm property of n1Mn1T}). It produces eight dangling Majorana zero modes localized at the $0$D block $\mu$ (Fig.~\ref{fig:spinless D4 2chain tau1}). These modes transform under the $0$D block symmetry group $G_f$ as:
\begin{equation}
    \begin{aligned}
    T: & ~ \left\{ \begin{aligned}
    &  i \mapsto -i, \\
       &  \gamma_{j, \sigma} \mapsto \gamma_{j, \sigma},\end{aligned}\right.\\
    M: & ~ \left\{ \begin{aligned}
&\gamma_{j,A} \mapsto \gamma_{6-j,A},\\
& \gamma_{j,B} \mapsto -\gamma_{6-j,B},\\
    \end{aligned}\right.\\
    C_4: & \quad \gamma_{j, \sigma} \mapsto \gamma_{j+1, \sigma},
\end{aligned}
\end{equation}
where \( \sigma = A, B \) label the two chains. The subscript $j$  is understood to be modulo $4$, indicating the position of each mode.

It is convenient to introduce complex fermions \begin{equation}c_j=\frac{1}{2}(\gamma_{j,A}+i\gamma_{j,B}),~j=1,2,3,4.\end{equation} Their symmetry transformations are then
\begin{equation}
    \begin{aligned}
   T:& (c_1,c_2,c_3,c_4) \mapsto (c^\dagger_1,c^\dagger_2,c^\dagger_3,c^\dagger_4)\\
   M:&(c_1,c_2,c_3,c_4) \mapsto (c^\dagger_1,c^\dagger_4,c^\dagger_3,c^\dagger_2)\\
   C_4:&(c_1,c_2,c_3,c_4) \mapsto (c_2,c_3,c_4,c_1)
\end{aligned}\label{eq:spinless D4 fermion sym n1mn1t}
\end{equation}

To gap out the gapless modes at the $0$D block \( \mu \), we first consider the Hubbard interaction (\( U > 0 \)):
\begin{equation}
H_U = U \sum_{j=1,2}  \left(c_j^\dagger c_j - \frac{1}{2}\right) \left(c_{j+2}^\dagger c_{j+2} - \frac{1}{2} \right).
\end{equation}
The complex fermion occupation numbers $(n_1, n_3)$, $(n_2, n_4)$ result in a four-fold ground-state degeneracy, which can be interpreted as two spin-\( 1/2 \) degrees of freedom:
\begin{equation}
    \tau_{ab}^\mu = 
\begin{pmatrix}
    c_a^\dagger & c_b^\dagger
\end{pmatrix}
\sigma^\mu
\begin{pmatrix}
    c_a \\ c_b
\end{pmatrix}, \label{eq:definition of spin12 degree}
\end{equation}
 where $\sigma^\mu$ $(\mu=x,y,z)$ are Pauli matrices. We can derive the symmetry properties of these spins from Eq.~\eqref{eq:spinless D4 fermion sym n1mn1t}:
\begin{equation}
\begin{aligned}
    T:& \begin{aligned}
       & (\tau_{13}^x,\tau_{13}^y,\tau_{13}^z) \mapsto (-\tau_{13}^x,-\tau_{13}^y,-\tau_{13}^z)\\
       &(\tau_{24}^x,\tau_{24}^y,\tau_{24}^z) \mapsto (-\tau_{24}^x,-\tau_{24}^y,-\tau_{24}^z)\\
    \end{aligned} \\
    M:& \begin{aligned}
       & (\tau_{13}^x,\tau_{13}^y,\tau_{13}^z) \mapsto (-\tau_{13}^x,\tau_{13}^y,-\tau_{13}^z)\\
       &(\tau_{24}^x,\tau_{24}^y,\tau_{24}^z) \mapsto (-\tau_{24}^x,-\tau_{24}^y,\tau_{24}^z)\\
        \end{aligned} \\
    C_4:& \begin{aligned}
       & (\tau_{13}^x,\tau_{13}^y,\tau_{13}^z) \mapsto (\tau_{24}^x,\tau_{24}^y,\tau_{24}^z)\\
       &(\tau_{24}^x,\tau_{24}^y,\tau_{24}^z) \mapsto (\tau_{13}^x,-\tau_{13}^y,-\tau_{13}^z)\\
        \end{aligned} \\        
\end{aligned}
\end{equation}
It can be verified that the following symmetric interaction term lifts the ground-state degeneracy at \( \mu \) ($J>0$):
\begin{equation}
H_{J}=J \bigl( \tau_{1 3}^{x} \tau_{2 4}^{x}+\tau_{1 3}^{y} \tau_{2 4}^{z}-\tau_{1 3}^{z} \tau_{2 4}^{y} \bigr).
\end{equation}
The symmetry property of $H_J$ can also be readily identified by expressing it in terms of Majorana operators as follows:
\begin{equation}
\begin{aligned} {{{H_{J}=}}} & {{} {{{}-\frac{J} {4} \left( \gamma_{1,A} \gamma_{3,B}-\gamma_{1,B} \gamma_{3,A} \right) \left( \gamma_{2,A} \gamma_{4,B}-\gamma_{2,B} \gamma_{4,A} \right)}}} \\ {{{}}} & {{} {{} {{}+\frac{J} {4} \left( \gamma_{1,A} \gamma_{3,A}+\gamma_{1,B} \gamma_{3,B} \right) \left( \gamma_{2,A} \gamma_{2,B}-\gamma_{4,A} \gamma_{4,B} \right)}}} \\ {{{}}} & {{} {{} {{}-\frac{J} {4} \left( \gamma_{1,A} \gamma_{1,B}-\gamma_{3,A} \gamma_{3,B} \right) \left( \gamma_{2,A} \gamma_{4,A}+\gamma_{2,B} \gamma_{4,B} \right)}}} \\ \end{aligned} 
\end{equation}

Therefore, the decoration of  \( n_1^T+ n_1^M \) FSPT phase on the $1$D block \( \tau_1 \) (or \( \tau_2 \)) satisfies the no-open-edge condition, leading to a classification \( \mathbb{Z}_4 \).
It is crucial to note that this obstruction-free decoration, or more precisely, the crystalline superconductor, realizes an intrinsic interacting FSPT phase that cannot be achieved in free-fermion systems. The intrinsic nature is evident, as any potential mass terms (of the form $i\gamma_{j,\sigma} \gamma_{k,\sigma^\prime}$ or their combinations) would necessarily break the time-reversal symmetry, preventing the system from opening a gap.

In summary, all obstruction-free $1$D block-states can be generated by stacking the following elementary decorations:
\begin{enumerate}
    \item Two independent \( n_1^T+ n_1^M \) FSPT phases: one on \( \tau_1 \) and one on \( \tau_2 \) blocks;
    \item Two independent \( (-1)^{n_1^T \cup n_1^T} \) BSPT phases: one on \( \tau_1 \) and one on \( \tau_2 \) blocks;
    \item  Two independent \( (-1)^{n_1^M \cup n_1^M} \) BSPT phases: one on \( \tau_1 \) and one on \( \tau_2 \) blocks.
\end{enumerate}
These states form the following group:
\begin{equation}
\{\mathrm{OFBS}\}^{1D} = \mathbb{Z}_4^2 \times \mathbb{Z}_2^2.
\end{equation}
The group structure here can be viewed as the quotient
\[
(\mathbb Z_4^2 \times \mathbb Z_2^4)/\mathbb Z_2^2.
\]
This is because, for each block type \(\tau_i\) (\(i=1,2\)), stacking two copies of the FSPT decoration on \(\tau_i\) yields precisely the BSPT decoration on \(\tau_i\). Consequently, among the decorations listed above, two are redundant: they constitute the \(\mathbb Z_2^2\) subgroup that is being quotiented out.

Next, we explore $2$D block-state decorations. All $2$D blocks have an on-site symmetry group \( G_f = \mathbb{Z}_2^f \times \mathbb{Z}_2^T \). It is well known that no non-trivial 2D FSPT phases can be protected by this symmetry~\cite{Wang2020}. Therefore, there are no  obstruction-free decorations for $2$D blocks, i.e., 
\begin{equation}
\{\mathrm{OFBS}\}^{2D} = \mathbb{Z}_1.
\end{equation}
 
\begin{figure}[tb]
    \centering
     \includegraphics[width=0.5\textwidth]{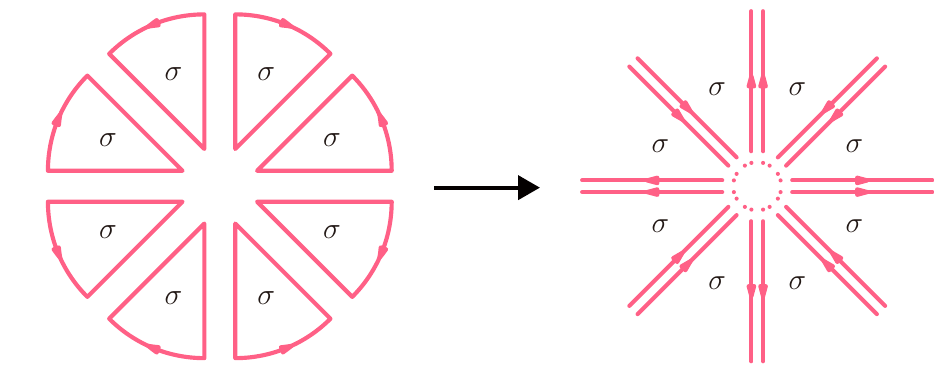}
    \caption{Deformation of the 2D ``Majorana-bubble'' construction. Left panel: Symmetric decoration of a ``Majorana bubble'' within each $2$D block. Each solid, oriented fan-shaped region represents a Majorana chain with anti-periodic boundary conditions (the arrows indicate the pairing direction; the anti-PBC is not shown explicitly). Right panel: Upon enlarging the ``Majorana bubbles'', the decorations can be smoothly deformed into effective $1$D block-states on $\tau_1$ and $\tau_2$.}    
    \label{fig:deformation of $2$D bubble decoration}
\end{figure}
With all obstruction-free block-states identified, we now discuss all possible trivializations. First, we consider the $2$D bubble equivalence by decorating a ``Majorana bubble" on each $2$D block \( \sigma \). Each bubble consists of a Majorana chain with anti-periodic boundary condition, and is therefore topologically trivial.
These 2D decorations can then be deformed into double Majorana chains living on the $1$D blocks \( \tau_1 \) and \( \tau_2 \) simultaneously (see Fig.~\ref{fig:deformation of $2$D bubble decoration}).

Since the two chains originate from adjacent bubbles on opposite sides of the reflection axis, the resulting 1D decoration inherits the same reflection action as the original bubble configuration:
\begin{equation}
M:\quad
\left\{
\begin{aligned}
\gamma^{L}_{j,A} &\mapsto \gamma^{L}_{j,B},\quad &\gamma^{L}_{j,B} &\mapsto \gamma^{L}_{j,A},\\
\gamma^{R}_{j,A} &\mapsto \gamma^{R}_{j,B},\quad &\gamma^{R}_{j,B} &\mapsto \gamma^{R}_{j,A}.
\end{aligned}
\right.
\end{equation}
The time-reversal action is defined by
\begin{equation}
T:\quad
\left\{
\begin{aligned}
&i \mapsto -i,\\
&\gamma^{L}_{j,\sigma} \mapsto \gamma^{L}_{j,\sigma},\quad
\gamma^{R}_{j,\sigma} \mapsto -\gamma^{R}_{j,\sigma},
\end{aligned}
\right.
\end{equation}
which is compatible with a single Majorana chain.

We now apply the same procedure used at the beginning of this section: stack two copies of the above two-chain configuration and determine the induced (projective) representation on the boundary Hilbert space. One finds
\begin{equation}
M^2=-1,\quad T^2=-1,\quad MT=TM.
\end{equation}
Hence, the symmetry-fractionalization pattern of the resulting 1D decorated state agrees with that in Eq.~\eqref{eq:n1mn1t relation}. It follows that the effective $1$D block-state realizes the \(n_1^{M}+n_1^{T}\) phase. Consequently, the trivialization group for $1$D block-states is
\begin{equation}
  \{\mathrm{TBS}\}^{1D}=\mathbb{Z}_4. 
\end{equation}

We now examine the effect of $2$D bubbles on $0$D block-states. Specifically, since (as we will show later) only the nontrivial fermion parity eigenstates of $0$D blocks remain robust against trivialization by $1$D bubbles, it suffices to ask whether $2$D bubbles can change the fermion parity carried by a $0$D block. It has been observed that the 2D Majorana bubbles leave double Majorana chains on all $1$D blocks. Consequently, this results in $16$ Majorana modes at the $0$D block \( \mu \), corresponding to the edge modes of the double Majorana chains on the $1$D blocks.

These Majorana modes cannot form Majorana chains with periodic boundary conditions (PBC) surrounding the $0$D blocks, which would have odd fermion parity. This is because Majorana chains are known to be incompatible with reflection symmetry~\cite{2Dcrystalline}, and any Majorana chain with PBC surrounding a $0$D block must necessarily cross at least one reflection axis. Consequently, the net effect of the 2D Majorana bubble is to trivialize the 1D FSPT decorations on all $1$D blocks, without altering the $0$D block-states.
Thus, $2$D bubbles contribute to the trivialization group \( \{\mathrm{TBS}\}^{1D} = \mathbb{Z}_4 \) in total, with its generator representing the \( n_1^M+ n_1^T \) FSPT state decorations on both $\tau_1$ and  $\tau_2$.

Next, we consider the $1$D bubble equivalences and decorate a pair of complex fermions on each $1$D block \( \tau_1 \) or \( \tau_2 \). Near each $0$D block, four complex fermions contributed by the adjacent $1$D bubbles form an atomic insulator,
\begin{equation}
\ket{TR} = c_1^\dagger c_2^\dagger c_3^\dagger c_4^\dagger \ket{0}.
\end{equation} Their symmetry transformations under \( G_f \) are:
\begin{equation}
\begin{aligned}
    {M}: & \quad c_j^\dagger \mapsto c_{6-j}^\dagger, \\
    {C_4}: & \quad c_j^\dagger \mapsto c_{j+1}^\dagger,
\end{aligned}
\end{equation}
where \( j \)  is understood modulo  $4$.

As established in the preceding analysis, the \( \mathbb{Z}_2^T \) symmetry group does not contribute to a non-trivial classification for 0D FSPT phases. We therefore restrict our attention to the \( M \) and \( C_4 \) symmetry transformations, which exhibit the following characteristic action on the state $\ket{TR}$:
\begin{equation}
  {C_4} \ket{TR} = -\ket{TR}, \quad {M} \ket{TR} = -\ket{TR}. \label{eq:D4 spinless $1$D bubble}  
\end{equation}

Alternatively, one may decorate both $1$D blocks \( \tau_1 \) and \( \tau_2 \) simultaneously. This configuration generates eight complex fermion modes localized near each $0$D block, forming an atomic insulator described by:
\begin{equation}
\ket{TR} = c_1^\dagger c_2^\dagger c_3^\dagger c_4^\dagger {c_1^\dagger}^\prime {c_2^\dagger}^\prime {c_3^\dagger}^\prime {c_4^\dagger}^\prime \ket{0},
\end{equation} where the prime notation distinguishes fermions from different $1$D blocks. The symmetry transformations now yield:
\begin{equation}
  {C_4} \ket{TR} = \ket{TR}, \quad {M} \ket{TR} = -\ket{TR}. \label{eq:D4 spinless $1$D bubbleilmu}  
\end{equation}
Eq.~\eqref{eq:D4 spinless $1$D bubbleilmu} and Eq.~\eqref{eq:D4 spinless $1$D bubble} indicate that $1$D bubbles can independently generate non-trivial eigenstates of \( M \) and \( C_4 \).
This leads to the $\mathbb{Z}_2^2$ structure of the trivialization group for $0$D block-states: \begin{equation}
     \{\mathrm{TBS}\}^{0D} = \mathbb{Z}_2^2.
\end{equation}

With all non-trivial block-states identified, we now systematically analyze the group structure of the ultimate classification. Crucially, this classification is not simply given by the direct product of non-trivial obstruction-free states $\{\mathrm{OFBS}\}^{nD}/{\{\mathrm{TBS}\}^{nD}}$ across different dimension $nD$, but may involve non-trivial group extensions. 
Physically, such extensions reflect whether stacking higher-dimensional block-states can be consistently identified with, or adiabatically deformed into, lower-dimensional ones. In the present spinless setting, however, there are no nontrivial $2$D block-states, so the only possible extension to consider is the one relating 0D and 1D data.

Recall that, after incorporating the bubble equivalences, the only remaining nontrivial $1$D block-states are the \((-1)^{n_1^T\cup n_1^T}\) and \((-1)^{n_1^M\cup n_1^M}\) phases realized on the \(\tau_1\) or \(\tau_2\) $1$D blocks, each carrying a \(\mathbb{Z}_2\) classification, together with the \(n_1^M+n_1^T\) phase on \(\tau_1\) or \(\tau_2\), which carries a \(\mathbb{Z}_4\) classification. All other sectors are trivialized. To investigate possible stacking effects, we first consider decorating four identical copies of 1D FSPT states with $n_1^T+n_1^M$ phase on $\tau_1$, which belong to trivial 1D FSPT phase. This configuration generates 32 dangling Majorana modes at $0$D block $\mu$.
Furthermore, since bubble equivalence enforces that only the non-trivial fermion parity eigenstate persists at $0$D blocks, the central question reduces to whether these trivial $1$D block-states can be adiabatically deformed into a non-trivial fermion parity eigenstate at the $0$D blocks.
These decorations cannot be deformed into a Majorana chain (with periodic boundary conditions) surrounding the $0$D block to alter the corresponding fermion parity, as Majorana chains are incompatible with reflection symmetry.
Subsequently, at each $0$D block \( \mu \), these 32 Majorana modes can be treated as 16 complex fermions, forming an atomic insulator state:
\begin{equation}
\ket{\phi} = c_1^\dagger c_2^\dagger \dots c_{16}^\dagger \ket{0},
\end{equation}
which indeed has even fermion parity. Since the only obstruction-free and non-trivial 0D state has odd parity, no further extensions are possible. Similarly, decorating two copies of $(-1)^{n_1^T\cup n_1^T}$ or $(-1)^{n_1^M\cup n_1^M}$ phases on the $1$D blocks also admits no non-trivial extension.

In summary, all independent non-trivial block-states with different dimensions are classified as follows:
\begin{equation}
\begin{aligned}
         & E^{\mathrm{2D}}=\{\mathrm{OFBS}\}^{\mathrm{2D}}=\mathbb{Z}_1                       \\
         & E^{\mathrm{1D}}=\{\mathrm{OFBS}\}^{\mathrm{1D}}/\{\mathrm{TBS}\}^{\mathrm{1D}}=\mathbb{Z}_4 \times \mathbb{Z}^2_2 \\
         & E^{\mathrm{0D}}=\{\mathrm{OFBS}\}^{\mathrm{0D}}/\{\mathrm{TBS}\}^{\mathrm{0D}}=\mathbb{Z}_2.
    \end{aligned}
\end{equation}
The ultimate classification is:
\begin{equation}
\mathcal{G}_0 = \mathbb{Z}_4 \times \mathbb{Z}_2^3.
\end{equation}

\subsubsection{ Spinful fermions}\label{sec:D4 spinful discussion}

For the spinful system, the $0$D block \( \mu \) possesses the total symmetry group \begin{equation}
    G_f = G_b\times_{\omega_2} \mathbb{Z}_2^f=(D_4 \times \mathbb{Z}_2^T) \times_{\omega_2} \mathbb{Z}_2^f,
\end{equation}where ``\( \times_{\omega_2} \)''  denotes a nontrivial extension of the physical symmetry group $G_b$ by fermion parity \( \mathbb{Z}_2^f \). The extension is specified by a 2-cocycle $\omega_2 \in \mathcal{H}^2(G_b, \mathbb{Z}_2)$,  defined previously in Eq.~\eqref{2-cocycle}. The corresponding  $0$D block FSPT phases are characterized by:
\begin{equation}
\begin{aligned}
    & n_0 \in \mathcal{H}^0(G_b, \mathbb{Z}_2^f) = \mathbb{Z}_2, \\
    & \nu_1 \in \mathcal{H}^1(G_b, U_T(1)) = \mathbb{Z}_2 \times \mathbb{Z}_2,
\end{aligned}
\end{equation}
subject to the consistency condition $d\nu_1=(-1)^{\omega_2\cup n_0}$~\cite{Wang2020}.
For $n_0 = 1$, the condition fails explicitly due to the nontrivial invariant:
\begin{equation}
(-1)^{\omega_2 \cup n_0}(T, T) = -1, \quad (T \in \mathbb{Z}_2^T),
\end{equation}
Therefore, $0$D block-states exhibit a $\mathbb{Z}_2 \times \mathbb{Z}_2$ classification. 
The two $\mathbb{Z}_2$ factors in the classification correspond to distinct 1D linear irreducible representations: the first $\mathbb{Z}_2$ reflects the reflection eigenvalues $\pm 1$ of $M$ , while the second $\mathbb{Z}_2$ represents the rotation eigenvalues $\pm 1$ of $C_4$.
The obstruction-free $0$D block-states form the following group:
\begin{equation}
    \{OFBS \}^{0D}=\mathbb{Z}_2^2.
\end{equation}

All one-dimensional blocks possess an identical on-site symmetry group \begin{equation}
    G_f = G_b \times_{\omega_2} \mathbb{Z}_2^f=(\mathbb{Z}_4^{Mf} \times \mathbb{Z}_4^{fT})/\mathbb{Z}_2^f.
\end{equation} The physical subgroup $G_b$ takes the form $G_b = \mathbb{Z}_2^M \times \mathbb{Z}_2^T$, where $M$ and $T$ denote reflection and time-reversal symmetries, respectively. The group extension is determined by
\begin{equation}\omega_2 = n_1^M \cup n_1^M + n_1^T \cup n_1^T,\end{equation}
where $n_1^M$ and $n_1^T$ represent non-trivial 1-cocycles in $\mathcal{H}^1(G_b,\mathbb{Z}_2^f)$. These cocycles are defined in accordance with Eq.~\eqref{eq:definition of n1M}. The classification data for one-dimensional block FSPT phases are characterized as follows:
\begin{align}
     & n_1 \in \mathcal{H}^1(G_b, \mathbb{Z}_2^f) = \mathbb{Z}_2 \times \mathbb{Z}_2 \notag \\
     & \nu_2 \in \mathcal{H}^2(G_b, U_T(1)) = \mathbb{Z}_2 \times \mathbb{Z}_2
\end{align} And they also should  obey the consistency equations listed in formula Eq.~\eqref{eq:equation of 1D FSPT}.
The non-trivial representative cocycles for $n_1$ are  $n_1^T$ and $n_1^M$. 
For \( \nu_2 \), the non-trivial representatives are:
\begin{equation}
(-1)^{n_1^T \cup n_1^M} \quad \text{and} \quad (-1)^{n_1^T \cup n_1^T},
\end{equation}
or equivalently:
\begin{equation}
(-1)^{n_1^M \cup n_1^M} \quad \text{and} \quad (-1)^{n_1^T \cup n_1^T}.
\end{equation}  
It can be verified that \( \mathrm{d} \nu_2(n_1^T) = (-1)^{\omega_2 \cup n_1^T} \) is a trivial 3-cocycle, whereas \( \mathrm{d} \nu_2(n_1^M) = (-1)^{\omega_2 \cup n_1^M} \) is non-trivial. Consequently, only \( n_1 = n_1^T \) is obstruction-free. Furthermore,  the trivialization group is \begin{equation} \Gamma = (-1)^{\omega_2} = (-1)^{n_1^T \cup n_1^T + n_1^M \cup n_1^T}, \end{equation}  implying that \( \nu_2 = (-1)^{n_1^T \cup n_1^T} \) and \( \nu_2 = (-1)^{n_1^M \cup n_1^T} \) are non-trivial and  equivalent.
 Moreover, it can be verified that \( (n_1, \nu_2) \) satisfies the twist stacking law~\cite{ren2023stacking}:
\begin{equation}
(n_1^T, 1) \boxplus (n_1^T, 1) = (0, (-1)^{n_1^T \cup n_1^T}) . \label{eq:D4 spinful stacking law}
\end{equation} 

Therefore, the one-dimensional FSPT systems are classified by $\mathbb{Z}_4$, where the generator (root state) corresponds to the non-trivial $n_1(T) = 1$ state. We denote this phase as the $n_1^T$ FSPT phase. The stacking of two $n_1^T$ FSPT phases produces a non-trivial BSPT phase $(-1)^{n_1^T \cup n_1^T}$ (or equivalently, phase $(-1)^{n_1^M \cup n_1^T}$).

\begin{figure}[tb]
    \centering
    \includegraphics[width=0.48\textwidth]{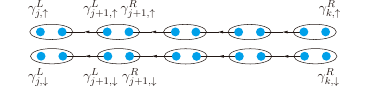}
    \caption{The 1D $n_1^T$ FSPT phase protected by $G_f=(\mathbb{Z}_4^{Mf} \times \mathbb{Z}_4^{fT})/\mathbb{Z}_2^f$, realized as two Majorana chains with opposite spin species. The illustration follows Fig.\ref{four chain definition}. } 
    \label{figure:twochain model}
\end{figure}

A physical realization of these phases has been identified in free fermionic systems, where the root state manifests as the ground state of two Majorana chains with opposite spin species. 
The Hamiltonian for this root state can be expressed as ($j \in \mathbb{Z}$ and $\sigma=\uparrow,\downarrow$):
\begin{equation}
    H=it\sum \gamma^R_{j,\sigma}\gamma^L_{j+1,\sigma},\label{form:twochain}
\end{equation}
where $j$ denotes the sites and $\uparrow,\downarrow$ label two chains, as depicted in Fig.~\ref{figure:twochain model}.

The symmetry transformations on the Majorana fermions \( \gamma_{i, \sigma} \) that leave the Hamiltonian in Eq.~\eqref{form:twochain} invariant are specified as follows:
\begin{equation}
    \begin{aligned}
    {T}: & \quad \left\{
            \begin{aligned}
            & i \mapsto -i, \\
            & \gamma^L_{j, \uparrow} \mapsto -\gamma^L_{j, \downarrow}, \quad \gamma^L_{j, \downarrow} \mapsto \gamma^L_{j, \uparrow} \\
            & \gamma^R_{j, \uparrow} \mapsto \gamma^R_{j, \downarrow}, \quad \gamma^R_{j, \downarrow} \mapsto -\gamma^R_{j, \uparrow}
            \end{aligned}
            \right ., \\
   {M}: & \quad \left\{
            \begin{aligned}
            & \gamma^L_{j, \uparrow} \mapsto -\gamma^L_{j, \downarrow}, \quad \gamma^L_{j, \downarrow} \mapsto \gamma^L_{j, \uparrow} \\
            & \gamma^R_{j, \uparrow} \mapsto -\gamma^R_{j, \downarrow}, \quad \gamma^R_{j, \downarrow} \mapsto \gamma^R_{j, \uparrow}
            \end{aligned}
            \right.,
    \end{aligned} 
    \label{eq:definition root phase}
\end{equation}
These transformations respect the multiplication relations within the symmetry group $G_f$. For instance, consider the action of $T^2$ on $\gamma^L_{j,\uparrow}$:
\begin{equation}
T^2\gamma^L_{j,\uparrow}T^{-2}=-\gamma^L_{j,\uparrow}=P_f \gamma^L_{j,\uparrow} P_f^{-1}
\end{equation}
This identity reveals that $T^2$ and $P_f$ act identically on the Majorana operators, establishing the group relation $T^2=P_f$.

Next, we examine whether this model realizes the \( n_1^T \) root phase.
As in the spinless case, we study the group representation of $G_f$ on the boundary Hilbert space for the doubled $n_1^T$ phase. Stacking two copies of the $n_1^T$ phase yields a BSPT phase,  which can be implemented using four Majorana chains. 

Introducing the complex fermion operators:
\begin{equation}
c_j = \frac{1}{2}(\gamma^L_{j, \uparrow} + i\gamma^L_{j, \downarrow}), \quad j = 1,2,
\end{equation}
we construct the four-dimensional Hilbert space on the boundary spanned by the states $\ket{0}$, $c_1^\dagger\ket{0}$, $c_2^\dagger\ket{0}$, and $c_1^\dagger c_2^\dagger\ket{0}$. The projective representation of $G_f$ on this space satisfies the algebraic relations:
\begin{equation}
T^2 = P_f, \quad (MT)^2 = -I,
\end{equation}
with the corresponding factor system:
\begin{equation}
\omega_2(MT, MT) = -1, \quad \omega_2(T, T) = 1.
\end{equation}

These results demonstrate that the four-Majorana-chain system realizes the $ (-1)^{n_1^M \cup n_1^T}$  BSPT phase. Through the stacking law~Eq.\ref{eq:D4 spinful stacking law}, we conclusively identify our model as realizing the $n_1^T$ root state.

For $1$D blocks, a single Majorana chain cannot be decorated due to the lack of on-site symmetry actions satisfying \( T^2 = -1 \). This implies that the obstruction function of the 1D invertible phase, \( O_1(n_0) = \omega_2 n_0 \), is non-trivial~\cite{wang2021domain}.

Next, we explore $2$D block-state decorations.  
All $2$D blocks share the same total symmetry group \begin{equation}
  G_f = \mathbb{Z}_4^{fT}.  
\end{equation}  It is well known that,  with symmetry group \( G_f  \), there exists a non-trivial 2D class-DIII topological superconductor (TSC) with a \( \mathbb{Z}_2 \) topological classification~\cite{grouptopo}. A 2D DIII TSC can be viewed  as a stacking of the  $p_x+ip_y$ and $p_x-ip_y$ states~\cite{xiaoliang}. The only reliable mass term, \( im\chi^\uparrow \chi^\downarrow \), breaks time-reversal symmetry (TRS), indicating that TRS protects the gapless helical edge. The non-trivial edge states in this system are represented by two counter-propagating chiral Majorana edge modes, $\chi^\uparrow$and $\chi^\downarrow$.

Next, we discuss the types of decorations on the blocks that satisfy the no-open-edge condition.
All zero-dimensional
decorations automatically meet this requirement, as they lack gapless edges and instead exhibit gapped bulk properties. So the obstruction-free $0$D block-states form the group:
\begin{equation}
    \{\mathrm{OFBS}\}^{0D} = \mathbb{Z}_2^2.
\end{equation}

\begin{figure}[tb]
    \centering
   \includegraphics[width=0.3\textwidth]{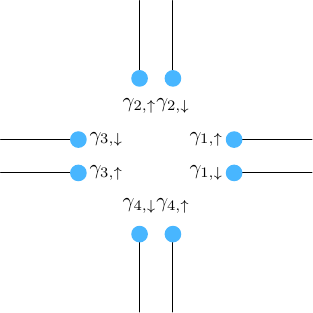}
    \caption{Dangling Majorana modes $\gamma_{i,\uparrow}$ and $\gamma_{i,\downarrow}$ ($i=1,2,3,4$) arising from the 1D FSPT decoration of the $\tau_1$ (or $\tau_2$) blocks on $\mu$.}
    \label{figure:tau3mu3}
\end{figure}

We now consider the 1D $n_1^T$ FSPT decorations (realized by two Majorana chains) on the $1$D blocks \( \tau_1 \) or \( \tau_2 \), which yield eight dangling Majorana fermions.
As shown in Fig.~\ref{figure:tau3mu3}, the $0$D block $\mu$ hosts eight dangling Majorana zero modes that transform under the on-site symmetry group $G_f$ according to the following representation:
\begin{equation}
    \begin{aligned}
    {T}:~ & 
    \begin{cases}
        i \mapsto -i, \\
        (\gamma_{j,\uparrow}, \gamma_{j,\downarrow}) \mapsto (-\gamma_{j,\downarrow}, \gamma_{j,\uparrow}), & j=1,2,3,4
    \end{cases} \\
    {C_4}:~ & 
    \begin{cases}
        (\gamma_{j,\uparrow}, \gamma_{j,\downarrow}) \mapsto (\gamma_{j+1,\uparrow}, \gamma_{j+1,\downarrow}), & j=1,2,3 \\
        (\gamma_{4,\uparrow}, \gamma_{4,\downarrow}) \mapsto (-\gamma_{1,\uparrow}, -\gamma_{1,\downarrow})
    \end{cases}  \\
    {M}:~ & 
    \begin{cases}
        (\gamma_{1,\uparrow}, \gamma_{1,\downarrow}) \mapsto (-\gamma_{1,\downarrow}, \gamma_{1,\uparrow}) \\
        (\gamma_{2,\uparrow}, \gamma_{2,\downarrow}) \mapsto (\gamma_{4,\downarrow}, -\gamma_{4,\uparrow}) \\
        (\gamma_{3,\uparrow}, \gamma_{3,\downarrow}) \mapsto (\gamma_{3,\downarrow}, -\gamma_{3,\uparrow}) \\
        (\gamma_{4,\uparrow}, \gamma_{4,\downarrow}) \mapsto (\gamma_{2,\downarrow}, -\gamma_{2,\uparrow})
    \end{cases}
\end{aligned}    \label{eq:22}
\end{equation}
The integer subscript $j$ labels the position of each Majorana mode, with all indices taken modulo 4 to reflect the periodic boundary conditions.
Note that the reflection operator \( M^\prime = M {C_4}^2 \), whose reflection axis is horizontal, acts on the edge \( (2, 4) \) in the same way as \( M \) acts on the edge \( (1, 3) \).

We now introduce the following mass terms that couple the Majorana modes:
\begin{equation}
H = i \sum_{j=1}^{2} \left( \gamma_{j, \uparrow} \gamma_{j+2, \uparrow} - \gamma_{j, \downarrow} \gamma_{j+2, \downarrow} \right) \label{eq:D4 spinful mass term 1D}
\end{equation}
Through direct computation, we verify that this Hamiltonian is invariant under all symmetry transformations specified in Eq.~\eqref{eq:22}. Moreover, the Hamiltonian (Eq.~\eqref{eq:D4 spinful mass term 1D}) possesses a unique fully gapped ground state, as required for a well-defined topological phase.

Consequently, the obstruction-free 1D decorations contribute to the classification group structure:
\begin{equation}
    \{\mathrm{OFBS}\}^{\mathrm{1D}} = \mathbb{Z}_4 \times \mathbb{Z}_4,
\end{equation}
where each $\mathbb{Z}_4$ factor counts the stacking number (mod $4$) of the $n_1^T$ FSPT decoration on the $\tau_1$ or $\tau_2$ blocks, respectively.

\begin{figure}[tb]
    \centering
    \includegraphics[width=0.2\textwidth]{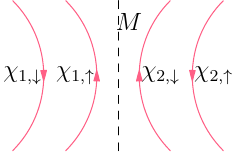}
        \caption{Chiral edge Majorana modes of two DIII topological superconductors related by reflection operator (whose reflection axis is represented by  dashed line ). Arrows of curves refer to chirality of Majorana modes.   } \label{one_DIII_layer_1}
\end{figure}

Next, we consider the block-state decoration on the two-dimensional block $\sigma$.
When the 2D DIII TSC is decorated on each $2$D block \( \sigma \), there are four chiral Majorana modes on each reflection axis, as shown in Fig.~\ref{one_DIII_layer_1}. These Majorana modes satisfy the following symmetry transformations ($j=1, 2$):
\begin{align}
        {T}:~   & i \mapsto -i , \quad \chi_{j,\uparrow} \mapsto -\chi_{j,\downarrow}, \quad \chi_{j,\downarrow} \mapsto \chi_{j,\uparrow},                 \notag \\
        {M}:~ &  \chi_{j,\uparrow} \mapsto -\chi_{3-j,\downarrow}, \quad \chi_{j,\downarrow} \mapsto \chi_{3-j,\uparrow} 
\end{align} 
Mass terms can be added along each reflection axis as follows:
\begin{align}
        H_{\tau}&\equiv H_0+H_m \notag\\
        &=i\int dx \, \sum_{j=1,2} (-1)^{j-1} (\chi_{j,\uparrow}  \partial \chi_{j,\uparrow}-\chi_{j,\downarrow } \partial \chi_{j,\downarrow}) \notag\\ 
        &+ i\int dx \, m(\chi_{1,\uparrow} \chi_{2,\uparrow}-\chi_{1,\downarrow} \chi_{2,\downarrow})   
\end{align} to gap out  dangling Majorana modes. Mass terms of this form can be consistently defined on all $1$D blocks while preserving the full 2D symmetry group $G_f$.
One readily checks that the two collinear $1$D blocks related by \(C_2\) admit symmetry-compatible mass terms with no relative sign. Hence the effective 1D theory has no nontrivial mass domain wall, implying that the $0$D block carries no protected zero mode and the no-open-edge condition is satisfied.
Therefore, the 2D DIII TSC decoration is obstruction-free and forms the group $\{ OFBS\}^{2D}=\mathbb{Z}_2$.

\begin{figure}[tb]
    \centering
 \includegraphics[width=0.5\textwidth]{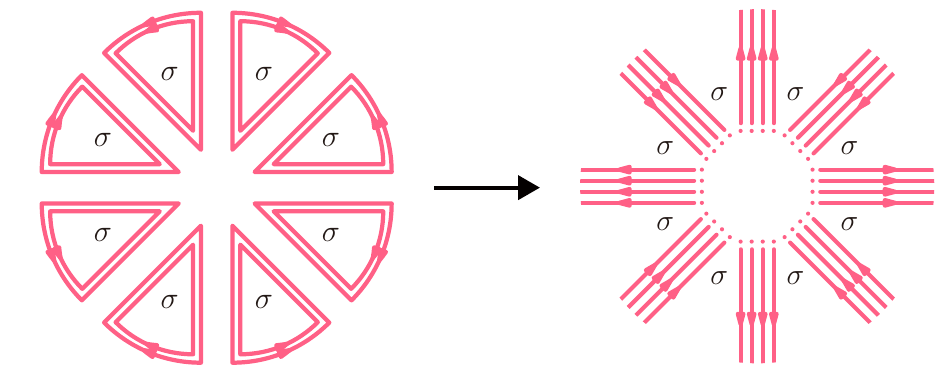}

    \caption{Deformation of the 2D ``Majorana bubble'' construction. Left panel: Decorate two ``Majorana bubble'' in each $2$D block in a symmetric way. Each solid oriented fan shape expresses a Majorana chain with anti-PBC (the arrows indicate the pairing direction; the anti-PBC is not shown explicitly). Right panel:   Upon enlarging the ``Majorana bubbles'', the decorations can be smoothly deformed into effective $1$D block-states on $\tau_1$ and $\tau_2$.}
    \label{figure:2Dbubble}
\end{figure}
With all obstruction-free block-states identified, we now discuss all possible trivializations. First, we consider the $2$D bubble equivalence.
Each $2$D bubble consists of two Majorana chains, as the bubbles must preserve the on-site symmetry \( T^2 = -1 \). Furthermore, these bubbles must be decorated symmetrically. 
The $2$D bubbles are illustrated in Fig.~\ref{figure:2Dbubble}. By enlarging the Majorana bubbles, they can be adiabatically deformed into $1$D block-states \( \ket{\mathbf{X}} \) on both \( \tau_1 \) and \( \tau_2 \), with an effective on-site symmetry derived from non-local space-time symmetry.

\begin{figure}[tb]
    \centering
    \includegraphics[height=0.15\textwidth]{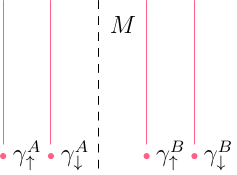}
   
    \caption{The $1$D block-state $\ket{\mathbf{X}}$ from shrink of $2$D bubbles (each bubble consists of two Majorana chains). Dashed line imply reflection axis.  }
    \label{fig:bubble_on-site_symmetry}
\end{figure}

However, the $2$D bubbles do not contribute to non-trivial 1D FSPT phases on the $1$D blocks. The symmetry transformations of the Majorana fermions in the block-state \( \ket{\mathbf{X}} \) are chosen as follows (see Fig.~\ref{fig:bubble_on-site_symmetry}):
\begin{equation}
    \begin{aligned}
        {T}:~ & i \mapsto -i                                                                                                  \\
                 & \gamma_{\uparrow}^{A} \mapsto -\gamma_{\downarrow}^{A},\quad\gamma_{\downarrow}^{A} \mapsto \gamma_{\uparrow}^{A}  \\
                 & \gamma_{\uparrow}^{B} \mapsto -\gamma_{\downarrow}^{B},\quad\gamma_{\downarrow}^{B} \mapsto \gamma_{\uparrow}^{B}  \\
        {M}:~ & \gamma_{\uparrow}^{A} \mapsto -\gamma_{\downarrow}^{B},\quad\gamma_{\downarrow}^{A} \mapsto \gamma_{\uparrow}^{B}  \\
                 & \gamma_{\uparrow}^{B} \mapsto -\gamma_{\downarrow}^{A},\quad\gamma_{\downarrow}^{B} \mapsto \gamma_{\uparrow}^{A},
    \end{aligned}
\end{equation}
Note that the symmetry transformation properties of \( M \) here differ from those of the 1D FSPT phase described in Eq.~\eqref{eq:definition root phase}. This distinction arises because operation \( M \) in this context is derived from the non-local reflection transformation associated with the bubble.
For the four Majorana edge modes associated with the $1$D block-state \( \ket{\mathbf{X}} \), an interaction term can be introduced as:
\begin{equation}
\gamma^A_{\uparrow} \gamma^A_{\downarrow} \gamma^B_{\uparrow} \gamma^B_{\downarrow} + i \gamma^A_{\uparrow} \gamma^B_{\uparrow} - i \gamma^A_{\downarrow} \gamma^B_{\downarrow},
\end{equation}
which adheres to the aforementioned symmetries. This interaction results in a unique ground state. Consequently, the on-site symmetry derived from space-time symmetry cannot protect a non-trivial 1D FSPT phase over the four Majorana chains. Therefore, there are no trivializations for 1D obstruction-free states, and the trivialization group is trivial: \begin{equation}
    \{\mathrm{TBS}\}^{1D} = \mathbb{Z}_1 .
\end{equation}  

Then we examine whether $2$D bubbles alter the $0$D block-state. For $2$D bubbles, although they can be deformed into odd-parity Majorana chains around
$\mu$~\cite{2Dcrystalline}, such configurations are explicitly prohibited in our framework since the 0D obstruction-free classification only admits even-parity 0D states.

Next, we consider $1$D bubbles by decorating a pair of complex fermions on each $1$D block. Since \( \mathbb{Z}_2^T \) does not contribute to non-trivial classification, we only consider the \( M \) and \( C_4 \) symmetry transformations.
Near each $0$D block $\mu$, there are four complex fermions contributed from adjacent bubbles, which form the following atomic insulator: $\ket{TR}=c_1^\dagger c_2^\dagger c_3^\dagger c_4^\dagger \ket{0}$. The symmetry transformations of these complex fermions are: \begin{equation}\begin{aligned}
    C_4:&~c_1^\dagger \mapsto c_2^\dagger \mapsto c_3^\dagger \mapsto c_4^\dagger \mapsto  -c_1^\dagger,\\
    M:&~c_1^\dagger \mapsto ic_1^\dagger,c_3^\dagger \mapsto -ic_3^\dagger,c_2^\dagger \mapsto -i c_4^\dagger,c_4^\dagger \mapsto -i c^\dagger_2.
\end{aligned} \end{equation}
The corresponding symmetry properties are:
\begin{equation}
\begin{aligned}
    {C_4} \ket{TR}=- c_2^\dagger c_3^\dagger c_4^\dagger c_1^\dagger \ket{0}=\ket{TR} \\
    {M}\ket{TR}=-c_1^\dagger c_4^\dagger c_3^\dagger c_2^\dagger \ket{0}=\ket{TR}
\end{aligned}
\end{equation}
Thus, at $0$D blocks \( \mu \), the $1$D bubble construction on  \( \tau_1 \) or \( \tau_2 \)  does not alter the eigenvalues of \( M \) and \( C_4 \).
We can also decorate a pair of complex fermions on each $1$D block $\tau_1$ and $\tau_2$ simultaneously. Near each $0$D block, there are eight complex fermions contributed from adjacent bubbles forming the following atomic insulator: \begin{equation}\ket{TR}=c_1^\dagger c_2^\dagger c_3^\dagger c_4^\dagger {c_1^\dagger}^\prime {c_2^\dagger}^\prime {c_3^\dagger}^\prime {c_4^\dagger}^\prime\ket{0}.\end{equation} 
The corresponding symmetry properties are:
\begin{equation}
    {C_4} \ket{TR}=\ket{TR}, \quad{M}\ket{TR}= \ket{TR}. 
\end{equation}
These results indicate that the $1$D bubbles do not contribute to non-trivial eigenstates of \( M \) and \( C_4 \).
Thus, there are no trivializations for the 0D FSPT states, and they form the group \begin{equation}
    \{\mathrm{TBS}\}^{0D} = \mathbb{Z}_1.
\end{equation}

With all non-trivial block-states identified, we now examine the group structure of the ultimate classification.
In Sec.~\ref{sec:classification strategy}, we proposed that the final classification group structure \( \mathcal{G}_0 \) of the two-dimensional FSPT phases is a central extension involving \( E^{0\mathrm{D}}, E^{1\mathrm{D}}, \) and \( E^{2\mathrm{D}} \):
\begin{equation}
\begin{gathered}
1 \rightarrow E^{0\mathrm{D}} \rightarrow E^{\leq 1\mathrm{D}} \rightarrow E^{1\mathrm{D}} \rightarrow 1, \\
1 \rightarrow E^{\leq 1\mathrm{D}} \rightarrow \mathcal{G}_0 \rightarrow E^{2\mathrm{D}} \rightarrow 1.
\end{gathered}
\end{equation}
Physically, this extension implies that  stacking of two \( n \)-dimensional block-states may results in a \( (n-1) \)-dimensional block-state.

We find that the 1D obstruction-free and trivialization-free block-state cannot extend to 0D obstruction-free and trivialization-free block-state in our study. For a $2n$-Majorana chains decoration on $1$D blocks $\tau_1$ (or $\tau_2$),
the 0D state at $0$D block $\mu$, extended by them, can be written as follows:
\begin{equation}
\ket{\Theta}^{0D}
=\Big(\prod_{j\in J} a_j^\dagger\Big)
\Big(\prod_{k\in K} a_k^\dagger\Big)
\Big(\prod_{l\in L} a_l^\dagger\Big)
\Big(\prod_{m\in M} a_m^\dagger\Big),
\end{equation}
where the fermion creation operators $a^\dagger$ are not necessarily the original microscopic fermions $c^\dagger$. The index sets $J,K,L,M\subset\{1,\dots,2n\}$ label four groups of modes that are permuted into one another by the $C_4$ rotation; for each orbit we order the labels as $(j,k,l,m)$ according to the action of $C_4$.
The symmetry actions on these operators are
\begin{equation}
\begin{aligned}
C_4:&\quad a_j^\dagger \mapsto a_k^\dagger \mapsto a_l^\dagger \mapsto a_m^\dagger \mapsto -a_j^\dagger,\\
M:&\quad a_j^\dagger \mapsto ia_j^\dagger,\qquad a_l^\dagger \mapsto -ia_l^\dagger,\\
&\quad a_k^\dagger \mapsto ia_m^\dagger,\qquad a_m^\dagger \mapsto ia_k^\dagger .
\end{aligned}
\end{equation}
The corresponding symmetry properties are:
\begin{equation}
    C_4 \ket{\Theta}^{0D}=\ket{\Theta}^{0D}, M\ket{\Theta}^{0D} = \ket{\Theta}^{0D},
\end{equation} for any integer $n$, which implies that $\ket{\Theta}^{0D}$ is trivial. Therefore, there is no extension between 1D and 0D obstruction-free block-states.

The decoration of $2$D block-states should extend to all $1$D blocks if there exists an extension.
We assert that the extension does exist and the extended $1$D block-states are non-trivial. 
Specifically, a trivial 2D decoration consisting of two copies of the 2D class-DIII TSC on all $2$D blocks results in non-trivial $n_1^T$ FSPT phases on all $1$D blocks. Notably, similar conclusions have been reached through entirely different approaches~\cite{Yaohong,Nonperturbativeconstraints}.

\begin{figure}[bt]
    \centering
    \includegraphics[width=0.33\textwidth]{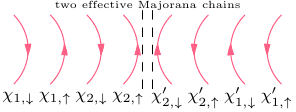}

    \caption{Extension problem for spinful systems with reflection and time-reversal symmetries: two copies of class-DIII topological superconductors on each side of the reflection axis are, without breaking any symmetry, equivalent to a nontrivial 1D FSPT state realized as two effective Majorana chains.}
        \label{figure:2 effective kitaev chain}
\end{figure}

To demonstrate this, we first examine a simplified scenario involving only time-reversal symmetry and a single reflection symmetry. Remarkably, we will show that nontrivial one-dimensional edge states can emerge even in the absence of any rotational subgroup.

We adopt a geometry illustrated in Fig.~\ref{figure:2 effective kitaev chain}, which consists of two $2$D blocks and one $1$D block. This provides a cell decomposition for the reflection group $\mathbb{Z}_2^M$. Accordingly, the total symmetry group of the $2$D block is \begin{equation}
    G_f = \mathbb{Z}_4^{fT},
\end{equation} while the symmetry group of the $1$D block is \begin{equation}
    G_f =  (\mathbb{Z}_4^{Mf} \times \mathbb{Z}_4^{fT}) / \mathbb{Z}_2^f.
\end{equation} This cell decomposition can therefore be viewed as a sub-decomposition of that for any dihedral group $D_n$. 

We decorate each $2$D block \( \sigma \) with two copies of DIII TSC,  which together form a trivial phase.  The resulting $1$D block hosts eight chiral Majorana modes ($\chi_{j,\uparrow},\chi_{j,\downarrow},\chi^\prime_{j,\uparrow},\chi^\prime_{j,\downarrow}$ with $j=1,2$), described by the edge Hamiltonian $H_{free}=H_{free}^{12}+{H^{12}_{free}}^\prime$ 
 \begin{align}
   H_{free}^{12}=& \int~dx~\Psi_{12}^T (i\nu\sigma_0\otimes\sigma_z\partial_x)\Psi_{12},\\
      {H_{free}^{12}}^\prime=& \int~dx~{\Psi_{12}^\prime}^T (-i\nu\sigma_0\otimes\sigma_z\partial_x)\Psi_{12}^\prime,\\
  & ~(\nu>0,x>0 ) \notag
 \end{align} with 
 \begin{equation}
  \begin{aligned}
  \Psi_{12}=(\chi_{1,\uparrow}(x),\chi_{1,\downarrow}(x),\chi_{2,\uparrow}(x),\chi_{2,\downarrow}(x))^T,\\\Psi^\prime_{12}=(\chi^\prime_{1,\uparrow}(x),\chi^\prime_{1,\downarrow}(x),\chi^\prime_{2,\uparrow}(x),\chi^\prime_{2,\downarrow}(x))^T.  
 \end{aligned}     
 \end{equation}
The integral is performed over the $1$D block, with the $x$-axis chosen to align with it.
The overall minus sign in ${H^{12}_{free}}^\prime$ is required for $H_{free}$ to be reflection invariant. Concretely,  we define the symmetry action as follows,
\begin{equation}
  \begin{aligned}
  T:&\quad i \to -i,  \\
   &\quad \chi_{j,\uparrow}(x) \to -\chi_{j,\downarrow}(x),~\chi_{j,\downarrow}(x) \to \chi_{j,\uparrow}(x)\\
      &\quad \text{and the same for}~\chi^\prime(x)  \\
  M:&\quad\chi_{j,\uparrow}(x) \to -\chi^\prime_{j,\downarrow}(x),~\chi_{j,\downarrow}(x) \to \chi^\prime_{j,\uparrow}(x), \\
  &\quad\chi^\prime_{j,\uparrow}(x) \to -\chi_{j,\downarrow}(x),~\chi^\prime_{j,\downarrow}(x) \to \chi_{j,\uparrow}(x) 
\end{aligned} \label{eq:D4 extension 1D 12 free}
\end{equation} with $j=1,2$.

We now introduce mass terms to gap these Majorana modes.  We require the mass term to be a sum of two parts,  each built solely from  chiral Majorana fermions originating from the same 2D decoration. In other words, terms such as $im\chi_{1,\uparrow} \chi^\prime_{1,\uparrow}$ should be excluded. This ensures a condition for addressing the extension problem, that is, that each 2D decorated state is already trivial on its own.

Under this restriction, the only symmetric mass term takes the form
\begin{align}
  H_{mass}=\int~dx~&(im\chi_{1,\uparrow} \chi_{2,\downarrow}-im\chi_{2,\uparrow}\chi_{1,\downarrow})\notag\\
  -&(im\chi^\prime_{1,\downarrow}\chi^\prime_{2,\uparrow} -im\chi^\prime_{2,\downarrow} \chi^\prime_{1,\uparrow}) \label{eq:D4 extension 1D 12 mass}
\end{align}
Here we have written the mass terms in standard form \( i m \chi^+ \chi^- \), where \( \chi^\pm \) denotes Majorana operators with positive/negative velocity relative to the \( x \)-direction. 
Here, each pair of (standard form) terms in $H_{mass}$ connected by the reflection operation exhibits a relative minus sign, with a total of two negative signs. For instance,\begin{equation}
    M:~im\chi_{1,\uparrow} \chi_{2,\downarrow} \to -im\chi^\prime_{1,\downarrow}\chi^\prime_{2,\uparrow}.
\end{equation}
Following  Ref.~\cite{Wang2020}, the number of  these minus signs equals the number of effective Majorana chains. 
Therefore, although the 2D decoration is trivial, it leaves two effective Majorana chains on the $1$D block.

In summary, the $1$D block-states induced by reflection-related block-states are equivalent to two effective Majorana chains. 
And it can be verified that the two-Majorana-chains state must belong to the non-trivial root state phase $n_1^T$ if they satisfy symmetry group $G_f = (\mathbb{Z}_4^{Mf} \times \mathbb{Z}_4^{fT})/\mathbb{Z}_2^f$ (even if their symmetry action differs from Eq.~\eqref{eq:definition root phase}).

Crucially, this conclusion remains robust when we restore the full $D_4$ point group symmetry. By incorporating the rotational symmetries, this extension correspondence naturally applies to all other 1D blocks in the system.

Ultimately, this demonstrates that a trivial $2$D block decoration inherently extends to non-trivial $1$D $n_1^T$ FSPT states on all $1$D blocks under the full symmetry group. This non-trivial extension fundamentally alters the final topological classification, manifesting mathematically as the following group extension:
\begin{equation}
1 \to \mathbb{Z}_4 \to \mathbb{Z}_8 \to \mathbb{Z}_2 \to 1,
\end{equation}
where \(\mathbb{Z}_4\) represents the decoration of the pair of Majorana chains on \(\tau_1\) and \(\tau_2\).

In summary, all independent non-trivial block-states with different dimensions are classified as follows:
\begin{align}
    \begin{aligned}
         & E^{\mathrm{2D}}=\{\mathrm{OFBS}\}^{\mathrm{2D}}=\mathbb{Z}_2   \\
         & E^{\mathrm{1D}}=\{\mathrm{OFBS}\}^{\mathrm{1D}}/\{\mathrm{TBS}\}^{\mathrm{1D}}=\mathbb{Z}_4^2   \\
         & E^{\mathrm{0D}}=\{\mathrm{OFBS}\}^{\mathrm{0D}}/\{\mathrm{TBS}\}^{\mathrm{0D}}=\mathbb{Z}_2^2
    \end{aligned}.
\end{align}
The ultimate classification $\mathcal{G}_{1/2}$  is $\mathbb{Z}_2^2 \times\mathbb{Z}_4 \times \mathbb{Z}_8$, as determined by the group extension discussed above.

\subsection{$D_2$ and $D_6$}
The cell decompositions for the $D_2$ and $D_6$ symmetry groups are presented in panels (a) and (d) of Figure~\ref{Dn cell}, respectively. We take $C_2$, $C_6$ (the $2\pi/n$ rotation, $n=2,6$) and $M_{\tau_j}$ (reflection about the $\tau_j$-block axis) as the generators of the  groups $D_2$ and $D_6$. 
And the cell decomposition remains invariant under the action of these generators.

\subsubsection{Spinless fermions} \label{sec:D2 D6 spinless}

 For the dihedral groups \( D_n \) ($n=2,6$), the $0$D blocks have physical symmetry group \( G_b = D_n \times \mathbb{Z}_2^T \). As in  the $D_4$ case, the corresponding 0D FSPT phases are characterized by \begin{equation}
    \mathcal{H}^0(G_b, \mathbb{Z}_2^f) \times \mathcal{H}^1(G_b, U_T(1))=\mathbb{Z}_2\times \mathbb{Z}_2^2. 
 \end{equation} 
Because the $\mathbb{Z}_2^T\subset G_b$ factor does not yield nontrivial cocycles in this classification, the irreducible representations can be labeled by the $\pm 1$ eigenvalues of $P_f$, $C_n$, and $M$.

Next, we consider the $1$D block-state decorations. For each $1$D block \( \tau_1 \) or \( \tau_2 \), the total symmetry group is \begin{equation}
    G_f = \mathbb{Z}_2^M \times \mathbb{Z}_2^T \times \mathbb{Z}_2^f
\end{equation} Thus, there are two possible $1$D block-states: the single Majorana chain and the 1D FSPT state. 

More precisely, the 1D FSPT phases form a \( \mathbb{Z}_4 \times \mathbb{Z}_4 \) classification. Each $\mathbb{Z}_4$ generator can be realized by two Majorana chains, corresponding to the $n_1^T$ and $n_1^M$ phases, respectively, as established in Sec.~\ref{sec:D4 spinless fermion}.

Finally, as discussed above, there is no root phase arising from $2$D block decorations, since every $2$D block has symmetry $G_f=\mathbb{Z}_2^T\times \mathbb{Z}_2^f$.

With all candidate states available for decoration, we proceed to examine which block-state decorations satisfy the no-open-edge conditions.  All 0D decorations automatically satisfy the no-open-edge conditions. The obstruction-free $0$D block-states therefore form the group:
\begin{equation}
\{\mathrm{OFBS}\}^{\mathrm{0D}} = \mathbb{Z}_2^3.
\end{equation}

Decorating $1$D blocks \(\tau_1\) (or \(\tau_2\)) with a single Majorana chain yields $n$ dangling Majorana modes $\gamma_1,\gamma_2,\dots,\gamma_n$ at the $0$D block \( \mu \). Near \( \mu \), these Majorana modes exhibit the following rotation symmetry (with all indices understood modulo $n$):
\begin{equation}
    C_n: \gamma_j \mapsto \gamma_{j+1}.
\end{equation}

The local fermion-parity operator and its symmetry properties are given by:
\begin{equation}
  P_f = i^{n/2}\gamma_1 \gamma_2 \dots \gamma_n,~ C_n:  P_f  \mapsto -P_f. 
\end{equation}
As a result, these Majorana modes break the fermion parity. Therefore, the decoration of a Majorana chain on the $1$D block \( \tau_1 \) or \( \tau_2 \) does not contribute to a non-trivial crystalline TSC, as it violates the no-open-edge condition.

Now, consider the decoration of single Majorana chains on both \( \tau_1 \) and \( \tau_2 \), leaving four dangling Majorana modes at the $0$D block \( \mu \). Although the local fermion-parity operator commutes with the rotation operator $C_n$, it anti-commutes with the reflection operator $M$. The fermion-parity operator transforms under $M$ as follows: \begin{equation}\begin{aligned}
   C_2:&P_f=\prod_{j=1}^{4} \gamma_j \mapsto \gamma_1 \gamma_2 \gamma_4 \gamma_3 = -P_f,\\
    C_6:&  P_f=\prod_{j=1}^{12} \gamma_j \mapsto \gamma_1 \gamma_6 \gamma_5 \gamma_4 \gamma_3 \gamma_2 \gamma_{12} \gamma_{11} \gamma_{10} \gamma_{9} \gamma_8 \gamma_7  = -P_f.
\end{aligned}\end{equation}  Here, we adopt the convention that the Majorana operators $\gamma_1, \gamma_2,  \dots ,\gamma_n$ originate from $\tau_1$, while $\gamma_{n+1}, \gamma_{n+2},  \dots ,\gamma_{2n}$ originate from $\tau_2$. The indices are arranged in ascending order, corresponding to the rotational sequence.  This anti-commutation under $M$ indicates that the Majorana chain decoration is incompatible with the $D_n$  symmetry ($n=2,6$), and thus remains forbidden.

\begin{figure}[bt]
    \centering
\includegraphics[width=0.3\textwidth]{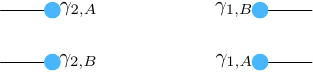}%width for pictures about dangling modes 
    \caption{ 1D FSPT decoration on $1$D blocks labeled by $\tau_1$ , which leaves $4$ dangling Majorana modes $\gamma_{1,A},\gamma_{1,B},\gamma_{2,A},\gamma_{2,B}$  at  $0$D block $\mu$.}
    \label{fig:spinless D2 2chain tau1}
\end{figure}

Now we consider decorating the $1$D blocks $\tau_1$ with a 1D FSPT phase (realized by two Majorana chains), as shown in Fig.~\ref{fig:spinless D2 2chain tau1}. We begin with the point group $D_2$.

At the $0$D block $\mu$, four Majorana zero modes remain dangling. They transform under the on-site symmetry group $G_f=D_2\times \mathbb{Z}_2^T \times \mathbb{Z}_2^f$ according to the symmetry actions specified below. For $n_1^T$ phase, we have
\begin{align}
    n_1^T:&\left\{\begin{aligned}
        T: &~ i \mapsto -i,\\
        &~\gamma_{j,\sigma} \mapsto \gamma_{j,\sigma}, \\
    M: & ~\gamma_{j,\sigma} \mapsto \gamma_{j,\sigma},\\
    C_2:& ~\gamma_{j,\sigma} \mapsto \gamma_{j+1,\sigma},
    \end{aligned}\right.  \label{eq:D2 n1T}
\end{align} where $j=1,2$ and $\sigma,\sigma^\prime = A,B$ with $\sigma \neq \sigma^\prime$.
The position indices $j$ are understood modulo $2$.

The Hilbert space spanned by these dangling Majorana modes furnishes a representation of $G_f$, whose symmetry operators obey the algebra
\begin{equation}
T^2=-1,\quad (MT)^2=-1,\quad (MC_2T)^2=1 .
\label{eq:D2 T tau1 algebra}
\end{equation}
Importantly, $\omega_2(T,T)$ and $\omega_2(MT,MT)$ are invariant under 2-coboundary equivalence, where $\omega_2(g_1,g_2)\in  \mathcal{H}^2(G_f,U_T(1))$ for $g_1,g_2\in G_f$. The relations $T^2=-1$ and $(MT)^2=-1$ therefore imply
$$\omega_2(T,T)=-1,\quad \omega_2(MT,MT)=-1,$$
showing that the symmetry acts projectively in a nontrivial way. Consequently, this decoration is obstructed.

\begin{figure}[bt]
    \centering
  \includegraphics[width=0.08\textwidth]{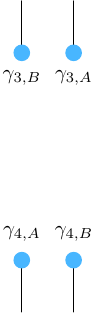}%width for pictures about dangling modes 
    \caption{ 1D FSPT decoration on $1$D blocks labeled by $\tau_2$ , which leaves $4$ dangling Majorana modes $\gamma_{3,A},\gamma_{3,B},\gamma_{4,A},\gamma_{4,B}$  at  $0$D block $\mu$.}
    \label{fig:spinless D2 2chain tau2}
\end{figure}

The same reasoning applies to decorating the $1$D block \(\tau_2\) with the 1D FSPT phase. In this case, there are again four dangling Majorana zero modes at the $0$D block \(\mu\) (see Fig.~\ref{fig:spinless D2 2chain tau2}), which span a new local Hilbert space. Their symmetry transformations agree with Eq.~\eqref{eq:D2 n1T}, except that the reflection action is modified to
\begin{equation}
M:\ \gamma_{3,\sigma} \leftrightarrow \gamma_{4,\sigma}.
\end{equation}
Keeping the same reflection axis and the same sense of rotation as above, one finds that on this Hilbert space the symmetry operators satisfy an algebra that includes
\begin{equation}
T^2=-1,\quad (MT)^2=1,\quad (MC_2T)^2=-1 .
\label{eq:D2 T tau2 algebra}
\end{equation}
Because the corresponding cohomology invariants \(\omega_2(T,T)\) and \(\omega_2(MC_2T,MC_2T)\) take nontrivial values, \(G_f\) acts projectively, i.e., the representation is a nontrivial projective representation. Moreover, this projective representation lies in a different cohomology class from the one obtained for \(\tau_1\) in the previous paragraph.

Since the projective representation classification for \(G_f\) is \( \mathcal{H}^2(G_f,U_T(1))=\mathbb{Z}_2^7\), the direct sum of two inequivalent nontrivial projective representations cannot be smoothly deformed into a trivial (linear) representation. Therefore, simultaneously decorating both \(\tau_1\) and \(\tau_2\) with the \(n_1^T\) 1D FSPT phase is also obstructed.

For 1D FSPT decoration of phase $n_1^M$  on $\tau_1$,  the symmetry actions are
\begin{align}
    n_1^M:&\left\{\begin{aligned}
    T: & ~i \mapsto -i,\\
 & ~\gamma_{j,A} \mapsto \gamma_{j,A},~\gamma_{j,B} \mapsto -\gamma_{j,B},\\
M:   & ~\gamma_{j,A} \mapsto \gamma_{j,A},~\gamma_{j,B} \mapsto -\gamma_{j,B},\\
C_2:& ~\gamma_{j,\sigma} \mapsto \gamma_{j+1,\sigma}.
\end{aligned}\right. \label{eq:D2 n1M}
\end{align} where $ j=1,2 \pmod{2},~\sigma=A,B$.
This choice uses the same reflection axis and rotation convention as in the $n_1^T$ cases discussed above.  For the Hilbert space formed by four dangling modes on $\mu$, the symmetry operators of $G_f$ satisfy
\begin{align}
    T^2=1,(MT)^2=-1,(MC_2T)^2=1. \label{eq:D2 M tau1}
\end{align}

Similarly, decorating $\tau_2$ with the $n_1^M$ phase yields a (generally different) projective action at $\mu$, characterized by
\begin{align}
T^2=1,\quad (MT)^2=1,\quad (MC_2T)^2=-1 .
\label{eq:D2 M tau2}
\end{align}

Therefore, the 1D FSPT decorations of the $n_1^M$ phase on $\tau_1$ (or on $\tau_2$), as well as the simultaneous decoration on both $\tau_1$ and $\tau_2$, are all obstructed.

 Similar to discussions in Sec.~\ref{sec:D4 spinless fermion}, the decorations on block $\tau_1$ (or $\tau_2$) with the following BSPT phases is obstruction-free:
 \begin{equation}
 (-1)^{n_1^T \cup n_1^T} \quad \text{or} \quad (-1)^{n_1^M \cup n_1^M}.
 \end{equation}
 And following Ref.~\cite{Song_2020}, let $[\alpha],[\beta] \in  \mathcal{H}^2(\mathbb{Z}_2^M\times \mathbb{Z}_2^T \times \mathbb{Z}_2^f,U_T(1))=\mathbb{Z}_2^4$ denote the wavefunctions of these two BSPT phases. Since  rotation (or reflection) operations do not act on the local degrees of freedom of each wavefunction, we have \begin{equation}[\alpha]+C_2[\alpha]=2[\alpha]=0,\end{equation} with an analogous relation holding for $[\beta]$. 
Therefore,  decorations of the BSPT phase \( (-1)^{n_1^T \cup n_1^T} \) or \( (-1)^{n_1^M \cup n_1^M} \)  on a $1$D block \( \tau_1 \) or \( \tau_2 \) satisfy the no-open-edge condition and are obstruction-free.

By contrast, for 1D FSPT decorations, the projective representations induced by the two phases $n_1^T$ and $n_1^M$ belong to different cohomology classes. As a result, decorating $\tau_1$ (or $\tau_2$) with two distinct phases (each contributing two Majorana chains) cannot be combined into a trivial (linear) representation, and such mixed decorations are obstructed.

\begin{figure}[bt]
    \centering
\includegraphics[width=0.3\textwidth]{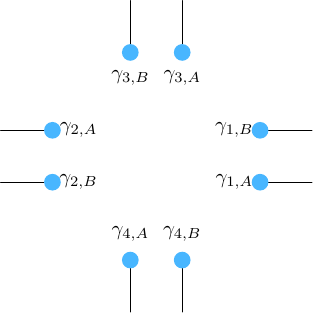}%width for pictures about dangling modes 
    \caption{ 1D FSPT decoration on $1$D blocks labeled by $\tau_1$ and $\tau_2$, which leaves $8$ dangling Majorana modes $\gamma_{1,A},\gamma_{1,B},\dots,\gamma_{4,A},\gamma_{4,B}$  at  $0$D block $\mu$.}
    \label{fig:spinless D2 2chain tau1tau2}
\end{figure}

However, due to the symmetry operator algebras discussed above, the decorations of the phase $n_1^T + n_1^M$ on $\tau_1$ contribute to a group representation satisfying $T^2=-1$ (while other relations are trivial) on the Hilbert space spanned by dangling edge modes, while those on $\tau_2$ also contribute to a group representation satisfying $T^2=-1$ (while other relations are trivial). Consequently, decorations of 1D FSPT phase $n_1^T + n_1^M$ on $\tau_1$ (or $\tau_2$) are obstructed.

In contrast, the direct sum of these representations contributed by $\tau_1$ and $\tau_2$ can be reduced to a trivial representation. This implies that decorating $\tau_1$ and $\tau_2$  simultaneously with the phase $n_1^T + n_1^M$ is obstruction-free.  In the following, we will explicitly construct the interaction terms that gap out the corresponding edge modes.

This decoration contributes eight dangling Majorana modes at block $\mu$ (see Fig.~\ref{fig:spinless D2 2chain tau1tau2}), and these modes obey symmetry properties as follows:
\begin{equation}
    \begin{aligned}
    T: & \quad i \mapsto -i, \\
       & \quad \gamma_{j, \sigma} \mapsto \gamma_{j, \sigma}, \\
    M: & ~ \left\{ \begin{aligned}
&\gamma_{1,A} \mapsto \gamma_{1,A},~~~\gamma_{2,A} \mapsto \gamma_{2,A},\\
& \gamma_{1,B} \mapsto-\gamma_{1,B},\gamma_{2,B}\mapsto-\gamma_{2,B},\\
&\gamma_{3,A} \leftrightarrow \gamma_{4,A},~~~ \gamma_{3,B} \leftrightarrow-\gamma_{4,B},
    \end{aligned}\right.\\
    C_2: & \quad \gamma_{1, \sigma} \leftrightarrow \gamma_{2, \sigma},~~~~~\gamma_{3, \sigma} \leftrightarrow \gamma_{4, \sigma},
\end{aligned}
\end{equation}
where $j=1,2,3,4$ and \( \sigma = A, B \). Introduce complex fermions by \begin{equation}c_j=\frac{1}{2}(\gamma_{j,A}+i\gamma_{j,B}),~j=1,2,3,4.\end{equation} Then they transform under symmetry group as follows:
\begin{equation}
    \begin{aligned}
   T:& (c_1,c_2,c_3,c_4) \mapsto (c^\dagger_1,c^\dagger_2,c^\dagger_3,c^\dagger_4)\\
   M:&(c_1,c_2,c_3,c_4) \mapsto (c^\dagger_1,c^\dagger_2,c^\dagger_4,c^\dagger_3)\\
   C_2:&(c_1,c_2,c_3,c_4) \mapsto (c_2,c_1,c_4,c_3)
\end{aligned}\label{eq:spinless D2 fermion sym n1mn1t}
\end{equation}

To gap out the gapless modes at the $0$D block \( \mu \), we first consider the Hubbard interaction (\( U > 0 \)):
\begin{equation}
H_U = U \sum_{j=1,3}  \left(c_j^\dagger c_j - \frac{1}{2}\right) \left(c_{j+1}^\dagger c_{j+1} - \frac{1}{2} \right).
\end{equation}
The complex fermion occupation numbers $(n_1, n_2)$, $(n_3, n_4)$ result in a four-fold ground-state degeneracy, which can be interpreted as two spin-\( 1/2 \) degrees of freedom (Eq.~\eqref{eq:definition of spin12 degree}). We can derive the symmetry properties of these spins from Eq.~\eqref{eq:spinless D2 fermion sym n1mn1t}:
\begin{equation}
    \begin{aligned}
    T:& \begin{aligned}
       & (\tau_{12}^x,\tau_{12}^y,\tau_{12}^z) \mapsto (-\tau_{12}^x,-\tau_{12}^y,-\tau_{12}^z)\\
       &(\tau_{34}^x,\tau_{34}^y,\tau_{34}^z) \mapsto (-\tau_{34}^x,-\tau_{34}^y,-\tau_{34}^z)\\
    \end{aligned} \\
    M:& \begin{aligned}
       & (\tau_{12}^x,\tau_{12}^y,\tau_{12}^z) \mapsto (-\tau_{12}^x,\tau_{12}^y,-\tau_{12}^z)\\
       &(\tau_{34}^x,\tau_{34}^y,\tau_{34}^z) \mapsto (-\tau_{34}^x,-\tau_{34}^y,\tau_{34}^z)\\
        \end{aligned} \\
    C_2:& \begin{aligned}
       & (\tau_{12}^x,\tau_{12}^y,\tau_{12}^z) \mapsto (\tau_{12}^x,-\tau_{12}^y,-\tau_{12}^z)\\
       &(\tau_{34}^x,\tau_{34}^y,\tau_{34}^z) \mapsto (\tau_{34}^x,-\tau_{34}^y,-\tau_{34}^z)\\
        \end{aligned}       
\end{aligned}
\end{equation}
It can be verified that the following interaction term remains invariant under the symmetry and lifts the ground-state degeneracy at \( \mu \) ($J>0$):
\begin{equation}
H_{J}=J \bigl( \tau_{1 2}^{x} \tau_{3 4}^{x}+\tau_{1 2}^{y} \tau_{3 4}^{z}+\tau_{1 2}^{z} \tau_{3 4}^{y} \bigr).
\end{equation}

Therefore, the decoration of two Majorana chains corresponding to the \( n_1^T+ n_1^M \) FSPT phase on the $1$D blocks \( \tau_1 \) and \( \tau_2 \) simultaneously satisfies the no-open-edge condition and is obstruction-free. Similarly to the conclusions in Sec.~\ref{sec:D4 spinless fermion}, this decorated state realizes an intrinsic interacting FSPT phase.  However, as we demonstrate through the bubble equivalence analysis below, this crystalline FSPT phase is actually trivial.

In summary, all obstruction-free $1$D block-states can be generated by stacking the following elementary decorations:
\begin{enumerate}
    \item \( n_1^T+ n_1^M \) FSPT phase on both  \( \tau_1 \) and \( \tau_2 \) blocks;
    \item Two independent \( (-1)^{n_1^T \cup n_1^T} \) BSPT phases: one on \( \tau_1 \) and one on \( \tau_2 \) blocks;
    \item  Two independent \( (-1)^{n_1^M \cup n_1^M} \) BSPT phases: one on \( \tau_1 \) and one on \( \tau_2 \) blocks.
\end{enumerate}
These states form the following group:  
\begin{equation}
\{\mathrm{OFBS}\}^{1D} = \mathbb{Z}_4 \times \mathbb{Z}_2^3,
\end{equation}  
where the \( \mathbb{Z}_4 \) factor represents the \( n_1^T+ n_1^M \) FSPT phase, and the three \( \mathbb{Z}_2 \) factors correspond to: the \( (-1)^{n_1^T \cup n_1^T} \) BSPT phase on \( \tau_1, \tau_2 \), and the \( (-1)^{n_1^M \cup n_1^M} \) BSPT phase on \( \tau_1 \), respectively.

The group structure can be written as the quotient
\[
(\mathbb Z_4 \times \mathbb Z_2^4)/\mathbb Z_2 .
\]
Indeed, stacking two copies of the FSPT decorations on \(\tau_1\) and \(\tau_2\) reproduces the BSPT decoration \((-1)^{n_1^M\cup n_1^M+n_1^T\cup n_1^T}\) on \(\tau_1\) and \(\tau_2\). Hence one linear combination of the above decorations is redundant, forming the \(\mathbb Z_2\) subgroup that is modded out.

For the point group $D_6$, the conclusions remain unchanged.  The obstruction-free $1$D block-states are the same as in the $D_2$ case, and they form the group
\begin{equation}
  \{\mathrm{OFBS}\}^{1D} = \mathbb{Z}_4 \times \mathbb{Z}_2^3.  
\end{equation} The reasoning used to determine which $1$D block-states satisfy the no-open-edge condition is also essentially identical. Without going into too much detail here, we will only briefly state the key details. 

For $n_1^T$ phase, twelve dangling Majorana zero modes from 1D FSPT decoration on $\tau_1$ preserve symmetry actions as following:
\begin{align}
    n_1^T:&\left\{\begin{aligned}
        T: &~ i \mapsto -i,\\
        &~\gamma_{j,\sigma} \mapsto \gamma_{j,\sigma}, \\
    M: & ~\gamma_{1,\sigma} \mapsto \gamma_{1,\sigma},~\gamma_{4,\sigma} \mapsto \gamma_{4,\sigma},\\
    & ~\gamma_{2,\sigma} \leftrightarrow \gamma_{6,\sigma},~\gamma_{3,\sigma} \leftrightarrow\gamma_{5,\sigma}, \\
    C_6:& ~\gamma_{j,\sigma} \mapsto \gamma_{j+1,\sigma},
    \end{aligned}\right.  \label{eq:D6 n1T}
\end{align} where $j=1,2,3,\dots,6$ and $\sigma,\sigma^\prime = A,B$ with $\sigma \neq \sigma^\prime$.
The subscripts of the operators indicate their respective positions and should mod $6$. Symmetry operators in the local Hilbert space satisfy the same algebra as Eq.~\eqref{eq:D2 T tau1 algebra}. Also, dangling modes from decoration on $\tau_2$ construct a representation of the same algebra relation as Eq.~\eqref{eq:D2 T tau2 algebra}. 

For 1D FSPT decoration of phase $n_1^M$  on $\tau_1$, symmetry properties of their dangling modes are as follows:
\begin{align}
    n_1^M:&\left\{\begin{aligned}
    T: & ~i \mapsto -i,\\
    & ~\gamma_{j,A} \mapsto \gamma_{j,A},~\gamma_{j,B} \mapsto -\gamma_{j,B},\\
    M: & ~\gamma_{1,A} \mapsto \gamma_{1,A},~\gamma_{4,A} \mapsto \gamma_{4,A},\\
    & ~\gamma_{1,B} \mapsto -\gamma_{1,B},~\gamma_{4,B} \mapsto -\gamma_{4,B},\\
    & ~\gamma_{2,A} \leftrightarrow \gamma_{6,A},~\gamma_{3,A} \leftrightarrow\gamma_{5,A}, \\
    & ~\gamma_{2,B} \leftrightarrow -\gamma_{6,B},~\gamma_{3,B} \leftrightarrow -\gamma_{5,B}, \\
C_6:& ~\gamma_{j,\sigma} \mapsto \gamma_{j+1,\sigma},
\end{aligned}\right.\label{eq:D6 n1M}
\end{align} where $j=1,2,3,\dots,6$ and $\sigma,\sigma^\prime = A,B,~\sigma \neq \sigma^\prime$. The subscripts of the operators should mod $6$. It can be verified that symmetry operators in the local Hilbert space satisfy the same algebra as Eq.~\eqref{eq:D2 M tau1} and Eq.~\eqref{eq:D2 M tau2}.

According to these algebras, we can conclude that 1D FSPT decoration of phase $n_1^T$ or $n_1^M$ on $\tau_1$ (or $\tau_2$) are obstructed, while the decorations on block $\tau_1$ (or $\tau_2$) corresponding to the following BSPT phases are free from obstruction:
\begin{equation}
(-1)^{n_1^T \cup n_1^T} \quad \text{or} \quad (-1)^{n_1^M \cup n_1^M}.
\end{equation}
Following Ref.~\cite{Song_2020}, we denote the wavefunctions of these two BSPT phases by $[\alpha],[\beta] \in  \mathcal{H}^2(\mathbb{Z}_2^M\times \mathbb{Z}_2^T \times \mathbb{Z}_2^f,U_T(1))=\mathbb{Z}_2^4$. Since rotational (or reflectional) operations preserve the local degrees of freedom of each wavefunction, we have
\begin{equation}[\alpha]+C_6[\alpha]+C_6^2[\alpha]+\dots+C_6^5[\alpha]=6[\alpha]=0,\end{equation}
with an analogous relation holding for $[\beta]$. Thus, BSPT phase decorations are obstruction-free.

\begin{figure}[tb]
    \centering
    \includegraphics[width=0.4\textwidth]{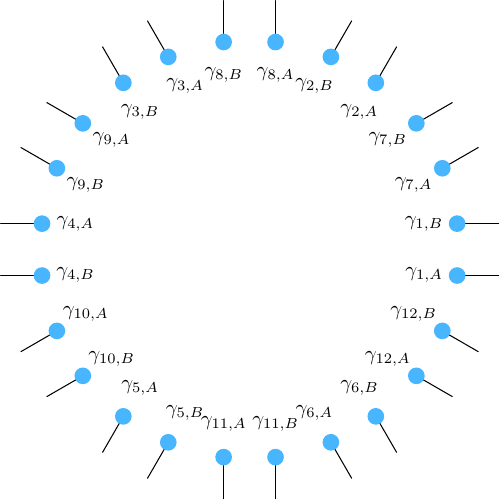}
    \caption{1D FSPT decoration on  $1$D blocks labeled by \( \tau_1 \) and \( \tau_2 \), leaving 24 dangling Majorana modes at a $0$D block \( \mu \).}
    \label{fig:spinless D6 2chain tau1tau2}
\end{figure}

For the same reasons discussed in the case $D_2$, the decorations of the  phase $n_1^T + n_1^M$ on $\tau_1$ and $\tau_2$, simultaneously, are obstruction-free, as they yield a trivial representation on the Hilbert space spanned by the dangling edge modes. We will explicitly present the interaction terms that open the corresponding gapless modes in the following. 

This decoration contributes 24 dangling Majorana modes at block $\mu$ (see Fig.~\ref{fig:spinless D6 2chain tau1tau2}), and these modes obey symmetry properties as follows:
\begin{equation}
    \begin{aligned}
    T: &\begin{cases} 
     i \mapsto -i, \\
         \gamma_{j, \sigma} \mapsto \gamma_{j, \sigma},\quad j=1,2,\dots,12  
    \end{cases} \\ 
    M: &\begin{cases} 
    \begin{aligned}
    &\gamma_{j,A} \mapsto \gamma_{8-j,A}, \\
    &\gamma_{j,B} \mapsto -\gamma_{8-j,B}, 
    \end{aligned}
    & j=1,2,\dots,6 \, \text{(mod 6)}, \\
    \begin{aligned}
    &\gamma_{j,A} \mapsto \gamma_{19-j,A}, \\
    &\gamma_{j,B} \mapsto -\gamma_{19-j,B},
    \end{aligned}
    & j=7,8,\dots,12, 
    \end{cases} \\
    C_6: & \quad \gamma_{j, \sigma} \mapsto \gamma_{j+1, \sigma}, \quad j=1,2,\dots,12 .
    \end{aligned}
\end{equation}
where \( \sigma = A, B \). Define complex fermions by \begin{equation}c_j=\frac{1}{2}(\gamma_{j,A}+i\gamma_{j,B}),~j=1,2,\dots,12.\end{equation} Then they transform under symmetry group as follows:
\begin{equation}
    \begin{aligned}
   T:~& (c_1,c_2,c_3,c_4,c_5,c_6) \mapsto (c^\dagger_1,c^\dagger_2,c^\dagger_3,c^\dagger_4,c^\dagger_5,c^\dagger_6)\\
   ~& (c_7,c_8,c_9,c_{10},c_{11},c_{12}) \mapsto (c^\dagger_7,c^\dagger_8,c^\dagger_9,c^\dagger_{10},c^\dagger_{11},c^\dagger_{12})\\
   M:~&(c_1,c_2,c_3,c_4,c_5,c_6) \mapsto (c^\dagger_1,c^\dagger_6,c^\dagger_5,c^\dagger_4,c^\dagger_3,c^\dagger_2)\\
   ~& (c_7,c_8,c_9,c_{10},c_{11},c_{12}) \mapsto (c^\dagger_{12},c^\dagger_{11},c^\dagger_{10},c^\dagger_9,c^\dagger_{8},c^\dagger_7) \\
   C_6:~&(c_1,c_2,c_3,c_4,c_5,c_6) \mapsto (c_2,c_3,c_4,c_5,c_6,c_1)\\
   ~&(c_7,c_8,c_9,c_{10},c_{11},c_{12}) \mapsto (c_8,c_9,c_{10},c_{11},c_{12},c_7)
\end{aligned}\label{eq:spinless D6 fermion sym n1mn1t}
\end{equation}

To gap out the gapless modes at the $0$D block \( \mu \), we first consider the Hubbard interaction (\( U > 0 \)):
\begin{equation}
H_U = U \sum_{j=1,2,3,7,8,9}  \left(c_j^\dagger c_j - \frac{1}{2}\right) \left(c_{j+3}^\dagger c_{j+3} - \frac{1}{2} \right).
\end{equation}
The complex fermion occupation numbers $(n_j, n_{j+3})$ ($j=1,2,3,7,8,9$) result in a $64$-fold ground-state degeneracy, which can be interpreted as six spin-\( 1/2 \) degrees of freedom. Symmetry properties of these spins from Eq.~\eqref{eq:spinless D6 fermion sym n1mn1t} are:
\begin{equation}
\begin{aligned}
    T:~& \begin{aligned}
       & (\tau_{a,b}^x,\tau_{a,b}^y,\tau_{a,b}^z) \mapsto (-\tau_{a,b}^x,-\tau_{a,b}^y,-\tau_{a,b}^z),
    \end{aligned} \\
    M:& \left\{\begin{aligned}
       & (\tau_{1,4}^x,\tau_{1,4}^y,\tau_{1,4}^z) \mapsto (-\tau_{1,4}^x,\tau_{1,4}^y,-\tau_{1,4}^z)\\
       &(\tau_{2,5}^x,\tau_{2,5}^y,\tau_{2,5}^z) \leftrightarrow (-\tau_{3,6}^x,-\tau_{3,6}^y,\tau_{3,6}^z)\\
        &(\tau_{8,11}^x,\tau_{8,11}^y,\tau_{8,11}^z)\mapsto (-\tau_{8,11}^x,-\tau_{8,11}^y,\tau_{8,11}^z)\\
        &(\tau_{7,10}^x,\tau_{7 ,10}^y,\tau_{7 ,10}^z) \leftrightarrow (-\tau_{9,12}^x,-\tau_{9,12}^y,\tau_{9,12}^z)
        \end{aligned}\right. \\
    C_6:&\left \{ \begin{aligned}
       & (\tau_{1,4}^x,\tau_{1,4}^y,\tau_{1,4}^z) \mapsto (\tau_{2,5}^x,\tau_{2,5}^y,\tau_{2,5}^z)\\
       &(\tau_{2,5}^x,\tau_{2,5}^y,\tau_{2,5}^z) \mapsto (\tau_{3,6}^x,\tau_{3,6}^y,\tau_{3,6}^z)\\
       & (\tau_{3,6}^x,\tau_{3,6}^y,\tau_{3,6}^z) \mapsto (\tau_{1,4}^x,-\tau_{1,4}^y,-\tau_{1,4}^z)\\
        & (\tau_{7,10}^x,\tau_{7,10}^y,\tau_{7,10}^z) \mapsto (\tau_{8,11}^x,\tau_{8,11}^y,\tau_{8,11}^z)\\
       &(\tau_{8,11}^x,\tau_{8,11}^y,\tau_{8,11}^z) \mapsto (\tau_{9,12}^x,\tau_{9,12}^y,\tau_{9,12}^z)\\
       & (\tau_{9,12}^x,\tau_{9,12}^y,\tau_{9,12}^z) \mapsto (\tau_{7,10}^x,-\tau_{7,10}^y,-\tau_{7,10}^z)
        \end{aligned} \right . ,      
\end{aligned}
\end{equation} where $\tau_{a,b}$ denote the spin degrees of freedom (Eq.~\eqref{eq:definition of spin12 degree}).
It can be verified that the following interaction terms remain invariant under the symmetry and lift the ground-state degeneracy at \( \mu \) ($J>0$):
\begin{equation}\begin{aligned}
    H_{J}=&J \bigl( \tau_{1,4}^{x} \tau_{8,11}^{x}+\tau_{1,4}^{y} \tau_{8,11}^{z}+\tau_{1,4}^{z} \tau_{8,11}^{y}\bigr)\\
    &+J \bigl( \tau_{2,5}^{x} \tau_{9,12}^{x}+\tau_{2,5}^{y} \tau_{9,12}^{z}+\tau_{2,5}^{z} \tau_{9,12}^{y}\bigr)\\
    &+J \bigl( \tau_{3,6}^{x} \tau_{7,10}^{x}-\tau_{3,6}^{y} \tau_{7,10}^{z}-\tau_{3,6}^{z} \tau_{7,10}^{y}\bigr) .
\end{aligned}
\end{equation}

Therefore, the decoration of 1D FSPT corresponding to the phase \( n_1^T+ n_1^M \)  on the $1$D blocks \( \tau_1 \) and \( \tau_2 \) simultaneously satisfies the no-open-edge condition and is obstruction-free. 
This decorated state also realizes an intrinsic interacting FSPT phase since all possible mass terms break time-reversal symmetry. However, bubble equivalence analysis reveals that this crystalline FSPT phase is trivial.

Now  we turn to discuss bubble equivalences. Following the same approach as in the spinless $D_4$ case (Sec.~\ref{sec:D4 spinless fermion}), we find that for $D_2$ and $D_6$, $2$D bubbles induce nontrivial 1D FSPT phases on both $\tau_1$ and $\tau_2$, specifically the phase $n_1^M+n_1^T$. As a result, these bubbles trivialize a subgroup
$$\{\mathrm{TBS}\}^{1D}=\mathbb{Z}_4$$
within the obstruction-free classification. Equivalently, any obstruction-free decoration by the phase $n_1^M+n_1^T$ simultaneously on $\tau_1$ and $\tau_2$ is trivial.

2D bubbles cannot generate a closed Majorana chain encircling the $0$D block \(\mu\), since such a closed Majorana chain is incompatible with reflection symmetry~\cite{2Dcrystalline}. Consequently, $2$D bubbles do not affect the $0$D block-states.

Next, consider the $1$D bubble equivalences. We decorate a pair of complex fermions on each $1$D block \( \tau_1 \) (or \( \tau_2 \)).
Near each $0$D block, there are \( n \) complex fermions contributed from adjacent bubbles that form the following atomic insulator ($n=2,6$):
\begin{equation}
\ket{TR} = c_1^\dagger c_2^\dagger c_3^\dagger \dots c_n^\dagger.
\end{equation}
Since the subgroup \( \mathbb{Z}_2^T \) does not contribute to a non-trivial classification, we only consider the symmetry transformation properties of \( \ket{TR} \) under \( C_n \) and \( M \).
For the even dihedral group \( D_n \) ($n=2,6$), we define the symmetry transformations of complex fermions as follows:
\begin{equation}
    \begin{aligned}
        {C_n}:~ & c^\dagger_j \mapsto c^\dagger_{j+1},~j=1,2,\dots,n~\pmod{n} \\
       {M}:~   & c^\dagger_j \mapsto c^\dagger_{n+2-j},~j=1,2,\dots,n~\pmod{n}
    \end{aligned}.
    \end{equation}
The decoration of $1$D bubbles on a $1$D block \( \tau_1 \) (or \( \tau_2 \)) yields states \( \ket{TR} \) that satisfy \begin{equation}
 C_n \ket{TR} = -\ket{TR} ,~  M \ket{TR} = \ket{TR}.   
\end{equation}  Similarly, the decoration on both \( \tau_1 \) and \( \tau_2 \) produces states \( \ket{TR} \) that satisfy \begin{equation}
     C_n \ket{TR} = \ket{TR},~  M \ket{TR} = -\ket{TR}.
\end{equation} 

Thus, the decoration of $1$D bubbles provides two independent eigenstates of \( M \) and \( C_n \), contributing to the trivialization group \begin{equation}
   \{\mathrm{TBS}\}^{\mathrm{0D}}=\mathbb{Z}_2^2. 
\end{equation}

The only possible extension is from \(1\)D block states to \(0\)D block states. However, dangling zero modes from trivial \(1\)D decorations cannot form a closed Majorana chain around the \(0\)D point \(\mu\) to realize an odd-parity \(0\)D state, since such a configuration is incompatible with reflection symmetry~\cite{2Dcrystalline}. Consequently, because all nontrivial obstruction-free \(0\)D states have odd parity, no extension from \(1\)D to \(0\)D block states is allowed.

In summary, all independent non-trivial block-states with different dimensions are classified as follows:
\begin{equation}
\begin{aligned}
         & E^{\mathrm{2D}}=\{\mathrm{OFBS}\}^{\mathrm{2D}}=\mathbb{Z}_1                       \\
         & E^{\mathrm{1D}}=\{\mathrm{OFBS}\}^{\mathrm{1D}}/\{\mathrm{TBS}\}^{\mathrm{1D}}= \mathbb{Z}_2^3 \\
         & E^{\mathrm{0D}}=\{\mathrm{OFBS}\}^{\mathrm{0D}}/\{\mathrm{TBS}\}^{\mathrm{0D}}=\mathbb{Z}_2.
    \end{aligned}
\end{equation}
The ultimate classification is:
\begin{equation}
\mathcal{G}_0 =  \mathbb{Z}_2^4.
\end{equation}
This result applies to both $D_2$ and $D_6$.

\subsubsection{Spinful fermions}\label{sec:D2 D6 spinful}

We first consider the decoration of $0$D block-states. For group \( D_n \) ($n=2,6$) , $0$D blocks have a physical symmetry group \( G_b = D_n \times \mathbb{Z}_2^T \), and the corresponding 0D FSPT phases are characterized by the data \( (n_0, \nu_1) \):
\begin{equation}
    \begin{aligned}
        & n_0 \in \mathcal{H}^0(G_b, \mathbb{Z}_2^f) = \mathbb{Z}_2, \\
        & \nu_1 \in \mathcal{H}^1(G_b, U_T(1)) = \mathbb{Z}_2 \times \mathbb{Z}_{2}.
    \end{aligned}
\end{equation}

As in Sec.~\ref{sec:D4 spinful discussion}, there are no non-trivial cocycles contributed from the subgroup \( \mathbb{Z}_2^T \subset G_b \), and only \( n_0 = 0 \) is obstruction-free. Therefore, the 0D FSPT classification group is \( \mathbb{Z}_2 \times \mathbb{Z}_{2} \), where each element of the group can be labeled by the eigenvalues \( \pm 1 \) of the operators \( C_n \) and \( M \).

Next, we consider the decoration of $1$D block-states. All $1$D blocks possess the physical symmetry group \begin{equation}
   G_b = \mathbb{Z}_2^M \times \mathbb{Z}_2^T. 
\end{equation} 
Thus, there are two possible $1$D block-states: the single Majorana chain and the 1D FSPT state. 

Specifically, the 1D FSPT state has a \( \mathbb{Z}_4\) classification, where the generator can be realized by two Majorana chains (Eq.~\eqref{form:twochain}) and is referred to  \( n_1^T \) phase. This conclusion has already been established in  Sec.~\ref{sec:D4 spinful discussion}. 
However, as concluded in the previous section, a single Majorana chain cannot be decorated here, due to the absence of an on-site symmetry action satisfying $T^2=-1$.

For $2$D blocks, the proper root phase is the class-DIII topological superconductor, as in Sec.~\ref{sec:D4 spinful discussion}. And the corresponding 2D FSPT phases have a $\mathbb{Z}_2$ classification.

\begin{figure}[tb]
    \centering
    \includegraphics[width=0.3\textwidth]{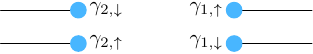}
    \caption{Dangling Majorana modes \(\gamma_{i,\uparrow}\) and \(\gamma_{i,\downarrow}\) (\(i=1,2\)) from the 1D FSPT decoration of the \(\tau_1\) (or \(\tau_2\)) blocks on \(\mu\) for the group \(D_2\).}
        \label{fig:Dn_spinless_no_open_edge_condition_D2}
\end{figure}

We now identify all block-state decorations that satisfy the no-open-edge conditions. First, all $0$D block-states naturally satisfy the required conditions. So the obstruction-free $0$D block-states form the group:
$\{\mathrm{OFBS}\}^{0D} = \mathbb{Z}_2^2.$

Next, we consider the decoration of $1$D block-states. 
For dihedral groups $D_2$ and $D_6$, decorating the $1$D block $\tau_1$ (or $\tau_2$) with the $n_1^T$ FSPT phase—realized by two Majorana chains—is obstruction-free. We now proceed to prove this statement.

\begin{figure}[tb]
    \centering
   \includegraphics[width=0.3\textwidth]{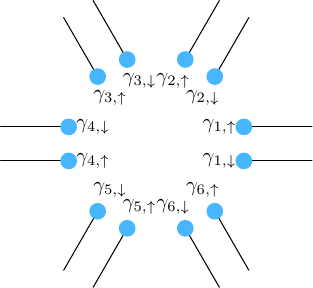}
    \caption{Dangling Majorana modes \(\gamma_{i,\uparrow}\) and \(\gamma_{i,\downarrow}\) (\(i=1-6\)) from the 1D FSPT decoration of the \(\tau_1\) (or \(\tau_2\)) blocks on \(\mu\) for the group \(D_6\).}
        \label{fig:Dn_spinless_no_open_edge_condition_D6}
\end{figure}

The resulting dangling Majorana (see Figs.~\ref{fig:Dn_spinless_no_open_edge_condition_D2},\ref{fig:Dn_spinless_no_open_edge_condition_D6}) contributed by the two-Majorana-chain decorations on \( \tau_1 \) (or \( \tau_2 \)) have the following symmetry properties. 

For the group \( D_2 \), the symmetry transformations are:
\begin{equation}
  \begin{aligned}
    {T}:~ & i \mapsto -i,  \\
             & (\gamma_{j, \uparrow}, \gamma_{j, \downarrow}) \mapsto (-\gamma_{j, \downarrow}, \gamma_{j, \uparrow}),  \\
    {C_2}:~ & (\gamma_{1, \uparrow}, \gamma_{2, \uparrow}) \mapsto (\gamma_{2, \uparrow}, -\gamma_{1, \uparrow}),  \\
               & (\gamma_{1, \downarrow}, \gamma_{2, \downarrow}) \mapsto (\gamma_{2, \downarrow}, -\gamma_{1, \downarrow}), \\
    {M}:~ & (\gamma_{1, \uparrow}, \gamma_{2, \uparrow}) \mapsto (-\gamma_{1, \downarrow}, \gamma_{2, \downarrow}),  \\
             & (\gamma_{1, \downarrow}, \gamma_{2, \downarrow}) \mapsto (\gamma_{1, \uparrow}, -\gamma_{2, \uparrow}).
\end{aligned}  
\end{equation}
We can straightforwardly construct symmetry-preserving interaction terms to open the spectrum gap for the dangling Majorana modes. These terms are given by:
\begin{align}
 i (\gamma_{1, \uparrow} \gamma_{2, \uparrow} - \gamma_{1, \downarrow} \gamma_{2, \downarrow})
\end{align}

For the group \( D_6 \), the symmetry transformations are:
\begin{align}
    {T}:~ & \left\{ \begin{aligned}
             & i \mapsto -i, \\
             & (\gamma_{j, \uparrow}, \gamma_{j, \downarrow}) \mapsto (-\gamma_{j, \downarrow}, \gamma_{j, \uparrow}), \quad j = 1, 2, \dots, 6,
    \end{aligned} \right. \notag \\
    {C_6}:~ & \left\{ \begin{aligned}
               & (\gamma_{j, \uparrow}, \gamma_{j, \downarrow}) \mapsto (\gamma_{j+1, \uparrow}, \gamma_{j+1, \downarrow}), \quad j = 1, 2, \dots, 5, \\
               & (\gamma_{6, \uparrow}, \gamma_{6, \downarrow}) \mapsto (-\gamma_{1, \downarrow}, -\gamma_{1, \downarrow}),
    \end{aligned} \right. \notag \\
    {M}:~ & \left\{ \begin{aligned}
             & (\gamma_{1, \uparrow}, \gamma_{2, \uparrow}, \gamma_{3, \uparrow}, \gamma_{4, \uparrow}, \gamma_{5, \uparrow}, \gamma_{6, \uparrow}) \\
             & \mapsto (-\gamma_{1, \downarrow}, \gamma_{6, \downarrow}, \gamma_{5, \downarrow}, \gamma_{4, \downarrow}, \gamma_{3, \downarrow}, \gamma_{2, \downarrow}), \\
             & (\gamma_{1, \downarrow}, \gamma_{2, \downarrow}, \gamma_{3, \downarrow}, \gamma_{4, \downarrow}, \gamma_{5, \downarrow}, \gamma_{6, \downarrow}) \\
             & \mapsto (\gamma_{1, \uparrow}, -\gamma_{6, \uparrow}, -\gamma_{5, \uparrow}, -\gamma_{4, \uparrow}, -\gamma_{3, \uparrow}, -\gamma_{2, \uparrow}).
    \end{aligned} \right.
\end{align}

We can also construct symmetry-preserving interaction terms as follows to open the gap for the dangling Majorana modes:
\begin{align}
     i \sum_{j=1,2,3} (\gamma_{j, \uparrow} \gamma_{j+3, \uparrow} - \gamma_{j, \downarrow} \gamma_{j+3, \downarrow}).
\end{align}

Therefore, $n_1^T$ FSPT decorations on $\tau_1$ (or $\tau_2$) are obstruction-free and form the group \begin{equation}
  \{ \mathrm{OFBS} \}^{1D} = \mathbb{Z}_4^2,   
\end{equation}  for the groups $D_2$ and $D_6$.

Next, we consider the block-state decoration on a two-dimensional block \( \sigma \). When a DIII topological superconductor (TSC) is decorated on each $2$D block \( \sigma \), four chiral Majorana modes appear along each reflection axis \( \tau_i \). As in the \( D_4 \) case, symmetric mass terms can be consistently defined on all $1$D blocks.
 The dangling modes on the $0$D blocks can be fully gapped, ensuring that the decoration of the 2D DIII TSC is obstruction-free. Consequently, it contributes to the classification group \begin{equation}
    \{ \mathrm{OFBS} \}^{2D} = \mathbb{Z}_2.
\end{equation}

We now discuss bubble equivalences. 
First, we decorate a pair of complex fermions on each $1$D block \( \tau_1 \) (or \( \tau_2 \)). Since \( \mathbb{Z}_2^T \) does not contribute to a non-trivial classification, we only need to consider the symmetry transformations \( M \) and \( C_n \) ($n=2,6$). Near each $0$D block, there are \( n \) complex fermions contributed from adjacent bubbles, forming the following atomic insulator:
\begin{equation}
\ket{TR} = c_1^\dagger c_2^\dagger \dots c_n^\dagger .
\end{equation}
For even dihedral groups \( D_n \) ($n=2,6$), complex fermions transform under the group \( G_f \) as follows:
\begin{equation}
    \begin{aligned}
        {C_n}: & c^\dagger_j \mapsto c^\dagger_{j+1}, \quad c^\dagger_n \mapsto -c^\dagger_1,~j=1,2,\dots,n\pmod{n} \\
        {M}:& \\
       &\text{for $D_2$:} ~  c_1^\dagger \mapsto i c^\dagger_1, \quad c^\dagger_2 \mapsto -i c^\dagger_{2} \\
       &\text{for $D_6$:} ~   c_1^\dagger \mapsto i c^\dagger_1, \quad c^\dagger_j \mapsto -i c^\dagger_{8-j},~j=2,\dots,6  
    \end{aligned}
\end{equation}

Then, the state \( \ket{TR} \) obeys the following symmetry properties:
\begin{equation}
    {C_n} \ket{TR} = \ket{TR}, \quad {M} \ket{TR} = \ket{TR}.
\end{equation}
Similarly, the decoration of $1$D bubbles on both $1$D blocks \( \tau_1 \) and \( \tau_2 \) produces states \( \ket{TR} \) that satisfy \begin{equation}
     {C_n} \ket{TR} = \ket{TR},~ {M} \ket{TR} = \ket{TR} .
\end{equation}
Thus, for dihedral groups $D_2$ and $D_6$, $1$D bubbles only contribute to a trivial classification group \begin{equation}
  \{\mathrm{TBS}\}^{0D} = \mathbb{Z}_1.  
\end{equation}

For reasons same to those discussed in Sec.~\ref{sec:D4 spinful discussion}, $2$D bubbles result in a one-dimensional state that is not protected by time-reversal symmetry. Consequently, there is no trivialization group for 1D obstruction-free states, i.e. \( \{\mathrm{TBS}\}^{1D} = \mathbb{Z}_1 \).

With all non-trivial block-states identified, we now examine the group structure of the ultimate classification. We find that the 1D obstruction-free and trivialization-free block-state cannot extend to 0D obstruction-free and trivialization-free block-state, following the analysis in Sec.~\ref{sec:D4 spinful discussion}. Then consider extension problem between 2D and 1D. 
We decorate a trivial block-state, i.e.  two copies of DIII TSC on each $2$D block $\sigma$.
 On each $1$D block, we can then add mass terms for the chiral Majorana fermions (see Fig.~\ref{figure:2 effective kitaev chain})  contributed from nearby $2$D block-states.  As concluded in Sec.~\ref{sec:D4 spinful discussion}, each pair of terms connected by the reflection operation exhibits a relative negative sign, with a total of two negative signs. According to Ref.~\cite{Wang2020}, the number of such negative signs indicates the number of effective Majorana chains. 
This implies that mass terms on each $1$D block contribute non-trivial two Majorana chains and there exists a group extension in the final classification:
\begin{equation}
1 \mapsto \mathbb{Z}_4 \mapsto \mathbb{Z}_8 \mapsto \mathbb{Z}_2 \mapsto 1,
\end{equation}
where \( \mathbb{Z}_4 \) represents the decoration of the pair of Majorana chains on \( \tau_1 \) and \( \tau_2 \) simultaneously.

In summary, all independent non-trivial block-states with different dimensions are classified as follows:
\begin{align}
    \begin{aligned}
         & E^{\mathrm{2D}}=\{\mathrm{OFBS}\}^{\mathrm{2D}}=\mathbb{Z}_2   \\
         & E^{\mathrm{1D}}=\{\mathrm{OFBS}\}^{\mathrm{1D}}/\{\mathrm{TBS}\}^{\mathrm{1D}}=\mathbb{Z}_4^2   \\
         & E^{\mathrm{0D}}=\{\mathrm{OFBS}\}^{\mathrm{0D}}/\{\mathrm{TBS}\}^{\mathrm{0D}}=\mathbb{Z}_2^2
    \end{aligned}.
\end{align}
The ultimate classification $\mathcal{G}_{1/2}$  is $\mathbb{Z}_2^2 \times\mathbb{Z}_4 \times \mathbb{Z}_8$, as determined by the group extension discussed above.  This result applies to both $D_2$ and $D_6$.

\subsection{$D_3$}
The cell decomposition of the group $D_3$ is shown in panel (b) of Fig.~\ref{Dn cell}. We take $C_3$ (the $2\pi/3$ rotation) and $M_{\tau_j}$ (reflection about the $\tau_j$-block axis) as the generators of the $D_3$ group. And the cell decomposition remains invariant under the action of these generators.
\subsubsection{Spinless fermions} \label{sec:D3 spinless}
For the \( D_3 \) group, the $0$D blocks have a physical symmetry group \( G_b = D_3 \times \mathbb{Z}_2^T \). The corresponding 0D FSPT phases are characterized by the data \begin{equation}
   \mathcal{H}^0(G_b, \mathbb{Z}_2^f) \times \mathcal{H}^1(G_b, U_T(1))=\mathbb{Z}_2\times \mathbb{Z}_2, 
\end{equation}   where each \( \mathbb{Z}_2 \) can be labeled by the eigenvalues \( \pm 1 \) of \( P_f \) and \( M \).

Next, we consider $1$D block-state decorations. For $1$D block \( \tau_1 \) (or \( \tau_2 \)), the total symmetry group is  \begin{equation}
    G_f = \mathbb{Z}_2^M \times \mathbb{Z}_2^T \times \mathbb{Z}_2^f.
\end{equation}  Thus, there are two possible $1$D block-state candidates: the single Majorana chain and the 1D FSPT state. 
Specifically, the 1D FSPT state has a \( \mathbb{Z}_4 \times \mathbb{Z}_4 \) classification, with each group generator realized by two Majorana chains, referred to as the \( n_1^T \) phase and the \( n_1^M \) phase, respectively. This conclusion has already been established in Sec.~\ref{sec:D4 spinful discussion}. 

Furthermore, as discussed earlier, there is no root phase for $2$D block-states, as all $2$D blocks exhibit the symmetry \( G_f = \mathbb{Z}_2^T \times \mathbb{Z}_2^f \). Therefore, there are no  obstruction-free decorations for $2$D blocks, i.e., 
\begin{equation}
\{\mathrm{OFBS}\}^{2D} = \mathbb{Z}_1.
\end{equation}

With all the states now available for decoration, we proceed to examine which block-state decorations satisfy the no-open-edge conditions.  All zero-dimensional decorations inherently satisfy the no-open-edge conditions. So the obstruction-free $0$D block-states form the group:
\begin{equation}
    \{\mathrm{OFBS}\}^{0D} = \mathbb{Z}_2^2.
\end{equation}

\begin{figure}[tb]
    \centering
    \includegraphics[width=0.3\textwidth]{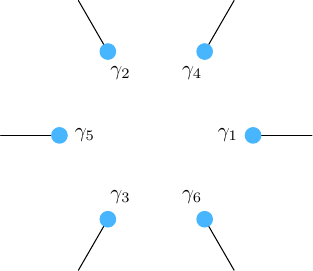}
    \caption{Single-Majorana-chain decorations on $1$D blocks labeled by \( \tau_1 \) and \( \tau_2 \), leaving six dangling Majorana modes at a $0$D block \( \mu \).}
    \label{D3_single_chain_tau1_and_tau2}
\end{figure}

\begin{figure}[tb]
    \centering
      \includegraphics[width=0.3\textwidth]{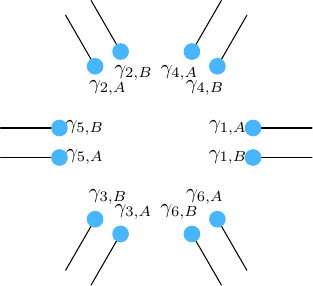}
    \caption{1D $n_1^T$ (or $n_1^M$) FSPT decoration on both $\tau_1$ and $\tau_2$ blocks, which leaves $12$ dangling Majorana modes at  $0$D block $\mu$.}
    \label{D3 2chain tau1 and tau2}
\end{figure}

A single Majorana chain decoration on $\tau_1$ (or $\tau_2$) cannot satisfy the no-open-edge condition due to an odd number of dangling zero modes at $\mu$. 
Consider decorating single Majorana chains on both types of $1$D blocks $\tau_1$ and $\tau_2$, which results in six dangling Majorana modes $\gamma_1,\gamma_2,\gamma_3,\dots,\gamma_6$ at the $0$D block $\mu$ (see Fig.~\ref{D3_single_chain_tau1_and_tau2}). Since the two types of $1$D blocks are independent, we can decorate single Majorana chains with opposite pairing directions on $\tau_1$ and $\tau_2$. Consequently, two types of dangling modes, $\gamma_j^L$ and $\gamma_j^R$, necessarily coexist at $\mu$, with distinct time-reversal symmetry actions. Near $\mu$, these Majorana modes transform under symmetries as follows ($\sigma=A,B$):
\begin{equation}
\begin{aligned}
T:&\left\{
\begin{aligned}
    ~ & i \mapsto -i, \\
    ~ & \gamma_j \mapsto \gamma_j, ~~~\quad j = 1,2,3, \\
    ~ & \gamma_j \mapsto -\gamma_j, \quad j = 4,5,6,
\end{aligned}
\right.\\
M:&~ \gamma_1 \mapsto \gamma_{1}, ~\gamma_5 \mapsto \gamma_{5},~ \gamma_2 \leftrightarrow \gamma_{3},~ \gamma_4 \leftrightarrow \gamma_{6}\\
C_3:& ~\gamma_1 \to \gamma_2 \to \gamma_3 \to \gamma_1,\quad 
       \gamma_4 \to \gamma_5 \to \gamma_6 \to \gamma_4.
\label{eq:D3 spinless sym prop of single}
\end{aligned}
\end{equation}

It can be verified that the following interaction term can lift the degeneracy of the ground state:
\begin{equation}
i (\gamma_1 \gamma_5 + \gamma_2 \gamma_6 + \gamma_3 \gamma_4),
\end{equation}
which is invariant under the symmetry transformations in Eq.~\eqref{eq:D3 spinless sym prop of single}. So, such a decoration is obstruction-free.

The decoration of \( n_1^T \) FSPT (via two Majorana chains) on $1$D blocks \( \tau_1 \) (or \( \tau_2 \)) is obstructed. Now we proceed to show that. 
This decoration contributes six dangling zero modes at $\mu$.  However, the local fermion parity $P_f=\prod_{j=1}^3i\gamma_{j,A}\gamma_{j,B}$ is violated by time-reversal symmetry, i.e. \begin{equation}
    T:P_f \to i\gamma_{1,A}\gamma_{1,B}\gamma_{2,A}\gamma_{2,B}\gamma_{3,A}\gamma_{3,B} = -P_f,
\end{equation} Thus, such decoration cannot satisfy the no-open-edge condition. 

Similarly, the decoration of  \( n_1^M \) FSPT phase on the $1$D blocks \( \tau_1 \) (or \( \tau_2 \)) is also obstructed. Because the local fermion parity is violated by reflection symmetry, i.e. \begin{equation}
  M:P_f\to-i\gamma_{1,B}\gamma_{1,A}\gamma_{3,B}\gamma_{3,A}\gamma_{2,B}\gamma_{2,A} = -P_f,  
\end{equation} such a decoration cannot satisfy the no-open-edge condition.

However, decorating 1D FSPT phases with opposite pairings on $\tau_1$ and $\tau_2$ will satisfy the no-open-edge condition, following the same strategy as above. We will present this conclusion explicitly by constructing symmetry-preserving mass terms that gap out all dangling modes.

1D FSPT phases with opposite pairing directions on $\tau_1$ and $\tau_2$, resulting in two types of dangling modes $\gamma_j^L$ and $\gamma_j^R$ at $\mu$ (see Fig.~\ref{D3 2chain tau1 and tau2}) with distinct time-reversal symmetry actions. 
For \( n_1^T \) phase, the decoration realized via two Majorana chains leaves twelve dangling Majorana modes. These modes transform under $0$D block symmetry as ($\sigma=A,B$):
\begin{equation}
\begin{aligned}
T:&\left\{
\begin{aligned}
    ~ & i \mapsto -i,\\
    ~ & \gamma_{j, \sigma} \mapsto \gamma_{j, \sigma}, ~~~\quad j = 1,2,3\\
    ~ & \gamma_{j, \sigma} \mapsto -\gamma_{j, \sigma},\quad j = 4,5,6
\end{aligned}
\right.\\
C_3:& ~\gamma_{1, \sigma} \to \gamma_{2, \sigma} \to \gamma_{3, \sigma} \to \gamma_{1, \sigma},~
       \gamma_{4, \sigma} \to \gamma_{5, \sigma} \to \gamma_{6, \sigma} \to \gamma_{4, \sigma}\\
M:&\left\{
\begin{aligned}
    ~ & \gamma_{j, \sigma} \mapsto \gamma_{j, \sigma}, ~~~\quad j = 1,5,\\
    ~ & \gamma_{j, \sigma} \leftrightarrow \gamma_{j^\prime, \sigma}, \quad (j, j^\prime) = (2, 3), (4, 6)
\end{aligned}
\right.
\end{aligned}
\end{equation}

A symmetric interaction term can be introduced to gap out the dangling modes, resulting in a non-degenerate gapped ground state:
\begin{align}
    & i \sum_{\sigma = A,B} (\gamma_{1, \sigma} \gamma_{5, \sigma} + \gamma_{2, \sigma} \gamma_{6, \sigma} + \gamma_{3, \sigma} \gamma_{4, \sigma}) 
    \label{eq:D3 spinless unique GS}
\end{align}
For \( n_1^M \) phase, the decoration leaves twelve dangling Majorana modes with the following symmetry properties ($\sigma=A,B$):
\begin{equation}
\begin{aligned}
    T:&\left\{ \begin{aligned}
        ~ & i \mapsto -i, \\
    ~&\gamma_{j, A} \mapsto \gamma_{j, A},\quad \gamma_{j, B} \mapsto -\gamma_{j, B},\quad j = 1,2,3,  \\
    ~ &   \gamma_{j, A} \mapsto -\gamma_{j, A},\quad \gamma_{j, B} \mapsto \gamma_{j, B} ,\quad j = 4,5,6,
    \end{aligned} \right. \\
C_3:& ~\gamma_{1, \sigma} \to \gamma_{2, \sigma} \to \gamma_{3, \sigma} \to \gamma_{1, \sigma},~
       \gamma_{4, \sigma} \to \gamma_{5, \sigma} \to \gamma_{6, \sigma} \to \gamma_{4, \sigma}\\
    M:&\ \left\{~ \begin{aligned}
        & \gamma_{j, A} \mapsto \gamma_{j, A},\quad \gamma_{j, B} \mapsto -\gamma_{j, B}, \quad j = 1,5, \\
        & \gamma_{j, A} \mapsto \gamma_{j^\prime, A},\quad \gamma_{j, B} \mapsto -\gamma_{j^\prime, B} \quad (j, j^\prime) = (2, 3), (4, 6), 
    \end{aligned}\right.
\end{aligned}
\end{equation}
The same mass term in Eq.~\eqref{eq:D3 spinless unique GS} can be applied here to gap out the dangling boundary modes, ensuring a non-degenerate gapped ground state.

In summary, all obstruction-free $1$D block-states can be generated by stacking the following elementary decorations:
\begin{enumerate}
    \item Single Majorana chains decorated on both \(\tau_1\) and \(\tau_2\) blocks;
    \item 1D FSPT phases \(n_1^T\) decorated on both \(\tau_1\) and \(\tau_2\) blocks;
    \item 1D FSPT phases  \(n_1^M\) decorated on both \(\tau_1\) and \(\tau_2\) blocks.
\end{enumerate}
These states form the following group:
\begin{equation}
\{\mathrm{OFBS}\}^{1D} = \mathbb{Z}_4 \times \mathbb{Z}_8,
\end{equation} where the \(\mathbb{Z}_4\) factor is generated by the 1D FSPT phase \(n_1^M\) decoration, while the \(\mathbb{Z}_8\) is generated by the single Majorana chain decoration. 
This $\mathbb{Z}_8$ classification arises because stacking two individual Majorana chains yields a 1D FSPT $n_1^T$ phase. This corresponds to the 1D BDI-class TSC phases, as previously established in Eq.~\eqref{eq:Z8}.

We now discuss bubble equivalences. As mentioned in Sec.~\ref{sec:D4 spinless fermion}, $2$D bubbles contribute to 1D non-trivial FSPT phases (belonging to the \( n_1^M+ n_1^T \) phase) and trivialize a \( \mathbb{Z}_4 \) subgroup from the obstruction-free classification. In other words, all  obstruction-free decorations of the \( n_1^M+ n_1^T \) phase on both blocks $\tau_1$ and $\tau_2$ are trivial. 

Moreover, as we will show below, all $0$D block-states are already trivialized by $1$D bubbles. Therefore, we do not consider the effect of $2$D bubbles on the $0$D block-states.
Thus, $2$D bubbles contribute to the trivialization group \begin{equation}
    \{\mathrm{TBS}\}^{1D} = \mathbb{Z}_4
\end{equation}   in total, with its generator representing the \( n_1^M+ n_1^T \) FSPT state decorations on $\tau_1$ and $\tau_2$  simultaneously.

Next, we consider the $1$D bubble equivalences. We decorate a pair of complex fermions on each $1$D block \( \tau_1 \) (or \( \tau_2 \)). Near each $0$D block, there are \( 3 \) complex fermions contributed from adjacent bubbles that form the following atomic insulator:
\begin{equation}
\ket{TR} = c_1^\dagger c_2^\dagger c_3^\dagger.
\end{equation}
Since the subgroup \( \mathbb{Z}_2^T \) does not contribute to a non-trivial classification, we only consider the symmetry transformation properties of \( \ket{TR} \) under \( M \) and \( P_f \).
For the odd dihedral group \( D_3 \), we define the symmetry transformations of complex fermions as follows:
\begin{equation}
    \begin{aligned}
        {M}:~   & c^\dagger_1 \mapsto c^\dagger_1, \quad c^\dagger_2 \mapsto c^\dagger_3, \quad c^\dagger_3 \mapsto c^\dagger_2,
    \end{aligned}
\end{equation}
Then the state \( \ket{TR} \) satisfies the symmetry properties: \begin{equation}
  P_f \ket{TR} = -\ket{TR},~ M \ket{TR} = -\ket{TR}.  
\end{equation}  Moreover, the decoration of $1$D bubbles on both \( \tau_1 \) and \( \tau_2 \) does not introduce any non-trivial $0$D block-states. 

However, we can instead decorate the $\tau_1$ (or $\tau_2$) with bosonic $1$D bubbles built from bosonic operators. This produces a trivial state $\ket{TR}$ on each block $\mu$,
\begin{equation}
\ket{TR}=b_1^\dagger b_2^\dagger b_3^\dagger ,
\end{equation}
where $b_i^\dagger$ are mutually commuting bosonic creation operators, each carrying reflection eigenvalue $-1$. Under the symmetry actions, $\ket{TR}$ obeys \begin{equation}
    P_f\ket{TR}=\ket{TR},~M\ket{TR}=-\ket{TR}.
\end{equation}

Therefore, the decoration of $1$D bubbles provides two independent eigenstates of \( M \) and \( P_f \), contributing to the trivialization group \begin{equation}
    \{\mathrm{TBS}\}^{\mathrm{0D}}=\mathbb{Z}_2^2.
\end{equation}

Among all possible extensions, the only potentially relevant one would be from 1D to 0D. However, this issue does not arise here: all $0$D block-states are already trivialized by $1$D bubbles. Therefore, we do not need to further consider any 1D-to-0D extension.

In summary, all independent non-trivial block-states with different dimensions are classified as follows:
\begin{equation}
\begin{aligned}
         & E^{\mathrm{2D}}=\{\mathrm{OFBS}\}^{\mathrm{2D}}=\mathbb{Z}_1                       \\
         & E^{\mathrm{1D}}=\{\mathrm{OFBS}\}^{\mathrm{1D}}/\{\mathrm{TBS}\}^{\mathrm{1D}}= \mathbb{Z}_8 \\
         & E^{\mathrm{0D}}=\{\mathrm{OFBS}\}^{\mathrm{0D}}/\{\mathrm{TBS}\}^{\mathrm{0D}}=\mathbb{Z}_1.
    \end{aligned}
\end{equation}
The ultimate classification is:
\begin{equation}
\mathcal{G}_0 =  \mathbb{Z}_8.
\end{equation}

\subsubsection{Spinful fermions} \label{sec:D3 spinful}
For spinful systems, we first consider the decoration of $0$D block-states. For \( D_3 \) group, $0$D blocks have a physical symmetry group \( G_b = D_3 \times \mathbb{Z}_2^T \), and the corresponding 0D FSPT phase is characterized by the data \( (n_0, \nu_1) \):
\begin{equation}
    \begin{aligned}
        & n_0 \in \mathcal{H}^0(G_b, \mathbb{Z}_2^f) = \mathbb{Z}_2, \\
        & \nu_1 \in \mathcal{H}^1(G_b, U_T(1)) = \mathbb{Z}_2 
    \end{aligned}
\end{equation}

Similar to the discussion in Sec.~\ref{sec:D4 spinful discussion}, there are no non-trivial cocycles contributed by the subgroup \( \mathbb{Z}_2^T \subset G_b \), and only \( n_0 = 0 \) is obstruction-free. Consequently, the 0D FSPT classification group is \( \mathbb{Z}_2\), where each element of the group can be labeled by the eigenvalues \( \pm 1 \) of the operator \( M \). And the obstruction-free $0$D block-states form the group:
\begin{equation}
  \{\mathrm{OFBS}\}^{0D} = \mathbb{Z}_2.  
\end{equation}

Next, we consider decorating $1$D block-states. All $1$D blocks have the physical symmetry group $G_b = \mathbb{Z}_2^M \times \mathbb{Z}_2^T$, leading to two possible states:
\begin{itemize}
    \item A single Majorana chain.
    \item A 1D FSPT state with a $\mathbb{Z}_4$ classification (realized by two Majorana chains, as in Eq.~\eqref{form:twochain}), referred to as the $n_1^T$ phase (Sec.~\ref{sec:D4 spinful discussion}).
\end{itemize}

However, a single Majorana chain cannot be decorated, as established earlier, due to the lack of valid on-site symmetry actions satisfying $T^2 = -1$.

\begin{figure}[tb]
    \centering
   \includegraphics[width=0.3\textwidth]{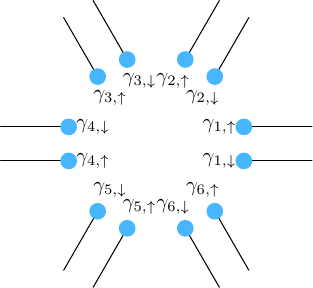}
    \caption{Dangling Majorana modes \(\gamma_{i,\uparrow}\) and \(\gamma_{i,\downarrow}\) (\(i=1-6\)) arising from the 1D FSPT decoration of the \(\tau_1\) and \(\tau_2\) blocks on \(\mu\) for the group \(D_3\).}
            \label{fig:Dn spinless no open edge condition D3 tau1 and tau2}
\end{figure}

Decoration of $n_1^T$ phases on $\tau_1$ (or $\tau_2$) remains obstructed.  Specifically, the decorations of two, four, or six Majorana chains on the $1$D block \( \tau_1 \) (or \( \tau_2 \)) are obstructed, as \( 3 \times n \) (for \( n = 2, 4, 6 \)) is not a multiple of 8. The detailed justification for this conclusion will be provided in Sec.~\ref{sec:appendix spinless Cn}.

However, if we simultaneously decorate 1D $n_1^T$ FSPT phase, realized by two Majorana chains, on both $1$D blocks \( \tau_1 \) and \( \tau_2 \), we leave 12 dangling Majorana zero modes (see Fig.~\ref{fig:Dn spinless no open edge condition D3 tau1 and tau2}) at the $0$D block \( \mu \). In this case, we can construct a symmetry-preserving mass term to open a spectral gap and obtain a unique ground state:
\begin{equation}
     i \sum_{j=1}^{3} \left( \gamma_{j, \uparrow} \gamma_{j+3, \uparrow} - \gamma_{j, \downarrow} \gamma_{j+3, \downarrow} \right)
\end{equation}
where \( j = 1, 3, 5 \) and \( \sigma = \uparrow, \downarrow \).
This mass term is invariant under the following symmetry transformations:
\begin{align}
    {T}:~ & i \mapsto -i, \notag \\
             & (\gamma_{j, \uparrow}, \gamma_{j, \downarrow}) \mapsto (-\gamma_{j, \downarrow}, \gamma_{j, \uparrow}), \quad 1 \leq j \leq 6, \notag \\
    {C_3}:~ & (\gamma_{j, \uparrow}, \gamma_{5, \uparrow}, \gamma_{6, \uparrow}) \mapsto (\gamma_{j+2, \uparrow}, -\gamma_{1, \uparrow}, -\gamma_{2, \uparrow}), \notag \\
               & (\gamma_{j, \downarrow}, \gamma_{5, \downarrow}, \gamma_{6, \downarrow}) \mapsto (\gamma_{j+2, \downarrow}, -\gamma_{1, \downarrow}, -\gamma_{2, \downarrow}), \quad 1 \leq j < 5, \notag \\
    {M}:~ & (\gamma_{1, \uparrow}, \gamma_{2, \uparrow}, \gamma_{6, \uparrow}) \mapsto (-\gamma_{1, \downarrow}, \gamma_{6, \downarrow}, \gamma_{2, \downarrow}), \notag \\
             & (\gamma_{3, \uparrow}, \gamma_{4, \uparrow}, \gamma_{5, \uparrow}) \mapsto (\gamma_{5, \downarrow}, \gamma_{4, \downarrow}, \gamma_{3, \downarrow}), \notag \\
             & (\gamma_{1, \downarrow}, \gamma_{2, \downarrow}, \gamma_{6, \downarrow}) \mapsto (\gamma_{1, \uparrow}, -\gamma_{6, \uparrow}, -\gamma_{2, \uparrow}), \notag \\
             & (\gamma_{3, \downarrow}, \gamma_{4, \downarrow}, \gamma_{5, \downarrow}) \mapsto (-\gamma_{5, \uparrow}, -\gamma_{4, \uparrow}, -\gamma_{3, \uparrow}).
\end{align}

Therefore, for the \( D_3 \) group, the simultaneous decoration of 1D $n_1^T$ phases on both \( \tau_1 \) and \( \tau_2 \) is the only obstruction-free scheme, which contributes to a classification group \begin{equation}
    \{ \mathrm{OFBS} \}^{1D} = \mathbb{Z}_4.
\end{equation} 

Next, we consider the block-state decoration on a two-dimensional block \( \sigma \). For the $2$D blocks, the proper root phase is identified as the class DIII topological superconductor, which aligns with the analysis presented in Sec.~\ref{sec:D4 spinful discussion}.
When a DIII topological superconductor (TSC) is decorated on each $2$D block \( \sigma \), four chiral Majorana modes appear along each reflection axis \( \tau_i \). As in the \( D_4 \) case, symmetric mass terms can be consistently defined on all $1$D blocks. 
The dangling modes on the $0$D blocks can be fully gapped, ensuring that the decoration of the 2D DIII TSC is obstruction-free. Consequently, it contributes to the classification group: \begin{equation}
    \{ \mathrm{OFBS} \}^{2D} = \mathbb{Z}_2.
\end{equation}

We now discuss bubble equivalences. 
First, we decorate a pair of complex fermions on each $1$D block \( \tau_1 \) (or \( \tau_2 \)). Since \( \mathbb{Z}_2^T \) does not contribute to a non-trivial classification, we only need to consider the symmetry transformation \( M \). Near each $0$D block, there are \( 3 \) complex fermions contributed from adjacent bubbles, forming the following atomic insulator:
\begin{equation}
\ket{TR} = c_1^\dagger c_2^\dagger c_3^\dagger.
\end{equation}
For odd dihedral groups \( D_3 \), complex fermions transform under the group \( G_f \) as follows:
\begin{equation}
    \begin{aligned}
        {C_3}:~ & c^\dagger_1 \mapsto c^\dagger_{2}, \quad c^\dagger_2 \mapsto c^\dagger_{3}, \quad c^\dagger_3 \mapsto -c^\dagger_1,  \\
        {M}:~   & c_1^\dagger \mapsto  c^\dagger_1, \quad c^\dagger_2 \mapsto - c^\dagger_{3}, \quad c^\dagger_3 \mapsto - c^\dagger_{2}
    \end{aligned}
\end{equation}

Then, the $1$D bubbles on the $1$D block \( \tau_1 \) (or \( \tau_2 \)) contribute a non-trivial state \( \ket{TR} \) that satisfies \( C_3 \ket{TR} = -\ket{TR} \) and \( M \ket{TR} = -\ket{TR} \), while the $1$D bubbles on both \( \tau_1 \) and \( \tau_2 \) contribute to a trivial state. 
Therefore, for \( D_3 \), $1$D bubbles contribute to a non-trivial classification group \begin{equation}
    \{\mathrm{TBS}\}^{0D} = \mathbb{Z}_2.
\end{equation} 

For analogous reasons to those detailed in Sec.~\ref{sec:D4 spinful discussion}, $2$D bubbles result in a 1D state \( \ket{\mathbf{X}} \) that lacks protection under time-reversal symmetry.
 Consequently, there is no trivialization group for 1D obstruction-free states. Since all $0$D block-states are already trivialized by $1$D bubbles, we do not consider the effect of $2$D bubbles on the $0$D block-states.  Thus, $2$D bubbles contribute to the trivialization group \begin{equation}
    \{\mathrm{TBS}\}^{1D} = \mathbb{Z}_1. \end{equation}

Then we now consider the extension problem. As in Sec.~\ref{sec:D4 spinful discussion}, the chiral Majorana modes (see Fig.~\ref{figure:2 effective kitaev chain}) arising from 2D trivial decorations admit mass terms on each reflection axis. We obtain two effective Majorana chains because of the relative minus sign of the mass terms.
This implies that there exists a group extension in the final classification:
\begin{equation}
1 \mapsto \mathbb{Z}_4 \mapsto \mathbb{Z}_8 \mapsto \mathbb{Z}_2 \mapsto 1,
\end{equation}
where \( \mathbb{Z}_4 \) represents the decoration of the pair of Majorana chains on \( \tau_1 \) and \( \tau_2 \) simultaneously.
However, we find no extension from 1D block-states to 0D block-states in this case.

In summary, all independent non-trivial block-states with different dimensions are classified as follows:
\begin{align}
    \begin{aligned}
         & E^{\mathrm{2D}}=\{\mathrm{OFBS}\}^{\mathrm{2D}}=\mathbb{Z}_2   \\
         & E^{\mathrm{1D}}=\{\mathrm{OFBS}\}^{\mathrm{1D}}/\{\mathrm{TBS}\}^{\mathrm{1D}}=\mathbb{Z}_4   \\
         & E^{\mathrm{0D}}=\{\mathrm{OFBS}\}^{\mathrm{0D}}/\{\mathrm{TBS}\}^{\mathrm{0D}}=\mathbb{Z}_1
    \end{aligned}.
\end{align}
The ultimate classification $\mathcal{G}_{1/2}$  is $ \mathbb{Z}_8$, as determined by the group extension discussed above.

\subsection{\( C_n \) } 

\subsubsection{Spinless fermions} \label{sec:appendix spinless Cn}
In this section, we detail the real-space construction of point group $C_n$ ($n=2,3,4,6$) and time-reversal symmetry-protected FSPT phases in 2D interacting spinless fermionic systems. The cell decomposition is illustrated in Fig.~\ref{Cn cell}. We alternately use $C_n$ to denote either the $n$-fold rotation group or its generator (the $2\pi/n$ rotation operator).

\paragraph{Block-State Decoration}

We first investigate the $0$D block-state decorations. For a general \( C_n \) group, the $0$D block \( \mu \) has a physical symmetry group \( G_b = \mathbb{Z}_n \times \mathbb{Z}_2^T \), and the corresponding 0D FSPT is characterized by the data \( (n_0, \nu_1) \):
\begin{equation}
    \begin{aligned}
         & n_0 \in \mathcal{H}^0(G_b, \mathbb{Z}_2^f) = \mathbb{Z}_2, \\
         & \nu_1 \in \mathcal{H}^1(G_b, U_T(1)) = \mathbb{Z}_{(2, n)},
    \end{aligned}
\end{equation}
where \( (m, n) \) denotes the greatest common divisor of \( m \) and \( n \).
The non-trivial cocycle of \( \nu_1 \) is \( \nu_1(C_n^{n/2}) = -1 \) when \( n \) is an even number. There are no non-trivial cocycles contributed from the subgroup \( \mathbb{Z}_2^T \subset G_b \). 
So the 0D FSPT classification group is simply the direct product of \( \mathcal{H}^0(G_b, \mathbb{Z}_2^f) \) and \( \mathcal{H}^1(G_b, U_T(1)) \). For an even-fold cyclic group, this results in \( \mathbb{Z}_2^2 \). Each \( \mathbb{Z}_2 \) can be labeled by the eigenvalues \( \pm 1 \) of \( P_f \) and \( C_n \). For an odd-fold cyclic group, the classification group is \( \mathbb{Z}_2 \). Each \( \mathbb{Z}_2 \) can be labeled by the eigenvalue \( \pm 1 \) of \( P_f \).

Next, we consider the $1$D block-state decoration. All $1$D blocks exhibit the symmetry \( G_f = \mathbb{Z}_2^T \times \mathbb{Z}_2^f \), and it is well known that the classification group of the corresponding 1D FSPT phases is \( \mathbb{Z}_4 \)~\cite{Wang2020}. However, a single Majorana chain is also a proper candidate to decorate. In fact, all the proper block-states constitute 1D class BDI topological superconductor~\cite{Fidkowski_2010}, each consisting of $v$ Majorana chains. Under interactions, they exhibit a $\mathbb{Z}_8$ classification, with $v=1,2,\ldots,8$.
The symmetry properties of these Majorana chains are given as follows (integer \( j \) denotes sites):
\begin{align}
    T: & \quad i \mapsto -i, \notag \\
       & \quad \gamma^L_j \mapsto \gamma^L_j, \quad \gamma^R_j \mapsto -\gamma^R_j. \label{eq:symm property spinless Cn 2chain}
\end{align}

Additionally, there is no root phase for $2$D block-states, since all $2$D blocks exhibit the symmetry group \( G_f = \mathbb{Z}_2^T \times \mathbb{Z}_2^f \), consistent with the analysis in Ref.~\cite{Wang2020}.

Now that we have identified all the states available for decoration, we proceed to determine which block-state decorations satisfy the no-open-edge conditions. 

All zero-dimensional block-state decorations naturally satisfy the required conditions. Consequently, the obstruction-free $0$D block-states constitute the groups:
\begin{equation}
    \{\mathrm{OFBS}\}^{\mathrm{0D}} = 
    \begin{cases}
        \mathbb{Z}_2^2 & \text{for even-order cyclic groups}, \\
        \mathbb{Z}_2                      & \text{for odd-order cyclic groups}.
    \end{cases}
\end{equation}

Next, we consider $1$D block-state decorations. Recall that the block-state candidates  admit a $\mathbb{Z}_8$ classification under time-reversal symmetry.  Therefore, for $1$D block-states to be obstruction-free when built from $nv$  Majorana chains, $nv$ must be a multiple of 8~\cite{RotationSPT}. The remaining task is to determine whether there exist $C_n$-invariant interaction terms ($n=2,3,4,6$) capable of gapping out the Majorana zero modes at the rotation center. 

 For $n=4$, decorations with $v=2,4,6,8$  Majorana chains on the $1$D block $\tau$  are time-reversal invariant.  For $n=2,6$, decorations with $v=4,8$  Majorana chains on the $1$D block $\tau$  are time-reversal invariant.  However, for $v=3$, no proper time-reversal invariant decorations exist.

\begin{figure}[tb]
    \centering
    \includegraphics[width=0.3\textwidth]{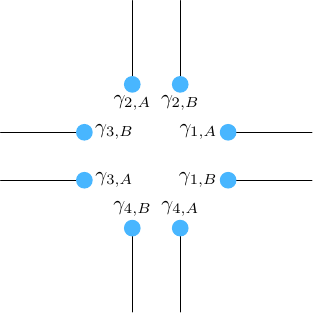}
    \caption{1D FSPT decoration on $1$D blocks labeled by $\tau$, which leaves $8$ dangling Majorana modes  at  $0$D block $\mu$.}
    \label{fig:Cn spinless 2chain}
    
\end{figure}

The explicit $C_4$-invariant interaction term to open the gap for the eight dangling Majorana fermions on \( \mu \) (see Fig.~\ref{fig:Cn spinless 2chain}) has been introduced by Meng and Wang~\cite{RotationSPT}. The 1D  block-state decorations with $v=2,4,6,8$ Majorana chains are obstruction-free and form the group \( \{\mathrm{OFBS}\}^{\mathrm{1D}}=\mathbb{Z}_4 \). Notably, this obstruction-free decoration represents an intrinsic interacting FSPT phase that cannot be realized in free-fermion systems. This follows from the fact that all possible mass terms (of the form $i\gamma_{j} \gamma_{k}$ or their combinations) break time-reversal symmetry.

For groups $C_2$ and $C_6$, the only possible decoration with $v=4$  Majorana chains belongs to BSPT phases. So it is straightforward to follow  Ref.~\cite{Song_2020} to identify whether such a decoration satisfies no-open-edge condition. We denote the wavefunction of such BSPT phases by $[\alpha] \in  \mathcal{H}^2( \mathbb{Z}_2^T \times \mathbb{Z}_2^f,U_T(1))=\mathbb{Z}_2^2$. Since the rotational operation $C_n$ preserves the local degrees of freedom of each wavefunction, we have
\begin{equation}[\alpha]+C_n[\alpha]+\dots+C_n^{n-1}[\alpha]=n[\alpha]=0,\end{equation} with $n=2,6$.  Consequently, only the $1$D block-state decorations with $v=4,8$ Majorana chains are obstruction-free and form the group \(\{\mathrm{OFBS}\}^{\mathrm{1D}}= \mathbb{Z}_2 \).

\paragraph{Bubble equivalence \& Extension \& Ultimate classification} \label{sec:Cn spinless bubble}

With all obstruction-free block-states, below we will discuss all possible trivializations.
First, we consider the $2$D bubble equivalences: we decorate a ``Majorana bubble" on each $2$D block \( \sigma \). Each bubble consists of a Majorana chain with anti-periodic boundary condition. These bubbles can then be deformed into two Majorana chains on each adjacent $1$D block. It is important to note that the two chains have opposite pairing directions, due to the rotational symmetry of the bubbles. Moreover, at each boundary of these two chains, two types of dangling modes, \( \gamma^L \) and \( \gamma^R \), must appear simultaneously. And time-reversal symmetry acts on them differently.

However, these double Majorana chains do not correspond to any non-trivial 1D FSPT phases. We denote the corresponding dangling zero modes contributed by a 1D block at $\mu$ as $\gamma_1^L$ and $\gamma_2^R$. Typically, their symmetry properties are manifested as follows:\begin{equation}
\begin{aligned}
    T:~ & i \mapsto -i, \\
        & \gamma_1^L \mapsto \gamma_1^L, \quad \gamma_2^R \mapsto -\gamma^R_2.
\end{aligned}\end{equation}
It is straightforward to introduce a mass term, such as:
\begin{equation}
H_I = i\gamma_1^L \gamma_2^R,
\end{equation}
to open a spectral gap for the dangling Majorana modes. 
As a result, these two Majorana chains belong to a trivial phase. Consequently,  one-dimensional obstruction-free block-states cannot be trivialized by two-dimensional bubbles for any point group $C_n$. Furthermore, the trivialization group in one dimension vanishes, i.e., \begin{equation}
    \{\mathrm{TBS}\}^{\mathrm{1D}}=\mathbb{Z}_1.
\end{equation}

2D bubbles indeed can contribute to a closed Majorana chain around 0D $\mu$ that has an odd fermion-parity. So, all odd fermion-parity 0D obstruction-free states should be trivialized.

Subsequently, we consider the $1$D bubble equivalences. We decorate a pair of complex fermions on each $1$D block $\tau$ for group $C_n$. 
Near each $0$D block, there are \( n \) complex fermions contributed from adjacent bubbles, which form the following atomic insulator:
\begin{equation}
\ket{TR} = c_1^\dagger c_2^\dagger c_3^\dagger \dots c_n^\dagger.
\end{equation}
Since the subgroup \( \mathbb{Z}_2^T \subset G_b \) does not contribute to a non-trivial classification, we only consider the symmetry transformation properties of \( \ket{TR} \) under \( C_n \) and \( P_f \) operations.
We observe that when \( n \) is odd:
\begin{equation}
{C_n} \ket{TR} = \ket{TR}, \quad {P_f} \ket{TR} = -\ket{TR}.
\end{equation}
However, when \( n \) is even:
\begin{equation}
{C_n} \ket{TR} = -\ket{TR}, \quad {P_f} \ket{TR} = \ket{TR}.
\end{equation}

Thus, the 0D trivial states contributed by 1D and $2$D bubbles form the group:
\begin{equation}
\{\mathrm{TBS}\}^{\mathrm{0D}} = \mathbb{Z}_2
\end{equation}
for all cyclic groups.
In conclusion, all non-trivial 0D FSPT decorations will be trivialized. Therefore, there is also no need to consider the extension problem here. 

To summarize, we present the classification groups $E^{nD}$ of all independent $n$-dimensional obstruction-free and non-trivial block-states for various cyclic point groups. For the $C_4$ group:
\begin{equation}
\begin{aligned}
     & E^{\mathrm{2D}}=\{\mathrm{OFBS}\}^{\mathrm{2D}}=\mathbb{Z}_1,\\
      & E^{\mathrm{1D}}=\{\mathrm{OFBS}\}^{\mathrm{1D}}/\{\mathrm{TBS}\}^{\mathrm{1D}}=\mathbb{Z}_4, \\&E^{\mathrm{0D}}=\{\mathrm{OFBS}\}^{\mathrm{0D}}/\{\mathrm{TBS}\}^{\mathrm{0D}}=\mathbb{Z}_1. 
\end{aligned}
  \end{equation}

For the $C_2$ and $C_6$ groups:
\begin{equation}
\begin{aligned}
    &E^{\mathrm{2D}}=\{\mathrm{OFBS}\}^{\mathrm{2D}}=\mathbb{Z}_1, \\&E^{\mathrm{1D}}=\{\mathrm{OFBS}\}^{\mathrm{1D}}/\{\mathrm{TBS}\}^{\mathrm{1D}}=\mathbb{Z}_2, \\&E^{\mathrm{0D}}=\{\mathrm{OFBS}\}^{\mathrm{0D}}/\{\mathrm{TBS}\}^{\mathrm{0D}}=\mathbb{Z}_1.    
\end{aligned}
 \end{equation}

For the $C_3$ group:
\begin{equation}
\begin{aligned}
 &E^{\mathrm{2D}}=\{\mathrm{OFBS}\}^{\mathrm{2D}}=\mathbb{Z}_1,\\ &E^{\mathrm{1D}}=\{\mathrm{OFBS}\}^{\mathrm{1D}}/\{\mathrm{TBS}\}^{\mathrm{1D}}=\mathbb{Z}_1, \\&E^{\mathrm{0D}}=\{\mathrm{OFBS}\}^{\mathrm{0D}}/\{\mathrm{TBS}\}^{\mathrm{0D}}=\mathbb{Z}_1.     
\end{aligned}
   \end{equation}

Notably, none of these cases exhibits a nontrivial group extension. The resulting classification groups are \(\mathcal{G}_{0}=\mathbb{Z}_4\) for \(C_4\), \(\mathcal{G}_{0}=\mathbb{Z}_2\) for \(C_2\) and \(C_6\), and \(\mathcal{G}_{0}=\mathbb{Z}_1\) for \(C_3\). A comprehensive classification for all cyclic point groups \(C_n\) is presented in Table~\ref{table II}.

\subsubsection{Spinful fermions} \label{sec:Cn spinfull}
In this section, we detail the real-space construction of point group $C_n$ ($n=2,3,4,6$) and time-reversal symmetry-protected FSPT phases in 2D interacting spinful fermionic systems. The cell decomposition is illustrated in Fig.~\ref{Cn cell}.
\paragraph{Block-State Decoration}

The $0$D block \( \mu \) has a physical symmetry group \( G_b = \mathbb{Z}_n \times \mathbb{Z}_2^T \), and the corresponding 0D FSPT phase is characterized by the data \( (n_0, \nu_1) \):
\begin{equation}
    \begin{aligned}
        & n_0 \in \mathcal{H}^0(G_b, \mathbb{Z}_2^f) = \mathbb{Z}_2, \\
        & \nu_1 \in \mathcal{H}^1(G_b, U_T(1)) = \mathbb{Z}_{(2, n)},
    \end{aligned}
\end{equation}
where \( (m, n) \) denotes the greatest common divisor of \( m \) and \( n \).
Similarly to the discussion in Sec.~\ref{sec:D4 spinful discussion}, there are no non-trivial cocycles contributed from the subgroup \( \mathbb{Z}_2^T \subset G_b \), and only \( n_0 = 0 \) is obstruction-free. Therefore, the 0D FSPT classification group is \( \mathbb{Z}_{(2, n)} \), where each group element can be labeled by the eigenvalues \( \pm 1 \) of \( C_n \).

\begin{figure}[tb]
    \centering
    \includegraphics[width=0.48 \textwidth]{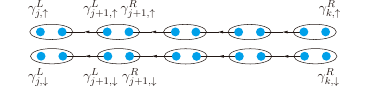}
    \caption{The 1D FSPT phase protected by \( G_f = \mathbb{Z}_4^{fT} \), realized as two copies of Majorana chains with opposite spin species.}
    \label{fig:Cn spinful root phase}
\end{figure}

Next, we consider the decoration of $1$D block-states.
In one-dimensional systems, all blocks possess an on-site symmetry group $G_f = \mathbb{Z}_4^{fT}$. As established in Ref.~\cite{Wang2020}, this symmetry leads to a $\mathbb{Z}_2$ classification of the corresponding FSPT phases. For the physical realization of these phases, we employ the two-Majorana-chains model proposed by Gu~\cite{PhysRevResearch.2.033290} as the root phase. The Hamiltonian of this model is given by:
\begin{equation}
    H = it \sum_{j,\sigma} \gamma^R_{j, \sigma} \gamma^L_{j+1, \sigma},
\end{equation} where $j$ labels the lattice sites and $\sigma = \uparrow,\downarrow$ represents the spin degrees of freedom, as illustrated in Fig.~\ref{fig:Cn spinful root phase}.
And this model exhibits the following symmetry transformation:
\begin{align}
    {T}:~ & i \mapsto -i, \notag \\
             & \gamma^L_{j, \uparrow} \mapsto -\gamma^L_{j, \downarrow}, \quad \gamma^L_{j, \downarrow} \mapsto \gamma^L_{j, \uparrow}, \notag \\
             & \gamma^R_{j, \uparrow} \mapsto \gamma^R_{j, \downarrow}, \quad \gamma^R_{j, \downarrow} \mapsto -\gamma^R_{j, \uparrow}. \label{eq:trs definition}
\end{align} 
Furthermore, we cannot decorate a single Majorana chain due to the absence of on-site symmetry actions that satisfy \( T^2 = -1 \). This leads to a non-trivial obstruction function for the 1D invertible phase, characterized by \( O_1(n_0) = \omega_2 n_0 \) \cite{wang2021domain}.

Finally, all $2$D blocks have a total symmetry group \( G_f = \mathbb{Z}_4^{fT} \). It is well known that there exists a non-trivial 2D class-DIII topological superconductor (TSC) with symmetry group \( G_f = \mathbb{Z}_4^{fT} \), characterized by a topological classification \( \mathbb{Z}_2 \).

Having identified all available states for decoration, we now explore which block-state decorations satisfy the no-open-edge conditions.

All $0$D block-state decorations naturally satisfy the required conditions. Consequently, the obstruction-free $0$D block-states constitute the groups:
\begin{equation}
    \{\mathrm{OFBS}\}^{\mathrm{0D}} = \mathbb{Z}_{(2,n)}
\end{equation} for cyclic groups $C_n$ ($n=2,3,4,6$).

Next, we consider the $1$D block-state decoration. For even cyclic groups \( C_n \), the decoration of 1D FSPT (realized by two Majorana chains) on a $1$D block \( \tau \) leaves \( n \) dangling Majorana modes in the $0$D block \( \mu \). These modes are labeled sequentially according to the order of rotation \( C_n \) as \( \gamma_1, \gamma_2, \ldots, \gamma_n \), and their symmetry properties are given by:\begin{equation}
\begin{aligned}
    T:~ & i \mapsto -i, \\
        ~& \gamma_{j, \uparrow} \mapsto -\gamma_{j, \downarrow}, \quad \gamma_{j, \downarrow} \mapsto \gamma_{j, \uparrow},  \\
    C_n:~ & \gamma_{1, \sigma} \mapsto \gamma_{2, \sigma}, \quad \gamma_{2, \sigma} \mapsto \gamma_{3, \sigma}, \quad \dots, \\
    ~& \gamma_{n-1, \sigma} \mapsto \gamma_{n, \sigma}, \quad  \gamma_{n, \sigma} \mapsto -\gamma_{1, \sigma},
\end{aligned}\end{equation}  where  $\sigma=\uparrow,\downarrow$.
It is straightforward to verify that the following interaction terms for each even cyclic group remain invariant under the symmetry transformations above:
\begin{equation}\begin{aligned}
    C_2:~ & i (\gamma_{1, \uparrow} \gamma_{2, \downarrow} + \gamma_{1, \downarrow} \gamma_{2, \uparrow}), \\
    C_4:~ & i \sum_{j=1,2} (\gamma_{j, \uparrow} \gamma_{j+2, \downarrow} + \gamma_{j, \downarrow} \gamma_{j+2, \uparrow}),  \\
    C_6:~ & i \sum_{j=1,2,3} (\gamma_{j, \uparrow} \gamma_{j+3, \downarrow} + \gamma_{j, \downarrow} \gamma_{j+3, \uparrow}).
\end{aligned}\end{equation}
However, for the \( C_3 \) point group, there is no obstruction-free decoration of the $1$D block-state.
A $C_3$-symmetric decoration contains a total of six Majorana chains. Yet, as established above, four Majorana chains correspond to a trivial phase in the FSPT classification.
Consequently, the boundary modes can only be symmetrically gapped out (yielding a trivial decoration) if the number of Majorana chains is a multiple of four.

Consequently, the obstruction-free $1$D block-states constitute the groups:
\begin{equation}
    \{\mathrm{OFBS}\}^{\mathrm{1D}} = 
    \begin{cases}
        \mathbb{Z}_2 & \text{for even-order cyclic groups}, \\
        \mathbb{Z}_1                      & \text{for odd-order cyclic groups},
    \end{cases}
\end{equation}
where the $\mathbb{Z}_2$ factor counts the stacking number (mod $2$) of the FSPT decoration on $1$D blocks.

Next, we consider the block-state decoration on a two-dimensional block \( \sigma \).  As concluded in Ref.~\cite{RotationSPT}, $p_x\pm ip_y$ superconductors are consistent with rotation symmetry group $C^n_n=-1$. Thus, the 2D DIII TSC decoration, i.e. stacking of $p_x\pm ip_y$ superconductors, is obstruction-free and forms a group \begin{equation}
    \{ \mathrm{OFBS} \}^{2D} = \mathbb{Z}_2.
\end{equation}

\paragraph{Bubble Equivalence \& Extension \& Ultimate Classification} \label{sec:extension cn spinful}

Here, we discuss the bubble equivalence. First, we decorate $1$D bubbles, each composed of a pair of complex fermions, on each $1$D block. Since the subgroup \( \mathbb{Z}_2^T \subset G_b \) does not contribute to non-trivial classifications, we only need to consider the \( C_n \) symmetry transformations.

Near each $0$D block, there are \( n \) complex fermions contributed from adjacent bubbles, forming the following atomic insulator:
\begin{equation}
\ket{TR} = c_1^\dagger c_2^\dagger c_3^\dagger \dots c_n^\dagger.
\end{equation}
We observe that when \( n \) is even, \( {C_n} \ket{TR} = \ket{TR} \), but when \( n \) is odd, \( {C_n} \ket{TR} = -\ket{TR} \). Consequently, the 0D trivial block-states from $1$D bubbles constitute the groups:
\begin{equation}
    \{\mathrm{TBS}\}^{\mathrm{0D}} = 
    \begin{cases}
        \mathbb{Z}_1 & \text{for even-order cyclic groups}, \\
        \mathbb{Z}_2                      & \text{for odd-order cyclic groups}.
    \end{cases}
\end{equation}
Thus, $1$D bubbles can trivialize all $0$D block-states for the \( C_n \) group with an odd \( n \), while having no effect on the \( C_n \) group with an even \( n \).

In two-dimensional systems, each $2$D bubble necessarily comprises two closed anti-PBC Majorana chains to maintain the on-site symmetry \( \mathbb{Z}_4^{fT} \). These chains are constrained by the time-reversal symmetry described in Eq.~\eqref{eq:trs definition}. Through an adiabatic enlargement process, these Majorana bubbles can be smoothly deformed into $1$D block-states, i.e. four Majorana chains, along each $1$D block while preserving time-reversal symmetry of Eq.~\eqref{eq:trs definition}. 
The chains originating from adjacent bubbles have opposite pairing directions, due to the rotational symmetry of the bubbles. Moreover, at each boundary of these $1$D block-states, two types of dangling modes, \( \gamma^L \) and \( \gamma^R \), must appear simultaneously. And time-reversal symmetry acts on them differently. These $1$D block-states are indeed topologically trivial which cannot be protected by time-reversal symmetry. Concretely, consider four dangling modes \begin{equation}
\gamma_{1,\uparrow}^L,\gamma_{1,\downarrow}^L,\gamma_{2,\uparrow}^R,\gamma_{2,\downarrow}^R,
\end{equation} where subscripts $1,2$ label their origin from different bubbles, and we assume they contribute $L$- and $R$-type modes, respectively. One can then write symmetric mass terms \begin{equation} i\gamma_{1,\uparrow}^L\gamma_{2,\uparrow}^R+i\gamma_{1,\downarrow}^L\gamma_{2,\downarrow}^R,
\end{equation} which are invariant under  both  time-reversal and rotation symmetry. Therefore,  $2$D bubbles do not contribute to any 1D non-trivial block-states, for any point group $C_n$. $2$D bubbles also have no effect on $0$D block-states (eigenstates of $C_n$). Therefore, $2$D bubbles do not contribute to any non-trivial trivialization group, and \begin{equation}
  \{\mathrm{TBS}\}^{\mathrm{1D}}=\mathbb{Z}_1.  
\end{equation}

We now turn to the extension problem. We find that there is no extension from 1D block-states to 0D block-states. We then follow the strategy employed in the analysis of the group \( D_4 \), and investigate the symmetry properties of the chiral edge modes arising from two-dimensional decorations.
We define the action of the $C_n$ symmetry on the chiral edge modes $\chi$ of DIII topological superconducting (TSC) decorations as follows:
\begin{align}
T:~&i\to-i \notag\\
&\chi^{(k)}_{j,\uparrow} \to -\chi^{(k)}_{j,\downarrow},~~~~~\chi^{(k)}_{j,\downarrow} \to \chi^{(k)}_{j,\uparrow}\notag\\
   C_n:~& \chi^{(k)}_{j,\sigma} \to  \chi^{(k+1)}_{j,\sigma},~~~~~1\leq k \leq n-1\notag\\
   &  \chi^{(n)}_{j,\sigma} \to  -\chi^{(1)}_{j,\sigma},~~~k=n
\end{align} Here, the superscript $(k)$, with $1\leq k \leq n$, labels the $2$D block to which the mode belongs, and the ordering of $k$ is defined by successive applications of the $C_n$ rotation operator. $j=1,2$ denotes the layer index within each $2$D block.

Under these symmetry constraints, the allowed symmetric mass terms, $H_{mass}$, for the Majorana fermions in the $k$-th and $(k+1)$-th blocks take the form:
\begin{align}
  \int~dx~&(im\chi^{(k)}_{1,\uparrow} \chi^{(k)}_{2,\downarrow}-im\chi^{(k)}_{2,\uparrow}\chi^{(k)}_{1,\downarrow})\notag\\
  +&(im\chi^{(k+1)}_{1,\uparrow} \chi^{(k+1)}_{2,\downarrow}-im\chi^{(k+1)}_{2,\uparrow}\chi^{(k+1)}_{1,\downarrow})\label{eq:cn extension 1D 12 mass}
\end{align}
where the mass terms have been written in the standard form \( i m \chi^+ \chi^- \), with \( \chi^\pm \) denoting Majorana modes with positive or negative velocity relative to the radial direction. Notably, pairs of mass terms related by the rotation symmetry do not acquire any relative minus sign. According to the criterion established in Ref.~\cite{Wang2020}, this implies that no non-trivial effective 1D states emerges and no extension from $2$D block-state decoration to $1$D block-state exists.

To summarize, we present the classification groups $E^{nD}$ of all independent $n$-dimensional obstruction-free and non-trivial block-states for various cyclic point groups. For the $C_4$ group:
\begin{equation}
\begin{aligned}
     & E^{\mathrm{2D}}=\{\mathrm{OFBS}\}^{\mathrm{2D}}=\mathbb{Z}_2,\\
      & E^{\mathrm{1D}}=\{\mathrm{OFBS}\}^{\mathrm{1D}}/\{\mathrm{TBS}\}^{\mathrm{1D}}=\mathbb{Z}_2, \\&E^{\mathrm{0D}}=\{\mathrm{OFBS}\}^{\mathrm{0D}}/\{\mathrm{TBS}\}^{\mathrm{0D}}=\mathbb{Z}_2. 
\end{aligned}
  \end{equation}

For the $C_2$ and $C_6$ groups:
\begin{equation}
\begin{aligned}
    &E^{\mathrm{2D}}=\{\mathrm{OFBS}\}^{\mathrm{2D}}=\mathbb{Z}_2, \\&E^{\mathrm{1D}}=\{\mathrm{OFBS}\}^{\mathrm{1D}}/\{\mathrm{TBS}\}^{\mathrm{1D}}=\mathbb{Z}_2, \\&E^{\mathrm{0D}}=\{\mathrm{OFBS}\}^{\mathrm{0D}}/\{\mathrm{TBS}\}^{\mathrm{0D}}=\mathbb{Z}_2.    
\end{aligned}
 \end{equation}

For the $C_3$ group:
\begin{equation}
\begin{aligned}
 &E^{\mathrm{2D}}=\{\mathrm{OFBS}\}^{\mathrm{2D}}=\mathbb{Z}_2,\\ &E^{\mathrm{1D}}=\{\mathrm{OFBS}\}^{\mathrm{1D}}/\{\mathrm{TBS}\}^{\mathrm{1D}}=\mathbb{Z}_1, \\&E^{\mathrm{0D}}=\{\mathrm{OFBS}\}^{\mathrm{0D}}/\{\mathrm{TBS}\}^{\mathrm{0D}}=\mathbb{Z}_1.     
\end{aligned}
   \end{equation}

Notably, none of these cases exhibits a nontrivial group extension. The resulting classification groups are \(\mathcal{G}_{1/2}=\mathbb{Z}_2^3\) for \(C_2\), \(C_4\), and \(C_6\), while for \(C_3\) we find \(\mathcal{G}_{1/2}=\mathbb{Z}_2\). The final classification for all cyclic groups \(C_n\) is summarized in Table~\ref{table I}.

\subsection{Higher-order  symmetry-protected topological phases
of interacting fermions }
Higher-order boundary states (HOBS)~\cite{Zhang_2013,Benalcazar_2017,Benalcazar_hing_2017,Schindler_2018,Zhang_2019,Zhang_2023}, which emerge in topological superconductors protected by crystalline symmetry, have attracted considerable attention. Within the framework of real-space construction, once we identify obstruction-free and non-trivial decoration phases, we can directly obtain the corresponding higher-order boundary states of each phase by implementing boundary truncation in real space~\cite{SongHao,Song_2020,zhang2022construction}. While there has been substantial progress~\cite{Rasmussen_2020,Dubinkin_2019} in classifying higher-order SPT phases of interacting bosons, the intrinsic interacting higher-order fermionic SPT phases remain less explored. In this work, we focus on these intrinsic interacting fermionic phases, which have not been thoroughly investigated. Without loss of generality, we restrict our attention to regular-polygon-shaped boundaries in the following discussion.

Based on the discussions in previous sections,  only spinless  systems with point groups $D_4$ and $C_4$ can host intrinsic interacting fermionic SPT phases. Notably, these phases originate from 1D FSPT block-states decoration. We conclude that such systems are the only viable candidates for hosting intrinsic interacting HOBS. Within the real-space construction framework, these $1$D block-states can generate 0D corner dangling modes when the boundary is ``cut off." In contrast, 0D fermionic block-state decorations do not produce dangling modes at the boundary.

\begin{figure}[bt]
    \centering
\includegraphics[width=0.45\textwidth]{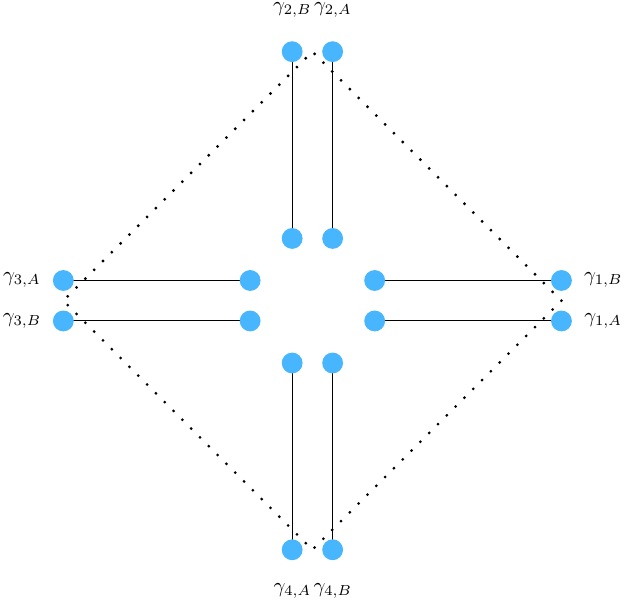}%width for pictures about dangling modes 
\caption{Higher-order 2D FSPT phases from 1D FSPT decorations  on $1$D blocks labeled by $\tau_1$. This configuration leaves eight dangling Majorana modes $\gamma_{1,A},\gamma_{1,B},\dots,\gamma_{4,A},\gamma_{4,B}$ at the corners of a square-shaped system. The dashed line indicates the system boundary.}
    \label{fig:HO spinless D4 2chain tau1}
\end{figure}

For point group $D_4$, the 1D $n_1^M+n_1^T$ FSPT decoration on block $\tau_1$ constructs an intrinsic FSPT phase, resulting in two dangling Majorana modes at each corner with square-shaped boundary conditions (Fig.~\ref{fig:HO spinless D4 2chain tau1}). These modes are symmetry-protected and remain robust under the local symmetry group \(G=\mathbb{Z}_2^M \times \mathbb{Z}_2^T \times \mathbb{Z}_2^f\), which includes reflection and time-reversal symmetries: the only allowed local mass term, \(i\,\gamma_{j,A}\gamma_{j,B}\), is odd under time reversal and therefore forbidden. A similar situation occurs for point group $C_4$, where the only possible mass terms for dangling modes also break time-reversal symmetry.

\section{CONSTRUCTION AND CLASSIFICATION OF SPACE GROUP AND TIME-REVERSAL SYMMETRY-PROTECTED TOPOLOGICAL SUPERCONDUCTORS}\label{sec:space group SPT}

In this section, we provide a detailed description of real-space constructions for crystalline TSCs protected by wallpaper groups and time-reversal symmetry in 2D interacting fermionic systems by analyzing several representative examples. It is well known that all 17 wallpaper groups can be categorized into five crystallographic systems, and their site symmetry groups can be derived from the International Tables for Crystallography~\cite{InternationalTablesforCrystallography}.
In particular, we apply the general framework of real-space construction, as outlined in Sec.~\ref{sec:classification strategy}, to investigate five representative cases belonging to different crystallographic systems:
\begin{enumerate}
    \item Square lattice: \( p4m \);
    \item Parallelogrammatic lattice: \( p2 \);
    \item Rhombic lattice: \( cmm \);
    \item Rectangle lattice: \( pgg \);
    \item Hexagonal lattice: \( p6m \).
\end{enumerate}
All other cases are provided in the appendix (App.~\ref{sec:other cases}). It can be observed that most of the classification results can be directly derived from the conclusions discussed in the context of point group- and time-reversal-symmetry protected TSCs. 
The classification results are summarized in Tables~\ref{plane group spinless} and~\ref{plane group spin-1/2}, for spinless and spinful fermions, respectively.

\subsection{ Square Lattice: $p4m$}

\begin{figure}[tb]
    \centering
    \includegraphics[width=0.45\textwidth]{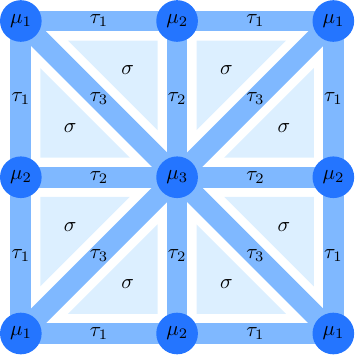}
    \caption{The \#11 2D space group $p4m$ and its  cell decomposition.}
    \label{p4m}
\end{figure}

For the square lattice, we illustrate the FSPT phase protected by the $p4m$ and time-reversal symmetry as an example. 
For the $2$D blocks $\sigma$, there is no nontrivial site symmetry, so the effective on-site symmetry reduces to $\mathbb{Z}_2^{T}$. For the $1$D blocks $\tau_1$, $\tau_2$, and $\tau_3$, the site symmetry group is the same, namely $\mathbb{Z}_2^{M}$ originating from reflection symmetry; hence their effective symmetry is $\mathbb{Z}_2^{M}\times \mathbb{Z}_2^{T}$. For the $0$D blocks $\mu_1$ and $\mu_3$, the effective symmetry is $\mathbb{Z}_4 \rtimes \mathbb{Z}_2 \times \mathbb{Z}_2^{T}$, inherited from the full $D_4$ site symmetry. For the $0$D block $\mu_2$, whose site symmetry is $D_2\subset D_4$, the effective symmetry is $\mathbb{Z}_2 \rtimes \mathbb{Z}_2 \times \mathbb{Z}_2^{T}$. The corresponding cell decomposition for $p4m$ is shown in Fig.~\ref{p4m}.

\subsubsection{Spinless fermions}
For spinless systems, the total symmetry group of each dimensional block is just the direct product of the physical symmetry group and the fermion-parity group $\mathbb{Z}_2^f$.

We begin by analyzing the decoration of $0$D block-states.
The following results, which were already established in our point group analysis (Sec.~\ref{sec:point_group}), are summarized here for completeness. The classification of $0$D block-states at each $\mu_j$ ($j = 1,2,3$) is determined by the 1D irreducible representations of $D_n \times \mathbb{Z}_2^T \times \mathbb{Z}_2^f$ ($n = 2,4$):
\begin{equation}
\mathcal{H}^1(D_n \times \mathbb{Z}_2^T \times \mathbb{Z}_2^f, U_T(1)) = \mathbb{Z}_2^3.
\end{equation}
Specifically:
\begin{itemize}
    \item For $\mu_1$ and $\mu_3$, the $\mathbb{Z}_2$ factors correspond to eigenvalues ($\pm 1$) of $M_{\tau_1}/M_{\tau_2}$, $M_{\tau_3}$, and $P_f$
    \item For $\mu_2$, they correspond to eigenvalues of $M_{\tau_1}$, $M_{\tau_2}$, and $P_f$
\end{itemize}

Since all $0$D block-states must satisfy the no-open-edge condition, the obstruction-free $0$D block-states form the group:
\begin{equation}
\{\mathrm{OFBS}\}^{0D} = \mathbb{Z}_2^9.
\end{equation}

As established in Sec.~\ref{sec:D4 spinless fermion}, all $1$D blocks here possess physical symmetry group $G_b = \mathbb{Z}_2 \times \mathbb{Z}_2^T$, leading to the FSPT classification: \( \mathbb{Z}_4 \times \mathbb{Z}_4 \). The generators of these $\mathbb{Z}_4$ factors correspond to the root phases $n_1^M$ and $n_1^T$, whose symmetry transformations were previously specified in Eqs.~\eqref{eq:symm property of n1T} and~\eqref{eq:symm property of n1M}. Additionally, following our earlier analysis, we may consider Majorana chain decorations as valid block-states.

We now determine which $1$D block-state decorations satisfy the no-open-edge condition. 

Consider decorating the $1$D block $\tau_1$. Such a 1D decoration generally produces dangling zero modes at its two endpoints, located on the $0$D blocks $\mu_1$ and $\mu_2$. As shown in previous sections, these dangling modes are not symmetry-protected under the corresponding $0$D block symmetries: they can be fully gapped out by adding appropriate (symmetry-allowed) interaction terms.
Concretely, for $\mu_1$ with physical symmetry $G_b = D_4 \times \mathbb{Z}_2^T$, the dangling modes can be gapped out by the following choices of decorations:
\begin{itemize}
\item a 1D FSPT decoration characterized by the phase $n_1^T + n_1^M$; or
\item 1D BSPT decorations characterized by $(-1)^{n_1^T \cup n_1^T}$ or $(-1)^{n_1^M \cup n_1^M}$.
\end{itemize}
For $\mu_2$ with $G_b = D_2 \times \mathbb{Z}_2^T$, the same BSPT decorations $(-1)^{n_1^T \cup n_1^T}$ and $(-1)^{n_1^M \cup n_1^M}$ likewise generate symmetry-allowed  interaction terms and gap out the dangling modes.

Therefore, only the 1D BSPT decorations ($(-1)^{n_1^T \cup n_1^T}$ or $(-1)^{n_1^M \cup n_1^M}$) on $\tau_1$ are obstruction-free, since they can simultaneously gap out all zero modes on both $\mu_1$ and $\mu_2$.

The analysis for decorating the $1$D block \(\tau_2\) is identical to that for \(\tau_1\). In particular, a 1D decoration on \(\tau_2\) produces dangling zero modes at its endpoints, and imposing the no-open-edge condition leads to the same obstruction-free choices of 1D decorations as for \(\tau_1\).

 When both $\tau_1$ and $\tau_2$ are decorated simultaneously, however, the dangling modes must be analyzed collectively at the shared $0$D blocks. Although the $n_1^T+n_1^M$ decoration on a single 1D block can leave unmatched modes at an endpoint, the modes contributed by $\tau_1$ and $\tau_2$ can pair up at $\mu_2$ (and similarly at $\mu_1$ and $\mu_3$) once both blocks are decorated, rendering the combined decoration obstruction-free. Indeed, our analysis in the $D_4$ and $D_2$ cases shows that only the 1D $n_1^T+n_1^M$ FSPT decoration can simultaneously gap out all zero modes at $\mu_1$, $\mu_2$, and $\mu_3$.
 We note that this decoration realizes an intrinsically interacting FSPT phase, corresponding to the intrinsic sector of $D_4$.

Finally, consider the decorations on block $\tau_3$. As established in Sec.~\ref{sec:D4 spinless fermion}, the zero modes at $\mu_1$ and $\mu_3$ under $D_4$ symmetry can be gapped out by decorating $\tau_3$ with block-states corresponding to either:
\begin{itemize}
    \item The $n_1^T + n_1^M$ FSPT phases, or
    \item The $(-1)^{n_1^T \cup n_1^T}$/$(-1)^{n_1^M \cup n_1^M}$ BSPT phases
\end{itemize}

Consequently, the group of all obstruction-free $1$D block-states is
\begin{equation}
\{\mathrm{OFBS}\}^{1\mathrm{D}}
= (\mathbb{Z}_4 \times \mathbb{Z}_2)\times(\mathbb{Z}_2 \times \mathbb{Z}_2)\times(\mathbb{Z}_4 \times \mathbb{Z}_2).
\end{equation}

It is convenient to view this group as a quotient of a larger ``free'' decoration group.  
Before imposing identifications among decorations, the obstruction-free decorations can be organized as
\begin{equation}
\tilde G^{1\mathrm{D}}
=
(\mathbb{Z}_2\times\mathbb{Z}_2)\times(\mathbb{Z}_2\times\mathbb{Z}_2)\times\mathbb{Z}_4\times(\mathbb{Z}_4\times\mathbb{Z}_2\times\mathbb{Z}_2).
\end{equation}
Here the first two $(\mathbb{Z}_2\times\mathbb{Z}_2)$ factors describe the BSPT decorations on $\tau_1$ and $\tau_2$, respectively; the $\mathbb{Z}_4$ factor records the FSPT decoration $n_1^T+n_1^M$ on $\tau_1$ and $\tau_2$; and the last factor $\mathbb{Z}_4\times\mathbb{Z}_2\times\mathbb{Z}_2$ encodes the independent FSPT and BSPT decorations on $\tau_3$.

However, $\tilde G^{1\mathrm{D}}$ contains redundancies. In particular, stacking two copies of the FSPT decoration $n_1^T+n_1^M$ on both $\tau_1$ and $\tau_2$ yields the BSPT decoration
\(
(-1)^{n_1^M\cup n_1^M+n_1^T\cup n_1^T}
\)
on both $\tau_1$ and $\tau_2$, which is already contained as a $\mathbb{Z}_2$ subgroup of the first two $(\mathbb{Z}_2\times\mathbb{Z}_2)$ factors. Likewise, stacking two copies of the $\mathbb{Z}_4$ FSPT decoration on $\tau_3$ produces an element in the $\mathbb{Z}_2$ subgroup inside the last $\mathbb{Z}_4\times\mathbb{Z}_2\times\mathbb{Z}_2$ factor. Therefore, the physical group $\{\mathrm{OFBS}\}^{1\mathrm{D}}$ is obtained from $\tilde G^{1\mathrm{D}}$ by quotienting out these overlap relations.

Elements of $\{\mathrm{OFBS}\}^{1\mathrm{D}}$ can be parameterized as
\begin{equation}
[mt_{12}, T_1, M_2, T_2, mt_3, T_3].
\end{equation}
Here $mt_{12}, mt_3 \in \{0,1,2,3\}$ and $M_j, T_j \in \{0,1\}$ label the decorated phases. Specifically,
\begin{itemize}
\item $mt_{ij}$ counts the number (mod $4$) of $n_1^M+n_1^T$ phases supported simultaneously on $\tau_i$ and $\tau_j$;
\item $mt_i$ counts the number (mod $4$) of $n_1^M+n_1^T$ phases supported on $\tau_i$;
\item $M_j$ records the presence or absence of a $(-1)^{\,n_1^M\cup n_1^M}$ phase on $\tau_j$;
\item $T_j$ records the presence or absence of a $(-1)^{\,n_1^T\cup n_1^T}$ phase on $\tau_j$.
\end{itemize}

The total symmetry group for the $2$D blocks is \( G_f = \mathbb{Z}_2^T \times \mathbb{Z}_2^f \). As concluded in Ref.~\cite{Wang2020}, there are no corresponding FSPT phases under this symmetry group. Consequently, the contribution of $2$D block decoration is classified as:
\begin{equation}
\{\mathrm{OFBS}\}^{2D} = \mathbb{Z}_1.
\end{equation}
and we will not further discuss the decoration of $2$D block-states in the following spinless cases.

Next, we consider bubble equivalence relations. First, we examine the $1$D bubble equivalences by decorating a pair of complex fermions on each $1$D block.
The $1$D bubble on the $1$D block \( \tau_1 \) contributes to a non-trivial state \begin{equation} \ket{\phi} = c_1^\dagger c_2^\dagger c_3^\dagger c_4^\dagger \ket{0} \end{equation} on the $0$D block \( \mu_1 \), which satisfies symmetry properties
\begin{equation} M_{\tau_1} \ket{\phi} = c_1^\dagger c_4^\dagger c_3^\dagger c_2^\dagger \ket{0} = -\ket{\phi}. \end{equation}
It also contributes to a non-trivial state \( \ket{\phi} = c_1^\dagger c_2^\dagger \ket{0} \) on the $0$D block \( \mu_2 \), which satisfies 
\begin{equation} M_{\tau_2} \ket{\phi} = c_2^\dagger c_1^\dagger \ket{0} = -\ket{\phi} .\end{equation} 
The remaining cases are analogous and can be summarized as follows:
\begin{enumerate}
    \item $1$D bubbles on $\tau_2$ induce $0$D block-states $\ket{\phi}$
    \begin{itemize}
        \item on $\mu_2$, satisfying $M_{\tau_1}\ket{\phi}=-\ket{\phi}$;
        \item on $\mu_3$, satisfying $M_{\tau_2}\ket{\phi}=-\ket{\phi}$.
    \end{itemize}
    \item $1$D bubbles on $\tau_3$ induce $0$D block-states $\ket{\phi}$
    \begin{itemize}
        \item on $\mu_1$, satisfying $M_{\tau_3}\ket{\phi}=-\ket{\phi}$;
        \item on $\mu_3$, satisfying $M_{\tau_3}\ket{\phi}=-\ket{\phi}$.
    \end{itemize}
\end{enumerate}
Thus, the $1$D bubbles form the trivialization group \begin{equation}
   \{\mathrm{TBS}\}^{\mathrm{0D}}= \mathbb{Z}_2^3 .
\end{equation}

Next, we consider the $2$D bubble equivalence: we decorate ``Majorana bubbles", i.e. Majorana chains with anti-periodic boundary conditions (anti-PBC) on each $2$D block, and enlarge all ``Majorana bubbles" near each $1$D block. Similarly to the conclusion in Sec.~\ref{sec:D4 spinless fermion}, 2D ``Majorana bubbles" can contribute to 1D non-trivial states, classified as \( \mathbb{Z}_4 \), on all $1$D blocks. This includes the \( n_1^M+ n_1^T \) and \( (-1)^{n_1^M \cup n_1^M} \oplus (-1)^{n_1^T \cup n_1^T} \) phases by stacking. 
Therefore, $2$D bubbles form the trivialization group: \begin{equation}\{\mathrm{TBS}\}^{\mathrm{1D}} = \mathbb{Z}_4 . \end{equation}

Therefore, all independent non-trivial block-states with different dimensions are classified as:
\begin{equation}
\begin{aligned}
    E^{\mathrm{2D}} &=\{\mathrm{OFBS}\}^{\mathrm{2D}}= \mathbb{Z}_1, \\
    E^{\mathrm{1D}} &= \{\mathrm{OFBS}\}^{\mathrm{1D}} / \{\mathrm{TBS}\}^{\mathrm{1D}} = \mathbb{Z}_4 \times \mathbb{Z}_2^4, \\
    E^{\mathrm{0D}} &= \{\mathrm{OFBS}\}^{\mathrm{0D}} / \{\mathrm{TBS}\}^{\mathrm{0D}} = \mathbb{Z}_2^6.
\end{aligned}
\end{equation}
 In fact, since the symmetries of the zero-dimensional blocks here belong to those we discussed earlier, and the cell decomposition nearby is consistent with the point group decomposition, there is no extension here. 
 Hence, the ultimate classification with an accurate group structure is: $\mathcal{G}_0 = \mathbb{Z}_4 \times \mathbb{Z}_2^{10}$.

\subsubsection{Spinful fermions}
We now extend our discussion to spinful fermions. For each $0$D block $\mu_j$ ($j=1,2,3$), the possible $0$D block-states are classified by the one-dimensional irreducible representations of the full symmetry group. Equivalently, they are captured by the first cohomology group (for $n=2,4$):
\begin{equation}
\mathcal{H}^1\big((D_n \times \mathbb{Z}_2^T)\times_{\omega_2}\mathbb{Z}_2^f,\; U_T(1)\big)=\mathbb{Z}_2^2.
\end{equation}
The $\omega_2$ appearing here and throughout Sec.~\ref{sec:space group SPT} is a 2-cocycle which satisfies the spinful conditions in Eq.~\eqref{2-cocycle}.
Specifically:
\begin{itemize}
    \item For $\mu_1$ and $\mu_3$, the $\mathbb{Z}_2$ factors correspond to eigenvalues ($\pm 1$) of $M$ and $C_4$
    \item For $\mu_2$, they correspond to eigenvalues ($\pm 1$) of $M$ and $C_2$
\end{itemize}
Since all $0$D block-states must satisfy the no-open-edge condition, the obstruction-free $0$D block-states form the group:
\begin{equation}
\{\mathrm{OFBS}\}^{0D} = \mathbb{Z}_2^6.
\end{equation}

Next, we discuss the decoration of $1$D block-states. In this case, the physical symmetry group for all $1$D blocks is \( G_b = \mathbb{Z}_2 \times \mathbb{Z}_2^T \). As discussed in Sec.~\ref{sec:D4 spinful discussion}, we cannot decorate an invertible topological order, i.e., a Majorana chain. However, we can decorate with 1D FSPT phases classified by \( \mathbb{Z}_4 \). Such a 1D FSPT phase features a root state consisting of double Majorana chains, with symmetry properties defined in Eq.~\eqref{eq:definition root phase}.

The 1D FSPT decoration on $\tau_1$ yields:
\begin{itemize}
    \item 8 Majorana modes at $\mu_1$
    \item 4 Majorana modes at $\mu_2$
\end{itemize}
As shown in Sec.~\ref{sec:D4 spinful discussion}, these modes can be gapped out under $G_b = D_4 \times \mathbb{Z}_2^T$ (for $\mu_1$) and $G_b = D_2 \times \mathbb{Z}_2^T$ (for $\mu_2$), making the decoration obstruction-free. Similar decorations on $\tau_2$ and $\tau_3$ are also obstruction-free, leading to the classification:
\begin{equation}
\{\mathrm{OFBS}\}^{\mathrm{1D}} = \mathbb{Z}_4^3,
\end{equation}
with group elements $[\mathbb{MT}_1, \mathbb{MT}_2, \mathbb{MT}_3]$, where $\mathbb{MT}_j \in \{0,1,2,3\}$ counts the 1D FSPT decorations on $\tau_j$.

The physical symmetry group for $2$D blocks is \( G_b = \mathbb{Z}_2^T \), and the corresponding $2$D block-states are classified as \( \mathbb{Z}_2 \), where the root state is the 2D DIII TSC state. Following the analysis in Sec.~\ref{sec:D4 spinful discussion}, the DIII TSC decoration on the $2$D blocks can always gap out the zero modes at each $1$D block. Therefore, the contribution of $2$D block decoration is classified as:
\begin{equation}
\{\mathrm{OFBS}\}^{2D} = \mathbb{Z}_2.
\end{equation}

Next, we examine bubble equivalence relations, starting with $1$D bubble equivalences by considering a decorated pair of complex fermions on each $1$D block. 
Recall from Sec.~\ref{sec:D4 spinful discussion} that $0$D blocks with site symmetry $D_4$ admit no trivialization via $1$D bubbles. In the present $p4m$ setting, this applies to the $0$D blocks $\mu_1$ and $\mu_3$. We therefore focus on the remaining $0$D block $\mu_2$, which has $D_2$ point-group symmetry.

The $1$D bubble decoration on the $1$D block $\tau_1$ (or $\tau_2$) generates a non-trivial state on the $0$D block $\mu_2$:
\begin{equation} \ket{\phi} = c_1^\dagger c_2^\dagger \ket{0} \end{equation}
This state transforms under symmetry operations as follows:
\begin{equation} {M} \ket{\phi} = ic_1^\dagger (-i)c_2^\dagger \ket{0} = \ket{\phi} \end{equation}
\begin{equation} {C_2} \ket{\phi} = -c_2^\dagger c_1^\dagger \ket{0} = \ket{\phi} \end{equation}
When decorating $1$D bubbles on both $1$D blocks $\tau_1$ and $\tau_2$, the resulting states $\ket{\phi}$ on $\mu_2$ exhibit the same transformation properties:
\begin{equation} {C_2} \ket{\phi} = \ket{\phi} \quad \text{and} \quad {M} \ket{\phi} = \ket{\phi} \end{equation}

In summary, our analysis reveals that the $1$D bubble decorations do not trivialize the $0$D block-states for systems with on-site physical symmetries $G_b = D_4 \times \mathbb{Z}_2^T$ and $G_b = D_2 \times \mathbb{Z}_2^T$. 

Furthermore, we decorate $2$D bubbles on each $2$D block, consisting of two closed anti-PBC Majorana chains, as the bubbles must preserve the on-site symmetry \( T^2 = -1 \). $2$D bubble decorations fail to trivialize $1$D block-states, since 1D states deformed from $2$D bubbles are gappable under reflection symmetry. We have demonstrated such a conclusion in Sec.~\ref{sec:D4 spinful discussion}. Note that the bubble equivalences around $0$D blocks $\mu_1$ and $\mu_3$ follow the same pattern as in the $D_4$ case, while those around $\mu_2$ align with the $D_2$ case (detailed analysis provided in Sec.~\ref{sec:D4 spinful discussion} and Sec.~\ref{sec:D2 D6 spinful}). Thus, $2$D bubbles also have no effect on $0$D blocks. Therefore, we conclude that the total trivialization groups are:
\begin{equation}
\{\mathrm{TBS}\}^{\mathrm{1D}}=\{\mathrm{TBS}\}^{\mathrm{0D}} = \mathbb{Z}_1
\end{equation}

Finally, we consider the extension problem. As discussed in Sec.~\ref{sec:D4 spinful discussion}, when only reflection symmetry is present, there exists an extension from $2$D block-states to $1$D block-states. Specifically, the decoration of two layers of DIII TSC states can adiabatically connect to the double Majorana chains decoration on all $1$D blocks. 
Because the local symmetry here, along with the corresponding cell decomposition, is the same as in the point-group analysis, there is no extension from 1D block-states to 0D block-states in this case.

Therefore, all independent non-trivial block-states with different dimensions are classified as:
\begin{equation}
\begin{aligned}
    E^{\mathrm{2D}} &=\{\mathrm{OFBS}\}^{\mathrm{2D}}= \mathbb{Z}_2, \\
    E^{\mathrm{1D}} &= \{\mathrm{OFBS}\}^{\mathrm{1D}} / \{\mathrm{TBS}\}^{\mathrm{1D}} = \mathbb{Z}_4^3, \\
    E^{\mathrm{0D}} &= \{\mathrm{OFBS}\}^{\mathrm{0D}} / \{\mathrm{TBS}\}^{\mathrm{0D}} = \mathbb{Z}_2^6.
\end{aligned}
\end{equation}
The final classification is:
\begin{equation}
\mathcal{G}_{1/2} = \mathbb{Z}_2^6 \times \mathbb{Z}_4^2 \times \mathbb{Z}_8,
\end{equation}
where the \( \mathbb{Z}_8 \) term is contributed by the extension of $2$D block-states to $1$D block-states.

\subsection{Parallelogram Lattice: \( p2 \)}

\begin{figure}[tb]
    \centering
    \includegraphics[width=0.42\textwidth]{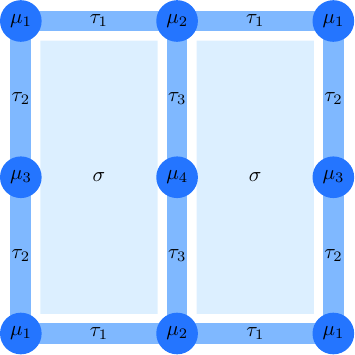}
    \caption{The \#2 2D space group $p2$ and its  cell decomposition.}
    \label{p2}
\end{figure}

For the parallelogram lattice, we consider the example of FSPT phases protected by \( p2 \) and time-reversal symmetry. The point group corresponding to \( p2 \) is the rotational group \( C_2 \). For 1D and $2$D blocks, there is no site symmetry, and the physical symmetry group is \( \mathbb{Z}_2^T \). However, the rotational subgroup \( C_2 \) acts internally on each $0$D block, making their physical symmetry group \( \mathbb{Z}_2 \times \mathbb{Z}_2^T \), as illustrated in Fig.~\ref{p2}.

\subsubsection{Spinless fermions} \label{sec:spinless p2}

For spinless systems, each $0$D block $\mu_j$ ($j=1,2,3,4$) can be decorated by a $0$D block-state classified by the distinct one-dimensional irreducible representations of the full onsite symmetry group. This classification is captured by
\begin{equation}
\mathcal{H}^1(\mathbb{Z}_2 \times \mathbb{Z}_2^T \times \mathbb{Z}_2^f, U_T(1)) = \mathbb{Z}_2^2,
\end{equation}
where each \( \mathbb{Z}_2 \) is labeled by the eigenvalue \( \pm 1 \) of \( C_2 \) or \( P_f \). Therefore, the obstruction-free $0$D block-states form the group:
\begin{equation}
\{\mathrm{OFBS}\}^{0D} = \mathbb{Z}_2^8.
\end{equation}

Next, we discuss the decoration of $1$D block-states. In this case, the total symmetry group for all $1$D blocks is \( G_f = \mathbb{Z}_2^T \times \mathbb{Z}_2^f \). As discussed in Sec.~\ref{sec:appendix spinless Cn}, the corresponding FSPT phases are classified by \( \mathbb{Z}_4 \), where the fourth-order group generator has a free-fermionic realization consisting of two Majorana chains with symmetry properties listed in Eq.~\eqref{eq:symm property spinless Cn 2chain}. 
In addition, we can also consider an invertible topological order, i.e., a Majorana chain.

Now we turn to the $1$D block-state decoration on the $1$D block $\tau_1$. Such a decoration leaves dangling Majorana zero modes at the adjacent $0$D blocks $\mu_1$ and $\mu_2$. Because the physical symmetry at $\mu_1/\mu_2$ is
$G_b=\mathbb{Z}_2\times\mathbb{Z}_2^T$,
and because the local cell decomposition around these sites reduces to the $C_2$ point-group case once $\tau_2/\tau_3$ are ignored, we can directly apply the $C_2$ analysis (see Sec.~\ref{sec:appendix spinless Cn}). It follows that the zero modes at $\mu_1$ can be gapped out only for 1D BSPT decorations, realized concretely as four Majorana chains.

The same reasoning applies to decorations on the $1$D blocks $\tau_2$ and $\tau_3$, which similarly leave Majorana zero modes at $\mu_3$ and $\mu_4$.

However, the cell decomposition near $\mu_1/\mu_2$ - when considering all adjacent $1$D blocks - is inconsistent with our $C_2$ point group analysis. This suggests an alternative decoration scheme where 1D FSPT decoration (implemented via two Majorana chains)  can be implemented simultaneously on $\tau_1$ together with $\tau_2$ (or, equivalently, with $\tau_3$).
Nevertheless, such two-chain decorations on $\tau_2/\tau_3$ fail to gap out the zero modes at $\mu_3/\mu_4$, as established in Sec.~\ref{sec:appendix spinless Cn}. This obstruction persists even when combined with decorations on $\tau_1$.

Consequently, only the 1D BSPT phase (realized via four Majorana chains) decorations on each $1$D block are obstruction-free. The complete classification of these states forms the group:
\begin{equation}
\{\mathrm{OFBS}\}^{\mathrm{1D}} = \mathbb{Z}_2^3,
\end{equation}
with group elements labeled as:
\begin{equation}
[\mathcal{T}_1, \mathcal{T}_2, \mathcal{T}_3], \quad \mathcal{T}_j \in \{0,1\}.
\end{equation}
Here, each $\mathcal{T}_j$ enumerates the four-Majorana-chain decorations on block $\tau_j$ ($j=1,2,3$), protected by the symmetry $G_f = \mathbb{Z}_2^T \times \mathbb{Z}_2^f$.

The total symmetry group for the $2$D blocks is \( G_f = \mathbb{Z}_2^T \times \mathbb{Z}_2^f \). As concluded in Ref.~\cite{Wang2020}, there are no corresponding FSPT phases under this symmetry group. Consequently, the contribution of $2$D block decoration is classified as:
\begin{equation}
\{\mathrm{OFBS}\}^{2D} = \mathbb{Z}_1.
\end{equation}

Next, we consider bubble equivalence relations. First, we examine the $1$D bubble equivalences by decorating a pair of complex fermions on each $1$D block. 
The $1$D bubble on the $1$D block \( \tau_1 \) contributes to a non-trivial state \( \ket{\phi} = c_1^\dagger c_2^\dagger \ket{0} \) on the $0$D block \( \mu_1 \), which satisfies \begin{equation} C_2 \ket{\phi} = c_2^\dagger c_1^\dagger \ket{0} = -\ket{\phi}, \end{equation} and a non-trivial state \( \ket{\phi} = c_1^\dagger c_2^\dagger \ket{0} \) on the $0$D block \( \mu_2 \), which satisfies \begin{equation} C_2 \ket{\phi} = c_2^\dagger c_1^\dagger \ket{0} = -\ket{\phi}. \end{equation} 
Similarly, the $1$D bubbles on $\tau_2$ and $\tau_3$ induce non-trivial 0D states on their adjacent $0$D blocks:
\begin{enumerate}
    \item \textbf{Bubble on $\tau_2$:} it contributes a non-trivial state $\ket{\phi}$ on $\mu_1$ and on $\mu_3$, both satisfying
    \begin{equation}
        C_2\ket{\phi}=-\ket{\phi}.
    \end{equation}
    \item \textbf{Bubble on $\tau_3$:} it contributes a non-trivial state $\ket{\phi}$ on $\mu_2$ and on $\mu_4$, both satisfying
    \begin{equation}
        C_2\ket{\phi}=-\ket{\phi}.
    \end{equation}
\end{enumerate}
Thus, the $1$D bubbles contribute to four independent $0$D block-states and form the trivialization group \( \mathbb{Z}_2^4 \). As mentioned in Sec.~\ref{sec:Cn spinless bubble}, $2$D bubbles decorations related by rotation symmetry do not contribute to non-trivial 1D FSPT states but can be used to trivialize the eigenstates of \( P_f \) associated with $\mu_j$ ($j=1,2,3,4$). Therefore, the total trivialization groups are:
\begin{equation}
\{\mathrm{TBS}\}^{\mathrm{1D}} = \mathbb{Z}_1,~\{\mathrm{TBS}\}^{\mathrm{0D}} = \mathbb{Z}_2^4.
\end{equation}

It is evident that there is no extension between 1D and $0$D block-states. Thus, all independent non-trivial block-states with different dimensions are classified as:
\begin{equation}
\begin{aligned}
    E^{\mathrm{2D}} &=\{\mathrm{OFBS}\}^{\mathrm{2D}}= \mathbb{Z}_1, \\
    E^{\mathrm{1D}} &= \{\mathrm{OFBS}\}^{\mathrm{1D}} / \{\mathrm{TBS}\}^{\mathrm{1D}} = \mathbb{Z}_2^3, \\
    E^{\mathrm{0D}} &= \{\mathrm{OFBS}\}^{\mathrm{0D}} / \{\mathrm{TBS}\}^{\mathrm{0D}} = \mathbb{Z}_2^4.
\end{aligned}
\end{equation}
Hence, the ultimate classification with an accurate group structure is:
\begin{equation}
\mathcal{G}_0 = \mathbb{Z}_2^7.
\end{equation}

\subsubsection{Spinful fermions}

In spinful systems, for all $0$D blocks \( \mu_j \) (\( j = 1, 2, 3, 4 \)), the corresponding $0$D block-state decorations are characterized by the distinct 1D irreducible representations of the total symmetry group, which are captured by:
\begin{equation}
\mathcal{H}^1\big((\mathbb{Z}_2 \times \mathbb{Z}_2^T) \times_{\omega_2} \mathbb{Z}_2^f, U_T(1)\big) = \mathbb{Z}_2,
\end{equation}
where each \( \mathbb{Z}_2 \) is labeled by the eigenvalue \( \pm 1 \) of \( C_2 \). Therefore, the obstruction-free $0$D block-states form the group:
\begin{equation}
\{\mathrm{OFBS}\}^{0D} = \mathbb{Z}_2^4.
\end{equation}

Next, we discuss the decoration of $1$D block-states. In this case, the physical symmetry group for all $1$D blocks is \( G_b = \mathbb{Z}_2^T \). As discussed in Sec.~\ref{sec:Cn spinfull}, decoration with an invertible topological order, i.e., a Majorana chain, is not possible; however, decorations with 1D FSPT phases, classified by \( \mathbb{Z}_2 \), are possible. Such 1D FSPT phases feature a root state consisting of double Majorana chains, with symmetry properties defined in Eq.~\eqref{eq:trs definition}.

As a direct consequence of our discussion of the $C_2$ case in Sec.~\ref{sec:Cn spinfull}, it is straightforward to conclude that the obstruction-free 1D decorations are classified by the group:
\begin{equation}
\{\mathrm{OFBS}\}^{\mathrm{1D}}= \mathbb{Z}_2^3,
\end{equation}
with elements labeled as triples:
\begin{equation}
[\mathbb{T}_1, \mathbb{T}_2, \mathbb{T}_3], \quad \mathbb{T}_j \in \{0,1\}.
\end{equation}
Each binary index $\mathbb{T}_j$ counts the double Majorana chain decorations on $1$D block $\tau_j$ ($j=1,2,3$), protected by the symmetry $G_f = \mathbb{Z}_2^T \times_{\omega_2} \mathbb{Z}_2^f$.

Following analogous arguments to the spinful case discussed earlier, the 2D class DIII topological superconductor decorations are obstruction-free. This yields a classification:
\begin{equation}
\{\mathrm{OFBS}\}^{\mathrm{2D}} = \mathbb{Z}_2,
\end{equation}
where the non-trivial element corresponds to the DIII TSC state decoration.

Next, we consider the bubble equivalence relations. The $1$D bubble on the $1$D block \( \tau_1 \) acts on the $0$D blocks \( \mu_1 \) and \( \mu_2 \), the bubble state on the $1$D block \( \tau_2 \) acts on the $0$D blocks \( \mu_1 \) and \( \mu_3 \), and the bubble state on the $1$D block \( \tau_3 \) acts on the $0$D blocks \( \mu_2 \) and \( \mu_4 \). 

However, as concluded in Sec.~\ref{sec:Cn spinfull}, the $1$D bubbles have no trivializing effect on the $0$D block-states with on-site physical symmetry \( G_b = \mathbb{Z}_2 \times \mathbb{Z}_2^T \). Similarly, the $2$D bubbles have no trivializing effect on either 0D or $1$D block-states.  Therefore, the total trivialization groups are:
\begin{equation}
\{\mathrm{TBS}\}^{\mathrm{1D}}=\{\mathrm{TBS}\}^{\mathrm{0D}}= \mathbb{Z}_1.
\end{equation}

Finally, we consider the extension problem. We find no extension from 1D block-states to 0D block-states in this case.
Since we are only dealing with cyclic point groups, as discussed in Sec.~\ref{sec:extension cn spinful},  the boundary Majorana modes for the 2D decoration do not contribute to non-trivial 1D phases.
Hence, there is no extension from $2$D block-states to $1$D block-states. 

Therefore, all independent non-trivial block-states with different dimensions are classified as:
\begin{equation}
\begin{aligned}
    E^{\mathrm{2D}} &=\{\mathrm{OFBS}\}^{\mathrm{2D}}= \mathbb{Z}_2, \\
    E^{\mathrm{1D}} &= \{\mathrm{OFBS}\}^{\mathrm{1D}} / \{\mathrm{TBS}\}^{\mathrm{1D}} = \mathbb{Z}_2^3, \\
    E^{\mathrm{0D}} &= \{\mathrm{OFBS}\}^{\mathrm{0D}} / \{\mathrm{TBS}\}^{\mathrm{0D}} = \mathbb{Z}_2^4.
\end{aligned}
\end{equation}
Thus, the final classification is:
\begin{equation}
\mathcal{G}_{1/2} =  \mathbb{Z}_2^8.
\end{equation}

\subsection{Rhombic Lattice: \( cmm \)}

\begin{figure}[tb]
    \centering
    \includegraphics[width=0.46\textwidth]{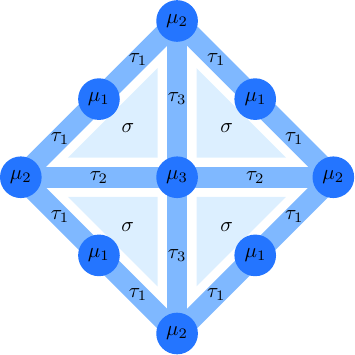}
    \caption{The \#9 2D space group $cmm$ and its  cell decomposition.}
    \label{cmm}
\end{figure}

For the rhombic lattice, we consider the example of FSPT phases protected by \( cmm \)  and time-reversal symmetry.  For the $2$D blocks \( \sigma \) and the $1$D block \( \tau_1 \), there is no site symmetry, and thus the physical symmetry group is \( \mathbb{Z}_2^T \). For the $1$D blocks \( \tau_2 \) and \( \tau_3 \), the reflection symmetry acts internally. For the $0$D block \( \mu_1 \), there is site symmetry \( C_2 \), and thus the physical symmetry group is \( \mathbb{Z}_2 \times \mathbb{Z}_2^T \). For the $0$D blocks \( \mu_2 \) and \( \mu_3 \), the \( D_2 \) symmetry acts internally, and the physical symmetry group is \( \mathbb{Z}_2 \ltimes \mathbb{Z}_2 \times \mathbb{Z}_2^T \). The cell decomposition for \( cmm \) is shown in Fig.~\ref{cmm}.

\subsubsection{Spinless fermions}

For spinless systems, we first consider the $0$D block-state decoration. For the $0$D block \( \mu_1 \), candidate states are characterized by the distinct 1D irreducible representations of the total symmetry group, which are captured by:
\begin{equation}
\mathcal{H}^1(\mathbb{Z}_2 \times \mathbb{Z}_2^T \times \mathbb{Z}_2^f, U_T(1)) = \mathbb{Z}_2^2,
\end{equation}
while for the $0$D blocks \( \mu_2 \) and \( \mu_3 \),
\begin{equation}
\mathcal{H}^1(D_2 \times \mathbb{Z}_2^T \times \mathbb{Z}_2^f, U_T(1)) = \mathbb{Z}_2^3.
\end{equation}
At each $0$D block labeled by \( \mu_1 \), the block-states can be labeled by the fermion-parity \( P_f \) and the rotation \( C_2 \) eigenvalue \( \pm 1 \), respectively. On the $0$D blocks \( \mu_2 \) and \( \mu_3 \), the block-states are labeled by the eigenvalues \( \pm 1 \) of \( C_2 \), \( M \), and \( P_f \). 
Therefore, the obstruction-free $0$D block-states form the group:
\begin{equation}
\{\mathrm{OFBS}\}^{0D} = \mathbb{Z}_2^8.
\end{equation}

Next, we discuss the decoration of $1$D block-states. The physical symmetry group for the $1$D block \( \tau_1 \) is \( G_b = \mathbb{Z}_2^T \), while for the $1$D blocks \( \tau_2 \) and \( \tau_3 \), it is \( G_b = \mathbb{Z}_2 \times \mathbb{Z}_2^T \).
As established in Sec.~\ref{sec:D4 spinless fermion},  the corresponding FSPT phases of the $1$D blocks \( \tau_2 \) and \( \tau_3 \) exhibit a classification: \( \mathbb{Z}_4 \times \mathbb{Z}_4 \). The generators of these $\mathbb{Z}_4$ factors correspond to the root phases $n_1^M$ and $n_1^T$, whose symmetry transformations were previously specified in Eqs.~\eqref{eq:symm property of n1T} and~\eqref{eq:symm property of n1M}.
The corresponding FSPT phases of the $1$D block \( \tau_1 \) have a \( \mathbb{Z}_4 \) classification, where the fourth-order group generator has a free-fermionic realization consisting of two Majorana chains with symmetry properties listed in Eq.~\eqref{eq:symm property spinless Cn 2chain}.
Furthermore, we can also consider an invertible topological order, i.e., a Majorana chain, as a block-state.

Specifically, the $0$D block \( \mu_1 \) has the physical symmetry group \( G_b = \mathbb{Z}_2 \times \mathbb{Z}_2^T \), and the cell decomposition near it is consistent with our analysis of the point group \( C_2 \). The corresponding decoration that can gap out the corresponding dangling zero modes at the $0$D block \( \mu_1 \) consists of 1D BSPT  decorations (four Majorana chains) on $\tau_1$. However, these decorations also leave dangling Majorana modes on $\mu_2$, which has physical symmetry \( G_b = D_2 \times \mathbb{Z}_2^T \). It should be noticed that, near $\mu_2$, the blocks $\tau_1$  do not align with cell decomposition of point group $D_2$. 
However, as we will demonstrate in Appendix.~\ref{sec:p4g D2 special}, the zero modes from 1D FSPT decorations (two Majorana chains), satisfying the symmetry properties in Eq.~\eqref{eq:D2 special symmetry}, already can exhibit a unique ground state. Consequently, zero modes from 1D BSPT phase decorations can also be gapped out.   Therefore, it follows directly that four-Majorana-chain decorations on $\tau_1$ are obstruction-free.

The $0$D block $\mu_3$ possesses physical symmetry group $G_b = D_2 \times \mathbb{Z}_2^T$, and its local cell decomposition is fully consistent with our $D_2$ point group analysis. Similarly, the $0$D block $\mu_2$ exhibits $D_2$ symmetry when we disregard $\tau_1$, with its cell decomposition also matching the $D_2$ point group structure. Within this framework, 1D BSPT decorations (realized as four-Majorana-chain configurations) of either $(-1)^{n_1^T \cup n_1^T}$ or $(-1)^{n_1^M \cup n_1^M}$ phases on $\tau_2$ (or $\tau_3$) satisfy the no-open-edge condition while permitting complete gapping of their $\mu_2$ (or $\mu_3$) boundary modes. Crucially, this obstruction-free property extends to the $n_1^M + n_1^T$ phase decorations when simultaneously implemented on both $\tau_2$ and $\tau_3$.

The cell decomposition near \( \mu_2 \) does not align with our analysis of the point group \( D_2 \), when considering the $1$D block $\tau_1$. This indicates it is possible to decorate the $1$D blocks \( \tau_1 \), \( \tau_2 \), and \( \tau_3 \) simultaneously with 1D FSPT decorations. However, this decoration on the $1$D block \( \tau_1 \) cannot gap out the zero modes at \( \mu_1 \). 

Therefore,  all obstruction-free $1$D block-states form the group:
\begin{equation}
\{\mathrm{OFBS}\}^{1D} = \mathbb{Z}_4 \times \mathbb{Z}_2^4,
\end{equation}
with group elements parameterized as:
\begin{equation}
[\mathcal{T}_1, mt_{23}, M_2, T_2, M_3].
\end{equation}
Here, the binary indices $M_j \in \{0,1\}$ and $T_j \in \{0,1\}$ enumerate decorations of the $(-1)^{n_1^M \cup n_1^M}$ and $(-1)^{n_1^T \cup n_1^T}$ phases on $\tau_j$ ($j=2,3$) respectively. Meanwhile, $\mathcal{T}_1 \in \{0,1,2,3\}$ counts 1D BSPT decorations on $\tau_1$ protected by $G_f = \mathbb{Z}_2^T \times \mathbb{Z}_2^f$. Finally, $mt_{ij}$ counts the total number of $n_1^M$ and $n_1^T$ phase decorations on $\tau_i$ and $\tau_j$.

The total symmetry group for the $2$D blocks is \( G_f = \mathbb{Z}_2^T \times \mathbb{Z}_2^f \). As concluded in Ref.~\cite{Wang2020}, there are no corresponding FSPT phases under this symmetry group. Consequently, the contribution of $2$D block decoration is classified as:
\begin{equation}
\{\mathrm{OFBS}\}^{2D} = \mathbb{Z}_1.
\end{equation}

Next, we consider bubble equivalence relations. First, we examine the $1$D bubble equivalences by decorating a pair of complex fermions on each $1$D block. 
The $1$D bubble on the $1$D block \( \tau_1 \) contributes to a non-trivial state \( \ket{\phi} = c_1^\dagger c_2^\dagger \ket{0} \) on the $0$D block \( \mu_1 \), which satisfies \begin{equation} C_2 \ket{\phi} = c_2^\dagger c_1^\dagger \ket{0} = -\ket{\phi}. \end{equation} 
Similarly, the $1$D bubbles on $\tau_2$ and $\tau_3$ induce the following 0D states:
\begin{enumerate}
    \item \textbf{Bubble on $\tau_2$:} it contributes a non-trivial state $\ket{\phi}$ on $\mu_1$ and on $\mu_3$, both satisfying
 \begin{equation}
     M_{\tau_3}\ket{\phi}=-\ket{\phi};
 \end{equation}
    \item \textbf{Bubble on $\tau_3$:} it contributes a non-trivial state $\ket{\phi}$ on $\mu_2$ and on $\mu_3$, both satisfying
 \begin{equation}
     M_{\tau_2}\ket{\phi}=-\ket{\phi}.
 \end{equation}
\end{enumerate}
Thus, the $1$D bubbles form the trivialization group: \begin{equation}
 \{\mathrm{TBS}\}^{\mathrm{0D}}= \mathbb{Z}_2^3  . 
\end{equation}

2D ``Majorana bubbles" can contribute to non-trivial 1D states on the $1$D blocks \( \tau_2 \) and \( \tau_3 \) but yield only trivial 1D states on \( \tau_1 \), as the bubbles near \( \mu_1 \) are related by rotation symmetry (see Sec.~\ref{sec:Cn spinless bubble}). These non-trivial states,  classified by \( \mathbb{Z}_4 \),   include the \( n_1^M+ n_1^T \) phase and the  \( (-1)^{n_1^M \cup n_1^M} \oplus (-1)^{n_1^T \cup n_1^T} \) phase obtained through stacking.
Hence,  the $2$D bubbles form the trivialization group:
\begin{equation}
\{\mathrm{TBS}\}^{\mathrm{1D}} = \mathbb{Z}_4.
\end{equation}

Therefore, all independent non-trivial block-states with different dimensions are classified as:
\begin{equation}
\begin{aligned}
    E^{\mathrm{2D}} &=\{\mathrm{OFBS}\}^{\mathrm{2D}}= \mathbb{Z}_1, \\
    E^{\mathrm{1D}} &= \{\mathrm{OFBS}\}^{\mathrm{1D}} / \{\mathrm{TBS}\}^{\mathrm{1D}} = \mathbb{Z}_2^4, \\
    E^{\mathrm{0D}} &= \{\mathrm{OFBS}\}^{\mathrm{0D}} / \{\mathrm{TBS}\}^{\mathrm{0D}} = \mathbb{Z}_2^5.
\end{aligned}
\end{equation}

It is evident that there is no extension between the 1D and $0$D block-states, and the ultimate classification is:
\begin{equation}
\mathcal{G}_0 =\mathbb{Z}_2^9.
\end{equation}

\subsubsection{Spinful fermions}

For spinful systems, we consider the decoration of the $0$D block-states. For the $0$D block \( \mu_1 \), the corresponding $0$D block-states are characterized by the distinct 1D irreducible representations of the total symmetry group, which are captured by:
\begin{equation}
\mathcal{H}^1\big((\mathbb{Z}_2 \times \mathbb{Z}_2^T) \times_{\omega_2} \mathbb{Z}_2^f, U_T(1)\big) = \mathbb{Z}_2,
\end{equation}
while for the $0$D blocks \( \mu_2 \) and \( \mu_3 \),
\begin{equation}
\mathcal{H}^1\big((D_2 \times \mathbb{Z}_2^T) \times_{\omega_2} \mathbb{Z}_2^f, U_T(1)\big) = \mathbb{Z}_2^2.
\end{equation}
At each $0$D block labeled by \( \mu_1 \), the block-states can be labeled by the rotation \( C_2 \) eigenvalue \( \pm 1 \). On the $0$D blocks \( \mu_2 \) and \( \mu_3 \), the block-states are labeled by the eigenvalues \( \pm 1 \) of \( C_2 \) and \( M \).
Therefore, the obstruction-free $0$D block-states form the group:
\begin{equation}
\{\mathrm{OFBS}\}^{0D} = \mathbb{Z}_2^5.
\end{equation}

Next, we discuss the decoration of $1$D block-states. The physical symmetry group for the $1$D block \( \tau_1 \) is \( G_b = \mathbb{Z}_2^T \), while for the $1$D blocks \( \tau_2 \) and \( \tau_3 \), it is \( G_b = \mathbb{Z}_2 \times \mathbb{Z}_2^T \). As discussed in Sec.~\ref{sec:Cn spinfull} and Sec.~\ref{sec:D4 spinful discussion}, decoration with an invertible topological order, such as the Majorana chain, is not possible, but decoration with 1D FSPT phase (implemented via double Majorana chains) is. The classification group of candidate states for \( \tau_1 \) is \( \mathbb{Z}_2 \), with symmetry properties defined in Eq.~\eqref{eq:trs definition}; while for \( \tau_2 \) and \( \tau_3 \), it is \( \mathbb{Z}_4 \), with symmetry properties defined in Eq.~\eqref{eq:definition root phase}.

The decoration of 1D FSPT phase on the $1$D block \( \tau_1 \) leaves four dangling Majorana fermion modes on the $0$D block \( \mu_1 \) and eight dangling Majorana fermion modes on the $0$D block \( \mu_2 \). Based on our analysis of the point group, these Majorana fermions can gap out the zero modes at the $0$D blocks with physical symmetry \( G_b = \mathbb{Z}_2 \times \mathbb{Z}_2^T \) and \( G_b = D_2 \times \mathbb{Z}_2^T \), making this decoration obstruction-free. The decorations on the other two $1$D blocks are similarly obstruction-free.

Therefore, the 1D decoration contributes to the classification group:
\begin{equation}
\{\mathrm{OFBS}\}^{1D} = \mathbb{Z}_2 \times \mathbb{Z}_4^2,
\end{equation}
where the group elements can be labeled as:
\begin{equation}
[\mathbb{T}_1, \mathbb{MT}_2, \mathbb{MT}_3].
\end{equation}
Here, \( \mathbb{MT}_j \in \{ 0, 1, 2, 3\} \) and \(\mathbb{T}_j \in \{ 0, 1\} \) represent the number of decorated 1D FSPT phases protected by $G_f=(\mathbb{Z}_2 \times \mathbb{Z}_2^T)\times_{\omega_2} \mathbb{Z}_2^f$ and $G_f= \mathbb{Z}_2^T\times_{\omega_2} \mathbb{Z}_2^f$ on the $1$D blocks \( \tau_j \) (\( j = 1, 2, 3 \)), respectively.

Following analogous arguments to the spinful case discussed earlier, the decoration of 2D DIII TSC states is always obstruction-free, contributing to a classification group: \( \{\mathrm{OFBS}\}^{2D} = \mathbb{Z}_2 \).

Next, we consider the bubble equivalence relations. The $1$D bubble on the block \( \tau_1 \) acts on the $0$D blocks \( \mu_1 \) and \( \mu_2 \), the $1$D bubble on the block \( \tau_2 \) acts on the $0$D blocks \( \mu_2 \) and \( \mu_3 \), and the $1$D bubble on the block \( \tau_3 \) also acts on the $0$D blocks \( \mu_2 \) and \( \mu_3 \). 

However, based on the conclusions drawn from point group cases, the $1$D bubble does not trivialize the $0$D block-states with physical symmetry groups \( G_b = \mathbb{Z}_2 \times \mathbb{Z}_2^T \) and \( G_b = D_2 \times \mathbb{Z}_2^T \). Similarly, the $2$D bubble does not have any trivializing effect on the 0D and $1$D block-states. Thus, the overall trivialization groups are:
\begin{equation}
\{\mathrm{TBS}\}^{\mathrm{1D}}=\{\mathrm{TBS}\}^{\mathrm{0D}}\ = \mathbb{Z}_1.
\end{equation}

Lastly, we consider the extension problem. We find no extension from 1D block-states to 0D block-states in this case. Here, we encounter both cyclic point groups and dihedral point groups. Based on the discussions in Sec.~\ref{sec:extension cn spinful} and Sec.~\ref{sec:D4 spinful discussion}, the decoration of $2$D block-states around \( \mu_1 \) (i.e., two layers of DIII topological superconductor states) does not contribute non-trivial $1$D block-states on the $1$D block \( \tau_1 \), since adjacent bubbles around $\tau_1$ are connected only by rotation symmetry.  In contrast, the decoration of $2$D block-states around \( \mu_3 \) does contribute non-trivial $1$D block-states on the $1$D blocks \( \tau_2 \) and \( \tau_3 \). 
This indicates that there exists an extension from $2$D block-states to $1$D block-states, resulting in the simultaneous decoration of double Majorana chains on the $1$D blocks \( \tau_2 \) and \( \tau_3 \).

Therefore, all independent non-trivial block-states with different dimensions are classified as:
\begin{equation}
\begin{aligned}
    E^{\mathrm{2D}} &=\{\mathrm{OFBS}\}^{\mathrm{2D}}= \mathbb{Z}_2, \\
    E^{\mathrm{1D}} &= \{\mathrm{OFBS}\}^{\mathrm{1D}} / \{\mathrm{TBS}\}^{\mathrm{1D}} =\mathbb{Z}_2 \times \mathbb{Z}_4^2, \\
    E^{\mathrm{0D}} &= \{\mathrm{OFBS}\}^{\mathrm{0D}} / \{\mathrm{TBS}\}^{\mathrm{0D}} = \mathbb{Z}_2^5.
\end{aligned}
\end{equation}
Thus, the final classification is:
\begin{equation}
\mathcal{G}_{1/2} = \mathbb{Z}_4 \times \mathbb{Z}_8 \times \mathbb{Z}_2^6,
\end{equation}
where the factor \( \mathbb{Z}_8 \) corresponds to the extension of $2$D block-states to $1$D block-states.

\subsection{Rectangular Lattice: \( pgg \)}

\begin{figure}[tb]
    \centering
    \includegraphics[width=0.46\textwidth]{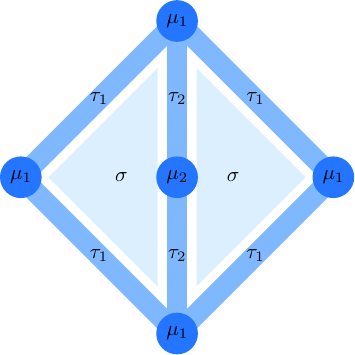}
    \caption{The \#8 2D space group $pgg$ and its cell decomposition.}
    \label{pgg}
\end{figure}

For the rectangular lattice, we consider \( pgg \) and time-reversal symmetry-protected crystal FSPTs as an example. The \( pgg \) wallpaper group is non-symmorphic, and its corresponding point group is the cyclic group \( C_2 \). For the $2$D block \( \sigma \) and the $1$D blocks \( \tau_1 \) and \( \tau_2 \), there is no site symmetry group, and the physical symmetry group is \( \mathbb{Z}_2^T \). For the $0$D blocks \( \mu_1 \) and \( \mu_2 \), the physical symmetry is \( \mathbb{Z}_2 \times \mathbb{Z}_2^T \), since the two-fold rotational symmetry \( C_2 \) acts internally on the $0$D blocks, as shown in Fig.~\ref{pgg}.

\begin{figure}[tb]
    \centering
    \includegraphics[width=0.3\textwidth]{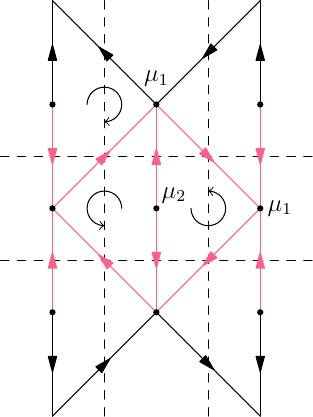}
    \caption{Symmetry  property  of the cell decomposition  for $pgg$: The red lines make up the unit cell of cell decomposition, while the  dashed lines indicate the glide operation, which involves reflecting across the dashed line followed by a translation of half a unit cell along the direction of the dashed line. The  arrows represent the pairing direction of the decorated Majorana chains, and the directed arcs denote the pairing direction of the decorated 2D Majorana chain bubble states.}
    \label{fig: pgg chiral}
\end{figure}

\subsubsection{Spinless fermions}

For the spinless system, we first consider the decoration of $0$D block-states. For all $0$D blocks \( \mu_1 \) and \( \mu_2 \), the corresponding $0$D block-states are characterized by the distinct 1D irreducible representations of the total symmetry group, which are captured by:
\begin{equation}
\mathcal{H}^1(\mathbb{Z}_2 \times \mathbb{Z}_2^T \times \mathbb{Z}_2^f, U_T(1)) = \mathbb{Z}_2^2,
\end{equation}
where group elements in each \( \mathbb{Z}_2 \) can be labeled by the eigenvalues of \( C_2 \) and \( P_f \) as \( \pm 1 \). Therefore, the obstruction-free $0$D block-states form the group:
\begin{equation}
\{\mathrm{OFBS}\}^{0D} = \mathbb{Z}_2^4.
\end{equation}

\begin{figure}[tb]
    \centering
    \includegraphics[width=0.3\textwidth]{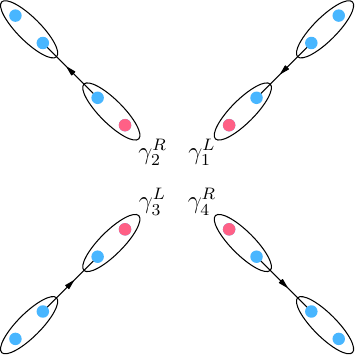}
    \caption{
   Single Majorana chain decoration on \(\tau_1\). The direction of the arrows represents the pairing direction of the Majorana fermions and the central origin represents the $0$D block \(\mu_1\). Due to the differing pairing directions of these four decorated states, there exist two types of dangling Majorana zero modes, \(\gamma^L\) and \(\gamma^R\), near \(\mu_1\), as defined in the Eq.~\eqref{eq:definition of Majorana chain}.}
\label{fig:pgg one chain on tau1}
\end{figure}

\begin{figure}[tb]
    \centering
    \includegraphics[width=0.35\textwidth]{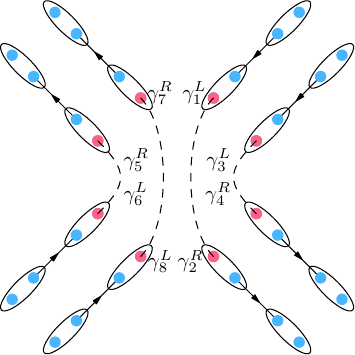}
    \caption{Decoration by two single Majorana chains on $\tau_1$.
    In the figure, the central origin represents the $0$D block \(\mu_1\), with black arrows indicating the two Majorana chain decoration on the four adjacent $1$D blocks $\tau_1$. We refer to Fig.~\ref{fig:two_single_Majorana_chain_deformation} to introduce the dashed lines: it depicts the same state, but with additional pairing that couples the dangling Majorana modes along the dashed-line direction. } 
    \label{fig:pgg two chain on tau1}
\end{figure}

Next, we discuss the decoration of $1$D block-states. The physical symmetry group for all $1$D blocks is given by \( G_b = \mathbb{Z}_2^T \). As mentioned previously, the corresponding FSPT phases are classified as \( \mathbb{Z}_4 \), with a root state model consisting of two Majorana chains, whose symmetry properties are detailed in Eq.~\eqref{eq:symm property spinless Cn 2chain}. Furthermore, we can incorporate an invertible topological order, namely, a single Majorana chain.

We now consider the $1$D block-state decoration on the $1$D block \( \tau_1 \). Unlike the point group discussion, although the local symmetry at the $0$D block $\mu_1$ here is \( C_2 \), this $0$D block is connected to four $1$D blocks in its vicinity. Additionally, due to the glide symmetry of this 2D space group, these decorated states will simultaneously contribute both types of zero modes, \( \gamma^L \) and \( \gamma^R \), on the $0$D block $\mu_1$  (see Fig.~\ref{fig: pgg chiral}). 

Thus, the decoration of a single Majorana chain on a $1$D block \( \tau_1 \) leaves four dangling Majorana modes \( \gamma_1^L, \gamma_2^R, \gamma_3^L, \gamma_4^R \) in each $0$D block \( \mu_1 \) (Fig.~\ref{fig:pgg one chain on tau1}). The symmetric interaction term
\begin{equation}
i\gamma_1^L\gamma_2^R + i\gamma_3^L\gamma_4^R
\end{equation}
can be introduced to gap out these zero modes. 

Next, we consider the 1D FSPT decoration of two Majorana chains on the $1$D block \( \tau_1 \), which leaves eight dangling Majorana modes \( \gamma_1^L, \gamma_2^R, \gamma_3^L, \gamma_4^R, \gamma_5^R, \gamma_6^L, \gamma_7^R, \gamma_8^L \) on each $0$D block \( \mu_1 \) (see Fig.~\ref{fig:pgg two chain on tau1}). The symmetry transformations of these Majorana modes are given by:
\begin{equation}
\begin{aligned}
    T:  & \quad i \mapsto -i, \\
        & \quad \gamma^L_j \mapsto \gamma^L_j, \quad \gamma^R_j \mapsto -\gamma^R_j, \quad 1 \leq j \leq 8, \\
    C_2: & \quad \gamma^L_1 \leftrightarrow \gamma^L_6, \quad \gamma^L_3 \leftrightarrow \gamma^L_8, \\
         & \quad \gamma^R_2 \leftrightarrow \gamma^R_5, \quad \gamma^R_4 \leftrightarrow \gamma^R_7.
\end{aligned}
\end{equation}

We introduce the symmetric interaction term:
\begin{equation}
i\gamma_1^L\gamma_2^R + i\gamma_3^L\gamma_4^R + i\gamma_6^L\gamma_5^R + i\gamma_8^L\gamma_7^R,
\end{equation}
to gap out these zero modes. Therefore, both the single- and two-Majorana-chain decorations on the 1D block $\tau_1$ are obstruction-free. By taking the single Majorana chain as the generator, these decorations form a $\mathbb{Z}_8$ classification, corresponding to a class-BDI topological superconductor.

Now, we consider the decoration on the $1$D block \( \tau_2 \), which is the same as in the point group discussion. In this case, only four-Majorana-chain decorations are obstruction-free, giving a \( \mathbb{Z}_2 \) classification. 

Therefore, the 1D decoration contributes to the classification group:
\begin{equation}
\{\mathrm{OFBS}\}^{1D} = \mathbb{Z}_8 \times \mathbb{Z}_2,
\end{equation}
where the group elements can be labeled as:
\begin{equation}
[\mathcal{M}_1, T_2].
\end{equation}
Here, \( \mathcal{M}_1 = 0, 1, 2, \dots, 7 \) and \( T_2 = 0, 1 \) represent the number of decorated 1D single Majorana chains or \( (-1)^{n_1^T \cup n_1^T} \) phase states on the $1$D blocks \( \tau_1 \) and \( \tau_2 \) , respectively.

The total symmetry group for the $2$D blocks is \( G_f = \mathbb{Z}_2^T \times \mathbb{Z}_2^f \). As concluded in Ref.~\cite{Wang2020}, there are no corresponding FSPT phases under this symmetry group. Consequently, the contribution of $2$D block decoration is classified as:
\begin{equation}
\{\mathrm{OFBS}\}^{2D} = \mathbb{Z}_1.
\end{equation}

Next, we consider bubble equivalence relations. The discussion of $2$D bubbles here differs from that of the point groups. Due to the glide symmetry of this 2D space group, decorating the $2$D bubbles in accordance with this symmetry (see Fig.~\ref{fig: pgg chiral}) results in a contribution of a 1D FSPT state on \( \tau_1 \), while on \( \tau_2 \), it yields a trivial state.

This distinction arises because, near \( \tau_1 \), the two Majorana chains deformed from the bubbles exhibit the same pairing direction and share the symmetry properties described in Eq.~\eqref{eq:symm property spinless Cn 2chain}. Consequently, this configuration leads to a non-trivial 1D two-Majorana-chains FSPT phase, as established in Sec.~\ref{sec:appendix spinless Cn}. Furthermore, stacking these bubbles ultimately contributes to 1D FSPT phases classified by $\mathbb{Z}_4$. In contrast, near \( \tau_2 \), the two Majorana chains have opposite pairing directions, allowing the straightforward introduction of an interaction term to gap out the dangling modes at the boundary. 
Thus, the $2$D bubbles form the trivialization group:
\begin{equation}
\{\mathrm{TBS}\}^{1D} = \mathbb{Z}_4.
\end{equation}

 Next, we consider the $1$D bubble equivalences by decorating a pair of complex fermions on each $1$D block. 
The $1$D bubble on the block \( \tau_1 \) contributes a trivial state 
\begin{equation}
\ket{\phi} = c_1^\dagger c_2^\dagger c_3^\dagger c_4^\dagger \ket{0}
\end{equation}
on the $0$D block \( \mu_1 \), satisfying 
\begin{equation}
C_2 \ket{\phi} = c_3^\dagger c_4^\dagger c_1^\dagger c_2^\dagger \ket{0} = \ket{\phi}.
\end{equation}

The $1$D bubble on the block \( \tau_2 \) contributes a non-trivial state 
\begin{equation}
\ket{\phi} = c_1^\dagger c_2^\dagger \ket{0}
\end{equation}
on the $0$D block \( \mu_1 \), satisfying 
\begin{equation}
C_2 \ket{\phi} = c_2^\dagger c_1^\dagger \ket{0} = -\ket{\phi},
\end{equation}
and also contributes a non-trivial state 
\begin{equation}
\ket{\phi} = c_1^\dagger c_2^\dagger \ket{0}
\end{equation}
on the $0$D block \( \mu_2 \), satisfying 
\begin{equation}
C_2 \ket{\phi} = c_2^\dagger c_1^\dagger \ket{0} = -\ket{\phi}.
\end{equation}

Thus, the $1$D bubbles form the trivialization group:
\begin{equation}
\{\mathrm{TBS}\}^{0D} = \mathbb{Z}_2.
\end{equation}

\begin{figure}[tb]
    \centering
    \includegraphics[width=0.35\textwidth]{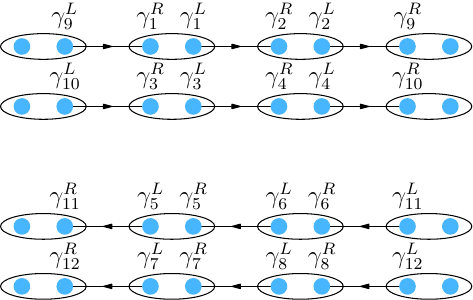}
    \caption{Two single Majorana chains decoration on $1$D block $\tau_1$}
    \label{fig:two_single_Majorana_chain_deformation}
\end{figure}

\begin{figure}[tb]
    \centering
    \includegraphics[width=0.35\textwidth]{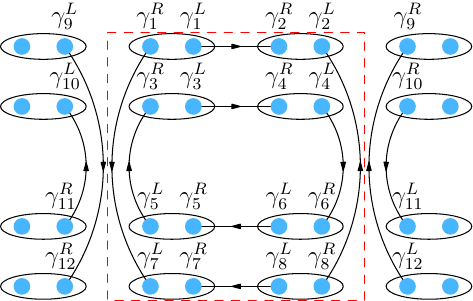}
    \caption{ The state adiabatically deformed from the double Majorana chain (see Fig.~\ref{fig:two_single_Majorana_chain_deformation}). Here the intersite entanglement only occurs in the $0$D block (red dashed region).}
    \label{fig:adiabatically}
\end{figure}

\begin{figure}[tb]
    \centering
    \includegraphics[width=0.43\textwidth]{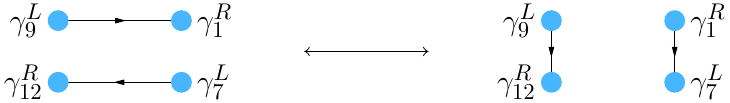}
     \caption{Adiabatic deformation of two coupling pairs of Majorana fermions.}
    \label{fig:pgg adiabatic pair}
\end{figure}

Finally, we consider the extension problem. The physical meaning of the group structure is whether the stacking of $1$D block-states extends to $0$D block-states. To investigate the possible stacking, we consider two identical $1$D block-states: we decorate two copies of single Majorana chain states (which are not trivialized by bubbles) on each $1$D blocks labeled by \( \tau_1 \). This decoration leaves eight dangling Majorana modes (highlighted in red in Fig.~\ref{fig:pgg two chain on tau1}) at each $0$D block labeled by \( \mu_1 \). 

These states can, in fact, be regarded as four copies of Majorana chains, as shown in Fig.~\ref{fig:two_single_Majorana_chain_deformation}. 
We will show that, without closing the energy gap or breaking the time-reversal symmetry, the double chain can be adiabatically deformed into a state where inter-site entanglement only exists between the two sites near the rotation center (see Fig.~\ref{fig:adiabatically}). Consider the following Hamiltonian~\cite{RotationSPT}:
\begin{equation}
H(\theta) = \sin \theta (-i\gamma_9^L \gamma_1^R + i\gamma_{12}^R \gamma_7^L) + \cos \theta (-i\gamma_1^R \gamma_7^L - i\gamma_9^L \gamma_{12}^R). \label{eq:adiabatically}
\end{equation}

When \( \theta = 0 \), the ground state corresponds to the one on the left in Fig.~\ref{fig:pgg adiabatic pair}; when \( \theta = \pi/2 \), the ground state corresponds to the one on the right in Fig.~\ref{fig:pgg adiabatic pair}.

The entire spectrum is independent of \( \theta \), and the ground state has energy \( -2 \). Applying this type of smooth deformation to the four copies of Majorana chains (see Fig.~\ref{fig:two_single_Majorana_chain_deformation}) in a \( C_2 \)-rotation symmetric manner results in the state shown in Fig.~\ref{fig:adiabatically}. Explicitly, we define the \( C_2 \) -rotation transformations as follows:
\begin{equation}
\gamma^h_1 \leftrightarrow \gamma^h_8, \quad \gamma^h_9 \leftrightarrow \gamma^h_{12}, \quad \gamma^h_3 \leftrightarrow \gamma^h_6, \quad \gamma^h_{10} \leftrightarrow \gamma^h_{11},
\end{equation}
where \( h = L, R \).

The state within the red dashed box can be regarded as a zero-dimensional block-state. Our calculations confirm that it is a non-trivial eigenstate of \( C_2 \) (but with even fermion parity). Accordingly, two Majorana chain decorations on the $1$D block \( \tau_1 \) stack into the root state of $0$D block-states.
Therefore, all independent non-trivial block-states with different dimensions are classified as:
\begin{equation}
\begin{aligned}
    E^{\mathrm{2D}} &=\{\mathrm{OFBS}\}^{\mathrm{2D}}= \mathbb{Z}_1, \\
    E^{\mathrm{1D}} &= \{\mathrm{OFBS}\}^{\mathrm{1D}} / \{\mathrm{TBS}\}^{\mathrm{1D}} = \mathbb{Z}_4 \times \mathbb{Z}_2^3, \\
    E^{\mathrm{0D}} &= \{\mathrm{OFBS}\}^{\mathrm{0D}} / \{\mathrm{TBS}\}^{\mathrm{0D}} = \mathbb{Z}_2^3.
\end{aligned}
\end{equation}

Hence, the ultimate classification with an accurate group structure is:
\begin{equation}
\mathcal{G}_0 = \mathbb{Z}_4 \times \mathbb{Z}_2^6,
\end{equation}
where \( \mathbb{Z}_4 \) comes from the extension of $1$D block-states to $0$D block-states.

\subsubsection{Spinful fermions}

For the spinful system, we first consider the decoration of $0$D block-states. For all $0$D blocks \( \mu_1 \) and \( \mu_2 \), the corresponding $0$D block-states are characterized by the distinct 1D irreducible representations of the total symmetry group, which are captured by:
\begin{equation}
\mathcal{H}^1\big((\mathbb{Z}_2 \times \mathbb{Z}_2^T) \times_{\omega_2} \mathbb{Z}_2^f, U_T(1)\big) = \mathbb{Z}_2,
\end{equation}
where the group elements in each \( \mathbb{Z}_2 \) can be labeled by the eigenvalues of \( C_2 \) as \( \pm 1 \). Therefore, the obstruction-free $0$D block-states form the group:
\begin{equation}
\{\mathrm{OFBS}\}^{0D} = \mathbb{Z}_2^2.
\end{equation}

Next, we discuss the decoration of $1$D block-states. At this point, the physical symmetry group of all $1$D blocks is \( G_b = \mathbb{Z}_2^T \). As discussed in Sec.~\ref{sec:Cn spinfull}, an invertible topological Majorana chain cannot be decorated, but  double Majorana chains can, which has a classification of \( \mathbb{Z}_2 \).

The decoration of  double Majorana chains on the $1$D block \( \tau_1 \) leaves eight dangling Majorana fermion modes on the $0$D block \( \mu_1 \). Similarly, the decoration on the $1$D block \( \tau_2 \) leaves four dangling Majorana fermion modes on both \( \mu_1 \) and \( \mu_2 \). Based on the point group discussion, these Majorana fermions can gap out the zero modes on the $0$D blocks with physical symmetry \( G_b = \mathbb{Z}_2 \times \mathbb{Z}_2^T \), making the decoration obstruction-free. 
Therefore, the contribution to the classification group of the 1D decoration is:
\begin{equation}
\{\mathrm{OFBS}\}^{1D} = \mathbb{Z}_2^2,
\end{equation}
where the group elements can be labeled as:
\begin{equation}
[\mathbb{T}_1,\mathbb{T}_2].
\end{equation}
Here, \( \mathbb{T}_j = 0, 1 \) represents the number of decorated double Majorana chain states, protected by  $G_f= \mathbb{Z}_2^T\times_{\omega_2} \mathbb{Z}_2^f$, on the $1$D blocks \( \tau_j \) (\( j = 1, 2 \)), respectively.

For the same reasons as in the previously discussed spinful example, the decoration of 2D DIII TSC states is always obstruction-free, contributing to a classification \begin{equation}
  \{\mathrm{OFBS}\}^{2D} = \mathbb{Z}_2  .
\end{equation}    

Next, we consider the bubble equivalence relations. The $1$D bubble on the block \( \tau_1 \) acts on the $0$D block \( \mu_1 \), and the $1$D bubble on the block \( \tau_2 \) acts on both \( \mu_1 \) and \( \mu_2 \). However, based on the conclusions drawn from Sec.~\ref{sec:Cn spinfull}, the $1$D bubbles do not trivialize the $0$D block-states with on-site physical symmetry \( G_b = \mathbb{Z}_2 \times \mathbb{Z}_2^T \). Similarly, the $2$D bubbles do not have any trivializing effect on either 0D or $1$D block-states. 
Therefore, the total trivialization groups are:
\begin{equation}
\{\mathrm{TBS}\}^{\mathrm{1D}} =\{\mathrm{TBS}\}^{\mathrm{0D}}= \mathbb{Z}_1.
\end{equation}

Lastly, we consider the extension problem. Since the \( pgg \) group involves a ``glide reflection" symmetry operation (for the decoration of 2D chiral fermions, \( \tau_1 \) is effectively a reflection axis), similar to the discussions in Sec.~\ref{sec:D4 spinful discussion}, the boundary Majorana mode mass terms of the 2D decorated state, satisfying the \( pgg \) group symmetry, acquire an additional negative sign under symmetry transformations. 
Thus, there exists an extension of the $2$D block-states to the $1$D block-states on the $1$D block \( \tau_1 \)$\tau_1$, however, we find no extension from 1D block-states to 0D block-states in this case.

Therefore, all independent non-trivial block-states with different dimensions are classified as:
\begin{equation}
\begin{aligned}
    E^{\mathrm{2D}} &=\{\mathrm{OFBS}\}^{\mathrm{2D}}= \mathbb{Z}_2, \\
    E^{\mathrm{1D}} &= \{\mathrm{OFBS}\}^{\mathrm{1D}} / \{\mathrm{TBS}\}^{\mathrm{1D}} = \mathbb{Z}_2^2, \\
    E^{\mathrm{0D}} &= \{\mathrm{OFBS}\}^{\mathrm{0D}} / \{\mathrm{TBS}\}^{\mathrm{0D}} = \mathbb{Z}_2^2.
\end{aligned}
\end{equation}
The final classification group is:
\begin{equation}
\mathcal{G}_{1/2} = \mathbb{Z}_2^3 \times \mathbb{Z}_4,
\end{equation} where the factor $\mathbb{Z}_4$ comes from the extension of the $2$D block-states to the $1$D block-states.

\subsection{Hexagonal Lattice: \( p6m \)}

For the hexagonal lattice, we demonstrate the crystalline TSC protected by $p6m$ and time-reversal symmetry as an example.
 For the $2$D block labeled as \( \sigma \), there is no site symmetry, and the physical symmetry group is \( \mathbb{Z}_2^T \). For any $1$D block, the physical symmetry is \( \mathbb{Z}_2^T \times \mathbb{Z}_2^M \), in which $\mathbb{Z}_2^M$ is attributed to the reflection symmetry. 

For the $0$D block \( \mu_1 \), the physical symmetry group is \( \mathbb{Z}_6 \rtimes \mathbb{Z}_2 \times \mathbb{Z}_2^T \), due to the internal action of the \( D_6 \) group. For the $0$D block \( \mu_2 \), the physical symmetry is \( \mathbb{Z}_2 \rtimes \mathbb{Z}_2 \times \mathbb{Z}_2^T \), due to the internal action of \( D_2 \subset D_6 \). For the $0$D block \( \mu_3 \), the physical symmetry is \( \mathbb{Z}_3 \rtimes \mathbb{Z}_2 \times \mathbb{Z}_2^T \), due to the internal action of \( D_3 \subset D_6 \). 

\begin{figure}[tb]
    \centering
    \includegraphics[width=0.46\textwidth]{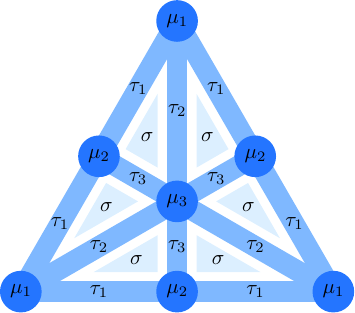}
     \caption{The \#17  2D space group $p6m$ and its  cell decomposition.}
    \label{p6m}
\end{figure}

The cell decomposition is shown in Fig.~\ref{p6m}.

 \subsubsection{Spinless fermions}

For spinless systems, we first consider the $0$D block-state decoration. For the $0$D blocks \( \mu_1 \) and \( \mu_2 \), candidate states can be characterized by the distinct 1D irreducible representations of the total symmetry group (\( n = 2, 6 \)), which are captured by:
\begin{equation}
\mathcal{H}^1(D_n \times \mathbb{Z}_2^T \times \mathbb{Z}_2^f, U_T(1)) = \mathbb{Z}_2^3,
\end{equation}
while for the $0$D block \( \mu_3 \),
\begin{equation}
\mathcal{H}^1(D_3 \times \mathbb{Z}_2^T \times \mathbb{Z}_2^f, U_T(1)) = \mathbb{Z}_2^2.
\end{equation}

At each $0$D block labeled by \( \mu_1 \) and \( \mu_2 \), the block-states can be labeled by the eigenvalues \( \pm 1 \) of \( C_6 / C_2 \), \( M \), and \( P_f \). Similarly, on the $0$D block \( \mu_3 \), the block-states are labeled by the eigenvalues \( \pm 1 \) of \( M \) and \( P_f \).

Therefore, the obstruction-free $0$D block-states form the group:
\begin{equation}
\{\mathrm{OFBS}\}^{0D} = \mathbb{Z}_2^8.
\end{equation}

Next, we discuss the decoration of $1$D block-states. The physical symmetry group for all $1$D blocks is given by \( G_b = \mathbb{Z}_2^M \times \mathbb{Z}_2^T \). 
As established in Sec.~\ref{sec:D4 spinless fermion}, the corresponding FSPT phases are classified as \( \mathbb{Z}_4 \times \mathbb{Z}_4 \). The generators of these $\mathbb{Z}_4$ factors correspond to the root phases $n_1^M$ and $n_1^T$, whose symmetry transformations were previously specified in Eqs.~\eqref{eq:symm property of n1T} and~\eqref{eq:symm property of n1M}.
 Furthermore, we can incorporate an invertible topological order, that is, a single Majorana chain. 

We now consider the $1$D block-state decoration on the $1$D block \( \tau_1 \). Based on our analysis of the point groups \( D_2 \) and \( D_6 \), the decoration of 1D BSPT (four Majorana chains) in the phase \( (-1)^{n_1^T \cup n_1^T} \) or \( (-1)^{n_1^M \cup n_1^M} \)  can gap out the zero modes at the $0$D blocks \( \mu_1 \) and \( \mu_2 \), making this decoration obstruction-free. This configuration is classified as \( \mathbb{Z}_2 \times \mathbb{Z}_2 \).

For the $1$D blocks \( \tau_2 \) and \( \tau_3 \), the only feasible decoration is a simultaneous decoration on both blocks. This is because the $0$D block \( \mu_3 \) (with point group \( D_3 \)) allows only a 1D FSPT decoration (two Majorana chains) in the \( n_1^M \) (or \( n_1^T \)) phase on \( \tau_2 \) and \( \tau_3 \) together to remain obstruction-free. Furthermore, since \( \tau_2 \) and \( \tau_3 \) are also connected to \( \mu_1 \) and \( \mu_2 \), and are restricted by the conclusions for the point groups \( D_2 \) and \( D_6 \), only the decoration of 1D BSPT decoration (four Majorana chains) in the  phase \( (-1)^{n_1^T \cup n_1^T} \) or \( (-1)^{n_1^M \cup n_1^M} \) on  \( \tau_2 \) and \( \tau_3 \) can simultaneously gap out the zero modes at \( \mu_1 \), \( \mu_2 \), and \( \mu_3 \). This configuration is also classified as \( \mathbb{Z}_2 \times \mathbb{Z}_2 \).

Moreover, decorating the 1D blocks \(\tau_1\), \(\tau_2\), and \(\tau_3\) simultaneously with the 1D FSPT phase \(n_1^{T}+n_1^{M}\) is also obstruction-free for the point groups \(D_2\) and \(D_6\). For \(D_3\), the same decoration lies in the subgroup of obstruction-free states. Therefore, this type of decoration contributes to the \(\mathbb{Z}_4\) classification.

Therefore, eliminating duplicate parts, the obstruction-free $1$D block-states form the group:
\begin{equation}
\{\mathrm{OFBS}\}^{1D} = \mathbb{Z}_4 \times \mathbb{Z}_2^3,
\end{equation}
where the group elements can be labeled as:
\begin{equation}
[mt_{123},T_1,  T_{23}, M_{23}].
\end{equation}
Here $M_{23}\in\{0,1\}$ and $T_{23}\in\{0,1\}$ denote, respectively, the number of decorated 1D phases $(-1)^{n_1^M\cup n_1^M}$ and $(-1)^{n_1^T\cup n_1^T}$ on the 1D blocks $\tau_2$ and $\tau_3$. Similarly, $mt_{123}$ denotes the number of decorated 1D FSPT phases $n_1^T+n_1^M$ placed simultaneously on the 1D blocks $\tau_1,\tau_2,\tau_3$. Finally, $T_1\in\{0,1\}$ denotes the number of decorated 1D phases $(-1)^{n_1^T\cup n_1^T}$ on the 1D block $\tau_1$.

The total symmetry group for the $2$D blocks is \( G_f = \mathbb{Z}_2^T \times \mathbb{Z}_2^f \). As concluded in Ref.~\cite{Wang2020}, there are no corresponding FSPT phases under this symmetry group. Consequently, the contribution of $2$D block decoration is classified as:
\begin{equation}
\{\mathrm{OFBS}\}^{2D} = \mathbb{Z}_1.
\end{equation}

Next, we consider bubble equivalence relations. First, we examine the $1$D bubble equivalences by decorating a pair of complex fermions on each $1$D block.
The $1$D bubble on the $1$D block \( \tau_1 \) contributes to a non-trivial state
\begin{equation}
\ket{\phi} = c_1^\dagger c_2^\dagger c_3^\dagger c_4^\dagger c_5^\dagger c_6^\dagger \ket{0}
\end{equation}
on the $0$D block \( \mu_1 \), which satisfies
\begin{equation}
M_{\tau_2} \ket{\phi} = c_2^\dagger c_1^\dagger c_6^\dagger c_5^\dagger c_4^\dagger c_3^\dagger \ket{0} = -\ket{\phi}.
\end{equation}

It also contributes a non-trivial state
\begin{equation}
\ket{\phi} = c_1^\dagger c_2^\dagger \ket{0}
\end{equation}
on the $0$D block \( \mu_2 \), which satisfies
\begin{equation}
M_{\tau_3} \ket{\phi} = c_2^\dagger c_1^\dagger \ket{0} = -\ket{\phi}.
\end{equation}

Similarly, the $1$D bubble on the $1$D block \( \tau_2 \) contributes to a non-trivial state \( \ket{\phi} \) on the $0$D block \( \mu_1 \), which satisfies
\begin{equation}
M_{\tau_1} \ket{\phi} = -\ket{\phi},
\end{equation}
and a non-trivial state \( \ket{\phi} \) on the $0$D block \( \mu_3 \), which satisfies
\begin{equation}
M_{\tau_2} \ket{\phi} = -\ket{\phi}, \quad M_{\tau_3} \ket{\phi} = -\ket{\phi}, \quad P_f \ket{\phi} = -\ket{\phi}.
\end{equation}

The $1$D bubble on the $1$D block \( \tau_3 \) contributes to a non-trivial state \( \ket{\phi} \) on the $0$D block \( \mu_2 \), which satisfies
\begin{equation}
M_{\tau_1} \ket{\phi} = -\ket{\phi},
\end{equation}
and a non-trivial state \( \ket{\phi} \) on the $0$D block \( \mu_3 \), which satisfies
\begin{equation}
M_{\tau_2} \ket{\phi} = -\ket{\phi}, \quad M_{\tau_3} \ket{\phi} = -\ket{\phi}, \quad P_f \ket{\phi} = -\ket{\phi}.
\end{equation}

We can also consider bosonic $1$D bubbles. The 1D bosonic bubble on the $1$D block \( \tau_2 \) contributes a non-trivial state \( \ket{\phi} \) on the $0$D block \( \mu_3 \), which satisfies
\begin{equation}
M_{\tau_2} \ket{\phi} = -\ket{\phi}, \quad M_{\tau_3} \ket{\phi} = -\ket{\phi}.
\end{equation}

Similarly, the 1D bosonic bubble on the $1$D block \( \tau_3 \) contributes a non-trivial state \( \ket{\phi} \) on the $0$D block \( \mu_3 \), which satisfies
\begin{equation}
M_{\tau_2} \ket{\phi} = -\ket{\phi}, \quad M_{\tau_3} \ket{\phi} = -\ket{\phi}.
\end{equation} Thus, all $1$D bubbles contribute to a $\mathbb{Z}_2^4$ classification.

2D ``Majorana bubbles" can contribute to 1D non-trivial states, whose classification is \( \mathbb{Z}_4 \), on all $1$D blocks, including \( n_1^M+ n_1^T \) and \( (-1)^{n_1^M \cup n_1^M} \oplus (-1)^{n_1^T \cup n_1^T} \) phases through stacking.

Therefore, the total trivialization group are:
\begin{equation}
\{\mathrm{TBS}\}^{\mathrm{1D}} = \mathbb{Z}_4  ,~\{\mathrm{TBS}\}^{\mathrm{0D}}=\mathbb{Z}_2^4.
\end{equation}

Therefore, all independent non-trivial block-states with different dimensions are classified as:
\begin{equation}
\begin{aligned}
    E^{\mathrm{2D}} &=\{\mathrm{OFBS}\}^{\mathrm{2D}}= \mathbb{Z}_1, \\
    E^{\mathrm{1D}} &= \{\mathrm{OFBS}\}^{\mathrm{1D}} / \{\mathrm{TBS}\}^{\mathrm{1D}} = \mathbb{Z}_2^3, \\
    E^{\mathrm{0D}} &= \{\mathrm{OFBS}\}^{\mathrm{0D}} / \{\mathrm{TBS}\}^{\mathrm{0D}} = \mathbb{Z}_2^4.
\end{aligned}
\end{equation}

It is evident that there is no extension between the 1D and $0$D block-states. Hence, the ultimate classification with an accurate group structure is:
\begin{equation}
\mathcal{G}_0 = \mathbb{Z}_2^{7}.
\end{equation}

\subsubsection{Spinful fermions}

For spinful systems, we first consider the $0$D block-state decoration. For the $0$D blocks \( \mu_1 \) and \( \mu_2 \), candidate states can be characterized by the distinct 1D irreducible representations of the total symmetry group (\( n = 2, 4 \)), which are captured by:
\begin{equation}
\mathcal{H}^1\big((D_n \times \mathbb{Z}_2^T) \times_{\omega_2} \mathbb{Z}_2^f, U_T(1)\big) = \mathbb{Z}_2^2,
\end{equation}
while for the $0$D block \( \mu_3 \),
\begin{equation}
\mathcal{H}^1\big(D_3 \times \mathbb{Z}_2^T \times_{\omega_2} \mathbb{Z}_2^f, U_T(1)\big) = \mathbb{Z}_2.
\end{equation}

At each $0$D block labeled by \( \mu_1 \) and \( \mu_2 \), the block-states can be labeled by the eigenvalues \( \pm 1 \) of \( C_6 / C_2 \) and \( M \). Similarly, on the $0$D block \( \mu_3 \), the block-states are labeled by the eigenvalues \( \pm 1 \) of \( M \).
Therefore, the obstruction-free $0$D block-states form the group:
\begin{equation}
\{\mathrm{OFBS}\}^{0D} = \mathbb{Z}_2^5.
\end{equation}

Next, we discuss the decoration of $1$D block-states. At this stage, the physical symmetry group for the $1$D blocks \( \tau_1 \), \( \tau_2 \), and \( \tau_3 \) is \( G_b = \mathbb{Z}_2 \times \mathbb{Z}_2^T \). As discussed earlier, an invertible topological Majorana chain cannot be decorated, but 1D FSPT decoration (implemented via double Majorana chains) can, which has a classification of \( \mathbb{Z}_4 \). The decoration of the 1D FSPT phase on the $1$D block \( \tau_1 \) leaves twelve dangling Majorana fermion modes on the $0$D block \( \mu_1 \) and four dangling Majorana fermion modes on the $0$D block \( \mu_2 \). Based on  discussions in Sec.~\ref{sec:D2 D6 spinful}, these Majorana fermions can gap out the zero modes on the $0$D blocks with physical symmetries \( G_b = D_6 \times \mathbb{Z}_2^T \) and \( G_b = D_2 \times \mathbb{Z}_2^T \), making this decoration obstruction-free.

The decoration of the 1D FSPT phase on the $1$D block \( \tau_2 \) leaves twelve dangling Majorana fermion modes on the $0$D block \( \mu_1 \) and six dangling Majorana fermion modes on the $0$D block \( \mu_3 \). The Majorana fermions on the $0$D block \( \mu_1 \) can gap out the zero modes, consistent with the previous case. However, based on our discussions of the point group, the Majorana fermions on the $0$D block \( \mu_3 \) cannot gap out the zero modes with physical symmetry \( G_b = D_3 \times \mathbb{Z}_2^T \).

The case for \( \tau_3 \) is similar to \( \tau_2 \): the Majorana fermions on the $0$D block \( \mu_1 \) can gap out the zero modes, but those on the $0$D block \( \mu_3 \) cannot. However, if we decorate the double Majorana chains on both \( \tau_2 \) and \( \tau_3 \), the zero modes on \( \mu_3 \) can also be gapped out, as discussed earlier. Thus, the contribution to the classification group from the 1D decoration is:
\begin{equation}
\{\mathrm{OFBS}\}^{1D} = \mathbb{Z}_4^2,
\end{equation}
where the group elements can be labeled as:
\begin{equation}
[\mathbb{MT}_1, \mathbb{MT}_2, \mathbb{MT}_3],\quad \text{with}~\mathbb{MT}_2=\mathbb{MT}_3
\end{equation}

Here, \( \mathbb{MT}_j \in \{ 0, 1, 2, 3\} \) represents the number of decorated 1D double Majorana chains, protected by $G_f=(\mathbb{Z}_2 \times \mathbb{Z}_2^T)\times_{\omega_2} \mathbb{Z}_2^f$, on the $1$D blocks \( \tau_j \) (\( j = 1, 2, 3 \)), respectively. According to the aforementioned discussions, a necessary condition for an obstruction-free block-state is \( \mathbb{MT}_2 = \mathbb{MT}_3 \).

Following analogous arguments to the spinful case discussed earlier, the decoration of 2D DIII TSC states is always obstruction-free, contributing to a \( \{\mathrm{OFBS}\}^{2D} = \mathbb{Z}_2 \) classification. 

Next, we consider the bubble equivalence relations. The $1$D bubble on the $1$D block \( \tau_1 \) acts on the $0$D blocks \( \mu_1 \) and \( \mu_2 \), the $1$D bubble on the $1$D block \( \tau_2 \) acts on the $0$D blocks \( \mu_1 \) and \( \mu_3 \), and the $1$D bubble on the $1$D block \( \tau_3 \) acts on the $0$D blocks \( \mu_2 \) and \( \mu_3 \).

However, as concluded in  Sec.~\ref{sec:D2 D6 spinful}, the $1$D bubbles do not trivialize the $0$D block-states with on-site physical symmetry \( G_b = D_2 \times \mathbb{Z}_2^T \) and \( G_b = D_6 \times \mathbb{Z}_2^T \), but they do trivialize the $0$D block-states with symmetry \( G_b = D_3 \times \mathbb{Z}_2^T \). Similarly, the $2$D bubbles have no trivializing effect on either 0D or $1$D block-states. 
Therefore, the total trivialization groups are:
\begin{equation}
\{\mathrm{TBS}\}^{\mathrm{0D}} = \mathbb{Z}_2,\{\mathrm{TBS}\}^{\mathrm{1D}}=\mathbb{Z}_1 .
\end{equation}

Lastly, we consider the extension problem. According to the discussions in Sec.~\ref{sec:D4 spinful discussion}, the $2$D block-states decorated on both sides of all $1$D blocks (i.e., two layers of DIII topological superconductors) contribute non-trivial $1$D block-states, since the dihedral point group is involved.
Thus, there is an extension from the $2$D block-states to the $1$D block-states, where double Majorana chains are decorated on the $1$D blocks \( \tau_1 \), \( \tau_2 \), and \( \tau_3 \). Furthermore, there is no extension from 1D block-states to 0D block-states in this case.

Therefore, all independent non-trivial block-states with different dimensions are classified as:
\begin{equation}
\begin{aligned}
    E^{\mathrm{2D}} &=\{\mathrm{OFBS}\}^{\mathrm{2D}}= \mathbb{Z}_2, \\
    E^{\mathrm{1D}} &= \{\mathrm{OFBS}\}^{\mathrm{1D}} / \{\mathrm{TBS}\}^{\mathrm{1D}} = \mathbb{Z}_4^2, \\
    E^{\mathrm{0D}} &= \{\mathrm{OFBS}\}^{\mathrm{0D}} / \{\mathrm{TBS}\}^{\mathrm{0D}} = \mathbb{Z}_2^4.
\end{aligned}
\end{equation}
As a result, the final classification is:
\begin{equation}
\mathcal{G}_{1/2} = \mathbb{Z}_2^4 \times \mathbb{Z}_8 \times \mathbb{Z}_4,
\end{equation}
where the contribution \( \mathbb{Z}_8 \) comes from the extension of the $2$D block-states to the $1$D block-states.

\section{CONCLUSIONS}
In conclusion, our study presents a rigorous and systematic framework for the construction and classification of 2D interacting fermionic symmetry-protected topological superconductors with crystalline symmetry (including point group and space group symmetries) and time-reversal symmetry. Through explicit real-space constructions, we have elucidated the intricate landscape of these fascinating quantum states, providing a comprehensive understanding of their topological properties. Our work extends beyond mere theoretical constructs; it offers a practical roadmap for experimental realization in real-world materials. Notably, the occurrence of time-reversal symmetric crystalline topological superconductors in materials like iron-based superconductors underscores the relevance and potential impact of our findings in the realm of quantum material.

Moreover, it is crucial to emphasize the versatility and applicability of the methodology proposed in our study. By extending this approach to systems with other internal symmetries, such as spin rotational symmetry and charge conservation symmetry, we open the door to a more holistic understanding of the ten-fold way classification with additional space group symmetries for 2D interacting fermion systems. This extension promises to enrich our insights into the interplay between different symmetries and their effects on the emergent topological phases, paving the way for a deeper exploration of novel quantum phenomena.

\begin{acknowledgments}
The authors thank Shangqiang Ning, Weicheng Ye, Wenjie Xi, and Qingrui Wang for their invaluable assistance with the mathematical details and for insightful discussions on the interpretation of the relevant physical ideas. This work is supported by funding from Hong Kong’s Research Grants Council (GRF No. 14307621 and CRF C7015-24G)
\end{acknowledgments}

\appendix

\section{Group extension} \label{sec:group extension}

Let \( G \) be a finite group, and let \( N \subset G \) be a normal subgroup. Then we can form the quotient group \( Q = G / N \). We say that \( G \) is an extension of \( Q \) by \( N \). Equivalently, the three groups \( N \), \( G \), and \( Q \) fit into the following short exact sequence:
\begin{equation}
    1 \to N \to G \to Q \to 1. \label{eq:short exact}
\end{equation}

This short exact sequence means that there is a surjective map \( \pi: G \to Q \), whose kernel is exactly \( N \).

The extension is said to be central if \( N \) lies in the center of \( G \), which also implies that \( N \) is an Abelian group. All extensions considered in this paper are central. The extension problem refers to determining \( G \) from \( N \) and \( Q \). For a central extension, the additional information needed to determine \( G \) is a group 2-cocycle \( \omega \in Z^2[Q, N] \). 

More explicitly, \( \omega \) is a function from \( Q \times Q \) to \( N \) that satisfies the 2-cocycle condition:
\begin{equation}
    \omega(q_1, q_2)\omega(q_1 q_2, q_3) = \omega(q_1, q_2 q_3)\omega(q_2, q_3),
\end{equation}
for all \( q_1, q_2, q_3 \in Q \).

Once \( \omega \) is given, \( G \) can be explicitly constructed as follows: we represent \( G \) set-wise as the Cartesian product
\begin{equation}
G = Q \times N = \{(q, n) \mid q \in Q, n \in N \},
\end{equation}
equipped with the following group multiplication law:
\begin{equation}
    (q_1, n_1) \times (q_2, n_2) = (q_1 q_2, n_1 n_2 \omega(q_1, q_2)).
\end{equation}

The 2-cocycle condition of \( \omega \) ensures that the multiplication is associative. The map \( \pi \) is defined as \( \pi((q, n)) = q \).

It can be shown that the group extension depends only on the cohomology class \( [\omega] \), and thus corresponds one-to-one with the group cohomology \( \mathcal{H}^2[Q, N] \). 
In particular, inequivalent extensions correspond to inequivalent short exact sequences~\ref{eq:short exact}, even though distinct extensions with fixed $Q$ and $N$ may yield isomorphic total groups $G$.

\section{Other cases of time-reversal symmetric crystalline topological superconductors } \label{sec:other cases}

 \subsection{$p1$}
 The on-site physical symmetry group for any block in $p1$ is $G_b = \mathbb{Z}_2^T$.

 \begin{figure}[bt]
     \centering
     \includegraphics[width=0.4\textwidth]{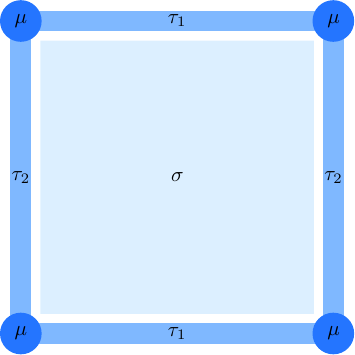}
     \caption{The \#1  2D space group $p1$ and its  cell decomposition.}
     \label{p1}
 \end{figure}

 \subsubsection{Spinless fermions}
  For $0$D block $\mu$, the corresponding $0$D block-states can be determined by the distinct 1D irreducible representations of the total symmetry group $G_f=\mathbb{Z}_2^T \times  \mathbb{Z}_2^f$, which are captured by:
\begin{equation}
    \mathcal{H}^1(\mathbb{Z}_2^T \times  \mathbb{Z}_2^f,U_T(1))=\mathbb{Z}_2,
\end{equation} where the two irreps are distinguished by the $P_f$ eigenvalue $\pm1$. Therefore, all obstruction-free $0$D block-states form the group \begin{equation}
  \{OFBS \}^{0D}=\mathbb{Z}_2.  
\end{equation}

\begin{figure}[tb]
    \centering
    \includegraphics[width=0.3\textwidth]{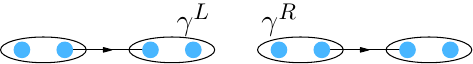}
\caption{Two single Majorana chains decorating the $1$D block \( \tau_1 \). At their junction, corresponding to the $0$D block \( \mu \), dangling Majorana modes \( \gamma^{L} \) and \( \gamma^{R} \) remain, which transform differently under the symmetry.}
    \label{fig:p1_gammaR}
\end{figure}

Next, we discuss the decoration of 1D block states. In this case, the total symmetry group relevant to each 1D block is $G_f=\mathbb{Z}_2^T\times \mathbb{Z}_2^f$. As reviewed in Sec.~\ref{sec:appendix spinless Cn}, one can decorate a 1D block by 1D FSPT phases classified by $\mathbb{Z}_4$; the root state is given by a double Majorana chain, whose symmetry actions are specified in Eq.~\eqref{eq:symm property spinless Cn 2chain}.

In addition, we may decorate the block with an invertible topological order, namely a single Majorana chain. Note that stacking two such single Majorana chains is equivalent to the \(\mathbb{Z}_4\) FSPT generator (the double Majorana chains). With both the single Majorana chain and the double-chain \(\mathbb{Z}_4\) generator available, all proper 1D block-state decorations can be viewed as class-BDI topological superconductors~\cite{Fidkowski_2010} consisting of \(v\) Majorana chains, which collapse to a \(\mathbb{Z}_8\) classification in the presence of interactions, i.e., \(v=1,2,\ldots,8\).

All $1$D blocks $\tau_1$ adjacent to a given  $0$D block  are related by translation symmetry. Consequently, decorating a single Majorana chain on  $1$D blocks $\tau_1$ generates two distinct types of dangling Majorana modes, denoted as $\gamma^L$ and $\gamma^R$ (Fig.~\ref{fig:p1_gammaR}), localized at the $0$D block $\mu$. Crucially, these two modes exhibit different symmetry properties, as characterized by Eq.~\eqref{eq:symm property spinless Cn 2chain}. The spectral gap for these modes can be opened by introducing the interaction term $i \gamma^L \gamma^R$.
Similarly, for decorating the double Majorana chain on the $1$D blocks \( \tau_1 \), we use the same form of interaction terms, such as \( i \gamma_1^L \gamma_2^R + i \gamma_3^L \gamma_4^R \), to gap out the zero modes. Therefore, both single Majorana chain and double Majorana chain decorations on the $1$D block \( \tau_1 \) are obstruction-free and admit  a \( \mathbb{Z}_8 \) classification. 
The same conclusion holds for the decoration on the $1$D blocks \( \tau_2 \).

Hence, all obstruction-free $1$D block-states form the group:
\begin{equation}
\{\mathrm{OFBS}\}^{1D} = \mathbb{Z}_8 \times \mathbb{Z}_8,
\end{equation}
where the group elements can be labeled as \( [\mathcal{M}_1, \mathcal{M}_2] \), with \( \mathcal{M}_j = 0, 1, 2, \dots,7 \)  representing the number of decorated single Majorana chain states on \( \tau_j \) (\( j = 1, 2 \)), respectively.

The total symmetry group for $2$D blocks is \( G_f = \mathbb{Z}_2^T \times \mathbb{Z}_2^f \), and there are no corresponding FSPT phases~\cite{Wang2020}. Therefore, the obstruction-free 2D  decorations are classified as:
\begin{equation}
\{\mathrm{OFBS}\}^{2D} = \mathbb{Z}_1.
\end{equation}

Next, we examine the bubble equivalence. As in Sec.~\ref{sec:appendix spinless Cn}, $2$D bubbles do not generate nontrivial 1D FSPT phases: adjacent bubbles around a $1$D block contribute Majorana chains with opposite pairing orientations and hence are gappable. Instead, they induce a nontrivial 0D \(P_f\)-eigenstate.
We then consider the $1$D bubble equivalences and decorate a pair of complex fermions on each $1$D block. However, $1$D bubbles can only contribute to 0D even fermion-parity block-states and thus have no impact on $0$D block-states. Consequently, the 0D trivialization originates solely from the 2D bubbles. In summary,  bubbles form the trivialization groups:
\begin{equation}
\{\mathrm{TBS}\}^{1D} = \mathbb{Z}_1, \quad \{\mathrm{TBS}\}^{0D} = \mathbb{Z}_2.
\end{equation}

There is no extension between 1D and $0$D block-states. Thus, all independent non-trivial block-states with different dimensions are classified by:
\begin{align}
    \begin{aligned}
        & E^{\mathrm{2D}} =\{\mathrm{OFBS}\}^{\mathrm{2D}}= \mathbb{Z}_1, \\
        & E^{\mathrm{1D}} = \{\mathrm{OFBS}\}^{\mathrm{1D}} / \{\mathrm{TBS}\}^{\mathrm{1D}} = \mathbb{Z}_8^2, \\
        & E^{\mathrm{0D}} = \{\mathrm{OFBS}\}^{\mathrm{0D}} / \{\mathrm{TBS}\}^{\mathrm{0D}} = \mathbb{Z}_1.
    \end{aligned}
\end{align}

Hence, the ultimate classification with an accurate group structure is:
\begin{equation}
\mathcal{G}_0 =  \mathbb{Z}_8^{2}.
\end{equation}

 \subsubsection{Spinful fermions}
For the $0$D block \( \mu \), the corresponding $0$D block-states can be determined by the 1D irreducible representation of the total symmetry group \( G_f = \mathbb{Z}_2^T \rtimes \mathbb{Z}_2^f = \mathbb{Z}_4^{fT} \), which are captured by:
\begin{equation}
    \mathcal{H}^1(\mathbb{Z}_4^{fT}, U_T(1)) = \mathbb{Z}_1,
\end{equation}
indicating that there are no non-trivial $0$D block-states.

According to Sec.~\ref{sec:Cn spinfull}, we can decorate with 1D FSPT phases classified by \( \mathbb{Z}_2 \). Such 1D FSPT phases feature a root state consisting of double Majorana chains, with symmetry properties defined in Eq.~\eqref{eq:trs definition}. 

\begin{figure}[tb]
    \centering
    \includegraphics[width=0.6\linewidth]{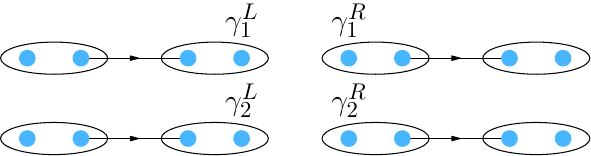}
    \caption{Two 1D FSPT phase decorations on the $1$D block \( \tau_1 \), leaving dangling Majorana modes \( \gamma_j^L \) and \( \gamma_j^R \) ($j=1,2$) at the $0$D block \( \mu \).}
    \label{fig:p1_tau1_two_chains}
\end{figure}

 All $1$D blocks $\tau_1$ adjacent to a given  $0$D block  are related by translation symmetry. Consequently, decorating a 1D FSPT phase on  $1$D blocks $\tau_1$ generates two distinct types of dangling Majorana modes (Fig.~\ref{fig:p1_tau1_two_chains}), denoted as $\gamma_j^L$ and $\gamma_j^R$ ($j=1,2$), localized at the $0$D block $\mu$. Crucially, these two modes exhibit different symmetry properties, as characterized by Eq.~\eqref{eq:trs definition}. The spectral gap for these modes can be opened by introducing the interaction term $i \gamma_1^L \gamma_1^R+i \gamma_2^L \gamma_2^R$. An analogous decoration procedure applied to the $1$D blocks $\tau_2$ similarly produces gappable Majorana modes.  Hence, the contribution to the classification from $1$D block-state decorations is:
\begin{equation}
\{\mathrm{OFBS}\}^{1D} = \mathbb{Z}_2^2,
\end{equation}
where the group elements can be labeled as:
\begin{equation}
[\mathbb{T}_1,\mathbb{T}_2].
\end{equation}
Here, \( \mathbb{T}_j = 0, 1 \) represents the number of decorated 1D FSPT phases (double Majorana chains) on the $1$D blocks \( \tau_j \) (\( j = 1, 2 \)), respectively.

The physical symmetry group for $2$D blocks is \( G_b = \mathbb{Z}_2^T \), and the corresponding $2$D block-states have a classification \( \mathbb{Z}_2 \), where the root state is the 2D DIII TSC state. The DIII TSC decoration on the $2$D blocks can always open a spectral gap at each $1$D block. Therefore, the contribution of $2$D block decorations is classified as:
\begin{equation}
\{\mathrm{OFBS}\}^{2D} = \mathbb{Z}_2.
\end{equation}

According to the results of Sec.~\ref{sec:Cn spinfull}, the $2$D bubble does not trivialize any 0D or $1$D block-states, since in this setting no symmetry can protect the induced effective 1D state. Moreover, considering $1$D bubbles is not meaningful here, because there are no non-trivial 0D decoration states to begin with.  Therefore,  bubbles form the trivialization groups:
\begin{equation}
\{\mathrm{TBS}\}^{1D} = \mathbb{Z}_1, \quad \{\mathrm{TBS}\}^{0D} = \mathbb{Z}_1.
\end{equation}

Lastly, we consider the extension problem. We find no extension from 1D block-states to 0D block-states in this case. The 2D decorations contribute chiral edge modes along each $1$D block; however, in a manner similar to the discussion in Sec.~\ref{sec:extension cn spinful}, the mass terms that gap the boundary Majorana modes of the 2D-decorated states do not pick up any additional sign under symmetry transformations. Therefore, there is no extension from $2$D block-states to $1$D block-states.

All independent non-trivial block-states with  different dimensions are classified by:
 \begin{align}
    \begin{aligned}
         & E^{\mathrm{2D}}=\{\mathrm{OFBS}\}^{\mathrm{2D}}=\mathbb{Z}_2  \\
         & E^{\mathrm{1D}}=\{\mathrm{OFBS}\}^{\mathrm{1D}}/\{\mathrm{TBS}\}^{\mathrm{1D}}=\mathbb{Z}_2^2   \\
         & E^{\mathrm{0D}}=\{\mathrm{OFBS}\}^{\mathrm{0D}}/\{\mathrm{TBS}\}^{\mathrm{0D}}=\mathbb{Z}_1,
    \end{aligned}
\end{align}  The final classification is $\mathbb{Z}_2^3$.

\subsection{$pm$}
\begin{figure}[tb]
    \centering
    \includegraphics[width=0.46\textwidth]{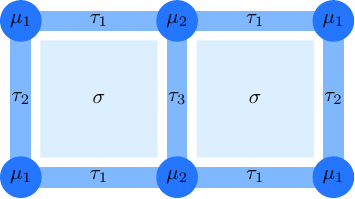}
    \caption{The \#3  2D space group $pm$ and its  cell decomposition.}
    \label{pm}
\end{figure}

For the 2D and $1$D blocks labeled by $\sigma$ and $\tau_1$, there is no site symmetry, and the physical symmetry group is $\mathbb{Z}_2^T$. For the $1$D blocks labeled by $\tau_2$ and $\tau_3$, the internal physical symmetry arises from the reflection symmetry and is given by $G_b=\mathbb{Z}_2 \times \mathbb{Z}_2^T$. All $0$D blocks possess the physical symmetry $G_b=\mathbb{Z}_2 \times \mathbb{Z}_2^T$ due to reflection symmetry, as shown in Fig.~\ref{pm}.

\subsubsection{Spinless fermions}
 As discussed in Sec.~\ref{sec:appendix spinless Cn}, the classification of 0D spinless FSPT states protected by the physical symmetry group $G_b=\mathbb{Z}_2 \times \mathbb{Z}_2^T$ is $\mathbb{Z}_2^2$, where the group elements are labeled by the eigenvalues $\pm 1$ of $M$ and $P_f$.
 There are two inequivalent 0D blocks, hence the obstruction-free $0$D block-states form the group $\{ \mathrm{OFBS} \}^{0D} = \mathbb{Z}_2^4$.

\begin{figure}[tb]
    \centering
    \includegraphics[width=0.25\textwidth]{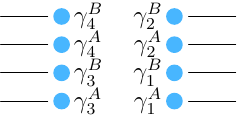}
      \caption{1D BSPT decoration (realized via four Majorana chains) on $\tau_1$  which leaves eight dangling Majorana fermions on $\mu_1$ (or $\mu_2$).}
    \label{fig:pm tau1 four chain}
\end{figure}

We now turn to the decoration of $1$D block-states. The total symmetry group for the $1$D block \( \tau_1 \) is \( G_f = \mathbb{Z}_2^T \times \mathbb{Z}_2^f \). As discussed in Sec.~\ref{sec:appendix spinless Cn}, the corresponding FSPT phases are classified by \( \mathbb{Z}_4 \), with the root state consisting of two Majorana chains satisfying the symmetry properties detailed in Eq.~\eqref{eq:symm property spinless Cn 2chain}. The total symmetry group for the $1$D blocks \( \tau_2 \) and \( \tau_3 \) is given by \( G_f = \mathbb{Z}_2^M \times \mathbb{Z}_2^T \times \mathbb{Z}_2^f \). As discussed in Sec.~\ref{sec:D4 spinless fermion}, the corresponding FSPT phases are classified by \( \mathbb{Z}_4 \times \mathbb{Z}_4 \), where the two fourth-order group generators correspond to the root phases \( n_1^M \) and \( n_1^T \) (their symmetry properties are listed in Eq.~\eqref{eq:symm property of n1T} and Eq.~\eqref{eq:symm property of n1M}).
Furthermore, we can also consider an invertible topological order, such as a Majorana chain, as a block-state. 

We now discuss the decoration of the $1$D block-states on the $1$D block \( \tau_1 \), which leaves dangling Majorana zero modes on the $0$D blocks \( \mu_1 \) and \( \mu_2 \) (see Fig.~\ref{fig:pm tau1 four chain}).  The case with only reflection symmetry has not been discussed previously. Only the decoration with 1D BSPT phase (realized via four Majorana chains) can remain obstruction-free, as this ensures that the number of dangling boundary modes is a multiple of 8 (similar to the discussion in Sec.~\ref{sec:appendix spinless Cn}). To open an energy gap in the zero modes while preserving symmetry, we introduce the interaction term:
\begin{equation}
(c^\dagger_1 c_1-\frac{1}{2})(c^\dagger_2 c_2-\frac{1}{2})+(c^\dagger_3 c_3-\frac{1}{2})(c^\dagger_4 c_4-\frac{1}{2})+\sum_{\sigma={x,y,z}}\tau_{12}^\sigma\tau_{34}^\sigma
\end{equation} 
Here we define the complex fermion operator $c_j$ as $(\gamma_{j,A}+i\gamma_{j,B})/2$. The operators $\tau^\sigma_{xy}$ are defined in Eq.~\eqref{eq:definition of spin12 degree}. This obstruction-free decoration contributes a $\mathbb{Z}_2$ classification. Notably, it realizes an intrinsically interacting FSPT phase that cannot be realized within free-fermion systems.

We now analyze the $1$D block-state decoration on \( \tau_2 \) and \( \tau_3 \), which contribute to the zero modes on \( \mu_1 \) and \( \mu_2 \), respectively. Different from decoration on $\tau_1$, each pair of $1$D block-states adjoining a $0$D block is related by the transformation symmetry operation. Consequently, similar to the \( p1 \) case, these decorated states simultaneously contribute both types of zero modes, \( \gamma^L \) and \( \gamma^R \), on the $0$D block where they intersect. Thus, the decoration of a single Majorana chain is free of obstructions, as it can open a spectral gap at both \( \mu_1 \) and \( \mu_2 \) by employing mass terms of the form \( i \gamma^L \gamma^R \). This obstruction-free property naturally extends to the decoration of general 1D FSPT phases.
Furthermore, since stacking two single Majorana chains is equivalent to decorating with two Majorana chains of the $n_1^T$ phase, the complete classification of obstruction-free $1$D block-states for either $\tau_2$ or $\tau_3$ forms the group $\mathbb{Z}_8 \times \mathbb{Z}_4$.

In conclusion, all 1D obstruction-free block-states form the group:
\begin{equation}
\{\mathrm{OFBS}\}^{1D} = \mathbb{Z}_8^2 \times \mathbb{Z}_4^2 \times \mathbb{Z}_2,
\end{equation}
where each group element is uniquely labeled by the quintuple:
\begin{equation}
[\mathcal{M}_2, m_2, \mathcal{M}_3, m_3, \mathcal{T}_1].
\end{equation}
The integer labels represent:
\begin{itemize}
\item $\mathcal{M}_j \in \{0,1,...,7\}$ counts the decorated single Majorana chains on $\tau_j$
\item $m_j \in \{0,1,2,3\}$ counts the decorated 1D FSPT phases (corresponding to the $n_1^M$ phase) on $\tau_j$
\item $\mathcal{T}_1 \in \{0,1\}$ counts the decorated 
 1D BSPT phases (realized via four Majorana chains) protected by $G_f = \mathbb{Z}_2^T \times \mathbb{Z}_2^f$ on $\tau_1$
\end{itemize}

The total symmetry group for the $2$D blocks is \( G_f = \mathbb{Z}_2^T \times \mathbb{Z}_2^f \). As concluded in Ref.~\cite{Wang2020}, there are no corresponding FSPT phases under this symmetry group. Consequently, the contribution of $2$D block decoration is classified as:
\begin{equation}
\{\mathrm{OFBS}\}^{2D} = \mathbb{Z}_1.
\end{equation}

Next, we consider bubble equivalence relations. First, we examine the $1$D bubble equivalences by decorating a pair of complex fermions on each $1$D block. 
The $1$D bubble on \( \tau_1 \) contributes a non-trivial state \( \ket{\phi} = c_1^\dagger c_2^\dagger \ket{0} \) on \( \mu_1 \), satisfying \begin{equation} M_{\tau_2} \ket{\phi} = c_2^\dagger c_1^\dagger \ket{0} = -\ket{\phi}, \end{equation} and a non-trivial state \( \ket{\phi} \) on \( \mu_2 \), satisfying \begin{equation}\ M_{\tau_3} \ket{\phi} = -\ket{\phi}. \end{equation} However, the $1$D bubbles on \( \tau_2 \) and \( \tau_3 \) contribute no non-trivial $0$D block-states. Thus, $1$D bubbles form the trivialization group \begin{equation}
\{\mathrm{TBS}\}^{0D}= \mathbb{Z}_2.
\end{equation}

\begin{figure}[tb]
    \centering
    \includegraphics[width=0.46\textwidth]{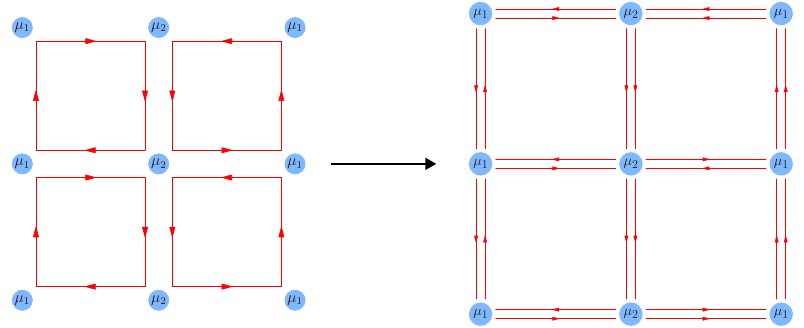}
    \caption{2D bubble decorations (left panel) and their adiabatic deformation into $1$D block-states (right panel). 
Red lines depict closed Majorana chains with anti-periodic boundary conditions (anti-PBC), while arrows indicate pairing directions (pairing near boundaries is omitted for clarity). }
    \label{fig:pm_bubbles}
\end{figure}

Next, we consider the $2$D bubble equivalence. We decorate Majorana chains with anti-periodic boundary conditions related by symmetry elements of the space group \( pm \) on all $2$D blocks and enlarge all ``Majorana bubbles'' near each $1$D block (see Fig.~\ref{fig:pm_bubbles}). 
Near the $1$D blocks \( \tau_2 \) and \( \tau_3 \), bubbles contribute to effective double Majorana chains with internal symmetry \( G_b = \mathbb{Z}_2^M \times \mathbb{Z}_2^T \). These chains, therefore, realize the \( n_1^M + n_1^T \) FSPT phase, analogous to the discussion in Sec.~\ref{sec:D4 spinless fermion}.  In contrast, near the $1$D block \(\tau_1\), the effective double Majorana chains exhibit distinct pairing directions, rendering them gappable. This results in a trivial phase, mirroring the behavior described in Sec.~\ref{sec:Cn spinless bubble}.  
Hence, $2$D bubbles form a trivialization group:
\begin{equation}
\{ \mathrm{TBS} \}^{1D} = \mathbb{Z}_4.
\end{equation}

It is evident that there is no extension between 1D and $0$D block-states. Thus, all independent non-trivial block-states with different dimensions are classified by:
\begin{align}
    \begin{aligned}
        & E^{\mathrm{2D}}=\{\mathrm{OFBS}\}^{\mathrm{2D}} = \mathbb{Z}_1, \\
        & E^{\mathrm{1D}} = \{\mathrm{OFBS}\}^{\mathrm{1D}} / \{\mathrm{TBS}\}^{\mathrm{1D}} = \mathbb{Z}_8^2 \times \mathbb{Z}_4 \times \mathbb{Z}_2, \\
        & E^{\mathrm{0D}} = \{\mathrm{OFBS}\}^{\mathrm{0D}} / \{\mathrm{TBS}\}^{\mathrm{0D}} = \mathbb{Z}_2^3.
    \end{aligned}
\end{align}

Hence, the ultimate classification with an accurate group structure is:
\begin{equation}
\mathcal{G}_0 = \mathbb{Z}_8^2 \times \mathbb{Z}_4 \times \mathbb{Z}_2^4.
\end{equation}

\subsubsection{Spinful fermions}

We now turn to discussing systems with spinful fermions. In Sec.~\ref{sec:Cn spinfull}, we demonstrated that the classification group of 0D spinful FSPT states protected by a physical symmetry group $G_b=\mathbb{Z}_2 \times \mathbb{Z}_2^T$ is \( \mathbb{Z}_2 \), with group elements labeled by the eigenvalues \( \pm 1 \) of \( M \). There are two inequivalent 0D blocks, hence obstruction-free $0$D block-states form the group:
\begin{equation}
\{ \mathrm{OFBS} \}^{0D} = \mathbb{Z}_2^2.
\end{equation}

Next, we discuss the decoration of $1$D block-states. The  physical symmetry group for the $1$D block \( \tau_1 \) is \( G_b = \mathbb{Z}_2^T \), while for \( \tau_2 \) and \( \tau_3 \), it is \( G_b = \mathbb{Z}_2 \times \mathbb{Z}_2^T \). Based on previous discussions, the invertible topological order of a single Majorana chain cannot be decorated, whereas  1D FSPT phases can. Specifically, we can decorate \( \tau_2 \) (or \( \tau_3 \)) with 1D FSPT phases classified by \( \mathbb{Z}_4 \). Such a 1D FSPT phase features a root state consisting of double Majorana chains, with symmetry properties defined in Eq.~\eqref{eq:definition root phase}. Similarly, we can decorate \( \tau_1 \) with 1D FSPT phases classified by \( \mathbb{Z}_2 \). This 1D FSPT phase also features a root state consisting of double Majorana chains, with symmetry properties defined in Eq.~\eqref{eq:trs definition}.

Decorating the 1D FSPT phases on the $1$D block \( \tau_1 \) leaves four dangling Majorana fermion modes on each of the $0$D blocks \( \mu_1 \) and \( \mu_2 \).  As established in Sec.~\ref{sec:D2 D6 spinful},  the dangling Majorana fermions arising from such a decoration can be gapped out under the $D_2$ symmetry at the 0D block. Consequently,  these decorations are also obstruction-free under the $\mathbb{Z}_2^M$ subgroup of \( D_2 \). The decorations of 1D FSPT phases on the other two $1$D blocks follow similarly and are also obstruction-free.

In summary, the complete classification of all obstruction-free $1$D block-states is given by the direct product group:
\begin{equation}
\{\mathrm{OFBS}\}^{1D} = \mathbb{Z}_2 \times \mathbb{Z}_4^2,
\end{equation}
where each group element is uniquely specified by the triplet:
\begin{equation}
[\mathbb{T}_1, \mathbb{MT}_2, \mathbb{MT}_3].
\end{equation}
The integer labels correspond to:
\begin{itemize}
\item $\mathbb{T}_1 \in \{0,1\}$ counts the number of decorated 1D FSPT phases on $\tau_1$ 
\item $\mathbb{MT}_j \in \{0,1,2,3\}$ counts the number of decorated 1D FSPT phases on $\tau_j$ ($j=2,3$)
\end{itemize}
This classification fully characterizes the possible $1$D block-state decorations while remaining obstruction-free.

Analogous to the previous spinful cases, 2D class DIII TSC state decorations are obstruction-free, yielding $\{\mathrm{OFBS}\}^{2D} = \mathbb{Z}_2$.

Next, we consider the bubble equivalence. First, we examine the $1$D bubble equivalences by decorating a pair of complex fermions on each $1$D block. The $1$D bubble on \( \tau_1 \) contributes an atomic product state \( \ket{\phi} = c_1^\dagger c_2^\dagger \ket{0} \) on \( \mu_1 \), satisfying \begin{equation} M_{\tau_2} \ket{\phi} = -c_2^\dagger c_1^\dagger \ket{0} = \ket{\phi}, \end{equation} and another product state \( \ket{\phi} \) on \( \mu_2 \), satisfying \begin{equation} M_{\tau_3} \ket{\phi} = \ket{\phi}. \end{equation} Similarly, the $1$D bubbles on \( \tau_2 \) and \( \tau_3 \) contribute no non-trivial $0$D block-states.

Next, we decorate each $2$D block \( \sigma \) with a $2$D bubble.  Each $2$D bubble consists of two closed anti-PBC Majorana chains, as the bubbles must preserve the on-site symmetry \( T^2 = -1 \). As concluded in Sec.~\ref{sec:D4 spinful discussion}, regardless of whether the effective Majorana chains deformed from the $2$D bubbles possess reflection symmetry or have opposite pairing directions, they have no trivializing effect on the $1$D block-states. Additionally, $2$D bubbles have no impact on $0$D block-states. It is  worth noting that in spinful cases, $2$D bubbles do not contribute any trivial states. Therefore, the total trivializing group is:
\begin{equation}
\{\mathrm{TBS}\}^{1D} = \mathbb{Z}_1, \quad \{\mathrm{TBS}\}^{0D} = \mathbb{Z}_1.
\end{equation}

Finally, we address the extension problem for the space group \( pm \), which involves reflection symmetry acting on the $1$D blocks \( \tau_2 \) and \( \tau_3 \). As demonstrated in Sec.~\ref{sec:D4 spinful discussion}, while two copies of 2D FSPT phases constitute a trivial bulk phase, they nevertheless induce effective nontrivial 1D FSPT phases on \( \tau_2 \) and \( \tau_3 \) due to reflection symmetry. This induces a natural extension from 2D to $1$D block-states. In contrast, the absence of reflection symmetry prevents the formation of any nontrivial 1D FSPT phases on $\tau_1$, leaving only trivial states. Moreover, we find no extension from 1D block-states to 0D block-states in this case.

All independent non-trivial block-states with different dimensions are classified by:
\begin{align}
    \begin{aligned}
        & E^{\mathrm{2D}} = \{\mathrm{OFBS}\}^{\mathrm{2D}} = \mathbb{Z}_2, \\
        & E^{\mathrm{1D}} = \{\mathrm{OFBS}\}^{\mathrm{1D}} / \{\mathrm{TBS}\}^{\mathrm{1D}} = \mathbb{Z}_4^2 \times \mathbb{Z}_2, \\
        & E^{\mathrm{0D}} = \{\mathrm{OFBS}\}^{\mathrm{0D}} / \{\mathrm{TBS}\}^{\mathrm{0D}} = \mathbb{Z}_2^2.
    \end{aligned}
\end{align}

The final classification is:
\begin{equation}
\mathcal{G}_{1/2} = \mathbb{Z}_2^3 \times \mathbb{Z}_4 \times \mathbb{Z}_8,
\end{equation}
where the \( \mathbb{Z}_8 \) term arises from the extension of $2$D block-states to $1$D block-states.

\subsection{$pg$}
\begin{figure}[tb]
    \centering
    \includegraphics[width=0.46\textwidth]{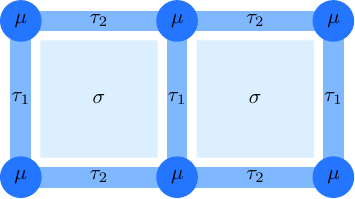}
    \caption{The \#4  2D space group $pg$ and its  cell decomposition.}
    \label{pg}
\end{figure}

For blocks of any dimension, there is no  site symmetry, and the physical symmetry group is $\mathbb{Z}_2^T$. In particular, the group $pg$ contains a ``glide reflection" operation, as shown in Fig.~\ref{pg}.

\subsubsection{Spinless fermions}

In the case of a spinless fermion system, the  physical symmetry group for all $0$D blocks \( \mu \) is \( G_b = \mathbb{Z}_2^T \). In Sec.~\ref{sec:appendix spinless Cn}, we demonstrated that the classification group of 0D FSPT states protected by this symmetry group is \( \mathbb{Z}_2 \), with group elements labeled by the eigenvalues \( \pm 1 \) of \( P_f \). Therefore, obstruction-free $0$D block-states form the group:
\begin{equation}
\{ \mathrm{OFBS} \}^{0D} = \mathbb{Z}_2.
\end{equation}

Next, we discuss the decoration of $1$D block-states. The physical symmetry group for all $1$D blocks is \( G_b = \mathbb{Z}_2^T \). As discussed in Sec.~\ref{sec:appendix spinless Cn}, the corresponding FSPT phases are classified by \( \mathbb{Z}_4 \), where the group generator corresponds to the root phase realized by two Majorana chains (their symmetry properties are listed in Eq.~\eqref{eq:symm property spinless Cn 2chain}).
Furthermore, we can also consider an invertible topological order, i.e., a Majorana chain, as a block-state. 

Now, we discuss the decoration of the $1$D block \( \tau_1 \), which leaves dangling Majorana zero modes on the $0$D block \( \mu \). Due to the glide symmetry of this 2D space group, these decorated states simultaneously contribute both types of zero modes, \( \gamma^L \) and \( \gamma^R \), on the $0$D blocks. Therefore, the single Majorana chain decoration is obstruction-free, as it can open a spectral gap at \( \mu \) using the \( i \gamma^L \gamma^R \) mass terms. This obstruction-free property naturally extends to the decoration of general 1D FSPT phases.
Furthermore, since stacking two single Majorana chains is equivalent to decorating with 1D FSPT phase realized via two Majorana chains, the total 1D obstruction-free block-states for \( \tau_1 \) form the group \( \mathbb{Z}_8 \). The situation for the decoration of the $1$D block \( \tau_2 \) is similar. 
Thus, all 1D obstruction-free block-states form the group:
\begin{equation}
\{\mathrm{OFBS}\}^{1D} = \mathbb{Z}_8^2,
\end{equation}
where the group elements can be labeled as:
\begin{equation}
[\mathcal{M}_1, \mathcal{M}_2  ].
\end{equation}
Here, \( \mathcal{M}_j = 0, 1, \dots,7 \) represents the number of decorated  Majorana chains  on the $1$D blocks \( \tau_j \) (\( j = 1, 2 \)), respectively. 

The total symmetry group for the $2$D blocks is \( G_f = \mathbb{Z}_2^T \times \mathbb{Z}_2^f \). As concluded in Ref.~\cite{Wang2020}, there are no corresponding FSPT phases under this symmetry group. Consequently, the contribution of $2$D block decoration is classified as:
\begin{equation}
\{\mathrm{OFBS}\}^{2D} = \mathbb{Z}_1.
\end{equation}

Next, we consider bubble equivalence relations. First, we examine the $1$D bubble equivalences by decorating a pair of complex fermions on each $1$D block. The 0D obstruction-free block-states only contain odd-parity states, but $1$D bubbles result in even fermion-parity 0D states. Thus, $1$D bubbles form the trivialization group:
\begin{equation}
\{\mathrm{TBS}\}^{0D} = \mathbb{Z}_1.
\end{equation}

Subsequently, we consider the $2$D bubble equivalences. Majorana chains with anti-periodic boundary conditions, related by the symmetry elements of the space group \( pg \), are decorated on all $2$D blocks. Effective 1D states are formed by enlarging all ``Majorana bubbles'' near each $1$D block. Similarly to Sec.~\ref{sec:Cn spinless bubble}, the effective 1D states near \(\tau_2\) are trivial, since the Majorana chains deformed from the bubbles have opposite pairing directions. This triviality can be verified by showing that their edge modes can be fully gapped out. In contrast, the effective 1D states near \( \tau_1 \) are 1D non-trivial FSPT states classified by \( \mathbb{Z}_4 \), as the Majorana chains deformed from the bubbles have the same pairing direction due to the glide symmetry of \( pg \). These 1D non-trivial FSPT states include the \( n_1^T \) phase and the \( (-1)^{n_1^T \cup n_1^T} \) phase through stacking. Therefore, the total trivialization group is given by:
\begin{equation}
\{\mathrm{TBS}\}^{1D} = \mathbb{Z}_4.
\end{equation}

It is evident that there is no extension between 1D and $0$D block-states. Thus, all independent non-trivial block-states with different dimensions are classified by:
\begin{align}
    \begin{aligned}
        & E^{\mathrm{2D}}= \{\mathrm{OFBS}\}^{\mathrm{2D}} = \mathbb{Z}_1, \\
        & E^{\mathrm{1D}} = \{\mathrm{OFBS}\}^{\mathrm{1D}} / \{\mathrm{TBS}\}^{\mathrm{1D}} = \mathbb{Z}_8 \times \mathbb{Z}_2, \\
        & E^{\mathrm{0D}} = \{\mathrm{OFBS}\}^{\mathrm{0D}} / \{\mathrm{TBS}\}^{\mathrm{0D}} = \mathbb{Z}_2.
    \end{aligned}
\end{align}

The ultimate classification is:
\begin{equation}
\mathcal{G}_0 = \mathbb{Z}_8 \times \mathbb{Z}_2^2.
\end{equation}

\subsubsection{Spinful fermions}

In the case of a spinful fermion system, the on-site physical symmetry group for all $0$D blocks \( \mu \) is \( G_b = \mathbb{Z}_2^T \). In Sec.~\ref{sec:Cn spinfull}, we demonstrated that the classification of 0D FSPT states protected by this symmetry group is \( \mathbb{Z}_1 \), meaning there is no non-trivial decoration ($\{\mathrm{OFBS}\}^{0D} = \mathbb{Z}_1$).

Next, we examine the decoration of $1$D block-states. The physical symmetry group for all $1$D blocks is \( G_b = \mathbb{Z}_2^T \). As discussed earlier, the invertible topological order of a single Majorana chain cannot be decorated, but the 1D FSPT phase can, with a classification of \( \mathbb{Z}_2 \). Decorating the 1D FSPT phase (realized via double Majorana chains) on the $1$D block \( \tau_1 \) leaves four dangling Majorana fermion modes on the $0$D blocks \( \mu \). Due to the glide symmetry, these decorated states simultaneously contribute both types of zero modes, \( \gamma^L \) and \( \gamma^R \), on the $0$D blocks. Therefore, the 1D FSPT phase decorations are obstruction-free, as they can open a spectral gap at \( \mu \) using mass term \( i \gamma^L \gamma^R \) . The decoration on the other $1$D blocks follows similarly, and all decorations are obstruction-free, contributing to the classification:
\begin{equation}
\{\mathrm{OFBS}\}^{1D} = \mathbb{Z}_2^2,
\end{equation}
where the group elements can be labeled as:
\begin{equation}
[\mathbb{T}_1,\mathbb{T}_2].
\end{equation}
Here, \( \mathbb{T}_j = 0, 1 \) represent the number of decorated 1D FSPT phases on the $1$D blocks \( \tau_j \) (\( j = 1, 2 \)), respectively.

For the same reasons as in the previously discussed spinful example, the decoration of 2D DIII TSC states is always obstruction-free, contributing to a classification group \( \{\mathrm{OFBS}\}^{2D} = \mathbb{Z}_2 \).

Next, we consider the bubble equivalence. Each $2$D block \( \sigma \) is decorated with a $2$D bubble, where each bubble consists of two closed anti-PBC Majorana chains. These bubbles are required to preserve the on-site symmetry \( T^2 = -1 \). By enlarging the bubbles, we obtain four Majorana chains near each $1$D block. This state shares the same symmetry properties as the 1D obstruction-free states decorated above and belongs to the trivial phase.

The $1$D bubbles on the $1$D blocks \( \tau_1 \) and \( \tau_2 \) act on \( \mu \). However, discussing $1$D bubbles in this context is unnecessary, as there are no non-trivial $0$D block-states. Also, there is no need to consider effect of the $2$D bubbles on the $0$D block-states. Therefore, the total trivialization groups are given by:
\begin{equation}
\{\mathrm{TBS}\}^{1D} = \mathbb{Z}_1, \quad \{\mathrm{TBS}\}^{0D} = \mathbb{Z}_1.
\end{equation}

Finally, we consider the extension problem. Since the space group \( pg \) involves glide  symmetry (for the decoration of 2D chiral fermions, \( \tau_1 \) or \( \tau_2 \) can be viewed as effective reflection axes), as discussed in Sec.~\ref{sec:D4 spinful discussion}, the mass term for the boundary Majorana modes of the 2D decorated states acquires an additional sign under the symmetry. Thus, there is an extension from $2$D block-states to $1$D block-states. But we find no extension from 1D block-states to 0D block-states in this case.

Thus, all independent non-trivial block-states with different dimensions are classified by:
\begin{align}
    \begin{aligned}
        & E^{\mathrm{2D}} = \{\mathrm{OFBS}\}^{\mathrm{2D}}= \mathbb{Z}_2, \\
        & E^{\mathrm{1D}} = \{\mathrm{OFBS}\}^{\mathrm{1D}} / \{\mathrm{TBS}\}^{\mathrm{1D}} = \mathbb{Z}_2^2, \\
        & E^{\mathrm{0D}} = \{\mathrm{OFBS}\}^{\mathrm{0D}} / \{\mathrm{TBS}\}^{\mathrm{0D}} = \mathbb{Z}_1.
    \end{aligned}
\end{align}

The ultimate classification is:
\begin{equation}
\mathcal{G}_0 = \mathbb{Z}_2 \times \mathbb{Z}_4.
\end{equation}

\subsection{$cm$}
\begin{figure}[tb]
    \centering
    \includegraphics[width=0.4\textwidth]{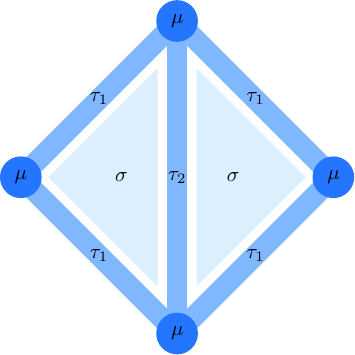}
    \caption{The \#5  2D space group $cm$ and its  cell decomposition.}
    \label{cm}
\end{figure}

There is no site symmetry on the $2$D block \( \sigma \) and the $1$D block \( \tau_1 \), with the physical symmetry group being \( \mathbb{Z}_2^T \). However, on the $1$D block \( \tau_2 \) and the $0$D block \( \mu \), there is a physical symmetry \( \mathbb{Z}_2 \times \mathbb{Z}_2^T \) due to the reflection symmetry, as shown in Fig.~\ref{cm}.

\begin{figure}[tb]
    \centering
    \includegraphics[width=0.3\textwidth]{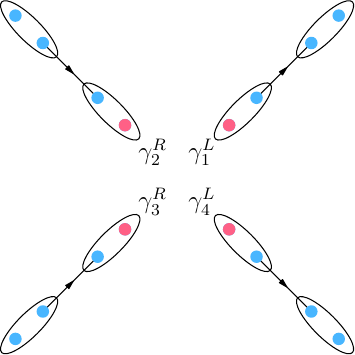}
    \caption{Single Majorana chain decorations on $1$D blocks $\tau_1$ near $0$D block $\mu$ and their dangling zero modes.}
    \label{fig:cm mu}
\end{figure}

\subsubsection{Spinless fermions}

For a spinless fermion system, the on-site physical symmetry group for all $0$D blocks \( \mu \) is \( G_b = \mathbb{Z}_2 \times \mathbb{Z}_2^T \). In earlier sections, we demonstrated that the classification of 0D FSPTs protected by this symmetry group is \( \mathbb{Z}_2^2 \), with group elements labeled by the eigenvalues \( \pm 1 \) of \( M \) and \( P_f \). Therefore, obstruction-free $0$D block-states form the group:
\begin{equation}
\{ \mathrm{OFBS} \}^{0D} = \mathbb{Z}_2^2.
\end{equation}

Next, we discuss the decoration of $1$D block-states. The physical symmetry group for the $1$D block \( \tau_1 \) is given by \( G_b = \mathbb{Z}_2^T \). As mentioned previously, the corresponding FSPT phases are classified by \( \mathbb{Z}_4 \), with a root state model consisting of two Majorana chains, whose symmetry properties are detailed in Eq.~\eqref{eq:symm property spinless Cn 2chain}. The physical symmetry group for the $1$D block \( \tau_2 \) is \( G_b = \mathbb{Z}_2 \times \mathbb{Z}_2^T \). As discussed in Sec.~\ref{sec:D4 spinless fermion}, the corresponding FSPT phases are classified by \( \mathbb{Z}_4 \times \mathbb{Z}_4 \), where the two fourth-order group generators correspond to the root phases \( n_1^M \) and \( n_1^T \) (their symmetry properties are listed in Eq.~\eqref{eq:symm property of n1T} and Eq.~\eqref{eq:symm property of n1M}).
Furthermore, we can consider an invertible topological order, i.e., a Majorana chain, as a block-state. 

Now, we discuss the decoration of the $1$D block-state on \( \tau_1 \). Due to the translation symmetry, these decorated states simultaneously contribute both types of zero modes, \( \gamma^L \) and \( \gamma^R \), on the $0$D block. The decoration of a single Majorana chain leaves four dangling Majorana zero modes (Fig.~\ref{fig:cm mu}). Using the interaction term \( i \gamma_1^L \gamma_3^R + i \gamma_2^R \gamma_4^L \), we can open the gap for the zero modes. Hence, the decoration of  single Majorana chains satisfies the no-open-edge condition and has the classification \( \mathbb{Z}_8 \). 

The decoration of the $1$D block-state on \( \tau_2 \) is similar to the decoration of the $1$D block-state on \( \tau_3 \) in the \( pm \) case, and it remains obstruction-free, contributing to a classification of \( \mathbb{Z}_8 \times \mathbb{Z}_4 \). Therefore, all obstruction-free $1$D block-states form the group:
\begin{equation}
\{ \mathrm{OFBS} \}^{1D} = \mathbb{Z}_8^2 \times \mathbb{Z}_4,
\end{equation}
where each group element is uniquely specified by the triplet:
\begin{equation}
[\mathcal{M}_1,\mathcal{M}_2, m_2 ].
\end{equation}
The integer labels correspond to:
\begin{itemize}
    \item $\mathcal{M}_j \in \{0,1,\ldots,7\}$ counts the number of decorated single Majorana chains on $\tau_j$ ($j=1,2$)
    \item $m_2 \in \{0,1,2,3\}$ counts the number of decorated $n_1^M$ FSPT phases  on $\tau_2$
\end{itemize}
This classification fully characterizes the possible $1$D block-state decorations while maintaining obstruction-free conditions.

The total symmetry group for the $2$D blocks is \( G_f = \mathbb{Z}_2^T \times \mathbb{Z}_2^f \). As concluded in Ref.~\cite{Wang2020}, there are no corresponding FSPT phases under this symmetry group. Consequently, the contribution of $2$D block decoration is classified as:
\begin{equation}
\{\mathrm{OFBS}\}^{2D} = \mathbb{Z}_1.
\end{equation}

Next, we consider bubble equivalence relations. First, we examine the $1$D bubble equivalences by decorating a pair of complex fermions on each $1$D block. The $1$D bubble on \( \tau_1 \) contributes to the trivial state \( \ket{TR} = c_1^\dagger c_2^\dagger c_3^\dagger c_4^\dagger \ket{0} \) on the $0$D block \( \mu \), satisfying \( M_{\tau_2} \ket{TR} = c_2^\dagger c_1^\dagger c_4^\dagger c_3^\dagger \ket{0} = \ket{TR} \). Here, $M_{\tau_2}$ represents the reflection operation with the axis coinciding with the $\tau_2$ block.  Similarly, the $1$D bubble on \( \tau_2 \) contributes to the trivial state \( \ket{TR} = c_1^\dagger c_2^\dagger \ket{0} \) on the $0$D block, satisfying \( M_{\tau_2} \ket{TR} = c_1^\dagger c_2^\dagger \ket{0} = \ket{TR} \). Therefore, $1$D bubbles have no effect on $0$D block-states and form the trivialization group:
\begin{equation}
\{\mathrm{TBS}\}^{0D} = \mathbb{Z}_1.
\end{equation}

According to the symmetry properties of $cm$ (translation and reflection symmetries),  2D ``Majorana bubbles" can contribute to 1D non-trivial states classified by \( \mathbb{Z}_4 \) on all $1$D blocks, including the \(  n_1^T \) and \(  (-1)^{n_1^T \cup n_1^T} \) phases through stacking, since the effective Majorana chains deformed from bubbles have the same pairing direction (similarly to the $pg$ case). 
Therefore, the  trivialization group for $1$D block-states is:
\begin{equation}
\{ \mathrm{TBS} \}^{1D} = \mathbb{Z}_4.
\end{equation}

It is evident that there is no extension between 1D and $0$D block-states. Thus, all independent non-trivial block-states with different dimensions are classified by:
\begin{align}
    \begin{aligned}
        & E^{\mathrm{2D}} = \{\mathrm{OFBS}\}^{\mathrm{2D}}= \mathbb{Z}_1, \\
        & E^{\mathrm{1D}} = \{ \mathrm{OFBS} \}^{\mathrm{1D}} / \{ \mathrm{TBS} \}^{\mathrm{1D}} = \mathbb{Z}_8^2, \\
        & E^{\mathrm{0D}} = \{ \mathrm{OFBS} \}^{\mathrm{0D}} / \{ \mathrm{TBS} \}^{\mathrm{0D}} = \mathbb{Z}_2^2.
    \end{aligned}
\end{align}

The ultimate classification is:
\begin{equation}
\mathcal{G}_0 = \mathbb{Z}_8^2 \times \mathbb{Z}_2^2.
\end{equation}

\subsubsection{Spinful fermions}

For a spinful fermion system, the on-site physical symmetry group for all $0$D blocks \( \mu \) is \( G_b = \mathbb{Z}_2 \times \mathbb{Z}_2^T \). In earlier sections, we demonstrated that the classification of 0D FSPTs protected by this symmetry group is \( \mathbb{Z}_2 \), with group elements labeled by the eigenvalues \( \pm 1 \) of \( M \). Therefore, obstruction-free $0$D block-states form the group:
\begin{equation}
\{ \mathrm{OFBS} \}^{0D} = \mathbb{Z}_2.
\end{equation}

Next, we examine the decoration of $1$D block-states. The physical symmetry group for the $1$D block \( \tau_1 \) is \( G_b = \mathbb{Z}_2^T \), and for \( \tau_2 \), it is \( G_b = \mathbb{Z}_2 \times \mathbb{Z}_2^T \). As discussed earlier, it is not possible to decorate the invertible topological order of a single Majorana chain, but the 1D FSPT phases can be proper candidates. 
Specifically, we can decorate \( \tau_2 \) with 1D FSPT phases classified by \( \mathbb{Z}_4 \). Such a 1D FSPT phase features a root state consisting of double Majorana chains, with symmetry properties defined in Eq.~\eqref{eq:definition root phase}. Similarly, we can decorate \( \tau_1 \) with 1D FSPT phases classified by \( \mathbb{Z}_2 \). This 1D FSPT phase also features a root state consisting of double Majorana chains, with symmetry properties defined in Eq.~\eqref{eq:trs definition}.

Decorating the double Majorana chains on the $1$D block \( \tau_1 \) leaves eight dangling Majorana fermion modes on the $0$D block \( \mu_1 \), while decorating on the $1$D block \( \tau_2 \) leaves four dangling Majorana fermion modes on the $0$D block \( \mu_1 \). The translation symmetry enforces the simultaneous appearance of both $\gamma^L$ and $\gamma^R$ types of zero modes at the $0$D block, analogous to the spinless scenario. Therefore, these decorations are all obstruction-free, as they can open a spectral gap at \( \mu \) using \( i \gamma^L \gamma^R \)-type mass terms.

The 1D obstruction-free decorations contribute to the classification group:
\begin{equation}
\{\mathrm{OFBS}\}^{1D} = \mathbb{Z}_2 \times \mathbb{Z}_4,
\end{equation}
where each group element is uniquely labeled by the pair:
\begin{equation}
[\mathbb{T}_1, \mathbb{MT}_2].
\end{equation}
The integer labels correspond to:
\begin{itemize}
    \item $\mathbb{T}_1 \in \{0,1\}$ counts the number of decorated 1D FSPT phases on $\tau_1$ protected by $\mathbb{Z}_4^{fT}$ symmetry
    \item $\mathbb{MT}_2 \in \{0,1,2,3\}$ counts the number of decorated 1D FSPT phases on $\tau_2$ protected by $G_f=(\mathbb{Z}_2 \times \mathbb{Z}_2^T)\times_{\omega_2} \mathbb{Z}_2^f$ symmetry
\end{itemize}
This classification fully characterizes all possible obstruction-free $1$D block-state decorations.

For the same reasons as in the previous spinful cases, the decoration of 2D DIII TSC states is always obstruction-free, contributing to a \( \{\mathrm{OFBS}\}^{2D} = \mathbb{Z}_2 \) classification.

Next, we study the bubble equivalence. Consider the \(1\)D bubbles obtained by decorating each \(1\)D block with a pair of complex fermions. The bubbles on $\tau_1$ induce the following state on the $0$D block $\mu$:
\[
\ket{TR}=c_1^\dagger c_2^\dagger c_3^\dagger c_4^\dagger\ket{0},
\]
which is invariant under the reflection \(M_{\tau_2}\) (about the \(\tau_2\) axis):
\[
M_{\tau_2}\ket{TR}=c_2^\dagger c_1^\dagger c_4^\dagger c_3^\dagger\ket{0}=\ket{TR}.
\]
Similarly, the bubbles on \(\tau_2\) induce another product state $\ket{TR}$ satisfying:
\[
\ket{TR}=c_1^\dagger c_2^\dagger\ket{0},
\quad
M_{\tau_2}\ket{TR}=\ket{TR}.
\]
Hence, \(1\)D bubbles do not affect the \(0\)D block states, and the trivialization group is
\begin{equation}
\{\mathrm{TBS}\}^{0D}=\mathbb{Z}_1.
\end{equation}

Furthermore, we decorate each $2$D block with $2$D bubbles consisting of two closed anti-PBC Majorana chains. Owing to the symmetry properties of \(cm\) (translation and reflection), these 2D ``Majorana bubbles" can induce $1$D block-states on all $1$D blocks with identical pairing directions. Along \(\tau_1\), four identical Majorana chains are generated; they realize the same symmetry action as the root phase protected by \(\mathbb{Z}_4^{fT}\) and therefore together form a trivial phase. The same analysis applies to the induced $1$D block-states on \(\tau_2\). Hence, $2$D bubbles do not contribute any nontrivial $1$D block-states. Moreover, they also have no trivializing effect on the $0$D block-states, similar to the $p4m$ case. Therefore, the total trivializing group is:
\begin{equation}
\{\mathrm{TBS}\}^{1D} = \mathbb{Z}_1, \quad \{\mathrm{TBS}\}^{0D} = \mathbb{Z}_1.
\end{equation}

Finally, we consider the extension problem. 
As demonstrated in Sec.~\ref{sec:D4 spinful discussion}, while two copies of 2D FSPT phases constitute a trivial bulk phase, they nevertheless induce effective nontrivial 1D FSPT phases on \( \tau_2 \) due to reflection symmetry. This induces a natural extension from 2D to $1$D block-states. In contrast, the absence of reflection symmetry prevents the formation of any nontrivial 1D FSPT phases on $\tau_1$, leaving only trivial states.  However, we find no extension from 1D block-states to 0D block-states in this case.

Thus, all independent non-trivial block-states with different dimensions are classified by:
\begin{align}
    \begin{aligned}
        & E^{\mathrm{2D}} = \{\mathrm{OFBS}\}^{\mathrm{2D}}= \mathbb{Z}_2, \\
        & E^{\mathrm{1D}} = \{\mathrm{OFBS}\}^{\mathrm{1D}} / \{\mathrm{TBS}\}^{\mathrm{1D}} = \mathbb{Z}_4 \times \mathbb{Z}_2, \\
        & E^{\mathrm{0D}} = \{\mathrm{OFBS}\}^{\mathrm{0D}} / \{\mathrm{TBS}\}^{\mathrm{0D}} = \mathbb{Z}_2.
    \end{aligned}
\end{align}

The final classification is:
\begin{equation}
\mathcal{G}_{1/2} = \mathbb{Z}_2^2 \times \mathbb{Z}_8,
\end{equation}
where the \( \mathbb{Z}_8 \) term arises from the extension of the $2$D block-states to the $1$D block-states.

 \subsection{$pmm$}
 
\begin{figure}[bt]
    \centering
    \includegraphics[width=0.4\textwidth]{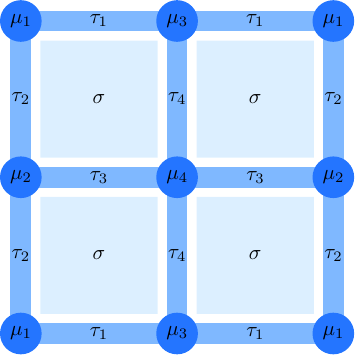}
    \caption{The \#6  2D space group $pmm$ and its  cell decomposition.}
    \label{pmm}
\end{figure}

The point group corresponding to this case is the dihedral group \( D_2 \). For the $2$D block \( \sigma \), there is no site symmetry, and the physical symmetry group is \( \mathbb{Z}_2^T \). For the $1$D blocks \( \tau_j \) (\( j = 1, 2, 3, 4 \)), there exists a reflection symmetry, and the physical symmetry group is \( \mathbb{Z}_2 \times \mathbb{Z}_2^T \). For the $0$D blocks \( \mu_k \) (\( k = 1, 2, 3, 4 \)), the physical symmetry is \( D_2 \times \mathbb{Z}_2^T \) due to the internal action of the \( D_2 \) symmetry, as shown in Fig.~\ref{pmm}.

\subsubsection{Spinless fermions}

For a spinless fermion system, the on-site physical symmetry group for all $0$D blocks \( \mu_1, \mu_2, \mu_3, \mu_4 \) is \( G_b = D_2 \times \mathbb{Z}_2^T \). In Sec.~\ref{sec:D2 D6 spinless}, we demonstrated that the classification of 0D FSPTs protected by this symmetry group is \( \mathbb{Z}_2^3 \), with group elements labeled by the eigenvalues \( \pm 1 \) of \( C_2 \), \( M \), and \( P_f \). Therefore, obstruction-free $0$D block-states at four independent blocks form the group:
\begin{equation}
\{ \mathrm{OFBS} \}^{0D} = \mathbb{Z}_2^{12}.
\end{equation}

The decoration of $1$D block-states is now examined. The physical symmetry group for all $1$D blocks is given by \( G_b = \mathbb{Z}_2 \times \mathbb{Z}_2^T \). As discussed in Sec.~\ref{sec:D4 spinless fermion}, the corresponding FSPT phases are classified by \( \mathbb{Z}_4 \times \mathbb{Z}_4 \), where the two fourth-order group generators correspond to the root phases \( n_1^M \) and \( n_1^T \) (their symmetry properties are listed in Eq.~\eqref{eq:symm property of n1T} and Eq.~\eqref{eq:symm property of n1M}). 
Additionally, we can incorporate an invertible topological order, i.e., a single Majorana chain. 

For instance, consider the $1$D block-state decoration on \( \tau_1 \), which leaves dangling Majorana zero modes on the $0$D blocks \( \mu_1 \) and \( \mu_3 \). Since the symmetries of the $0$D blocks here belong to those discussed earlier, we can directly apply the existing conclusions. The $0$D blocks \( \mu_1 \) and \( \mu_3 \) have the physical symmetry group \( G_b = D_2 \times \mathbb{Z}_2^T \), and the cell decomposition near them aligns with our analysis of the point group \( D_2 \). Thus, the decoration of 1D BSPT phases \( (-1)^{n_1^T \cup n_1^T} \) or \( (-1)^{n_1^M \cup n_1^M} \)  can open a spectral gap for the dangling zero modes at the $0$D blocks \( \mu_1 \) and \( \mu_3 \). The conclusions are similar for decorations on the $1$D blocks \( \tau_2 \) and \( \tau_3 \). 

Furthermore, for the physical symmetry group \(G_b = D_2 \times \mathbb{Z}_2^T\), an obstruction-free 1D decoration can also be realized by simultaneously decorating the 1D $n_1^T+ n_1^M$ FSPT phases associated with the two distinct $1$D block types of \(D_2\); in the present \(pmm\) setting, this corresponds precisely to decorating the same 1D $n_1^T+ n_1^M$ FSPT on all $1$D blocks at once.

Therefore, all obstruction-free $1$D block-states form the group:
\begin{equation}
\{ \mathrm{OFBS} \}^{1D} = \mathbb{Z}_4 \times \mathbb{Z}_2^7,
\end{equation}
where the group elements can be labeled by:
\begin{equation}
[mt_{1234},T_1, T_2, T_3, T_4, M_1, M_2, M_3].
\end{equation}
The integer labels \( mt_j \), \( M_j  \), and \( T_j \)  correspond to:
\begin{itemize}
    \item $ mt_{1234} \in \{0,1,2,3\}$ counts the number of decorated $n_1^M + n_1^T$ phase states (required to be equal across all $\tau_j$ by the obstruction-free condition);
    \item $T_j \in \{0,1\}$ counts the number of decorated $(-1)^{n_1^T \cup n_1^T}$ phase states on $\tau_j$ ($j=1,\ldots,4$);
    \item $M_j \in \{0,1\}$ counts the number of decorated $(-1)^{n_1^M \cup n_1^M}$ phase states on $\tau_j$ ($j=1,2,3$).
\end{itemize}
Note that the final classification group is not simply the direct product of all obstruction-free cases analyzed above. Rather, some of these cases are redundantly counted, so after removing the redundancies the remaining classification group is smaller.

The total symmetry group for the $2$D blocks is \( G_f = \mathbb{Z}_2^T \times \mathbb{Z}_2^f \). As concluded in Ref.~\cite{Wang2020}, there are no corresponding FSPT phases under this symmetry group. Consequently, the contribution of $2$D block decoration is classified as:
\begin{equation}
\{\mathrm{OFBS}\}^{2D} = \mathbb{Z}_1.
\end{equation}

We now turn to bubble equivalence relations. Similar to the \(D_2\) case discussed in Sec.~\ref{sec:D2 D6 spinless}, 2D ``Majorana bubbles" can induce nontrivial 1D states on all $1$D blocks. These induced states form a \(\mathbb{Z}_4\) classification and, under stacking, include the phases \(n_1^M + n_1^T\) and \(\,(-1)^{n_1^M \cup n_1^M+n_1^T \cup n_1^T}\).
Therefore, the $2$D bubbles form the trivialization group:
\begin{equation}
\{ \mathrm{TBS} \}^{1D} = \mathbb{Z}_4 .
\end{equation}

To evaluate whether the $2$D bubble can alter the fermion parity of the $0$D blocks, we observe that the 2D Majorana bubble leaves double Majorana chains on all $1$D blocks. Consequently, it leads to the presence of eight Majorana modes at the $0$D blocks \( \mu_j \), corresponding to the edge modes of the double Majorana chains on the $1$D blocks. These Majorana modes cannot be connected to Majorana chains with periodic boundary conditions (PBC) surrounding the $0$D blocks, which have odd fermion parity, since Majorana chains are incompatible with reflection symmetry~\cite{2Dcrystalline}.

The $1$D bubble equivalences are also examined by decorating a pair of complex fermions on each $1$D block. For instance, the $1$D bubble on \( \tau_1 \) contributes a non-trivial state \( \ket{\phi} = c_1^\dagger c_2^\dagger \ket{0} \) on \( \mu_1 \), satisfying \begin{equation} M_{\tau_2} \ket{\phi} = c_2^\dagger c_1^\dagger \ket{0} = -\ket{\phi}, \end{equation} and a non-trivial state \( \ket{\phi} = c_1^\dagger c_2^\dagger \ket{0} \) on \( \mu_3 \), satisfying \begin{equation} M_{\tau_4} \ket{\phi} = c_2^\dagger c_1^\dagger \ket{0} = -\ket{\phi} .\end{equation} Similarly, the $1$D bubbles on \( \tau_2, \tau_3, \tau_4 \) contribute non-trivial states in a similar fashion. Thus, the $1$D bubbles form the trivialization group:
\begin{equation}
\{ \mathrm{TBS} \}^{0D} = \mathbb{Z}_2^4.
\end{equation}

It is evident that there is no extension between 1D and $0$D block-states. Thus, all independent non-trivial block-states with different dimensions are classified by:
\begin{align}
    \begin{aligned}
        & E^{\mathrm{2D}}= \{\mathrm{OFBS}\}^{\mathrm{2D}} = \mathbb{Z}_1, \\
        & E^{\mathrm{1D}} = \{ \mathrm{OFBS} \}^{\mathrm{1D}} / \{ \mathrm{TBS} \}^{\mathrm{1D}} =  \mathbb{Z}_2^7, \\
        & E^{\mathrm{0D}} = \{ \mathrm{OFBS} \}^{\mathrm{0D}} / \{ \mathrm{TBS} \}^{\mathrm{0D}} = \mathbb{Z}_2^8.
    \end{aligned}
\end{align}

The ultimate classification is:
\begin{equation}
\mathcal{G}_0 =  \mathbb{Z}_2^{15}.
\end{equation}

 \subsubsection{Spinful fermions}

For a spinful fermion system, the on-site physical symmetry group for all $0$D blocks \( \mu_1, \mu_2, \mu_3, \mu_4 \) is \( G_b = D_2 \times \mathbb{Z}_2^T \). In Sec.~\ref{sec:D4 spinful discussion}, we demonstrated that the classification of 0D FSPTs protected by this symmetry group is \( \mathbb{Z}_2^2 \), with group elements labeled by the eigenvalues \( \pm 1 \) of \( C_2 \) and \( M \). Therefore, obstruction-free $0$D block-states form the group:
\begin{equation}
\{ \mathrm{OFBS} \}^{0D} = \mathbb{Z}_2^8.
\end{equation}

We now examine the decoration of $1$D block-states. The physical symmetry group for all $1$D blocks is \( G_b = \mathbb{Z}_2 \times \mathbb{Z}_2^T \). As discussed earlier, it is not possible to decorate the invertible topological order, i.e. a single Majorana chain. However, we can decorate with 1D FSPT phases classified by \( \mathbb{Z}_4 \). Such a 1D FSPT phase features a root state consisting of double Majorana chains, with symmetry properties defined in Eq.~\eqref{eq:definition root phase}.

For the $1$D block \( \tau_1 \), decorating the double Majorana chains leaves four dangling Majorana fermion modes on each of the $0$D blocks \( \mu_1 \) and \( \mu_3 \). As concluded in Sec.~\ref{sec:D2 D6 spinful}, these four Majorana fermions can open a spectral gap at the $0$D blocks with physical symmetry \( G_b = D_2 \times \mathbb{Z}_2^T \), making the decoration obstruction-free. The decorations on the three other $1$D blocks follow similarly, and all decorations are obstruction-free. 

Thus, the 1D obstruction-free decorations contribute to the classification group:
\begin{equation}
\{\mathrm{OFBS}\}^{1D} = \mathbb{Z}_4^4,
\end{equation}
where the group elements can be labeled as:
\begin{equation}
[\mathbb{MT}_1,\mathbb{MT}_2,\mathbb{MT}_3,\mathbb{MT}_4].
\end{equation}
Here,  \( \mathbb{MT}_j = 0, 1, 2, 3 \) represents the number of decorated 1D FSPT phases on the $1$D blocks \( \tau_j \) (\( j = 1, 2,3,4 \)), respectively.

For the same reasons as in the previously discussed spinful example, the decoration of 2D DIII TSC states is always obstruction-free, contributing to a classification group \( \{\mathrm{OFBS}\}^{2D} = \mathbb{Z}_2 \) . 

Turning to bubble equivalence, each $2$D block \( \sigma \) is decorated with a $2$D bubble, where each bubble consists of two closed anti-PBC Majorana chains. These bubbles are required to preserve the on-site symmetry \( T^2 = -1 \). By enlarging the bubbles, we obtain four Majorana chains near each $1$D block.  As concluded in Sec.~\ref{sec:D4 spinful discussion}, they are not protected by symmetry and belong to the trivial phase. Therefore, the $2$D bubbles form the trivialization group:
\begin{equation}
\{ \mathrm{TBS} \}^{1D} = \mathbb{Z}_1 .
\end{equation} 

The $1$D bubble on the $1$D block \( \tau_1 \) acts on the $0$D blocks \( \mu_1 \) and \( \mu_3 \). However, as concluded in Sec.~\ref{sec:D2 D6 spinful}, $1$D bubbles do not have a trivializing effect on $0$D blocks with on-site physical symmetry \( G_b = D_2 \times \mathbb{Z}_2^T \). The same holds for the three other $1$D bubbles. Additionally, $2$D bubbles have no trivializing effect on $0$D block-states. Therefore, the 0D trivial block-states form group:
\begin{equation}
\{\mathrm{TBS}\}^{0D} = \mathbb{Z}_1.
\end{equation}

We now consider the extension problem. This involves the dihedral point group \( D_2 \), which includes reflection symmetry. 
As demonstrated in Sec.~\ref{sec:D4 spinful discussion}, while two copies of 2D FSPT phases constitute a trivial bulk phase, they nevertheless induce effective nontrivial 1D FSPT phases on  all $1$D blocks due to reflection symmetry. 
Consequently, the $2$D block-states extend to the $1$D block-states, leading to the decoration of the double Majorana chain on \( \tau_1, \tau_2, \tau_3, \tau_4 \) blocks.
However, we find no extension from 1D block-states to 0D block-states in this case.

All independent non-trivial block-states with different dimensions are classified by:
\begin{align}
    \begin{aligned}
        & E^{\mathrm{2D}}= \{\mathrm{OFBS}\}^{\mathrm{2D}} = \mathbb{Z}_2, \\
        & E^{\mathrm{1D}} = \{\mathrm{OFBS}\}^{\mathrm{1D}} / \{\mathrm{TBS}\}^{\mathrm{1D}} = \mathbb{Z}_4^4, \\
        & E^{\mathrm{0D}} = \{\mathrm{OFBS}\}^{\mathrm{0D}} / \{\mathrm{TBS}\}^{\mathrm{0D}} = \mathbb{Z}_2^8.
    \end{aligned}
\end{align}

The final classification is:
\begin{equation}
\mathcal{G}_0 = \mathbb{Z}_8 \times \mathbb{Z}_4^3 \times \mathbb{Z}_2^8,
\end{equation}
where the \( \mathbb{Z}_8 \) term arises from the extension of the $2$D block-states to the $1$D block-states. 

\subsection{$pmg$}
\begin{figure}[tb]
    \centering
    \includegraphics[width=0.4\textwidth]{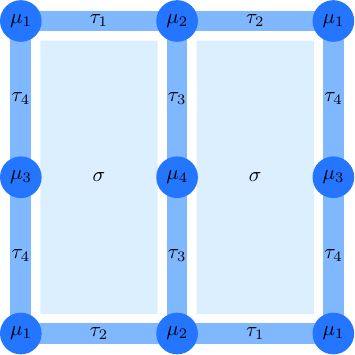}
    \caption{The \#7  2D space group $pmg$ and its  cell decomposition.}
    \label{pmg}
\end{figure}

 For the $2$D block \( \sigma \) and the $1$D blocks labeled \( \tau_3 \) and \( \tau_4 \), there is no site symmetry, and the physical symmetry group is \( \mathbb{Z}_2^T \). For the $1$D blocks labeled \( \tau_1 \) and \( \tau_2 \), and the $0$D blocks labeled \( \mu_1 \) and \( \mu_2 \), the physical symmetry is \( \mathbb{Z}_2 \times \mathbb{Z}_2^T \) due to the internal action of reflection symmetry along $\tau_1$ (or $\tau_2$). For the $0$D blocks \( \mu_3 \) and \( \mu_4 \), the physical symmetry is \( \mathbb{Z}_2 \times \mathbb{Z}_2^T \) due to the internal action of the two-fold rotation symmetry. Fig.~\ref{pmg} shows the cell decomposition.

\subsubsection{Spinless fermions}

For a spinless fermion system, the on-site physical symmetry group for the $0$D blocks \( \mu_1 \) and \( \mu_2 \) is \( G_b = \mathbb{Z}_2^M \times \mathbb{Z}_2^T \), while for \( \mu_3 \) and \( \mu_4 \), it is \( G_b = C_2 \times \mathbb{Z}_2^T \). In Sec.~\ref{sec:appendix spinless Cn}, we demonstrated that the classification of block-states protected by these symmetry groups for \( \mu_1 \) and \( \mu_2 \) is \( \mathbb{Z}_2^2 \), and they can be labeled by the eigenvalues \( \pm 1 \) of \( M \) and \( P_f \). Similarly, the classification for \( \mu_3 \) and \( \mu_4 \) is \( \mathbb{Z}_2^2 \), with group elements labeled by the eigenvalues \( \pm 1 \) of \( C_2 \) and \( P_f \). Therefore, obstruction-free $0$D block-states form the group:
\begin{equation}
\{ \mathrm{OFBS} \}^{0D} = \mathbb{Z}_2^8.
\end{equation}

\begin{figure}[tb]
    \centering
   \includegraphics[width=0.45\textwidth]{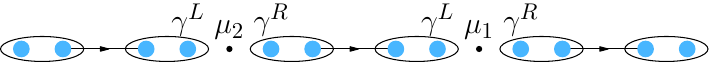}
    \caption{Single Majorana chain decorations on $1$D blocks $\tau_1$ and $\tau_2$ with the same pairing direction. These decorated states simultaneously contribute both types of zero modes, \( \gamma^L \) and \( \gamma^R \), on the $0$D blocks \( \mu_1 \) and \( \mu_2 \).}
    \label{fig:pmg tau1tau2}
\end{figure}

We now discuss the decoration of $1$D block-states. The physical symmetry group for the $1$D blocks \( \tau_1 \) and \( \tau_2 \) is given by \( G_b = \mathbb{Z}_2^M \times \mathbb{Z}_2^T \). As discussed in Sec.~\ref{sec:D4 spinless fermion}, the corresponding FSPT phases are classified by \( \mathbb{Z}_4 \times \mathbb{Z}_4 \), where the two fourth-order group generators correspond to the root phases \( n_1^M \) and \( n_1^T \) (their symmetry properties are listed in Eq.~\eqref{eq:symm property of n1T} and Eq.~\eqref{eq:symm property of n1M}). The physical symmetry group for the $1$D blocks \( \tau_3 \) and \( \tau_4 \) is \( G_b = \mathbb{Z}_2^T \). As mentioned in Sec.~\ref{sec:appendix spinless Cn}, the corresponding FSPT phases are classified by \( \mathbb{Z}_4 \), with a root state model that can consist of two Majorana chains, whose symmetry properties are detailed in Eq.~\eqref{eq:symm property spinless Cn 2chain}.
Additionally, we can incorporate an invertible topological order, i.e., a single Majorana chain. 

For the $1$D block-state decoration on \( \tau_1 \), we note that any dangling modes on \( \mu_1 \) or \( \mu_2 \) necessarily violate the no-open-edge condition. This follows because gapless edge modes in FSPT phases are, by definition, topologically protected and cannot be trivially gapped out. An analogous argument applies to the decorations on \( \tau_2 \).   Thus, we must decorate the $1$D block-states on both \( \tau_1 \) and \( \tau_2 \) to open the gap for zero modes on \( \mu_1 \) and \( \mu_2 \). To maximize the number of obstruction-free configurations, we decorate Majorana chains with the same pairing direction. These decorated states simultaneously contribute both types of zero modes, \( \gamma^L \) and \( \gamma^R \), on the $0$D blocks \( \mu_1 \) and \( \mu_2 \) (see Fig.~\ref{fig:pmg tau1tau2}). Similarly to the \( pm \) case, the decoration of a single Majorana chain and the 1D \( n_1^T \) or \( n_1^M \) FSPT phase is obstruction-free. Since stacking two single Majorana chains is equivalent to the 1D FSPT state of the phase \( n_1^T \) , the total 1D obstruction-free block-states for \( \tau_1 \) and \( \tau_2 \) form the group \( \mathbb{Z}_8 \times \mathbb{Z}_4 \).

The $0$D blocks \( \mu_3 \) and \( \mu_4 \) have the physical symmetry group \( G_b = C_2 \times \mathbb{Z}_2^T \), and the cell decomposition near them aligns with our analysis of the point group \( C_2 \). Thus, decorating 1D BSPT phases (realized via four Majorana chains) on the $1$D block \( \tau_4 \) (or \( \tau_3 \)) can open a spectral gap for the zero modes at the $0$D blocks \( \mu_3 \) (or \( \mu_4 \)). It is straightforward to verify that dangling modes at $\mu_1$ (or $\mu_2$) are also gappable. Therefore, all obstruction-free $1$D block-states form the group:
\begin{equation}
\{ \mathrm{OFBS} \}^{1D} = \mathbb{Z}_8 \times \mathbb{Z}_4 \times \mathbb{Z}_2 \times \mathbb{Z}_2,
\end{equation}
where the group elements can be labeled by:
\begin{equation}
[ \mathcal{M}_{12}, m_{12}, \mathcal{T}_3, \mathcal{T}_4 ],
\end{equation}
with the following physical interpretations:
\begin{itemize}
    \item $\mathcal{M}_{12}  \in \{0,1,\ldots,7\}$ counts the number of decorated single Majorana chains  on both $\tau_1$ and $\tau_2$;
    \item $m_{12} \in \{0,1,2,3\}$ counts the number of decorated $n_1^M$ FSPT phases on both $\tau_1$ and $\tau_2$;
    \item $\mathcal{T}_3, \mathcal{T}_4 \in \{0,1\}$ count the number of decorated 1D BSPT phases on $\tau_3$ and $\tau_4$ respectively.
\end{itemize}
This classification fully characterizes the possible obstruction-free $1$D block-state decorations under the given symmetry constraints.

The total symmetry group for the $2$D blocks is \( G_f = \mathbb{Z}_2^T \times \mathbb{Z}_2^f \). As concluded in Ref.~\cite{Wang2020}, there are no corresponding FSPT phases under this symmetry group. Consequently, the contribution of $2$D block decoration is classified as:
\begin{equation}
\{\mathrm{OFBS}\}^{2D} = \mathbb{Z}_1.
\end{equation}

We now turn to bubble equivalence relations. 
2D ``Majorana bubbles" can contribute to 1D non-trivial states, classified by \( \mathbb{Z}_4 \), on the $1$D blocks \( \tau_1 \) and \( \tau_2 \) (which have reflection symmetry), including \( n_1^M+ n_1^T \) and \( (-1)^{n_1^M \cup n_1^M} \oplus (-1)^{n_1^T \cup n_1^T} \) phases through stacking. However, only the \( (-1)^{n_1^M \cup n_1^M} \oplus (-1)^{n_1^T \cup n_1^T} \) phase decoration on all $1$D blocks belongs to \( \{ \mathrm{OFBS} \}^{1D} \), similar to the group \( pm \).

Then, we examine the $1$D bubble equivalences by decorating a pair of complex fermions on each $1$D block. For the $1$D block \( \tau_1 \), the $1$D bubble contributes to a non-trivial state \( \ket{\phi} = c_1^\dagger \ket{0} \) on the $0$D block \( \mu_1 \), which satisfies \( P_f \ket{\phi} = -\ket{\phi} \), and a non-trivial state \( \ket{\phi} = c_1^\dagger \ket{0} \) on the $0$D block \( \mu_2 \), which also satisfies \( P_f \ket{\phi} = -\ket{\phi} \). Similarly, the $1$D bubble on the $1$D block \( \tau_2 \) contributes to a non-trivial eigenstate of \( P_f \) on the $0$D blocks \( \mu_1 \) and \( \mu_2 \). 

For the $1$D block \( \tau_3 \), the $1$D bubble contributes a non-trivial state \( \ket{\phi} = c_1^\dagger c_2^\dagger \ket{0} \) on the $0$D block \( \mu_2 \), which satisfies \( M_{\tau_{1}} \ket{\phi} = -\ket{\phi} \), and a non-trivial state \( \ket{\phi} \) on the $0$D block \( \mu_4 \), which satisfies \( C_2 \ket{\phi} = -\ket{\phi} \). Similarly, the $1$D bubble on \( \tau_4 \) contributes a non-trivial state \( \ket{\phi} \) on the $0$D block \( \mu_1 \), satisfying \( M_{\tau_{1}} \ket{\phi} = -\ket{\phi} \), and a non-trivial state \( \ket{\phi} \) on the $0$D block \( \mu_3 \), satisfying \( C_2 \ket{\phi} = -\ket{\phi} \).

We also consider bosonic $1$D bubbles. For the $1$D block \( \tau_1 \), the bosonic bubble contributes a non-trivial state \( \ket{\phi} = a_1^\dagger \ket{0} \) on the $0$D block \( \mu_1 \), satisfying \( M_{\tau_{1}} \ket{\phi} = -\ket{\phi} \), and another non-trivial state \( \ket{\phi} = a_2^\dagger \ket{0} \) on the $0$D block \( \mu_2 \), which also satisfies \( M_{\tau_{1}} \ket{\phi} = -\ket{\phi} \). Similarly, for the $1$D block \( \tau_2 \), the bosonic bubble contributes a non-trivial state \( \ket{\phi} = a_1^\dagger \ket{0} \) on the $0$D block \( \mu_1 \), satisfying \( M_{\tau_{1}} \ket{\phi} = -\ket{\phi} \), and another non-trivial state \( \ket{\phi} = a_1^\dagger \ket{0} \) on the $0$D block \( \mu_2 \), satisfying \( M_{\tau_{1}} \ket{\phi} = -\ket{\phi} \).

Therefore, the total trivialization groups are:
\begin{equation}
\{\mathrm{TBS}\}^{1D} = \mathbb{Z}_4, \quad \{\mathrm{TBS}\}^{0D} = \mathbb{Z}_2^4.
\end{equation}

It is evident that there is no extension between 1D and $0$D block-states. Thus, all independent non-trivial block-states with different dimensions are classified by:
\begin{align}
    \begin{aligned}
        & E^{\mathrm{2D}}= \{\mathrm{OFBS}\}^{\mathrm{2D}} = \mathbb{Z}_1, \\
        & E^{\mathrm{1D}} = \{\mathrm{OFBS}\}^{\mathrm{1D}} / \{\mathrm{TBS}\}^{\mathrm{1D}} = \mathbb{Z}_8 \times \mathbb{Z}_2^2, \\
        & E^{\mathrm{0D}} = \{\mathrm{OFBS}\}^{\mathrm{0D}} / \{\mathrm{TBS}\}^{\mathrm{0D}} = \mathbb{Z}_2^4.
    \end{aligned}
\end{align}

The ultimate classification is:
\begin{equation}
\mathcal{G}_0 = \mathbb{Z}_8 \times \mathbb{Z}_2^6.
\end{equation}

\subsubsection{Spinful fermions}

For a spinful fermion system, the on-site physical symmetry group for the $0$D blocks \( \mu_1 \) and \( \mu_2 \) is \( G_b = \mathbb{Z}_2^M \times \mathbb{Z}_2^T \), while for \( \mu_3 \) and \( \mu_4 \), it is \( G_b = C_2 \times \mathbb{Z}_2^T \). In Sec.~\ref{sec:Cn spinfull}, we demonstrated that the classification of block-states protected by these symmetry groups for \( \mu_1 \) and \( \mu_2 \) is \( \mathbb{Z}_2 \), and they can be labeled by the eigenvalues \( \pm 1 \) of \( M \). Similarly, the classification for \( \mu_3 \) and \( \mu_4 \) is \( \mathbb{Z}_2 \), with group elements labeled by the eigenvalues \( \pm 1 \) of \( C_2 \). Therefore, obstruction-free $0$D block-states form the group:
\begin{equation}
\{ \mathrm{OFBS} \}^{0D} = \mathbb{Z}_2^4.
\end{equation}

Next, we examine the decoration of $1$D block-states. For the $1$D blocks \( \tau_1 \) and \( \tau_2 \), the physical symmetry group is \( G_b = \mathbb{Z}_2 \times \mathbb{Z}_2^T \), while for \( \tau_3 \) and \( \tau_4 \), it is \( G_b = \mathbb{Z}_2^T \). As discussed previously, it is not possible to decorate the invertible topological order, i.e. a single Majorana chain. However, the 1D FSPT phases, realized via double Majorana chains, can serve as a suitable candidate, with the classification for \( \tau_1 \) and \( \tau_2 \) being \( \mathbb{Z}_4 \), and for \( \tau_3 \) and \( \tau_4 \) being \( \mathbb{Z}_2 \). 

Decorating the 1D FSPT phase on $1$D blocks \( \tau_1 \) or \( \tau_2 \) leaves two dangling Majorana fermion modes on each of the $0$D blocks \( \mu_1 \) and \( \mu_2 \), which are protected and cannot be gapped out. However, if both \( \tau_1 \) and \( \tau_2 \) are decorated simultaneously (with the same pairing direction), these Majorana fermions can open a spectral gap at the $0$D blocks, making this decoration obstruction-free. Decorating the 1D FSPT phases on the $1$D block \( \tau_3 \) leaves four dangling Majorana fermion modes on each of the $0$D blocks \( \mu_2 \) and \( \mu_4 \). As concluded in Sec.~\ref{sec:Cn spinfull}, these Majorana fermions can open a spectral gap at the $0$D blocks with physical symmetry \( G_b = \mathbb{Z}_2 \times \mathbb{Z}_2^T \), making this decoration obstruction-free. 

Thus, the 1D decoration contributes to the classification group:
\begin{equation}
\{\mathrm{OFBS}\}^{1D} = \mathbb{Z}_2^2 \times \mathbb{Z}_4,
\end{equation}
where the group elements can be labeled as:
\begin{equation}
[\mathbb{MT}_{12}, \mathbb{T}_3, \mathbb{T}_4],
\end{equation}
The integer labels correspond to:
\begin{itemize}
    \item $\mathbb{MT}_{12} \in \{0,1,2,3\}$ counts the number of decorated 1D FSPT phases on $\tau_1$ and $\tau_2$, protected by $G_f = (\mathbb{Z}_2 \times \mathbb{Z}_2^T)\times_{\omega_2} \mathbb{Z}_2^f$ symmetry
    \item $\mathbb{T}_j \in \{0,1\}$ counts the number of decorated 1D FSPT phases on $\tau_j$ ($j=3,4$), protected by $G_f=\mathbb{Z}_4^{fT}$ symmetry
\end{itemize}
This classification fully characterizes the obstruction-free $1$D block-state decorations under the given symmetry constraints.

For the same reasons as in the previous spinful cases, the decoration of 2D DIII TSC states is always obstruction-free, contributing to a \( \{\mathrm{OFBS}\}^{2D} = \mathbb{Z}_2 \) classification.

We now analyze the bubble equivalence for the $1$D blocks. The $1$D bubbles on $\tau_1$ and $\tau_2$ act on the $0$D blocks $\mu_1$ and $\mu_2$, respectively, generating a non-trivial eigenstate of fermion parity $P_f$ that lies outside $\{\mathrm{OFBS}\}^{0D}$. However, simultaneous decoration of $1$D bubbles on both $\tau_1$ and $\tau_2$ produces an atomic insulator state $\ket{TR} = c_1^\dagger c_2^\dagger \ket{0}$ with the symmetry transformation property:
\begin{equation}
M\ket{TR} = -\ket{TR}, \quad \text{under} \quad M: c_1^\dagger \to ic_1^\dagger, c_2^\dagger \to ic_2^\dagger.
\end{equation} Consequently, the $1$D bubble decorations on $\tau_1$ and $\tau_2$ generate non-trivial states that admit a $\mathbb{Z}_2$ classification.

For the remaining $1$D blocks:
\begin{itemize}
    \item $1$D bubbles on $\tau_3$ act on $\mu_2$ and $\mu_4$
    \item $1$D bubbles on $\tau_4$ act on $\mu_1$ and $\mu_3$
\end{itemize}
As established in Sec.~\ref{sec:Cn spinfull}, these $1$D bubbles have no trivializing effect on $0$D blocks with two-fold rotation symmetry.

Furthermore, $2$D bubbles are similarly ineffective at trivializing either 0D or $1$D block-states. We therefore conclude that the total trivializing groups are:
\begin{equation}
\{\mathrm{TBS}\}^{1D} = \mathbb{Z}_1, \quad \{\mathrm{TBS}\}^{0D} = \mathbb{Z}_2.
\end{equation}

Finally, we consider the extension problem. We find no extension from 1D block-states to 0D block-states in this case. 
As demonstrated in Sec.~\ref{sec:D4 spinful discussion}, while two copies of 2D FSPT phases constitute a trivial bulk phase, they nevertheless induce effective nontrivial 1D FSPT phases on \( \tau_1 \) and \( \tau_2 \) due to reflection symmetry. This induces a natural extension from 2D to $1$D block-states. In contrast, the absence of reflection symmetry prevents the formation of any nontrivial 1D FSPT phases on $\tau_3$ and $\tau_4$, leaving only trivial states.

Thus, all independent non-trivial block-states with different dimensions are classified by:
\begin{align}
    \begin{aligned}
        & E^{\mathrm{2D}}= \{\mathrm{OFBS}\}^{\mathrm{2D}} = \mathbb{Z}_2, \\
        & E^{\mathrm{1D}} = \{\mathrm{OFBS}\}^{\mathrm{1D}} / \{\mathrm{TBS}\}^{\mathrm{1D}} = \mathbb{Z}_4 \times \mathbb{Z}_2^2, \\
        & E^{\mathrm{0D}} = \{\mathrm{OFBS}\}^{\mathrm{0D}} / \{\mathrm{TBS}\}^{\mathrm{0D}} = \mathbb{Z}_2^3.
    \end{aligned}
\end{align}

The final classification is:
\begin{equation}
\mathcal{G}_{1/2} = \mathbb{Z}_8 \times \mathbb{Z}_2^5,
\end{equation}
where the \( \mathbb{Z}_8 \) term arises from the extension of $2$D block-states to $1$D block-states.

 \subsection{$p4$}
 \begin{figure}[tb]
    \centering
    \includegraphics[width=0.42\textwidth]{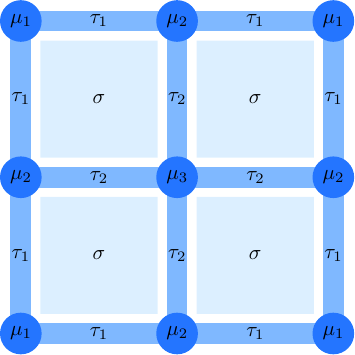}
    \caption{The \#10  2D space group $p4$ and its  cell decomposition.}
    \label{p4}
\end{figure}
 
For the $2$D block \( \sigma \) and the $1$D blocks \( \tau_1 \) and \( \tau_2 \), there is no site symmetry, and the physical symmetry group is \( \mathbb{Z}_2^T \). For the $0$D blocks labeled \( \mu_1 \) and \( \mu_3 \), the physical symmetry group is \( \mathbb{Z}_4 \times \mathbb{Z}_2^T \), due to the internal action of the rotation symmetry group \( C_4 \). For the $0$D block \( \mu_2 \), the physical symmetry group is \( \mathbb{Z}_2 \times \mathbb{Z}_2^T \), corresponding to the two-fold rotation subgroup \( C_2 \subset C_4 \). Fig.~\ref{p4} shows the cell decomposition.

\subsubsection{Spinless fermions}

For a spinless fermion system, the on-site physical symmetry group for the $0$D blocks \( \mu_1 \) and \( \mu_3 \) is \( G_b = \mathbb{Z}_4 \times \mathbb{Z}_2^T \), while for \( \mu_2 \), it is \( G_b = \mathbb{Z}_2 \times \mathbb{Z}_2^T \). In Sec.~\ref{sec:appendix spinless Cn}, we have discussed the classification of 0D FSPTs protected by these symmetry groups. The block-states on $0$D blocks \( \mu_1 \) and \( \mu_3 \) are classified by \( \mathbb{Z}_2^2 \), and can be labeled by the eigenvalues \( \pm 1 \) of \( C_4 \) and \( P_f \). Similarly, the classification of the block-state on the $0$D block \( \mu_2 \) is \( \mathbb{Z}_2^2 \), and can be labeled by the eigenvalues \( \pm 1 \) of \( C_2 \) and \( P_f \). Therefore, obstruction-free $0$D block-states form the group:
\begin{equation}
\{ \mathrm{OFBS} \}^{0D} = \mathbb{Z}_2^6.
\end{equation}

We now discuss the decoration of $1$D block-states. The physical symmetry group for all $1$D blocks is given by \( G_b = \mathbb{Z}_2^T \). As mentioned in Sec.~\ref{sec:appendix spinless Cn}, the corresponding FSPT phases are classified by \( \mathbb{Z}_4 \), with a root state model consisting of two Majorana chains, whose symmetry properties are detailed in Eq.~\eqref{eq:symm property spinless Cn 2chain}. Additionally, we can incorporate an invertible topological order, i.e., a single Majorana chain. 

Consider the $1$D block-state decoration on the $1$D block \(\tau_1\). From our analysis of the point groups \(C_4\) and \(C_2\), any 1D FSPT decoration can gap out the boundary modes at the $0$D block \(\mu_1\). However, at \(\mu_2\) only a restricted subset—namely, 1D BSPT decorations obtained by stacking root phases—can open a gap, and hence only these decorations are obstruction-free. Therefore, the obstruction-free 1D BSPT decoration on \(\tau_1\) contributes a \(\mathbb{Z}_2\) factor to the classification.

The decorations on \(\tau_2\) are analogous to those on \(\tau_1\). In addition, if one decorates \(\tau_1\) and \(\tau_2\) simultaneously by 1D FSPT phases, the resulting boundary modes at \(\mu_1\), \(\mu_2\), and \(\mu_3\) can all be gapped out. This yields a \(\mathbb{Z}_4\) contribution to the classification group. And this decoration realizes an intrinsically interacting FSPT phase, as it incorporates the intrinsically interacting decoration from the $C_4$ case.

Thus, all obstruction-free $1$D block-states form the group:
\begin{equation}
\{ \mathrm{OFBS} \}^{1D} = \mathbb{Z}_2  \times \mathbb{Z}_4 ,
\end{equation}
where the group elements can be labeled by:
\begin{equation}
[b\mathcal{T}_{12}, BM_2],
\end{equation}
with the following physical interpretations:
\begin{itemize}
    \item \( b\mathcal{T}_{12} = 0, 1, 2, 3 \) represent the number of decorated 1D FSPT phases on \( \tau_1 \) and $\tau_2$;
    \item  \( BM_2 = 0, 1 \)  represent the number of decorated 1D BSPT phases on \( \tau_2 \).
    \end{itemize}

The total symmetry group for the $2$D blocks is \( G_f = \mathbb{Z}_2^T \times \mathbb{Z}_2^f \). As concluded in Ref.~\cite{Wang2020}, there are no corresponding FSPT phases under this symmetry group. Consequently, the contribution of $2$D block decoration is classified as:
\begin{equation}
\{\mathrm{OFBS}\}^{2D} = \mathbb{Z}_1.
\end{equation}

We now turn to bubble equivalence relations. As mentioned in Sec.~\ref{sec:Cn spinless bubble}, the $2$D bubbles cannot contribute non-trivial 1D FSPT states, but they can be used to trivialize the eigenstate of \( P_f \). 

Then, we examine the $1$D bubble equivalences by decorating a pair of complex fermions on each $1$D block. For the $1$D block \( \tau_1 \), the $1$D bubble contributes a non-trivial state \( \ket{\phi} = c_1^\dagger c_2^\dagger c_3^\dagger c_4^\dagger \ket{0} \) on \( \mu_1 \), satisfying \( C_4 \ket{\phi} = c_2^\dagger c_3^\dagger c_4^\dagger c_1^\dagger \ket{0} = -\ket{\phi} \), and a non-trivial state \( \ket{\phi} = c_1^\dagger c_2^\dagger \ket{0} \) on \( \mu_2 \), satisfying \( C_2 \ket{\phi} = c_2^\dagger c_1^\dagger \ket{0} = -\ket{\phi} \). The $1$D bubble on \( \tau_2 \) has a similar effect on \( \mu_2 \) to the effect of bubbles on \( \tau_1 \) on \( \mu_1 \). Thus, $1$D bubbles form the trivialization group \( \mathbb{Z}_2^2 \). Therefore, the total trivialization groups are:
\begin{equation}
\{\mathrm{TBS}\}^{1D} = \mathbb{Z}_1, \quad \{\mathrm{TBS}\}^{0D} = \mathbb{Z}_2^3.
\end{equation}

It is evident that there is no extension between 1D and $0$D block-states. Thus, all independent non-trivial block-states with different dimensions are classified by:
\begin{align}
    \begin{aligned}
        & E^{\mathrm{2D}}= \{\mathrm{OFBS}\}^{\mathrm{2D}} = \mathbb{Z}_1, \\
        & E^{\mathrm{1D}} = \{\mathrm{OFBS}\}^{\mathrm{1D}} / \{\mathrm{TBS}\}^{\mathrm{1D}} = \mathbb{Z}_4 \times \mathbb{Z}_2, \\
        & E^{\mathrm{0D}} = \{\mathrm{OFBS}\}^{\mathrm{0D}} / \{\mathrm{TBS}\}^{\mathrm{0D}} = \mathbb{Z}_2^3.
    \end{aligned}
\end{align}

Hence, the ultimate classification with an accurate group structure is:
\begin{equation}
\mathcal{G}_0 = \mathbb{Z}_4 \times \mathbb{Z}_2^4.
\end{equation}

 \subsubsection{Spinful fermions}

For a spinful fermion system, the on-site physical symmetry group for the $0$D blocks \( \mu_1 \) and \( \mu_3 \) is \( G_b = \mathbb{Z}_4 \times \mathbb{Z}_2^T \), while for \( \mu_2 \), it is \( G_b = \mathbb{Z}_2 \times \mathbb{Z}_2^T \). In Sec.~\ref{sec:Cn spinfull}, we have discussed the classification of 0D FSPTs protected by these symmetry groups. The block-states on $0$D blocks \( \mu_1 \) and \( \mu_3 \) are classified by \( \mathbb{Z}_2 \), and can be labeled by the eigenvalues \( \pm 1 \) of \( C_4 \). Similarly, the classification of the block-state on the $0$D block \( \mu_2 \) is \( \mathbb{Z}_2 \), and can be labeled by the eigenvalues \( \pm 1 \) of \( C_2 \). Therefore, obstruction-free $0$D block-states form the group:
\begin{equation}
\{ \mathrm{OFBS} \}^{0D} = \mathbb{Z}_2^3.
\end{equation}

We now examine the decoration of the $1$D block-states. The physical symmetry group for all the $1$D blocks is \( G_b = \mathbb{Z}_2^T \). As discussed earlier, the invertible topological order of a single Majorana chain cannot be decorated, but the 1D FSPT phase (realized via double Majorana chains) can, and its classification is \( \mathbb{Z}_2 \). 

Decorating the 1D FSPT phase on the $1$D block \( \tau_1 \) leaves eight dangling Majorana fermion modes on the $0$D block \( \mu_1 \) and four on the $0$D block \( \mu_2 \). As concluded in Sec.~\ref{sec:Cn spinfull}, these Majorana fermions can open a spectral gap at the $0$D blocks with physical symmetries \( G_b = C_2 \times \mathbb{Z}_2^T \) and \( G_b = C_4 \times \mathbb{Z}_2^T \), making this decoration obstruction-free. The decorations on the remaining $1$D blocks follow a similar pattern and are also obstruction-free. Thus, the 1D decoration contributes to the classification group:
\begin{equation}
\{\mathrm{OFBS}\}^{1D} = \mathbb{Z}_2^2,
\end{equation}
where the group elements can be labeled as:
\begin{equation}
[\mathbb{T}_1,\mathbb{T}_2].
\end{equation}
Here, \( \mathbb{T}_j = 0, 1 \) represent the number of decorated 1D FSPT phases, protected by $\mathbb{Z}_4^{fT}$,  on the $1$D blocks \( \tau_j \) (\( j = 1, 2 \)), respectively.

Analogous to the previous spinful cases, 2D class DIII TSC state decorations are obstruction-free, yielding $\{\mathrm{OFBS}\}^{2D} = \mathbb{Z}_2$.

Turning to bubble equivalence, the $1$D bubble on the $1$D block \( \tau_1 \) acts on the $0$D blocks \( \mu_1 \) and \( \mu_2 \), while the $1$D bubble on the $1$D block \( \tau_2 \) acts on the $0$D blocks \( \mu_2 \) and \( \mu_3 \). As previously concluded in Sec.~\ref{sec:Cn spinfull}, $1$D bubbles have no trivializing effect on the $0$D blocks with on-site physical symmetries \( G_b = C_2 \times \mathbb{Z}_2^T \) and \( G_b = C_4 \times \mathbb{Z}_2^T \). Additionally, the $2$D bubbles have no trivializing effect on either 0D or $1$D block-states. Therefore, the total trivializing group is:
\begin{equation}
\{\mathrm{TBS}\}^{1D} = \mathbb{Z}_1, \quad \{\mathrm{TBS}\}^{0D} = \mathbb{Z}_1.
\end{equation}

Finally, we consider the extension problem. We find no extension from 1D block-states to 0D block-states in this case. Since this case involves only the cyclic point group \( C_4 \), according to the discussion in Sec.~\ref{sec:extension cn spinful}, the mass term of the boundary Majorana modes under 2D decoration does not acquire an additional minus sign under symmetry, so there is no extension of $2$D block-states to $1$D block-states.

Thus, all independent non-trivial block-states with different dimensions are classified by:
\begin{align}
    \begin{aligned}
        & E^{\mathrm{2D}}= \{\mathrm{OFBS}\}^{\mathrm{2D}} = \mathbb{Z}_2, \\
        & E^{\mathrm{1D}} = \{\mathrm{OFBS}\}^{\mathrm{1D}} / \{\mathrm{TBS}\}^{\mathrm{1D}} = \mathbb{Z}_2^2, \\
        & E^{\mathrm{0D}} = \{\mathrm{OFBS}\}^{\mathrm{0D}} / \{\mathrm{TBS}\}^{\mathrm{0D}} = \mathbb{Z}_2^3.
    \end{aligned}
\end{align}

The final classification is:
\begin{equation}
\mathcal{G}_{1/2} = \mathbb{Z}_2^6.
\end{equation}

\subsection{$p4g$}
\begin{figure}[tb]
    \centering
    \includegraphics[width=0.46\textwidth]{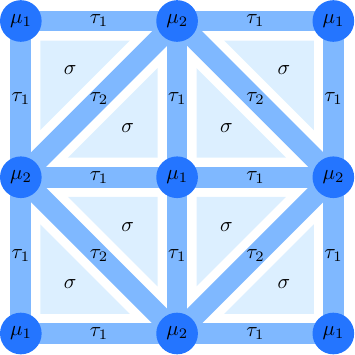}
    \caption{The \#12  2D space group $p4g$ and its  cell decomposition.}
    \label{p4g}
\end{figure}

For the $2$D block $\sigma$ and the $1$D block $\tau_1$, there is no site symmetry, and the physical symmetry group is $\mathbb{Z}_2^T$. For the $1$D block $\tau_2$, the internal action of the reflection symmetry gives the physical symmetry group $\mathbb{Z}_2 \times \mathbb{Z}_2^T$. For the $0$D block $\mu_1$, the physical symmetry group is $\mathbb{Z}_4 \times \mathbb{Z}_2^T$ due to the internal action of the symmetry $C_4$. For the $0$D block $\mu_2$, the physical symmetry group is $\mathbb{Z}_2 \rtimes \mathbb{Z}_2 \times \mathbb{Z}_2^T$, corresponding to the internal symmetry $D_2 \subset D_4$. See Fig.~\ref{p4g}.

\subsubsection{Spinless fermions} \label{sec:p4g D2 special}

For a spinless fermion system, the on-site physical symmetry group for the $0$D block \( \mu_1 \) is \( G_b = \mathbb{Z}_4 \times \mathbb{Z}_2^T \), while for \( \mu_2 \), it is \( G_b = D_2 \times \mathbb{Z}_2^T \). In Sec.~\ref{sec:appendix spinless Cn} and Sec.~\ref{sec:D2 D6 spinless}, we have discussed the classification of 0D FSPTs protected by these symmetry groups. The block-state on the $0$D block \( \mu_1 \) is classified by \( \mathbb{Z}_2^2 \), and it can be labeled by the eigenvalues \( \pm 1 \) of \( C_4 \) and \( P_f \). The classification of the block-state on the $0$D block \( \mu_2 \) is \( \mathbb{Z}_2^3 \), and it can be characterized by the eigenvalues \( \pm 1 \) of \( C_2 \), \( M \), and \( P_f \). Therefore, obstruction-free $0$D block-states form the group:
\begin{equation}
\{\mathrm{OFBS}\}^{0D} = \mathbb{Z}_2^5.
\end{equation}

\begin{figure}[tb]
    \centering

    \includegraphics[width=0.3\textwidth]{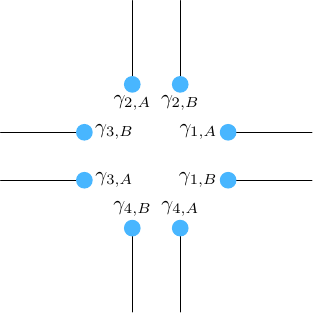}
    \caption{1D FSPT phase (via two Majorana chains) decorations on the $1$D block \( \tau_1 \), leaving 8 dangling Majorana modes at the $0$D block \( \mu_2 \).}
\label{fig:p4g 2chain}
\end{figure}

We now discuss the decoration of $1$D block-states. The physical symmetry group for the $1$D block \( \tau_1 \) is given by \( G_b = \mathbb{Z}_2^T \). As mentioned in Sec.~\ref{sec:appendix spinless Cn}, the corresponding FSPT phases are classified by \( \mathbb{Z}_4 \), with a root state model consisting of two Majorana chains, whose symmetry properties are detailed in Eq.~\eqref{eq:symm property spinless Cn 2chain}. The physical symmetry group for the $1$D block \( \tau_2 \) is \( G_b = \mathbb{Z}_2 \times \mathbb{Z}_2^T \). As discussed in Sec.~\ref{sec:D4 spinless fermion}, the corresponding FSPT phases are classified by \( \mathbb{Z}_4 \times \mathbb{Z}_4 \), where the two fourth-order group generators correspond to the root phases \( n_1^M \) and \( n_1^T \) (their symmetry properties are listed in Eq.~\eqref{eq:symm property of n1T} and Eq.~\eqref{eq:symm property of n1M}). Additionally, we can incorporate an invertible topological order, i.e., a single Majorana chain.

Consider the $1$D block-state decoration on the $1$D block \( \tau_1 \). Based on our analysis of the point group, the decoration of 1D FSPT phases can open a spectral gap at the $0$D block \( \mu_1 \). This decoration  also leaves eight dangling Majorana modes, $\gamma_{j,A/B}$ (see Fig.~\ref{fig:p4g 2chain}) on the $0$D block \( \mu_2 \). The cell decomposition near $\mu_2$ does not align with the point group $D_2$ case, though the site symmetry group of $\mu_2$ is just $D_2$. They possess the symmetry properties:
\begin{align}
    T:&\quad i \mapsto -i, \quad \gamma_{j,h} \mapsto \gamma_{j,h}, \notag \\
    M:&\quad \gamma_{1,h} \leftrightarrow \gamma_{2,h}, \quad \gamma_{3,h} \leftrightarrow \gamma_{4,h}, \notag \\
    C_2:&\quad \gamma_{1,h} \leftrightarrow \gamma_{3,h}, \quad \gamma_{2,h} \leftrightarrow \gamma_{4,h}, \label{eq:D2 special symmetry}
\end{align}
where \( h = A,B \) and \( j = 1, 2, 3, 4 \). It can be verified that the interaction term:
\begin{equation}
\begin{aligned}
     &U \Big[ (n_1 - \frac{1}{2})(n_2 - \frac{1}{2}) 
     + (n_3 - \frac{1}{2})(n_4 - \frac{1}{2}) \Big] \quad \\
     &+ J \sum_{j=x,y,z} \tau^j_{12} \tau^j_{34},
\end{aligned}
\end{equation}
where \( c_j \) and \( \tau_{ab}^\mu \) retain their forms defined in previous sections, provides a gapped  symmetric unique ground state. Thus, the 1D FSPT phase decoration on \( \tau_1 \) satisfies the no-open-edge condition.
It should be noted that this decoration realizes an intrinsically interacting FSPT phase, which can be viewed as a combination of the intrinsic sectors of \(C_4\) and \(D_2\) (although the $D_2$ component is trivial when only $D_2$ symmetry is present).

Next, consider $1$D block-state decorations on \(\tau_2\). As discussed in Sec.~\ref{sec:D2 D6 spinless} for the \(D_2\) case, the no-open-edge condition can be satisfied in two ways: (i) decorating each of the two inequivalent $1$D blocks by a 1D BSPT root phase, \( (-1)^{n_1^M\cup n_1^M}\) or equivalently \( (-1)^{n_1^T\cup n_1^T}\), which are both obstruction-free; and (ii) decorating the two inequivalent $1$D blocks simultaneously by the 1D FSPT phase \(n_1^T+n_1^M\).

In the present \(p4g\) example, the arrangement of the \(\tau_2\) blocks around the \(\mu_2\) is the same as the 1D-block configuration analyzed for \(D_2\), except that there is only a single symmetry-inequivalent type of $1$D block. Consequently, the obstruction-free decorations on \(\tau_2\) are generated by decorating \(n_1^T+n_1^M\) and by decorating the BSPT root phase \( (-1)^{n_1^M\cup n_1^M}\) (whose generated group also contains \( (-1)^{n_1^T\cup n_1^T}\) phase). These obstruction-free decorations therefore contribute a \(\mathbb{Z}_4\times \mathbb{Z}_2\) factor to the classification.

Therefore, all obstruction-free $1$D block-states form the group:
\begin{equation}
\{\mathrm{OFBS}\}^{1D} = \mathbb{Z}_4^2 \times \mathbb{Z}_2 ,
\end{equation}
where the group elements can be labeled by:
\begin{equation}
[b\mathcal{T}_1, mt_2, M_2],
\end{equation}
These group elements are integers and  correspond to:
\begin{itemize}
    \item $b\mathcal{T}_1 \in \{0,1,2,3\}$ counts the number of decorated 1D FSPT phases on $\tau_1$
    \item $mt_2 \in \{0,1,2,3\}$ counts the number of decorated $n_1^T + n_1^M$  FSPT phases on $\tau_2$
    \item $M_2 \in \{0,1\}$ counts the number of decorated $(-1)^{n_1^M \cup n_1^M}$ BSPT phases on $\tau_2$
\end{itemize}
This classification fully characterizes all possible obstruction-free $1$D block-state decorations under the given symmetry constraints.

The total symmetry group for the $2$D blocks is \( G_f = \mathbb{Z}_2^T \times \mathbb{Z}_2^f \). As concluded in Ref.~\cite{Wang2020}, there are no corresponding FSPT phases under this symmetry group. Consequently, the contribution of $2$D block decoration is classified as:
\begin{equation}
\{\mathrm{OFBS}\}^{2D} = \mathbb{Z}_1.
\end{equation}

Next, we consider bubble equivalence relations. Similarly to the cases in \( p4m \), $2$D bubbles, referred to as ``Majorana bubbles," can contribute to 1D non-trivial states, whose classification is \( \mathbb{Z}_4 \), on the $1$D block \( \tau_2 \). This includes the \( n_1^T + n_1^M \) and \( (-1)^{n_1^M \cup n_1^M} \oplus (-1)^{n_1^T \cup n_1^T} \) phases through stacking. 
Thus, $2$D bubbles contribute to a trivialization group:
\begin{equation}
\{\mathrm{TBS}\}^{1D} = \mathbb{Z}_4.
\end{equation}

We now examine the $1$D bubble equivalences by decorating a pair of complex fermions on each $1$D block. For the $1$D block \( \tau_1 \), the $1$D bubble contributes a non-trivial state \( \ket{\phi} = c_1^\dagger c_2^\dagger c_3^\dagger c_4^\dagger \ket{0} \) on \( \mu_1 \), satisfying:
\begin{equation}
C_4 \ket{\phi} = c_2^\dagger c_3^\dagger c_4^\dagger c_1^\dagger \ket{0} = -\ket{\phi},
\end{equation}
and a trivial state \( \ket{\phi} = c_1^\dagger c_2^\dagger c_3^\dagger c_4^\dagger \ket{0} \) on \( \mu_2 \), satisfying:
\begin{equation}
C_2 \ket{\phi} = c_3^\dagger c_4^\dagger c_1^\dagger c_2^\dagger \ket{0} = \ket{\phi}, \quad M\ket{\phi} = c_2^\dagger c_1^\dagger c_4^\dagger c_3^\dagger \ket{0} = \ket{\phi}.
\end{equation}
Similarly, the $1$D bubble on \( \tau_2 \) contributes a non-trivial state \( \ket{\phi} \) on \( \mu_2 \), satisfying:
\begin{equation}
M_{\tau_{2}} \ket{\phi} = -\ket{\phi}, \quad C_2 \ket{\phi} = \ket{\phi}.
\end{equation}
Thus, $1$D bubbles form the trivialization group:
\begin{equation}
\{\mathrm{TBS}\}^{0D} = \mathbb{Z}_2^2.
\end{equation}

It is evident that there is no extension between 1D and $0$D block-states. Thus, all independent non-trivial block-states with different dimensions are classified by:
\begin{align}
    \begin{aligned}
        & E^{\mathrm{2D}}= \{\mathrm{OFBS}\}^{\mathrm{2D}} = \mathbb{Z}_1, \\
        & E^{\mathrm{1D}} = \{\mathrm{OFBS}\}^{\mathrm{1D}} / \{\mathrm{TBS}\}^{\mathrm{1D}} = \mathbb{Z}_4 \times \mathbb{Z}_2, \\
        & E^{\mathrm{0D}} = \{\mathrm{OFBS}\}^{\mathrm{0D}} / \{\mathrm{TBS}\}^{\mathrm{0D}} = \mathbb{Z}_2^3.
    \end{aligned}
\end{align}

The ultimate classification is:
\begin{equation}
\mathcal{G}_0 = \mathbb{Z}_4 \times \mathbb{Z}_2^4.
\end{equation}

\subsubsection{Spinful fermions}

For a spinful fermion system, the on-site physical symmetry group for the $0$D block \( \mu_1 \) is \( G_b = \mathbb{Z}_4 \times \mathbb{Z}_2^T \), while for \( \mu_2 \), it is \( G_b = D_2 \times \mathbb{Z}_2^T \). In Sec.~\ref{sec:Cn spinfull} and Sec.~\ref{sec:D2 D6 spinful}, we have discussed the classification of 0D FSPTs protected by these symmetry groups. The block-state on the $0$D block \( \mu_1 \) is classified by \( \mathbb{Z}_2 \), and it can be labeled by the eigenvalues \( \pm 1 \) of \( C_4 \). The classification of the block-state on the $0$D block \( \mu_2 \) is \( \mathbb{Z}_2^2 \), and it can be labeled by the eigenvalues \( \pm 1 \) of \( C_2 \) and \( M \) (reflection symmetry). Therefore, obstruction-free $0$D block-states form the group:
\begin{equation}
\{\mathrm{OFBS}\}^{0D} = \mathbb{Z}_2^3.
\end{equation}

We now examine the decoration of $1$D block-states. The physical symmetry group for the $1$D block \( \tau_1 \) is \( G_b = \mathbb{Z}_2^T \), and for \( \tau_2 \), it is \( G_b = \mathbb{Z}_2 \times \mathbb{Z}_2^T \). As discussed previously, the invertible topological order, i.e., the single Majorana chain,  cannot be decorated, but the 1D FSPT phases can. The classification for \( \tau_1 \) is \( \mathbb{Z}_2 \), while for \( \tau_2 \), it is \( \mathbb{Z}_4 \). 

Decorating the 1D FSPT phase on the $1$D block \( \tau_1 \) leaves eight dangling Majorana fermion modes on the $0$D blocks \( \mu_1 \) and \( \mu_2 \). As concluded in previous sections, these Majorana fermions can open a spectral gap at $0$D blocks with physical symmetries \( G_b = C_4 \times \mathbb{Z}_2^T \) and \( G_b = D_2 \times \mathbb{Z}_2^T \), making this decoration obstruction-free. The decoration on the $1$D block \( \tau_2 \) follows a similar pattern and is also obstruction-free. Thus, the 1D decoration contributes to the classification group:
\begin{equation}
\{\mathrm{OFBS}\}^{1D} = \mathbb{Z}_2 \times \mathbb{Z}_4,
\end{equation}
These group elements are integers and correspond to:
\begin{itemize}
    \item $\mathbb{T}_1 \in \{0,1\}$ counts the number of decorated 1D FSPT phases on $\tau_1$, protected by $\mathbb{Z}_4^{fT}$ symmetry
    \item $\mathbb{MT}_2 \in \{0,1,2,3\}$ counts the number of decorated 1D FSPT phases on $\tau_2$, protected by $G_f=(\mathbb{Z}_2 \times \mathbb{Z}_2^T)\times_{\omega_2} \mathbb{Z}_2^f$ symmetry
\end{itemize}
This classification fully characterizes the obstruction-free $1$D block-state decorations under the given symmetry constraints.

For the same reasons as in the previously discussed spinful example, the decoration of 2D DIII TSC states is always obstruction-free, contributing to a classification group \( \{\mathrm{OFBS}\}^{2D} = \mathbb{Z}_2 \). 

Next, we consider the bubble equivalence. The $1$D bubble on the $1$D block \( \tau_1 \) acts on the $0$D blocks \( \mu_1 \) and \( \mu_2 \), while the $1$D bubble on the $1$D block \( \tau_2 \) acts on the $0$D block \( \mu_2 \). As previously concluded, the $1$D bubble has no trivializing effect on the $0$D blocks with on-site physical symmetries \( G_b = C_4 \times \mathbb{Z}_2^T \) and \( G_b = D_2 \times \mathbb{Z}_2^T \). Additionally, the $2$D bubbles have no trivializing effect on either 0D or $1$D block-states. Therefore, the total trivializing group is:
\begin{equation}
\{\mathrm{TBS}\}^{1D} = \mathbb{Z}_1, \quad \{\mathrm{TBS}\}^{0D} = \mathbb{Z}_1.
\end{equation}

Finally, we consider the extension problem. Here, both cyclic point groups and dihedral point groups are involved. 
As demonstrated in Sec.~\ref{sec:D4 spinful discussion} and Sec.~\ref{sec:extension cn spinful}, while two copies of 2D FSPT phases constitute a trivial bulk phase, they nevertheless induce effective nontrivial 1D FSPT phases on \( \tau_1 \) due to reflection symmetry. This induces a natural extension from 2D to $1$D block-states. In contrast, the absence of reflection symmetry prevents the formation of any nontrivial 1D FSPT phases on $\tau_2$, leaving only trivial states. But we find no extension from 1D block-states to 0D block-states in this case.

Thus, all independent non-trivial block-states with different dimensions are classified by:
\begin{align}
    \begin{aligned}
        & E^{\mathrm{2D}}= \{\mathrm{OFBS}\}^{\mathrm{2D}} = \mathbb{Z}_2, \\
        & E^{\mathrm{1D}} = \{\mathrm{OFBS}\}^{\mathrm{1D}} / \{\mathrm{TBS}\}^{\mathrm{1D}} = \mathbb{Z}_4 \times \mathbb{Z}_2, \\
        & E^{\mathrm{0D}} = \{\mathrm{OFBS}\}^{\mathrm{0D}} / \{\mathrm{TBS}\}^{\mathrm{0D}} = \mathbb{Z}_2^3.
    \end{aligned}
\end{align}

The final classification is:
\begin{equation}
\mathcal{G}_{1/2} = \mathbb{Z}_2^4 \times \mathbb{Z}_8,
\end{equation}
where the \( \mathbb{Z}_8 \) term comes from the extension of $2$D block-states to $1$D block-states.

 \subsection{$p3$}
 
\begin{figure}[tb]
 \centering
 \includegraphics[width=0.46\textwidth]{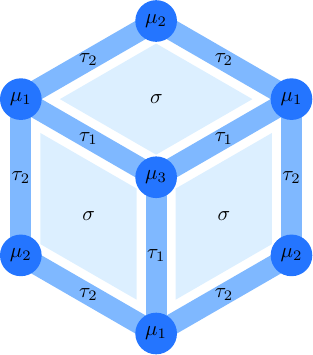}
 \caption{The \#13  2D space group $p3$ and its  cell decomposition.}
 \label{p3}
 \end{figure}

 The corresponding point group for this case is the rotational symmetry group $C_3$. For the $2$D block $\sigma$ and all $1$D blocks, there is no site symmetry, and the physical symmetry group is $\mathbb{Z}_2^T$. For the $0$D blocks $\mu_j$, where $j=1,2,3$, the physical symmetry group is $\mathbb{Z}_3 \times \mathbb{Z}_2^T$, due to the internal action of the rotational symmetry. See Fig.~\ref{p3}.

 \subsubsection{Spinless fermions}
 
 For a spinless fermion system, the on-site physical symmetry group for all $0$D blocks \( \mu_1, \mu_2, \mu_3 \) is \( G_b = C_3 \times \mathbb{Z}_2^T \). In Sec.~\ref{sec:appendix spinless Cn}, we demonstrated that the classification of 0D FSPTs protected by this symmetry group is \( \mathbb{Z}_2 \), labeled by the eigenvalues \( \pm 1 \) of \( P_f \). Therefore, obstruction-free $0$D block-states form the group:
\begin{equation}
\{\mathrm{OFBS}\}^{0D} = \mathbb{Z}_2^3.
\end{equation}

We now turn to the decoration of $1$D block-states. The physical symmetry group for all $1$D blocks is \( G_b = \mathbb{Z}_2^T \). As mentioned previously, the corresponding FSPT phases are classified by \( \mathbb{Z}_4 \), with a root state model consisting of two Majorana chains, whose symmetry properties are detailed in Eq.~\eqref{eq:symm property spinless Cn 2chain}. Additionally, we can incorporate an invertible topological order, i.e., a single Majorana chain. 

Consider the $1$D block-state decoration on the $1$D block \( \tau_1 \). Based on our analysis of the point group \( C_3 \), there is no obstruction-free decoration. The same conclusion applies to the decorations on the $1$D block \( \tau_2 \). Hence, all obstruction-free $1$D block-states form the group:
\begin{equation}
\{\mathrm{OFBS}\}^{1D} = \mathbb{Z}_1.
\end{equation}

The total symmetry group for the $2$D blocks is \( G_f = \mathbb{Z}_2^T \times \mathbb{Z}_2^f \). As concluded in Ref.~\cite{Wang2020}, there are no corresponding FSPT phases under this symmetry group. Consequently, the contribution of $2$D block decoration is classified as:
\begin{equation}
\{\mathrm{OFBS}\}^{2D} = \mathbb{Z}_1.
\end{equation}

Next, we examine bubble equivalence relations. First, we consider the $1$D bubble equivalences by decorating a pair of complex fermions on each $1$D block. For the $1$D block \( \tau_1 \), the $1$D bubble contributes a non-trivial state \( \ket{\phi} \) on \( \mu_1 \), satisfying \( P_f \ket{\phi} = -\ket{\phi} \), and another non-trivial state \( \ket{\phi} \) on \( \mu_3 \), also satisfying \( P_f \ket{\phi} = -\ket{\phi} \). Similarly, the $1$D bubble on the $1$D block \( \tau_2 \) contributes a non-trivial state \( \ket{\phi} \) on \( \mu_1 \), satisfying \( P_f \ket{\phi} = -\ket{\phi} \), and another non-trivial state \( \ket{\phi} \) on \( \mu_2 \), satisfying \( P_f \ket{\phi} = -\ket{\phi} \). Thus, $1$D bubbles form the trivialization group:
\begin{equation}
 \{\mathrm{TBS}\}^{0D}=\mathbb{Z}_2^2.  
\end{equation}

From Ref.~\cite{2Dcrystalline}, we know that $2$D bubbles have no effect on $1$D block-states and $0$D block-states. Therefore, all the trivialization groups are:
\begin{equation}
\{\mathrm{TBS}\}^{0D}=\mathbb{Z}_2^2, ~\{\mathrm{TBS}\}^{1D} = \mathbb{Z}_1.
\end{equation}

It is evident that there is no extension between 1D and $0$D block-states. Thus, all independent non-trivial block-states with different dimensions are classified by:
\begin{align}
    \begin{aligned}
        & E^{\mathrm{2D}}= \{\mathrm{OFBS}\}^{\mathrm{2D}} = \mathbb{Z}_1, \\
        & E^{\mathrm{1D}} = \{\mathrm{OFBS}\}^{\mathrm{1D}} / \{\mathrm{TBS}\}^{\mathrm{1D}} = \mathbb{Z}_1, \\
        & E^{\mathrm{0D}} = \{\mathrm{OFBS}\}^{\mathrm{0D}} / \{\mathrm{TBS}\}^{\mathrm{0D}} = \mathbb{Z}_2.
    \end{aligned}
\end{align}

Hence, the ultimate classification with an accurate group structure is:
\begin{equation}
\mathcal{G}_0 = \mathbb{Z}_2.
\end{equation}

 \subsubsection{Spinful fermions}

For a spinful fermion system, the on-site physical symmetry group for all $0$D blocks \( \mu_1, \mu_2, \mu_3 \) is \( G_b = C_3 \times \mathbb{Z}_2^T \). In Sec.~\ref{sec:Cn spinfull}, we demonstrated that the classification of 0D FSPTs protected by this symmetry group is trivial, i.e., \( \mathbb{Z}_1 \), meaning that no non-trivial decorated state exists. Thus, we have \( \{\mathrm{OFBS}\}^{0D} = \mathbb{Z}_1 \).

We now consider the decoration of $1$D block-states. For all $1$D blocks, the physical symmetry group is \( G_b = \mathbb{Z}_2^T \). As discussed earlier, the invertible topological order, i.e., the single Majorana chain, cannot be decorated, but the 1D FSPT phase (realized via double Majorana chains) can. This decoration has a classification of \( \mathbb{Z}_2 \). 

Decorating the 1D FSPT phase on the $1$D block \( \tau_1 \) leaves six dangling Majorana fermion modes on the $0$D blocks \( \mu_1 \) and \( \mu_3 \). As concluded in the discussion of the point group, these six Majorana fermions cannot open a spectral gap at the $0$D blocks with the physical symmetry \( G_b = \mathbb{Z}_3 \times \mathbb{Z}_2^T \). The same conclusion applies to the $1$D block \( \tau_2 \). 

If 1D FSPT phases are simultaneously decorated on both \( \tau_1 \) and \( \tau_2 \), the corresponding twelve dangling  Majorana fermions on the $0$D blocks \( \mu_1 \) and \( \mu_2 \) can open a spectral gap. However, the zero modes localized at \( \mu_3 \) remain gapless, rendering this decoration obstructed. Thus, the 1D decoration contributes to the classification group:
\begin{equation}
\{\mathrm{OFBS}\}^{1D} = \mathbb{Z}_1.
\end{equation}

For the same reasons as in the previously discussed spinful example, the decoration of 2D DIII TSC states is always obstruction-free, contributing to a \( \{\mathrm{OFBS}\}^{2D} = \mathbb{Z}_2 \) classification. 

We now turn to bubble equivalence. Discussing $1$D bubbles here is unnecessary, as no non-trivial 0D decorated state exists. Additionally, as we know from previous discussions, $2$D bubbles have no trivializing effect on either 0D or $1$D block-states. Therefore, the total trivializing group is:
\begin{equation}
\{\mathrm{TBS}\}^{\mathrm{0D}} = \mathbb{Z}_1, \{\mathrm{TBS}\}^{\mathrm{1D}} = \mathbb{Z}_1.
\end{equation}

Finally, we consider the extension problem. We find that there is no extension from 1D block-states to 0D block-states in this case. Since this case only involves the cyclic point group \( C_3 \), and as discussed in Sec.~\ref{sec:extension cn spinful}, the mass terms of the Majorana modes at the boundaries of 2D decorations do not generate any additional signs under symmetry. Therefore, there is no extension of $2$D block-states to $1$D block-states.

Thus, all independent non-trivial block-states with different dimensions are classified by:
\begin{align}
    \begin{aligned}
        & E^{\mathrm{2D}}= \{\mathrm{OFBS}\}^{\mathrm{2D}} = \mathbb{Z}_2, \\
        & E^{\mathrm{1D}} = \{\mathrm{OFBS}\}^{\mathrm{1D}} / \{\mathrm{TBS}\}^{\mathrm{1D}} = \mathbb{Z}_1, \\
        & E^{\mathrm{0D}} = \{\mathrm{OFBS}\}^{\mathrm{0D}} / \{\mathrm{TBS}\}^{\mathrm{0D}} = \mathbb{Z}_1.
    \end{aligned}
\end{align}
The final classification is:
\begin{equation}
\mathcal{G}_{1/2}= \mathbb{Z}_2.
\end{equation}

 \subsection{$p3m1$}
  \begin{figure}[tb]
     \centering
     \includegraphics[width=0.46\textwidth]{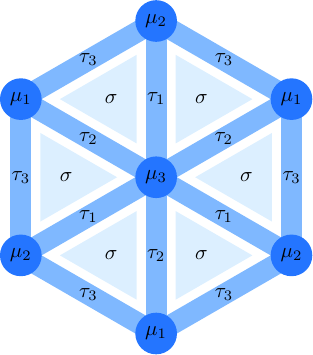}
     \caption{The \#14  2D space group $p3m1$ and its  cell decomposition.}
     \label{p3m1}
 \end{figure}
 
The corresponding point group for this case is the dihedral symmetry group \( D_3 \). For the $2$D block \( \sigma \), there is no site symmetry, and the physical symmetry group is \( \mathbb{Z}_2^T \). For the $1$D blocks \( \tau_j \) (\( j = 1, 2, 3 \)), the internal action of reflection symmetry gives the physical symmetry group \( \mathbb{Z}_2 \times \mathbb{Z}_2^T \). For the $0$D blocks \( \mu_j \) (\( j = 1, 2, 3 \)), the physical symmetry group is \( \mathbb{Z}_3 \rtimes \mathbb{Z}_2 \times \mathbb{Z}_2^T \), corresponding to the local action of the \( D_3 \) group. See Fig.~\ref{p3m1}.

 \subsubsection{Spinless fermions}
 
 For a spinless fermion system, the on-site physical symmetry group for all $0$D blocks \( \mu_1, \mu_2, \mu_3 \) is \( G_b = D_3 \times \mathbb{Z}_2^T \). In Sec.~\ref{sec:D3 spinless}, we demonstrated that the classification of 0D FSPTs protected by this symmetry group is \( \mathbb{Z}_2^2 \), and it can be labeled by the eigenvalues \( \pm 1 \) of the reflection symmetry \( M \) and \( P_f \). Therefore, obstruction-free $0$D block-states form the group:
\begin{equation}
\{\mathrm{OFBS}\}^{0D} = \mathbb{Z}_2^6.
\end{equation}

We now consider the decoration of $1$D block-states. The physical symmetry group for all $1$D blocks is \( G_b = \mathbb{Z}_2 \times \mathbb{Z}_2^T \). As discussed in Sec.~\ref{sec:D4 spinless fermion}, the corresponding FSPT phases are classified by \( \mathbb{Z}_4 \times \mathbb{Z}_4 \), where the two fourth-order group generators correspond to the root phases \( n_1^M \) and \( n_1^T \) (their symmetry properties are listed in Eq.~\eqref{eq:symm property of n1T} and Eq.~\eqref{eq:symm property of n1M}). Additionally, we can incorporate an invertible topological order, i.e., a single Majorana chain. 

Since the cell decomposition near each $0$D block aligns with the discussion for point group $D_3$ in Sec.~\ref{sec:D3 spinless}, only when all $1$D blocks are decorated either with a single Majorana chain or with \( n_1^T \) or \( n_1^M \) FSPT phases can the zero modes on all $0$D blocks open their gap.

Thus, all obstruction-free $1$D block-states form the group:
\begin{equation}
\{\mathrm{OFBS}\}^{1D} = \mathbb{Z}_8 \times \mathbb{Z}_4,
\end{equation}
where the group elements can be labeled by:
\begin{equation}
[\mathcal{M}_{123}, m_{123}],
\end{equation}
These group elements are integers and correspond to:
\begin{itemize}
    \item $\mathcal{M}_{123} \in \{0,1,\ldots,7\}$ counts the number of decorated single Majorana chains on all blocks $\tau_1,\tau_2,\tau_3$;
    \item $ m_{123} \in \{0,1,2,3\}$ counts the number of decorated $n_1^M$ FSPT phases on all blocks $\tau_1,\tau_2,\tau_3$.
\end{itemize}
This classification fully characterizes the possible obstruction-free $1$D block-state decorations under the given symmetry constraints.

The total symmetry group for the $2$D blocks is \( G_f = \mathbb{Z}_2^T \times \mathbb{Z}_2^f \). As concluded in Ref.~\cite{Wang2020}, there are no corresponding FSPT phases under this symmetry group. Consequently, the contribution of $2$D block decoration is classified as:
\begin{equation}
\{\mathrm{OFBS}\}^{2D} = \mathbb{Z}_1.
\end{equation}

Next, we examine bubble equivalence relations. We begin with $1$D bubble equivalences obtained by decorating a pair of complex fermions on each $1$D block. The induced 0D eigenstates are as follows:
\begin{enumerate}
\item \textbf{1D bubble on \(\tau_1\):}
  \begin{enumerate}
  \item on \(\mu_2\): a nontrivial eigenstate of \((M_{\tau_1},\, M_{\tau_3},\, P_f)\);
  \item on \(\mu_3\): a nontrivial eigenstate of \((M_{\tau_1},\, M_{\tau_2},\, P_f)\).
  \end{enumerate}

\item \textbf{1D bubble on \(\tau_2\):}
  \begin{enumerate}
  \item on \(\mu_1\): a nontrivial eigenstate of \((M_{\tau_2},\, M_{\tau_3},\, P_f)\);
  \item on \(\mu_3\): a nontrivial eigenstate of \((M_{\tau_1},\, M_{\tau_2},\, P_f)\).
  \end{enumerate}

\item \textbf{1D bubble on \(\tau_3\):}
  \begin{enumerate}
  \item on \(\mu_1\): a nontrivial eigenstate of \((M_{\tau_2},\, M_{\tau_3},\, P_f)\);
  \item on \(\mu_2\): a nontrivial eigenstate of \((M_{\tau_1},\, M_{\tau_3},\, P_f)\).
  \end{enumerate}
\end{enumerate}

If instead we decorate bosonic $1$D bubbles, they induce the same eigenstates, except that the \(P_f\) component is absent. Therefore, $1$D bubbles generate the trivialization group
$
\mathbb{Z}_2^4.
$

Additionally, 2D ``Majorana bubbles" can contribute to 1D non-trivial states, classified by \( \mathbb{Z}_4 \), on all $1$D blocks, including the \( n_1^M+ n_1^T \) and \( (-1)^{n_1^M \cup n_1^M} \oplus (-1)^{n_1^T \cup n_1^T} \) phases through stacking. However, only the \( (-1)^{n_1^M \cup n_1^M} \oplus (-1)^{n_1^T \cup n_1^T} \) phase decoration on all $1$D blocks belongs to \( \{\mathrm{OFBS}\}^{1D} \). Therefore, the total trivialization groups are:
\begin{equation}
\{\mathrm{TBS}\}^{1D} = \mathbb{Z}_4, \quad \{\mathrm{TBS}\}^{0D} = \mathbb{Z}_2^4.
\end{equation}

It is evident that there is no extension between 1D and $0$D block-states. Thus, all independent non-trivial block-states with different dimensions are classified by:
\begin{align}
    \begin{aligned}
        & E^{\mathrm{2D}} = \{\mathrm{OFBS}\}^{\mathrm{2D}}= \mathbb{Z}_1, \\
        & E^{\mathrm{1D}} = \{\mathrm{OFBS}\}^{\mathrm{1D}} / \{\mathrm{TBS}\}^{\mathrm{1D}} = \mathbb{Z}_8, \\
        & E^{\mathrm{0D}} = \{\mathrm{OFBS}\}^{\mathrm{0D}} / \{\mathrm{TBS}\}^{\mathrm{0D}} = \mathbb{Z}_2^2.
    \end{aligned}
\end{align}

Hence, the ultimate classification is:
\begin{equation}
\mathcal{G}_0 = \mathbb{Z}_8 \times \mathbb{Z}_2^2.
\end{equation}

 \subsubsection{Spinful fermions}

For a spinful fermion system, the on-site physical symmetry group for all $0$D blocks \( \mu_1, \mu_2, \mu_3 \) is \( G_b = D_3 \times \mathbb{Z}_2^T \). In Sec.~\ref{sec:D3 spinful}, we demonstrated that the classification of 0D FSPTs protected by this symmetry group is \( \mathbb{Z}_2 \), and it can be labeled by the eigenvalue \( \pm 1 \) of the reflection operator \( M \). Therefore, obstruction-free $0$D block-states form the group:
\begin{equation}
\{\mathrm{OFBS}\}^{0D} = \mathbb{Z}_2^3.
\end{equation}

We now consider the decoration of $1$D block-states. For all $1$D blocks, the physical symmetry group is \( G_b = \mathbb{Z}_2 \times \mathbb{Z}_2^T \). As previously discussed, the invertible topological order, i.e., the single Majorana chain,  cannot be decorated, but the 1D FSPT phase (realized via double Majorana chains) can be decorated, which has a \( \mathbb{Z}_4 \) classification. 

Decorating the 1D FSPT phase on the $1$D block \( \tau_1 \) leaves six dangling Majorana fermion modes on the $0$D blocks \( \mu_2 \) and \( \mu_3 \). As concluded in the point group discussion, these six Majorana fermions cannot open a spectral gap at the $0$D blocks with physical symmetry \( G_b = D_3 \times \mathbb{Z}_2^T \), so this decoration is obstructed. The same applies to the other two $1$D blocks. However, if the 1D FSPT phase is simultaneously decorated on all three blocks \( \tau_1, \tau_2, \tau_3 \), the twelve dangling Majorana fermions at each $0$D block can open a spectral gap. Thus, the 1D decoration contributes to the classification group:
\begin{equation}
\{\mathrm{OFBS}\}^{1D} = \mathbb{Z}_4,
\end{equation}
where the group elements can be labeled as:
\begin{equation}
[\mathbb{MT}_{123}].
\end{equation}
Here,  \( \mathbb{MT}_{123} = 0, 1, 2, 3 \) represent the number of decorated 1D FSPT phases on all $1$D blocks \( \tau_1,\tau_2,\tau_3 \).

For the same reasons as in the previously discussed spinful example, the decoration of 2D DIII TSC states is always obstruction-free, contributing to a \( \{\mathrm{OFBS}\}^{2D} = \mathbb{Z}_2 \) classification. 

Next, we consider the bubble equivalence. The $1$D bubble on the $1$D block \( \tau_1 \) acts on the $0$D blocks \( \mu_2 \) and \( \mu_3 \), while the $1$D bubble on \( \tau_2 \) acts on the $0$D blocks \( \mu_1 \) and \( \mu_3 \), and the $1$D bubble on \( \tau_3 \) acts on the $0$D blocks \( \mu_1 \) and \( \mu_2 \). As concluded in Sec.~\ref{sec:D3 spinful}, the $1$D bubble has a trivializing effect on the $0$D blocks with on-site physical symmetry \( G_b = D_3 \times \mathbb{Z}_2^T \). Starting from the trivial 0D decorated state \( (+, +, +) \) (block-states can be labeled by \( (\pm, \pm, \pm) \), where the three \( \pm \) represent the eigenvalues of the reflection generator on \( \mu_1, \mu_2, \mu_3 \), respectively), decorating with three independent numbers \( l_1, l_2, l_3 \) of $1$D bubbles on \( \tau_1, \tau_2, \tau_3 \), respectively, still leads to a trivial state, including:
\begin{equation}
\begin{cases}
(+, +, +), \\
(+, -, -), \\
(-, +, -), \\
(-, -, +).
\end{cases}
\end{equation}

Additionally, as we know, $2$D bubbles have no trivializing effect on either 0D or $1$D block-states. Thus, the total trivializing group is:
\begin{equation}
\{\mathrm{TBS}\}^{\mathrm{0D}} = \mathbb{Z}_2^2, \quad \{\mathrm{TBS}\}^{\mathrm{1D}} = \mathbb{Z}_1.
\end{equation}

Finally, we consider the extension problem. It should be noted that $p3m1$ involves dihedral point groups. According to the discussion in Sec.~\ref{sec:D4 spinful discussion}, the $2$D block-states decorating on both sides of all $1$D blocks (i.e., two layers of DIII topological superconductor states) contribute to non-trivial $1$D block-states, due to reflection symmetry. Therefore, the $2$D block-states extend to $1$D block-states, resulting in the simultaneous decoration of  double Majorana chains on the $1$D blocks \( \tau_1, \tau_2, \tau_3 \). However, we find no extension from 1D block-states to 0D block-states in this case.

Thus, all independent non-trivial block-states with different dimensions are classified by:
\begin{align}
    \begin{aligned}
        & E^{\mathrm{2D}} = \{\mathrm{OFBS}\}^{\mathrm{2D}}= \mathbb{Z}_2, \\
        & E^{\mathrm{1D}} = \{\mathrm{OFBS}\}^{\mathrm{1D}} / \{\mathrm{TBS}\}^{\mathrm{1D}} = \mathbb{Z}_4, \\
        & E^{\mathrm{0D}} = \{\mathrm{OFBS}\}^{\mathrm{0D}} / \{\mathrm{TBS}\}^{\mathrm{0D}} = \mathbb{Z}_2.
    \end{aligned}
\end{align}
Thus, the final classification is:
\begin{equation}
\mathcal{G}_{1/2} = \mathbb{Z}_2 \times \mathbb{Z}_8,
\end{equation}
where the \( \mathbb{Z}_8 \) term comes from the extension of $2$D block-states to $1$D block-states.

\begin{figure}[tb]
    \centering
    \includegraphics[width=0.46\textwidth]{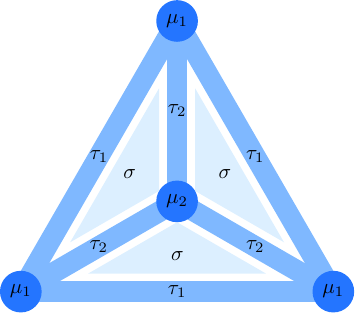}
    \caption{The \#15 The 2D space group \(p31m\) and its  cell decomposition.}
    \label{p31m}
\end{figure}

\subsection{$p31m$}

For the $2$D block \( \sigma \) and the $1$D block \( \tau_2 \), there is no local symmetry group, and the physical symmetry group is \( \mathbb{Z}_2^T \). For the $1$D block \( \tau_1 \), due to the action of reflection symmetry, the physical symmetry group is \( \mathbb{Z}_2 \times \mathbb{Z}_2^T \). For the $0$D block \( \mu_1 \), due to the local action of the group \( D_3 \), the physical symmetry group is \( \mathbb{Z}_3 \rtimes \mathbb{Z}_2 \times \mathbb{Z}_2^T \); and for the $0$D block \( \mu_2 \), due to the local action of the group \( C_3 \), the physical symmetry group is \( \mathbb{Z}_3 \times \mathbb{Z}_2^T \), as shown in Fig.~\ref{p31m}.

\subsubsection{Spinless fermions}

For a spinless fermion system, the on-site physical symmetry group of the $0$D block \( \mu_1 \) is \( G_b = D_3 \times \mathbb{Z}_2^T \), and for the $0$D block \( \mu_2 \), it is \( G_b = C_3 \times \mathbb{Z}_2^T \). As discussed in Sec.~\ref{sec:appendix spinless Cn} and Sec.~\ref{sec:D3 spinless}, the classification of 0D FSPTs protected by these symmetry groups yields a \( \mathbb{Z}_2^2 \) classification for the block-state on the $0$D block \( \mu_1 \), with group elements labeled by the eigenvalues \( \pm 1 \) of \( M \) and \( P_f \). For the $0$D block \( \mu_2 \), the classification is \( \mathbb{Z}_2 \), with group elements labeled by the eigenvalues \( \pm 1 \) of \( P_f \). Therefore, obstruction-free $0$D block-states form the group:
\begin{equation}
\{\mathrm{OFBS}\}^{0D} = \mathbb{Z}_2^3.
\end{equation}

We now discuss the decoration of $1$D block-states. The physical symmetry group for the $1$D block \( \tau_1 \) is \( G_b = \mathbb{Z}_2 \times \mathbb{Z}_2^T \). As discussed in Sec.~\ref{sec:D4 spinless fermion}, the corresponding FSPT phases are classified by \( \mathbb{Z}_4 \times \mathbb{Z}_4 \), where the two fourth-order group generators correspond to the root phases \( n_1^M \) and \( n_1^T \) (their symmetry properties are listed in Eq.~\eqref{eq:symm property of n1T} and Eq.~\eqref{eq:symm property of n1M}). The physical symmetry group for the $1$D block \( \tau_2 \) is \( G_b = \mathbb{Z}_2^T \). As mentioned previously, the corresponding FSPT phases are classified by \( \mathbb{Z}_4 \), with a root state model consisting of two Majorana chains, whose symmetry properties are detailed in Eq.~\eqref{eq:symm property spinless Cn 2chain}. Additionally, we can incorporate an invertible topological order, i.e., a single Majorana chain.

Consider the $1$D block-state decoration on the $1$D block \( \tau_1 \). Based on our analysis of the point group $D_3$ in the previous sections, the 1D  \( n_1^T \) or \( n_1^M \) FSPT phases decoration, as well as the single Majorana chain decoration, can open a spectral gap at the $0$D block \( \mu_1 \). This contributes to the classification \( \mathbb{Z}_8 \times \mathbb{Z}_4 \), making this decoration obstruction-free. Any decoration on \( \tau_2 \) is obstructed, as the dangling modes at \( \mu_2 \) cannot be gapped out. This is due to the discussion of the point group \( C_3 \) in Sec.~\ref{sec:appendix spinless Cn}.

Thus, all obstruction-free $1$D block-states form the group:
\begin{equation}
\{\mathrm{OFBS}\}^{1D} = \mathbb{Z}_8 \times \mathbb{Z}_4,
\end{equation}
where the group elements can be labeled by:
\begin{equation}
[\mathcal{M}_1, m_1],
\end{equation}
with \( \mathcal{M}_1 = 0, 1, 2, \dots, 6, 7 \) and \( m_1 = 0, 1, 2, 3 \) representing the number of decorated 1D single Majorana chains and  \( n_1^M \) FSPT phase states on \( \tau_1 \), respectively.

The total symmetry group for the $2$D blocks is \( G_f = \mathbb{Z}_2^T \times \mathbb{Z}_2^f \). As concluded in Ref.~\cite{Wang2020}, there are no corresponding FSPT phases under this symmetry group. Consequently, the contribution of $2$D block decoration is classified as:
\begin{equation}
\{\mathrm{OFBS}\}^{2D} = \mathbb{Z}_1.
\end{equation}

Next, we examine bubble equivalence relations. First, we consider the $1$D bubble equivalences by decorating a pair of complex fermions on each $1$D block. For the $1$D block \( \tau_1 \), the $1$D bubble contributes a trivial state \( \ket{\phi} \) on \( \mu_1 \), satisfying \( M \ket{\phi} = \ket{\phi}, P_f \ket{\phi} = \ket{\phi} \). The $1$D bubble on \( \tau_2 \) contributes a non-trivial state \( \ket{\phi} \) on \( \mu_2 \), satisfying \( P_f \ket{\phi} = -\ket{\phi} \). Thus, $1$D bubbles form the trivialization group $\mathbb{Z}_2$.

Additionally, 2D ``Majorana bubbles" can contribute to 1D non-trivial states, classified by \( \mathbb{Z}_4 \), on all $1$D blocks, including the \( n_1^M+ n_1^T \) and \( (-1)^{n_1^M \cup n_1^M} \oplus (-1)^{n_1^T \cup n_1^T} \) phases through stacking. However, only the \( (-1)^{n_1^M \cup n_1^M} \oplus (-1)^{n_1^T \cup n_1^T} \) phase decoration on all $1$D blocks belongs to \( \{\mathrm{OFBS}\}^{1D} \). And $2$D bubbles have no effect on $0$D block-states. Therefore, the total trivialization groups are:
\begin{equation}
\{\mathrm{TBS}\}^{1D} = \mathbb{Z}_4, \quad \{\mathrm{TBS}\}^{0D} = \mathbb{Z}_2.
\end{equation}

Since the symmetries of the $0$D blocks here align with those discussed earlier, and the cell decomposition nearby is consistent with the point group decomposition, there is no extension in this case. Thus, all independent non-trivial block-states with different dimensions are classified by:
\begin{align}
    \begin{aligned}
        & E^{\mathrm{2D}}= \{\mathrm{OFBS}\}^{\mathrm{2D}} = \mathbb{Z}_1, \\
        & E^{\mathrm{1D}} = \{\mathrm{OFBS}\}^{\mathrm{1D}} / \{\mathrm{TBS}\}^{\mathrm{1D}} =  \mathbb{Z}_8, \\
        & E^{\mathrm{0D}} = \{\mathrm{OFBS}\}^{\mathrm{0D}} / \{\mathrm{TBS}\}^{\mathrm{0D}} = \mathbb{Z}_2^2.
    \end{aligned}
\end{align}

The ultimate classification is:
\begin{equation}
\mathcal{G}_0 = \mathbb{Z}_8 \ \times \mathbb{Z}_2^2.
\end{equation}

\subsubsection{Spinful fermions}

For a spinful fermion system, the on-site physical symmetry group for the $0$D block \( \mu_1 \) is \( G_b = D_3 \times \mathbb{Z}_2^T \), and for the $0$D block \( \mu_2 \), it is \( G_b = C_3 \times \mathbb{Z}_2^T \). As discussed in Sec.~\ref{sec:Cn spinfull} and Sec.~\ref{sec:D3 spinful}, the classification of 0D FSPTs protected by these symmetry groups gives \( \mathbb{Z}_2 \) for the block-state on the $0$D block \( \mu_1 \), with group elements labeled by the eigenvalues \( \pm 1 \) of \( M \), while for the $0$D block \( \mu_2 \), the classification is \( \mathbb{Z}_1 \), meaning that no non-trivial decorations exist. Therefore, obstruction-free $0$D block-states form the group:
\begin{equation}
\{\mathrm{OFBS}\}^{0D} = \mathbb{Z}_2.
\end{equation}

We now consider the decoration of $1$D block-states. The physical symmetry group for the $1$D block \( \tau_2 \) is \( G_b = \mathbb{Z}_2^T \), and for \( \tau_1 \), it is \( G_b = \mathbb{Z}_2 \times \mathbb{Z}_2^T \). As discussed earlier, it is not possible to decorate an invertible topological Majorana chain, but  double Majorana chains can be decorated. The classification is \( \mathbb{Z}_2 \) for \( \tau_2 \) and \( \mathbb{Z}_4 \) for \( \tau_1 \). 

Decorating \( \tau_1 \) with  double Majorana chains leaves twelve dangling Majorana fermion modes on the $0$D block \( \mu_1 \), while decorating \( \tau_2 \) leaves six dangling Majorana fermion modes at both \( \mu_1 \) and \( \mu_2 \). As discussed in the point group section, the Majorana fermions at \( \mu_1 \) can gap out at the $0$D block with physical symmetry \( G_b = D_3 \times \mathbb{Z}_2^T \), while the Majorana fermions at \( \mu_2 \) cannot gap out at the $0$D block with physical symmetry \( G_b = C_3 \times \mathbb{Z}_2^T \). Thus, the decoration on \( \tau_1 \) is obstruction-free, while the decoration on \( \tau_2 \) is obstructed. Therefore, the 1D decoration contributes to the classification group:
\begin{equation}
\{\mathrm{OFBS}\}^{1D} = \mathbb{Z}_4,
\end{equation}
where the group elements can be labeled as:
\begin{equation}
[\mathbb{MT}_1].
\end{equation}
Here,  \( \mathbb{MT}_1 = 0, 1, 2, 3 \) represents the number of decorated 1D double Majorana chains, protected by $\mathbb{Z}_4^{fT}$, on the $1$D blocks \( \tau_1 \).

For the same reasons as in the previously discussed spinful example, the decoration of 2D DIII TSC states is always obstruction-free, contributing to a \( \{\mathrm{OFBS}\}^{2D} = \mathbb{Z}_2 \) classification. 

Next, we consider the bubble equivalence. We find that the 1D bubbles on either $\tau_1$ or $\tau_2$ do not contribute any non-trivial 0D states. Due to the properties of the cell decomposition of $p31m$, applying a 1D bubble decoration on $\tau_1$ or $\tau_2$ strictly corresponds to simultaneously decorating all 1D blocks in the $D_3$ case. Therefore, we can directly apply the conclusion derived in Section \ref{sec:D3 spinful}. Additionally, as discussed in Sec.~\ref{sec:Cn spinfull} and Sec.~\ref{sec:D3 spinful}, the $2$D bubbles have no trivializing effect on either 0D or $1$D block-states. Hence, the total trivializing group is:
\begin{equation}
\{\mathrm{TBS}\}^{1D} = \mathbb{Z}_1, \quad \{\mathrm{TBS}\}^{0D} = \mathbb{Z}_1.
\end{equation}

Finally, we consider the extension problem. This involves both cyclic and dihedral point groups. 
As demonstrated in Sec.~\ref{sec:D4 spinful discussion} and Sec.~\ref{sec:extension cn spinful}, while two copies of 2D FSPT phases constitute a trivial bulk phase, they nevertheless induce effective nontrivial 1D FSPT phases on \( \tau_1 \) due to reflection symmetry. This induces a natural extension from 2D to $1$D block-states. In contrast, the absence of reflection symmetry prevents the formation of any nontrivial 1D FSPT phases on $\tau_2$, leaving only trivial states. Because the local symmetry here, along with the local cell decomposition, is the same as in the point-group analysis, there is no extension from 1D block-states to 0D block-states in this case.

Thus, all independent non-trivial block-states with different dimensions are classified by:
\begin{align}
    \begin{aligned}
        & E^{\mathrm{2D}} = \{\mathrm{OFBS}\}^{\mathrm{2D}}= \mathbb{Z}_2, \\
        & E^{\mathrm{1D}} = \{\mathrm{OFBS}\}^{\mathrm{1D}} / \{\mathrm{TBS}\}^{\mathrm{1D}} = \mathbb{Z}_4, \\
        & E^{\mathrm{0D}} = \{\mathrm{OFBS}\}^{\mathrm{0D}} / \{\mathrm{TBS}\}^{\mathrm{0D}} = \mathbb{Z}_2.
    \end{aligned}
\end{align}

And the final classification is:
\begin{equation}
\mathcal{G}_{1/2} = \mathbb{Z}_2 \times \mathbb{Z}_8.
\end{equation}

\subsection{$p6$}

\begin{figure}[tb]
    \centering
    \includegraphics[width=0.46\textwidth]{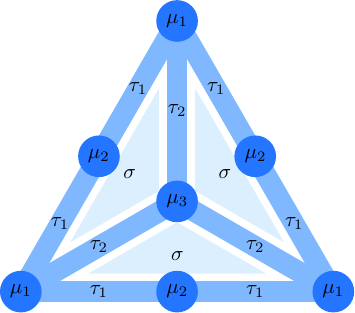}
    \caption{The \#16  2D space group $p6$ and its  cell decomposition.}
    \label{p6}
\end{figure}

For the $2$D block \( \sigma \) and the $1$D blocks \( \tau_1 \) and \( \tau_2 \), there is no site symmetry group, and the physical symmetry group is \( \mathbb{Z}_2^T \). For the $0$D block \( \mu_1 \), due to the effect of the sixfold rotation, the physical symmetry group is \( \mathbb{Z}_6 \times \mathbb{Z}_2^T \); for the $0$D block \( \mu_2 \), due to the action of the \( C_2 \) rotation, the physical symmetry group is \( \mathbb{Z}_2 \times \mathbb{Z}_2^T \); and for the $0$D block \( \mu_3 \), due to the action of the \( C_3 \) rotation, the physical symmetry group is \( \mathbb{Z}_3 \times \mathbb{Z}_2^T \), as shown in Fig.~\ref{p6}.

\subsubsection{Spinless fermions}

For the spinless fermion system, the on-site physical symmetry group \( G_b \) of the $0$D block \( \mu_1 \) is \( C_6 \times \mathbb{Z}_2^T \), for the $0$D block \( \mu_2 \), it is \( C_2 \times \mathbb{Z}_2^T \), and for the $0$D block \( \mu_3 \), it is \( C_3 \times \mathbb{Z}_2^T \). In Sec.~\ref{sec:appendix spinless Cn}, we discussed the 0D FSPTs protected by these symmetry groups. The classification of block-states on the $0$D block \( \mu_1 \) is \( \mathbb{Z}_2^2 \), labeled by the eigenvalues \( \pm 1 \) of \( C_6 \) and \( P_f \); the classification of block-states on \( \mu_2 \) is \( \mathbb{Z}_2^2 \), labeled by the eigenvalues \( \pm 1 \) of \( C_2 \) and \( P_f \); and the classification of block-states on \( \mu_3 \) is \( \mathbb{Z}_2 \), labeled by the eigenvalue \( \pm 1 \) of \( P_f \). Therefore, obstruction-free $0$D block-states form the group:
\begin{equation}
\{\mathrm{OFBS}\}^{0D} = \mathbb{Z}_2^5.
\end{equation}

We now discuss the decoration of $1$D block-states. The physical symmetry group for all $1$D blocks is \( G_b = \mathbb{Z}_2^T \). As mentioned previously, the corresponding FSPT phases are classified by \( \mathbb{Z}_4 \), with a root state model consisting of two Majorana chains, whose symmetry properties are detailed in Eq.~\eqref{eq:symm property spinless Cn 2chain}. Additionally, we can incorporate an invertible topological order, i.e., a single Majorana chain.

Consider the $1$D block-state decoration on the $1$D block \( \tau_1 \). Based on our analysis of the point groups \( C_6 \) and \( C_2 \), only the 1D BSPT phase (realized via four Majorana chains) decoration can open a spectral gap at the $0$D blocks \( \mu_1 \) and \( \mu_2 \), making this decoration obstruction-free. However, no obstruction-free $1$D block-states can be decorated on the $1$D block \( \tau_2 \). Based on the $C_3$ point group analysis in Sec.~\ref{sec:appendix spinless Cn}, corresponding dangling modes on the $0$D block \( \mu_3 \) are not gappable. 

Thus, all obstruction-free $1$D block-states form the group:
\begin{equation}
\{\mathrm{OFBS}\}^{1D} = \mathbb{Z}_2,
\end{equation}
where the group elements can be labeled by:
\begin{equation}
[\mathcal{T}_1],
\end{equation}
with \( \mathcal{T}_1 = 0, 1 \) representing the number of decorated 1D BSPT phases on \( \tau_1 \).

The total symmetry group for the $2$D blocks is \( G_f = \mathbb{Z}_2^T \times \mathbb{Z}_2^f \). As concluded in Ref.~\cite{Wang2020}, there are no corresponding FSPT phases under this symmetry group. Consequently, the contribution of $2$D block decoration is classified as:
\begin{equation}
\{\mathrm{OFBS}\}^{2D} = \mathbb{Z}_1.
\end{equation}

Next, we examine bubble equivalence relations. From Ref.~\cite{2Dcrystalline}, we know that $2$D bubbles have no effect on $1$D block-states but instead trivialize the eigenstate of \( P_f \) on $0$D blocks. Therefore, the  trivialization group of $1$D block-states is:
\begin{equation}
\{\mathrm{TBS}\}^{1D} = \mathbb{Z}_1.
\end{equation}
Then, we consider the $1$D bubble equivalences by decorating a pair of complex fermions on each $1$D block. For the $1$D block \( \tau_1 \), the $1$D bubble contributes a non-trivial eigenstate of \( C_6 \) on the $0$D block \( \mu_1 \), and a non-trivial eigenstate of \( C_2 \) on the $0$D block \( \mu_2 \). The $1$D bubble on \( \tau_2 \) contributes a non-trivial eigenstate of \( C_6 \) on the $0$D block \( \mu_1 \), and a non-trivial eigenstate of \( P_f \) on the $0$D block \( \mu_3 \). Thus, $1$D bubbles form the trivialization group:
\begin{equation}
\{\mathrm{TBS}\}^{0D}=\mathbb{Z}_2^3.
\end{equation}

Since there is no extension between 1D and $0$D block-states, all independent non-trivial block-states with different dimensions are classified by:
\begin{align}
    \begin{aligned}
        & E^{\mathrm{2D}} = \{\mathrm{OFBS}\}^{\mathrm{2D}}= \mathbb{Z}_1, \\
        & E^{\mathrm{1D}} = \{\mathrm{OFBS}\}^{\mathrm{1D}} / \{\mathrm{TBS}\}^{\mathrm{1D}} = \mathbb{Z}_2, \\
        & E^{\mathrm{0D}} = \{\mathrm{OFBS}\}^{\mathrm{0D}} / \{\mathrm{TBS}\}^{\mathrm{0D}} = \mathbb{Z}_2^2.
    \end{aligned}
\end{align}

The ultimate classification is:
\begin{equation}
\mathcal{G}_0 =  \mathbb{Z}_2^3.
\end{equation}

\subsubsection{Spinful fermions}

For the spinful fermion system, in Sec.~\ref{sec:Cn spinfull}, we discussed the 0D FSPTs protected by the corresponding physical symmetry groups \( G_b \). The classification of block-states on the $0$D block \( \mu_1 \) is \( \mathbb{Z}_2 \), labeled by the eigenvalue \( \pm 1 \) of \( C_6 \); the classification of block-states on \( \mu_2 \) is \( \mathbb{Z}_2 \), labeled by the eigenvalue \( \pm 1 \) of \( C_2 \); and the classification of block-states on \( \mu_3 \) is \( \mathbb{Z}_1 \), meaning that no non-trivial decoration exists. Therefore, obstruction-free $0$D block-states form the group:
\begin{equation}
\{\mathrm{OFBS}\}^{0D} = \mathbb{Z}_2^2.
\end{equation}

We now consider the decoration of $1$D block-states. The physical symmetry group \( G_b \) of all $1$D blocks is \( \mathbb{Z}_2^T \). Based on previous discussions, the invertible topological order, i.e., the single Majorana chain,  cannot be decorated, but the 1D FSPT phase (realized via double Majorana chains) can, with a classification group  \( \mathbb{Z}_4 \). Since the cell decomposition aligns with the discussion of cyclic groups, the conclusions in Sec.~\ref{sec:Cn spinfull} can be applied directly.

Decorating the 1D FSPT phase on \( \tau_1 \) leaves $12$ and $4$ dangling Majorana fermion modes on the $0$D blocks \( \mu_1 \) and \( \mu_2 \), respectively. As concluded in Sec.~\ref{sec:Cn spinfull}, these $12$ Majorana fermions on $\mu_1$ can open a spectral gap at the $0$D block with physical symmetry \( G_b = C_6 \times \mathbb{Z}_2^T \), and the $4$ Majorana fermions can open a spectral gap at the $0$D block with \( G_b = C_2 \times \mathbb{Z}_2^T \), making the decoration on \( \tau_1 \) obstruction-free.

Decorating the 1D FSPT phases on \( \tau_2 \) leaves $12$ and $6$ dangling Majorana fermion modes on the $0$D blocks \( \mu_1 \) and \( \mu_3 \), respectively. While the $12$ Majorana fermions  on $\mu_1$ can open a spectral gap, the $6$ Majorana fermions at the $0$D block \( \mu_3 \) with \( G_b = C_3 \times \mathbb{Z}_2^T \) cannot, making the decoration on \( \tau_2 \) obstructed. If 1D FSPT phases are simultaneously decorated on \( \tau_1, \tau_2 \), $6$ dangling Majorana modes remain on the $0$D block \( \mu_3 \), and they cannot open a spectral gap. Thus, all obstruction-free 1D decorations contribute to the classification group:
\begin{equation}
\{\mathrm{OFBS}\}^{1D} = \mathbb{Z}_2,
\end{equation}
where the group elements can be labeled as:
\begin{equation}
[\mathbb{T}_1].
\end{equation}
Here, \( \mathbb{T}_1 = 0, 1 \) represent the number of decorated 1D FSPT phases on the $1$D blocks \( \tau_1 \).

For the same reasons as in the previously discussed spinful example, the decoration of 2D DIII TSC states is always obstruction-free, contributing to a \( \{\mathrm{OFBS}\}^{2D} = \mathbb{Z}_2 \) classification. 

Next, we consider bubble equivalence. The $1$D bubble on \( \tau_1 \) acts on the $0$D blocks \( \mu_1 \) and \( \mu_2 \), while the $1$D bubble on \( \tau_2 \) acts on the $0$D blocks \( \mu_1 \) and \( \mu_3 \). As discussed earlier, the $1$D bubbles do not trivialize the $0$D block-states with physical symmetry \( G_b = C_6 \times \mathbb{Z}_2^T \) or \( G_b = C_2 \times \mathbb{Z}_2^T \). There is no need to consider the trivialization of \( \mu_3 \), as $\mu_3$ has no non-trivial $0$D block-state decoration. Additionally, it is known that $2$D bubbles do not trivialize the 0D or $1$D block-states. Therefore, the total trivialization groups are:
\begin{equation}
\{\mathrm{TBS}\}^{1D} = \mathbb{Z}_1, \quad \{\mathrm{TBS}\}^{0D} = \mathbb{Z}_1.
\end{equation}

Finally, we consider the extension problem. Since only the cyclic point groups are involved here, according to the discussion in Sec.~\ref{sec:extension cn spinful}, the mass term of the boundary Majorana modes of the 2D decoration does not introduce additional signs under symmetry. We also find no extension from 1D block-states to 0D block-states in this case. Consequently, no extension exists between the block-state decorations under the given symmetry constraints.

Thus, all independent non-trivial block-states with different dimensions are classified by:
\begin{align}
    \begin{aligned}
        & E^{\mathrm{2D}} = \{\mathrm{OFBS}\}^{\mathrm{2D}}= \mathbb{Z}_2, \\
        & E^{\mathrm{1D}} = \{\mathrm{OFBS}\}^{\mathrm{1D}} / \{\mathrm{TBS}\}^{\mathrm{1D}} = \mathbb{Z}_2, \\
        & E^{\mathrm{0D}} = \{\mathrm{OFBS}\}^{\mathrm{0D}} / \{\mathrm{TBS}\}^{\mathrm{0D}} = \mathbb{Z}_2^2.
    \end{aligned}
\end{align}
And the final classification is:
\begin{equation}
\mathcal{G}_{1/2} = \mathbb{Z}_2^4.
\end{equation}

% \nocite{*}

\bibliography{point_group}% Produces the bibliography via BibTeX.

@article{Wang2018,
  title = {Towards a Complete Classification of Symmetry-Protected Topological Phases for Interacting Fermions in Three Dimensions and a General Group Supercohomology Theory},
  author = {Wang, Qing-Rui and Gu, Zheng-Cheng},
  journal = {Phys. Rev. X},
  volume = {8},
  issue = {1},
  pages = {011055},
  numpages = {29},
  year = {2018},
  month = {Mar},
  publisher = {American Physical Society},
  doi = {10.1103/PhysRevX.8.011055},
  url = {https://link.aps.org/doi/10.1103/PhysRevX.8.011055}
}

@article{Wang2020,
  title = {Construction and Classification of Symmetry-Protected Topological Phases in Interacting Fermion Systems},
  author = {Wang, Qing-Rui and Gu, Zheng-Cheng},
  journal = {Phys. Rev. X},
  volume = {10},
  issue = {3},
  pages = {031055},
  numpages = {64},
  year = {2020},
  month = {Sep},
  publisher = {American Physical Society},
  doi = {10.1103/PhysRevX.10.031055},
  url = {https://link.aps.org/doi/10.1103/PhysRevX.10.031055}
}

@article{grouptopo,
  title     = {Classification of topological quantum matter with symmetries},
  author    = {Chiu, Ching-Kai and Teo, Jeffrey C. Y. and Schnyder, Andreas P. and Ryu, Shinsei},
  journal   = {Rev. Mod. Phys.},
  volume    = {88},
  issue     = {3},
  pages     = {035005},
  numpages  = {63},
  year      = {2016},
  month     = {Aug},
  publisher = {American Physical Society},
  doi       = {10.1103/RevModPhys.88.035005},
  url       = {https://link.aps.org/doi/10.1103/RevModPhys.88.035005}
}

@article{2Dcrystalline,
  title     = {Real-space construction of crystalline topological superconductors and insulators in 2D interacting fermionic systems},
  author    = {Zhang, Jian-Hao and Yang, Shuo and Qi, Yang and Gu, Zheng-Cheng},
  journal   = {Phys. Rev. Res.},
  volume    = {4},
  issue     = {3},
  pages     = {033081},
  numpages  = {33},
  year      = {2022},
  month     = {Jul},
  publisher = {American Physical Society},
  doi       = {10.1103/PhysRevResearch.4.033081},
  url       = {https://link.aps.org/doi/10.1103/PhysRevResearch.4.033081}
}

@article{pointgroup,
  title     = {Construction and classification of point-group symmetry-protected topological phases in two-dimensional interacting fermionic systems},
  author    = {Zhang, Jian-Hao and Wang, Qing-Rui and Yang, Shuo and Qi, Yang and Gu, Zheng-Cheng},
  journal   = {Phys. Rev. B},
  volume    = {101},
  issue     = {10},
  pages     = {100501},
  numpages  = {6},
  year      = {2020},
  month     = {Mar},
  publisher = {American Physical Society},
  doi       = {10.1103/PhysRevB.101.100501},
  url       = {https://link.aps.org/doi/10.1103/PhysRevB.101.100501}
}

@article{PhysRevResearch.2.033290,
  title     = {Fractionalized time reversal, parity, and charge conjugation symmetry in a topological superconductor: A possible origin of three generations of neutrinos and mass mixing},
  author    = {Gu, Zheng-Cheng},
  journal   = {Phys. Rev. Res.},
  volume    = {2},
  issue     = {3},
  pages     = {033290},
  numpages  = {21},
  year      = {2020},
  month     = {Aug},
  publisher = {American Physical Society},
  doi       = {10.1103/PhysRevResearch.2.033290},
  url       = {https://link.aps.org/doi/10.1103/PhysRevResearch.2.033290}
}

@article{InternationalTablesforCrystallography,
  number    = {3},
  pages     = {274-276},
  publisher = {International Union of Crystallography},
  title     = {International Tables for Crystallography, Volume A, Space-group symmetry. 6th edition. Edited by Mois I. Aroyo. Wiley, 2016. Pp. xxi + 873. Price GBP 295.00, EUR 354.00 (hardcover). ISBN 978‐0‐470‐97423‐0},
  volume    = {73},
  year      = {2017},
  author    = {Nespolo, Massimo},
  address   = {5 Abbey Square, Chester, Cheshire CH1 2HU, England},
  copyright = {International Union of Crystallography, 2017},
  issn      = {2053-2733},
  journal   = {Acta Crystallographica Section A}
}

@article{chenSymmetryProtectedTopological2013,
  title         = {Symmetry Protected Topological Orders and the Group Cohomology of Their Symmetry Group},
  author        = {Chen, Xie and Gu, Zheng-Cheng and Liu, Zheng-Xin and Wen, Xiao-Gang},
  year          = {2013},
  month         = apr,
  journal       = {Physical Review B},
  volume        = {87},
  number        = {15},
  eprint        = {1106.4772},
  primaryclass  = {cond-mat, physics:quant-ph},
  pages         = {155114},
  issn          = {1098-0121, 1550-235X},
  doi           = {10.1103/PhysRevB.87.155114},
  urldate       = {2023-01-27},
  archiveprefix = {arxiv},
  langid        = {english}
}

@article{Kitaev_2001,
  doi       = {10.1070/1063-7869/44/10S/S29},
  url       = {https://dx.doi.org/10.1070/1063-7869/44/10S/S29},
  year      = {2001},
  month     = {oct},
  publisher = {},
  volume    = {44},
  number    = {10S},
  pages     = {131},
  author    = {A Yu Kitaev},
  title     = {Unpaired Majorana fermions in quantum
               wires},
  journal   = {Physics-Uspekhi}
}

@misc{wang2021domain,
  title         = {Domain Wall Decorations, Anomalies and Spectral Sequences in Bosonic Topological Phases},
  author        = {Qing-Rui Wang and Shang-Qiang Ning and Meng Cheng},
  year          = {2021},
  eprint        = {2104.13233},
  archiveprefix = {arXiv},
  primaryclass  = {cond-mat.str-el}
}

@article{Chen_2014,
  title     = {Symmetry-protected topological phases from decorated domain walls},
  volume    = {5},
  issn      = {2041-1723},
  url       = {http://dx.doi.org/10.1038/ncomms4507},
  doi       = {10.1038/ncomms4507},
  number    = {1},
  journal   = {Nature Communications},
  publisher = {Springer Science and Business Media LLC},
  author    = {Chen, Xie and Lu, Yuan-Ming and Vishwanath, Ashvin},
  year      = {2014},
  month     = mar
}

@article{Chen_2011,
  title     = {Classification of gapped symmetric phases in one-dimensional spin systems},
  volume    = {83},
  issn      = {1550-235X},
  url       = {http://dx.doi.org/10.1103/PhysRevB.83.035107},
  doi       = {10.1103/physrevb.83.035107},
  number    = {3},
  journal   = {Physical Review B},
  publisher = {American Physical Society (APS)},
  author    = {Chen, Xie and Gu, Zheng-Cheng and Wen, Xiao-Gang},
  year      = {2011},
  month     = jan
}

@article{RotationSPT,
  title     = {Rotation symmetry-protected topological phases of fermions},
  author    = {Cheng, Meng and Wang, Chenjie},
  journal   = {Phys. Rev. B},
  volume    = {105},
  issue     = {19},
  pages     = {195154},
  numpages  = {29},
  year      = {2022},
  month     = {May},
  publisher = {American Physical Society},
  doi       = {10.1103/PhysRevB.105.195154},
  url       = {https://link.aps.org/doi/10.1103/PhysRevB.105.195154}
}

@article{Thorngren,
  title     = {Gauging Spatial Symmetries and the Classification of Topological Crystalline Phases},
  author    = {Thorngren, Ryan and Else, Dominic V.},
  journal   = {Phys. Rev. X},
  volume    = {8},
  issue     = {1},
  pages     = {011040},
  numpages  = {39},
  year      = {2018},
  month     = {Mar},
  publisher = {American Physical Society},
  doi       = {10.1103/PhysRevX.8.011040},
  url       = {https://link.aps.org/doi/10.1103/PhysRevX.8.011040}
}

@article{SongHao,
  title     = {Topological Phases Protected by Point Group Symmetry},
  author    = {Song, Hao and Huang, Sheng-Jie and Fu, Liang and Hermele, Michael},
  journal   = {Phys. Rev. X},
  volume    = {7},
  issue     = {1},
  pages     = {011020},
  numpages  = {23},
  year      = {2017},
  month     = {Feb},
  publisher = {American Physical Society},
  doi       = {10.1103/PhysRevX.7.011020},
  url       = {https://link.aps.org/doi/10.1103/PhysRevX.7.011020}
}

@misc{ren2023stacking,
  title         = {Stacking Group Structure of Fermionic Symmetry-Protected Topological Phases},
  author        = {Xing-Yu Ren and Shang-Qiang Ning and Yang Qi and Qing-Rui Wang and Zheng-Cheng Gu},
  year          = {2023},
  eprint        = {2310.19058},
  archiveprefix = {arXiv},
  primaryclass  = {cond-mat.str-el}
}

@article{Jiang2021,
  title     = {Generalized Lieb-Schultz-Mattis theorem on bosonic symmetry protected topological phases},
  author    = {Shenghan Jiang and Meng Cheng and Yang Qi and Yuan-Ming Lu},
  journal   = {SciPost Phys.},
  volume    = {11},
  pages     = {024},
  year      = {2021},
  publisher = {SciPost},
  doi       = {10.21468/SciPostPhys.11.2.024},
  url       = {https://scipost.org/10.21468/SciPostPhys.11.2.024}
}

@article{xiaoliang,
  title     = {Time-Reversal-Invariant Topological Superconductors and Superfluids in Two and Three Dimensions},
  author    = {Qi, Xiao-Liang and Hughes, Taylor L. and Raghu, S. and Zhang, Shou-Cheng},
  journal   = {Phys. Rev. Lett.},
  volume    = {102},
  issue     = {18},
  pages     = {187001},
  numpages  = {4},
  year      = {2009},
  month     = {May},
  publisher = {American Physical Society},
  doi       = {10.1103/PhysRevLett.102.187001},
  url       = {https://link.aps.org/doi/10.1103/PhysRevLett.102.187001}
}

@article{Nonperturbativeconstraints,
  title     = {Nonperturbative constraints from symmetry and chirality on Majorana zero modes and defect quantum numbers in (2+1) dimensions},
  author    = {Manjunath, Naren and Calvera, Vladimir and Barkeshli, Maissam},
  journal   = {Phys. Rev. B},
  volume    = {107},
  issue     = {16},
  pages     = {165126},
  numpages  = {43},
  year      = {2023},
  month     = {Apr},
  publisher = {American Physical Society},
  doi       = {10.1103/PhysRevB.107.165126},
  url       = {https://link.aps.org/doi/10.1103/PhysRevB.107.165126}
}

@article{Yaohong,
  title     = {Interaction effect on topological classification of superconductors in two dimensions},
  author    = {Yao, Hong and Ryu, Shinsei},
  journal   = {Phys. Rev. B},
  volume    = {88},
  issue     = {6},
  pages     = {064507},
  numpages  = {5},
  year      = {2013},
  month     = {Aug},
  publisher = {American Physical Society},
  doi       = {10.1103/PhysRevB.88.064507},
  url       = {https://link.aps.org/doi/10.1103/PhysRevB.88.064507}
}

@article{Song_2020,
  title     = {Real-space recipes for general topological crystalline states},
  volume    = {11},
  issn      = {2041-1723},
  url       = {http://dx.doi.org/10.1038/s41467-020-17685-5},
  doi       = {10.1038/s41467-020-17685-5},
  number    = {1},
  journal   = {Nature Communications},
  publisher = {Springer Science and Business Media LLC},
  author    = {Song, Zhida and Fang, Chen and Qi, Yang},
  year      = {2020},
  month     = aug
}

@article{Ginzburg:1950sr,
  author  = {Ginzburg, V. L. and Landau, L. D.},
  editor  = {ter Haar, D.},
  title   = {{On the Theory of superconductivity}},
  doi     = {10.1016/B978-0-08-010586-4.50035-3},
  journal = {Zh. Eksp. Teor. Fiz.},
  volume  = {20},
  pages   = {1064--1082},
  year    = {1950}
}

@article{LU,
  title     = {Local unitary transformation, long-range quantum entanglement, wave function renormalization, and topological order},
  author    = {Chen, Xie and Gu, Zheng-Cheng and Wen, Xiao-Gang},
  journal   = {Phys. Rev. B},
  volume    = {82},
  issue     = {15},
  pages     = {155138},
  numpages  = {28},
  year      = {2010},
  month     = {Oct},
  publisher = {American Physical Society},
  doi       = {10.1103/PhysRevB.82.155138},
  url       = {https://link.aps.org/doi/10.1103/PhysRevB.82.155138}
}

@article{TO,
  title     = {Vacuum degeneracy of chiral spin states in compactified space},
  author    = {Wen, X. G.},
  journal   = {Phys. Rev. B},
  volume    = {40},
  issue     = {10},
  pages     = {7387--7390},
  numpages  = {0},
  year      = {1989},
  month     = {Oct},
  publisher = {American Physical Society},
  doi       = {10.1103/PhysRevB.40.7387},
  url       = {https://link.aps.org/doi/10.1103/PhysRevB.40.7387}
}

@article{Gu_2009,
  title     = {Tensor-entanglement-filtering renormalization approach and symmetry-protected topological order},
  volume    = {80},
  issn      = {1550-235X},
  url       = {http://dx.doi.org/10.1103/PhysRevB.80.155131},
  doi       = {10.1103/physrevb.80.155131},
  number    = {15},
  journal   = {Physical Review B},
  publisher = {American Physical Society (APS)},
  author    = {Gu, Zheng-Cheng and Wen, Xiao-Gang},
  year      = {2009},
  month     = oct
}

@article{sigmaModel,
  title     = {Symmetry-protected topological orders for interacting fermions: Fermionic topological nonlinear $\ensuremath{\sigma}$ models and a special group supercohomology theory},
  author    = {Gu, Zheng-Cheng and Wen, Xiao-Gang},
  journal   = {Phys. Rev. B},
  volume    = {90},
  issue     = {11},
  pages     = {115141},
  numpages  = {59},
  year      = {2014},
  month     = {Sep},
  publisher = {American Physical Society},
  doi       = {10.1103/PhysRevB.90.115141},
  url       = {https://link.aps.org/doi/10.1103/PhysRevB.90.115141}
}

@article{Pollmann,
  title     = {Entanglement spectrum of a topological phase in one dimension},
  author    = {Pollmann, Frank and Turner, Ari M. and Berg, Erez and Oshikawa, Masaki},
  journal   = {Phys. Rev. B},
  volume    = {81},
  issue     = {6},
  pages     = {064439},
  numpages  = {10},
  year      = {2010},
  month     = {Feb},
  publisher = {American Physical Society},
  doi       = {10.1103/PhysRevB.81.064439},
  url       = {https://link.aps.org/doi/10.1103/PhysRevB.81.064439}
}

@misc{kapustin2014symmetry,
  title         = {Symmetry Protected Topological Phases, Anomalies, and Cobordisms: Beyond Group Cohomology},
  author        = {Anton Kapustin},
  year          = {2014},
  eprint        = {1403.1467},
  archiveprefix = {arXiv},
  primaryclass  = {cond-mat.str-el}
}

@article{Kapustin_2015,
  title     = {Fermionic symmetry protected topological phases and cobordisms},
  volume    = {2015},
  issn      = {1029-8479},
  url       = {http://dx.doi.org/10.1007/JHEP12(2015)052},
  doi       = {10.1007/jhep12(2015)052},
  number    = {12},
  journal   = {Journal of High Energy Physics},
  publisher = {Springer Science and Business Media LLC},
  author    = {Kapustin, Anton and Thorngren, Ryan and Turzillo, Alex and Wang, Zitao},
  year      = {2015},
  month     = dec,
  pages     = {1–21}
}

@misc{kapustin2015equivariant,
  title         = {Equivariant Topological Quantum Field Theory and Symmetry Protected Topological Phases},
  author        = {Anton Kapustin and Alex Turzillo},
  year          = {2015},
  eprint        = {1504.01830},
  archiveprefix = {arXiv},
  primaryclass  = {cond-mat.str-el}
}

@article{Freed_2021,
  title     = {Reflection positivity and invertible topological phases},
  volume    = {25},
  issn      = {1465-3060},
  url       = {http://dx.doi.org/10.2140/gt.2021.25.1165},
  doi       = {10.2140/gt.2021.25.1165},
  number    = {3},
  journal   = {Geometry and Topology},
  publisher = {Mathematical Sciences Publishers},
  author    = {Freed, Daniel S and Hopkins, Michael J},
  year      = {2021},
  month     = may,
  pages     = {1165–1330}
}

@article{TIreview,
  title     = {Colloquium: Topological insulators},
  author    = {Hasan, M. Z. and Kane, C. L.},
  journal   = {Rev. Mod. Phys.},
  volume    = {82},
  issue     = {4},
  pages     = {3045--3067},
  numpages  = {0},
  year      = {2010},
  month     = {Nov},
  publisher = {American Physical Society},
  doi       = {10.1103/RevModPhys.82.3045},
  url       = {https://link.aps.org/doi/10.1103/RevModPhys.82.3045}
}

@article{TSC,
  title     = {Topological insulators and superconductors},
  author    = {Qi, Xiao-Liang and Zhang, Shou-Cheng},
  journal   = {Rev. Mod. Phys.},
  volume    = {83},
  issue     = {4},
  pages     = {1057--1110},
  numpages  = {0},
  year      = {2011},
  month     = {Oct},
  publisher = {American Physical Society},
  doi       = {10.1103/RevModPhys.83.1057},
  url       = {https://link.aps.org/doi/10.1103/RevModPhys.83.1057}
}

@article{CryTI,
  title     = {Surface states and topological invariants in three-dimensional topological insulators: Application to ${\text{Bi}}_{1\ensuremath{-}x}{\text{Sb}}_{x}$},
  author    = {Teo, Jeffrey C. Y. and Fu, Liang and Kane, C. L.},
  journal   = {Phys. Rev. B},
  volume    = {78},
  issue     = {4},
  pages     = {045426},
  numpages  = {15},
  year      = {2008},
  month     = {Jul},
  publisher = {American Physical Society},
  doi       = {10.1103/PhysRevB.78.045426},
  url       = {https://link.aps.org/doi/10.1103/PhysRevB.78.045426}
}

@article{CryTI2,
  title     = {Topological Crystalline Insulators},
  author    = {Fu, Liang},
  journal   = {Phys. Rev. Lett.},
  volume    = {106},
  issue     = {10},
  pages     = {106802},
  numpages  = {4},
  year      = {2011},
  month     = {Mar},
  publisher = {American Physical Society},
  doi       = {10.1103/PhysRevLett.106.106802},
  url       = {https://link.aps.org/doi/10.1103/PhysRevLett.106.106802}
}

@misc{zhang2022construction,
  title         = {Construction and classification of crystalline topological superconductor and insulators in three-dimensional interacting fermion systems},
  author        = {Jian-Hao Zhang and Yang Qi and Zheng-Cheng Gu},
  year          = {2022},
  eprint        = {2204.13558},
  archiveprefix = {arXiv},
  primaryclass  = {cond-mat.str-el}
}

@misc{debray2021invertible,
  title         = {Invertible phases for mixed spatial symmetries and the fermionic crystalline equivalence principle},
  author        = {Arun Debray},
  year          = {2021},
  eprint        = {2102.02941},
  archiveprefix = {arXiv},
  primaryclass  = {math-ph}
}

@article{Fidkowski_2010,
  title     = {Effects of interactions on the topological classification of free fermion systems},
  volume    = {81},
  issn      = {1550-235X},
  url       = {http://dx.doi.org/10.1103/PhysRevB.81.134509},
  doi       = {10.1103/physrevb.81.134509},
  number    = {13},
  journal   = {Physical Review B},
  publisher = {American Physical Society (APS)},
  author    = {Fidkowski, Lukasz and Kitaev, Alexei},
  year      = {2010},
  month     = apr
}

@article{Dubinkin_2019,
  title     = {Higher-order bosonic topological phases in spin models},
  volume    = {99},
  issn      = {2469-9969},
  url       = {http://dx.doi.org/10.1103/PhysRevB.99.235132},
  doi       = {10.1103/physrevb.99.235132},
  number    = {23},
  journal   = {Physical Review B},
  publisher = {American Physical Society (APS)},
  author    = {Dubinkin, Oleg and Hughes, Taylor L.},
  year      = {2019},
  month     = jun
}

@article{Tanaka_2012,
  title     = {Experimental realization of a topological crystalline insulator in SnTe},
  volume    = {8},
  issn      = {1745-2481},
  url       = {http://dx.doi.org/10.1038/nphys2442},
  doi       = {10.1038/nphys2442},
  number    = {11},
  journal   = {Nature Physics},
  publisher = {Springer Science and Business Media LLC},
  author    = {Tanaka, Y. and Ren, Zhi and Sato, T. and Nakayama, K. and Souma, S. and Takahashi, T. and Segawa, Kouji and Ando, Yoichi},
  year      = {2012},
  month     = sep,
  pages     = {800–803}
}

@article{Dziawa_2012,
  title     = {Topological crystalline insulator states in {$\mathrm{Pb}_{1-x}\mathrm{Sn}_x\mathrm{Se}$}},
  volume    = {11},
  issn      = {1476-4660},
  url       = {http://dx.doi.org/10.1038/nmat3449},
  doi       = {10.1038/nmat3449},
  number    = {12},
  journal   = {Nature Materials},
  publisher = {Springer Science and Business Media LLC},
  author    = {Dziawa, P. and Kowalski, B. J. and Dybko, K. and Buczko, R. and Szczerbakow, A. and Szot, M. and Łusakowska, E. and Balasubramanian, T. and Wojek, B. M. and Berntsen, M. H. and Tjernberg, O. and Story, T.},
  year      = {2012},
  month     = sep,
  pages     = {1023–1027}
}

@article{Okada_2013,
  title     = {Observation of Dirac Node Formation and Mass Acquisition in a Topological Crystalline Insulator},
  volume    = {341},
  issn      = {1095-9203},
  url       = {http://dx.doi.org/10.1126/science.1239451},
  doi       = {10.1126/science.1239451},
  number    = {6153},
  journal   = {Science},
  publisher = {American Association for the Advancement of Science (AAAS)},
  author    = {Okada, Yoshinori and Serbyn, Maksym and Lin, Hsin and Walkup, Daniel and Zhou, Wenwen and Dhital, Chetan and Neupane, Madhab and Xu, Suyang and Wang, Yung Jui and Sankar, R. and Chou, Fangcheng and Bansil, Arun and Hasan, M. Zahid and Wilson, Stephen D. and Fu, Liang and Madhavan, Vidya},
  year      = {2013},
  month     = sep,
  pages     = {1496–1499}
}

@article{ma2017experimental,
  title     = {Experimental evidence of hourglass fermion in the candidate nonsymmorphic topological insulator KHgSb},
  author    = {Ma, Junzhang and Yi, Changjiang and Lv, Baiqing and Wang, Zhijun and Nie, Simin and Wang, Le and Kong, Lingyuan and Huang, Yaobo and Richard, Pierre and Zhang, Peng and others},
  journal   = {Science advances},
  volume    = {3},
  number    = {5},
  pages     = {e1602415},
  year      = {2017},
  publisher = {American Association for the Advancement of Science}
}

@article{defect_networks,
  title     = {Crystalline topological phases as defect networks},
  author    = {Else, Dominic V. and Thorngren, Ryan},
  journal   = {Phys. Rev. B},
  volume    = {99},
  issue     = {11},
  pages     = {115116},
  numpages  = {16},
  year      = {2019},
  month     = {Mar},
  publisher = {American Physical Society},
  doi       = {10.1103/PhysRevB.99.115116},
  url       = {https://link.aps.org/doi/10.1103/PhysRevB.99.115116}
}

@article{hsieh2012topological,
  title     = {Topological crystalline insulators in the SnTe material class},
  author    = {Hsieh, Timothy H and Lin, Hsin and Liu, Junwei and Duan, Wenhui and Bansil, Arun and Fu, Liang},
  journal   = {Nature communications},
  volume    = {3},
  number    = {1},
  pages     = {982},
  year      = {2012},
  publisher = {Nature Publishing Group UK London}
}

@article{Isobe_2015,
  title     = {Theory of interacting topological crystalline insulators},
  author    = {Isobe, Hiroki and Fu, Liang},
  journal   = {Phys. Rev. B},
  volume    = {92},
  issue     = {8},
  pages     = {081304},
  numpages  = {5},
  year      = {2015},
  month     = {Aug},
  publisher = {American Physical Society},
  doi       = {10.1103/PhysRevB.92.081304},
  url       = {https://link.aps.org/doi/10.1103/PhysRevB.92.081304}
}

@article{Zou_2018,
  title = {Bulk characterization of topological crystalline insulators: Stability under interactions and relations to symmetry enriched $U$(1) quantum spin liquids},
  author = {Zou, Liujun},
  journal = {Phys. Rev. B},
  volume = {97},
  issue = {4},
  pages = {045130},
  numpages = {11},
  year = {2018},
  month = {Jan},
  publisher = {American Physical Society},
  doi = {10.1103/PhysRevB.97.045130},
  url = {https://link.aps.org/doi/10.1103/PhysRevB.97.045130}
}

@article{Po_2017,
   title={Symmetry-based indicators of band topology in the 230 space groups},
   volume={8},
   ISSN={2041-1723},
   url={http://dx.doi.org/10.1038/s41467-017-00133-2},
   DOI={10.1038/s41467-017-00133-2},
   number={1},
   journal={Nature Communications},
   publisher={Springer Science and Business Media LLC},
   author={Po, Hoi Chun and Vishwanath, Ashvin and Watanabe, Haruki},
   year={2017},
   month=jun }

@article{Song_2020_beyond,
   title={Bosonic crystalline symmetry protected topological phases beyond the group cohomology proposal},
   volume={101},
   ISSN={2469-9969},
   url={http://dx.doi.org/10.1103/PhysRevB.101.165129},
   DOI={10.1103/physrevb.101.165129},
   number={16},
   journal={Physical Review B},
   publisher={American Physical Society (APS)},
   author={Song, Hao and Xiong, Charles Zhaoxi and Huang, Sheng-Jie},
   year={2020},
   month=apr }

@article{Jiang_2017,
   title={Anyon condensation and a generic tensor-network construction for symmetry-protected topological phases},
   volume={95},
   ISSN={2469-9969},
   url={http://dx.doi.org/10.1103/PhysRevB.95.125107},
   DOI={10.1103/physrevb.95.125107},
   number={12},
   journal={Physical Review B},
   publisher={American Physical Society (APS)},
   author={Jiang, Shenghan and Ran, Ying},
   year={2017},
   month=mar }

@article{Kruthoff_2017,
   title={Topological Classification of Crystalline Insulators through Band Structure Combinatorics},
   volume={7},
   ISSN={2160-3308},
   url={http://dx.doi.org/10.1103/PhysRevX.7.041069},
   DOI={10.1103/physrevx.7.041069},
   number={4},
   journal={Physical Review X},
   publisher={American Physical Society (APS)},
   author={Kruthoff, Jorrit and de Boer, Jan and van Wezel, Jasper and Kane, Charles L. and Slager, Robert-Jan},
   year={2017},
   month=dec }

@article{wire_2022,
  title = {Wire construction of class DIII topological crystalline superconductors in two dimensions},
  author = {Peng, Bingrui and Weng, Hongming and Fang, Chen},
  journal = {Phys. Rev. B},
  volume = {106},
  issue = {17},
  pages = {174512},
  numpages = {26},
  year = {2022},
  month = {Nov},
  publisher = {American Physical Society},
  doi = {10.1103/PhysRevB.106.174512},
  url = {https://link.aps.org/doi/10.1103/PhysRevB.106.174512}
}

@article{Shiozaki_2022,
   title={Atiyah-Hirzebruch spectral sequence in band topology: General formalism and topological invariants for 230 space groups},
   volume={106},
   ISSN={2469-9969},
   url={http://dx.doi.org/10.1103/PhysRevB.106.165103},
   DOI={10.1103/physrevb.106.165103},
   number={16},
   journal={Physical Review B},
   publisher={American Physical Society (APS)},
   author={Shiozaki, Ken and Sato, Masatoshi and Gomi, Kiyonori},
   year={2022},
   month=oct }

@article{Rasmussen_2020,
  title     = {Classification and construction of higher-order symmetry-protected topological phases of interacting bosons},
  volume    = {101},
  issn      = {2469-9969},
  url       = {http://dx.doi.org/10.1103/PhysRevB.101.085137},
  doi       = {10.1103/physrevb.101.085137},
  number    = {8},
  journal   = {Physical Review B},
  publisher = {American Physical Society (APS)},
  author    = {Rasmussen, Alex and Lu, Yuan-Ming},
  year      = {2020},
  month     = feb
}

@misc{rasmussen2018intrinsicallyinteractingtopologicalcrystalline,
      title={Intrinsically interacting topological crystalline insulators and superconductors}, 
      author={Alex Rasmussen and Yuan-Ming Lu},
      year={2018},
      eprint={1810.12317},
      archivePrefix={arXiv},
      primaryClass={cond-mat.str-el},
      url={https://arxiv.org/abs/1810.12317}, 
}

@article{Cheng_Lieb_2019,
   title={Fermionic Lieb-Schultz-Mattis theorems and weak symmetry-protected phases},
   volume={99},
   ISSN={2469-9969},
   url={http://dx.doi.org/10.1103/PhysRevB.99.075143},
   DOI={10.1103/physrevb.99.075143},
   number={7},
   journal={Physical Review B},
   publisher={American Physical Society (APS)},
   author={Cheng, Meng},
   year={2019},
   month=feb }

@article{Huang_Surface_2018,
   title={Surface field theories of point group symmetry protected topological phases},
   volume={97},
   ISSN={2469-9969},
   url={http://dx.doi.org/10.1103/PhysRevB.97.075145},
   DOI={10.1103/physrevb.97.075145},
   number={7},
   journal={Physical Review B},
   publisher={American Physical Society (APS)},
   author={Huang, Sheng-Jie and Hermele, Michael},
   year={2018},
   month=feb }

@article{ono_2021,
  title = {${\mathbb{Z}}_{2}$-enriched symmetry indicators for topological superconductors in the 1651 magnetic space groups},
  author = {Ono, Seishiro and Po, Hoi Chun and Shiozaki, Ken},
  journal = {Phys. Rev. Res.},
  volume = {3},
  issue = {2},
  pages = {023086},
  numpages = {13},
  year = {2021},
  month = {May},
  publisher = {American Physical Society},
  doi = {10.1103/PhysRevResearch.3.023086},
  url = {https://link.aps.org/doi/10.1103/PhysRevResearch.3.023086}
}

@article{Zhang_2013,
   title={Surface State Magnetization and Chiral Edge States on Topological Insulators},
   volume={110},
   ISSN={1079-7114},
   url={http://dx.doi.org/10.1103/PhysRevLett.110.046404},
   DOI={10.1103/physrevlett.110.046404},
   number={4},
   journal={Physical Review Letters},
   publisher={American Physical Society (APS)},
   author={Zhang, Fan and Kane, C. L. and Mele, E. J.},
   year={2013},
   month=jan }

@article{Benalcazar_2017,
   title={Quantized electric multipole insulators},
   volume={357},
   ISSN={1095-9203},
   url={http://dx.doi.org/10.1126/science.aah6442},
   DOI={10.1126/science.aah6442},
   number={6346},
   journal={Science},
   publisher={American Association for the Advancement of Science (AAAS)},
   author={Benalcazar, Wladimir A. and Bernevig, B. Andrei and Hughes, Taylor L.},
   year={2017},
   month=jul, pages={61–66} }

@article{Benalcazar_hing_2017,
   title={Electric multipole moments, topological multipole moment pumping, and chiral hinge states in crystalline insulators},
   volume={96},
   ISSN={2469-9969},
   url={http://dx.doi.org/10.1103/PhysRevB.96.245115},
   DOI={10.1103/physrevb.96.245115},
   number={24},
   journal={Physical Review B},
   publisher={American Physical Society (APS)},
   author={Benalcazar, Wladimir A. and Bernevig, B. Andrei and Hughes, Taylor L.},
   year={2017},
   month=dec }

@article{Schindler_2018,
   title={Higher-order topological insulators},
   volume={4},
   ISSN={2375-2548},
   url={http://dx.doi.org/10.1126/sciadv.aat0346},
   DOI={10.1126/sciadv.aat0346},
   number={6},
   journal={Science Advances},
   publisher={American Association for the Advancement of Science (AAAS)},
   author={Schindler, Frank and Cook, Ashley M. and Vergniory, Maia G. and Wang, Zhijun and Parkin, Stuart S. P. and Bernevig, B. Andrei and Neupert, Titus},
   year={2018},
   month=jun }

@article{Zhang_2019,
   title={Higher-Order Topology and Nodal Topological Superconductivity in Fe(Se,Te) Heterostructures},
   volume={123},
   ISSN={1079-7114},
   url={http://dx.doi.org/10.1103/PhysRevLett.123.167001},
   DOI={10.1103/physrevlett.123.167001},
   number={16},
   journal={Physical Review Letters},
   publisher={American Physical Society (APS)},
   author={Zhang, Rui-Xing and Cole, William S. and Wu, Xianxin and Das Sarma, S.},
   year={2019},
   month=oct }

@article{Zhang_2023,
   title={Kitaev building-block construction for inversion-protected higher-order topological superconductors},
   volume={108},
   ISSN={2469-9969},
   url={http://dx.doi.org/10.1103/PhysRevB.108.115146},
   DOI={10.1103/physrevb.108.115146},
   number={11},
   journal={Physical Review B},
   publisher={American Physical Society (APS)},
   author={Zhang, Rui-Xing and Sau, Jay D. and Das Sarma, S.},
   year={2023},
   month=sep }

@misc{lee2024crystallineequivalenttopologicalphasesmanybody,
      title={Crystalline-equivalent topological phases of many-body fermionic systems in one dimension}, 
      author={Chen-Shen Lee and Ken Shiozaki and Chang-Tse Hsieh},
      year={2024},
      eprint={2411.19268},
      archivePrefix={arXiv},
      primaryClass={cond-mat.str-el},
      url={https://arxiv.org/abs/2411.19268}, 
}

@misc{lee2024connectionfreefermioninteractingcrystalline,
      title={Connection between Free-Fermion and Interacting Crystalline Symmetry-Protected Topological Phases}, 
      author={Chen-Shen Lee and Ken Shiozaki and Chang-Tse Hsieh},
      year={2024},
      eprint={2411.19287},
      archivePrefix={arXiv},
      primaryClass={cond-mat.str-el},
      url={https://arxiv.org/abs/2411.19287}, 
}

@misc{zeng2018quantuminformationmeetsquantum,
      title={Quantum Information Meets Quantum Matter -- From Quantum Entanglement to Topological Phase in Many-Body Systems}, 
      author={Bei Zeng and Xie Chen and Duan-Lu Zhou and Xiao-Gang Wen},
      year={2018},
      eprint={1508.02595},
      archivePrefix={arXiv},
      primaryClass={cond-mat.str-el},
      url={https://arxiv.org/abs/1508.02595}, 
}

@article{Sato_2017,
   title={Topological superconductors: a review},
   volume={80},
   ISSN={1361-6633},
   url={http://dx.doi.org/10.1088/1361-6633/aa6ac7},
   DOI={10.1088/1361-6633/aa6ac7},
   number={7},
   journal={Reports on Progress in Physics},
   publisher={IOP Publishing},
   author={Sato, Masatoshi and Ando, Yoichi},
   year={2017},
   month=may, pages={076501} }

@article{FU_TSC_2010,
  title = {Odd-Parity Topological Superconductors: Theory and Application to ${\mathrm{Cu}}_{x}{\mathrm{Bi}}_{2}{\mathrm{Se}}_{3}$},
  author = {Fu, Liang and Berg, Erez},
  journal = {Phys. Rev. Lett.},
  volume = {105},
  issue = {9},
  pages = {097001},
  numpages = {4},
  year = {2010},
  month = {Aug},
  publisher = {American Physical Society},
  doi = {10.1103/PhysRevLett.105.097001},
  url = {https://link.aps.org/doi/10.1103/PhysRevLett.105.097001}
}

@article{Sasaki_2012,
  title = {Odd-Parity Pairing and Topological Superconductivity in a Strongly Spin-Orbit Coupled Semiconductor},
  author = {Sasaki, Satoshi and Ren, Zhi and Taskin, A. A. and Segawa, Kouji and Fu, Liang and Ando, Yoichi},
  journal = {Phys. Rev. Lett.},
  volume = {109},
  issue = {21},
  pages = {217004},
  numpages = {5},
  year = {2012},
  month = {Nov},
  publisher = {American Physical Society},
  doi = {10.1103/PhysRevLett.109.217004},
  url = {https://link.aps.org/doi/10.1103/PhysRevLett.109.217004}
}

@article{FU_proximity_2008,
  title = {Superconducting Proximity Effect and Majorana Fermions at the Surface of a Topological Insulator},
  author = {Fu, Liang and Kane, C. L.},
  journal = {Phys. Rev. Lett.},
  volume = {100},
  issue = {9},
  pages = {096407},
  numpages = {4},
  year = {2008},
  month = {Mar},
  publisher = {American Physical Society},
  doi = {10.1103/PhysRevLett.100.096407},
  url = {https://link.aps.org/doi/10.1103/PhysRevLett.100.096407}
}

\end{document}